%% file: NNMoE.tex
\documentclass[onecolumn,12pt]{article} 
 
\usepackage[utf8]{inputenc}

  \usepackage{setspace}
\usepackage{dsfont,amssymb,amsmath,graphics,graphicx,fancyhdr,float,verbatim,
ifthen,ifpdf,algorithmic,multirow,array,color, enumerate,theorem}
\usepackage[english]{babel}
\usepackage[perpage]{footmisc}
\usepackage[ruled]{algorithm}
\usepackage[subnum]{cases}

\renewcommand{\parallel}{|\!|}
 
\usepackage[margin=1.1in]{geometry}

\usepackage[authoryear, round]{natbib}
 \usepackage[
                  breaklinks = true,
                 colorlinks = true,
                 linkcolor = red,
                 urlcolor  = black, 
                 citecolor = blue,
                 anchorcolor = green,
                 ]{hyperref} 

\graphicspath{ 
{Results/Simulation-1-with-Outliers/NMoE/} 
{Results/Simulation-1/NMoE/}
{Results/Simulation-1/SNMoE/}
{Results/Simulation-1/TMoE/}
{Results/Simulation-1/STMoE/} 
{Results/Simulation-2/Bishop/}
{Results/Real-Data/Tone/} 
{Results/Real-Data-with-Outliers/Tone/}
{Results/Real-Data/TemperatureAnomaly/}
} 

\input{mymathsym.tex}

\date{}
\begin{document}
\title{Non-Normal Mixtures of Experts}
\author{Faicel Chamroukhi}
 \maketitle
\begin{center}
Aix Marseille Universit\'e, CNRS, ENSAM, LSIS, UMR 7296, 13397 Marseille, France\\
Universit\'e de Toulon, CNRS, LSIS, UMR 7296, 83957 La Garde, France\\
\href{mailto:chamroukhi@univ-tln.fr}{chamroukhi@univ-tln.fr}
\end{center}


%

\begin{abstract} 
Mixture of Experts (MoE) is a popular framework for modeling heterogeneity in data for regression, classification and clustering. For continuous data which we consider here in the context of regression and cluster analysis, MoE usually use normal experts, that is, expert components following  the Gaussian distribution. However, for a set of data containing a group or groups of observations with asymmetric behavior, heavy tails or atypical observations, the use of normal experts may be unsuitable and can unduly affect the fit of the MoE model.
In this paper,  we introduce new non-normal mixture of experts (NNMoE) which can deal with these issues regarding possibly skewed, heavy-tailed data and with outliers.  The proposed models are the skew-normal MoE and the robust $t$ MoE and skew $t$ MoE, respectively named SNMoE, TMoE and STMoE.
We develop dedicated expectation-maximization (EM) and expectation conditional maximization (ECM) algorithms  to estimate the parameters of the proposed models by monotonically maximizing the observed data log-likelihood.
We describe how the presented models can be used in prediction and in model-based clustering of regression data. 
Numerical experiments carried out on simulated data show the effectiveness and the robustness of the proposed models in terms modeling non-linear regression functions as well as in model-based clustering. 
Then, to show their usefulness for practical applications, the proposed models are applied to the real-world data of tone perception for musical data analysis, and the one of temperature anomalies for  the analysis of climate change data.
\end{abstract}
\vspace{-.1cm}
{\bf keywords:}
mixture of experts,
skew normal distribution, 
$t$ distribution, 
skew $t$ distribution, 
EM algorithm,
ECM algorithm,
non-linear regression,
model-based clustering
\vspace{-.6cm}
{\footnotesize \tableofcontents}

\section{Introduction}

Mixture of experts (MoE) introduced by \citep{jacobsME} are widely studied in statistics and machine learning. They consist in a fully conditional mixture model where both the mixing proportions, known as the gating functions, and the component densities, known as the experts, are conditional on some input covariates. MoE have been investigated, in their simple form, as well as in their hierarchical form \citet{jordanHME} (e.g Section 5.12 of \citet{McLachlanFMM}) for regression and model-based cluster and discriminant analyses and in different application domains. A complete review of the MoE models can be found in \citet{YukselWG12}. 
For continuous data, which we consider here in the context of non-linear regression and model-based cluster analysis, MoE usually use normal experts, that is, expert components following  the Gaussian distribution. Along this paper, We will call it the  normal mixture of experts, abbreviated as NMoE.
However, it is well-known that the normal distribution is sensitive to outliers. Moreover, for a set of data containing a group or groups of observations with heavy tails or asymmetric behavior, the use of normal experts may be unsuitable and can unduly affect the fit of the MoE model. 
In this paper, we attempt to overcome these limitations in MoE by proposing more adapted and robust mixture of experts models which can deal with possibly skewed, heavy-tailed and atypical data.

Recently, the problem of sensitivity of NMoE to outliers have been considered by \citet{Nguyen2014-MoLE} where the authors proposed  a Laplace mixture of linear experts (LMoLE) for a robust modeling of non-linear regression data. The model parameters are estimated by maximizing the observed-data likelihood via a minorization-maximization (MM) algorithm.  
Here, we propose alternative MoE models, by relaying on other non-normal distributions that generalize the normal distribution, that is, the skew-normal, $t$, and the skew-$t$ distributions.  We call these proposed NNMoE models, respectively, the skew-normal MoE (SNMoE), the $t$ MoE (TMoE), and the skew-$t$ MoE (STMoE).
Indeed, in these last years, the use of the skew normal distribution, firstly proposed by \citet{Azzalini1985, Azzalini1986},  
has been shown beneficial in dealing with asymmetric data in various theoretic and applied problems.
This has been studied in the finite mixture literature by namely  \citet{Lin07univSkewNMixture} for modeling asymmetric univariate data with the univariate skew-normal mixture.
On the other hand, the $t$ distribution provides a natural robust extension of the normal distribution to model data with possible outliers. This has been integrated to develop the $t$ mixture model proposed by \citet{Mclachlan98robustTmixture} for robust cluster analysis of multivariate data. Recently, \citet{Bai2012} proposed a robust mixture modeling in the regression context on univariate data, by using a univariate $t$-mixture model. 
Moreover, in many practical problems,  the robustness of $t$ mixtures may however be not sufficient in the presence of asymmetric observations. To deal with this issue, \citet{Lin07univSkewtMixture} proposed the univariate skew-$t$ mixture model which allows for accommodation of both skewness and thick tails in the data, by relying on the skew-$t$ distribution, introduced by \citet{AzzaliniAndCapitanio2003}. 
For the general multivariate case using $t$, skew-normal and skew-$t$ mixtures, one can refer to 
\citet{Mclachlan98robustTmixture,Peel2000robusTtmixture},
\citet{Pyne2009},
\citep{Lin2010SkewtMvMixture},
\citet{LeeAndMchLachlan13non-normal-mix},
\citet{LeeAndMcLachlan13skew},
\citet{LeeAndMcLachlan14-skewtmix}, and recently, the unifying framework for previous restricted and unrestricted skew-$t$ mixtures, using the CFUST distribution \citet{LeeAndMcLachlan15-CFUST}.
The inference in the previously described approaches is performed by maximum likelihood estimation via expectation-maximization (EM)  or extensions \citep{dlr,McLachlanEM2008}, in particular the  expectation conditional maximization (ECM) algorithm \citep{meng_and_rubin_ECM_93}. For the Bayesian framework,  \citet{Fruhwirth10BayesSkewMixtures} have considered the Bayesian inference for both the univariate  and the multivariate skew-normal and skew-$t$ mixtures.
%
For the regression context, the robust modeling of regression data has been studied namely by
\citet{Wei2012}  who considered a $t$-mixture model for regression analysis of univariate data, as well as by \citet{Bai2012} who relied on the M-estimate in mixture of linear regressions. 
%
In the same context of regression, \citet{Song2014} proposed the mixture of Laplace regressions, which has been then extended by \citet{Nguyen2014-MoLE} to the case of mixture of experts, by introducing the Laplace mixture of linear experts (LMoLE).
Recently, \citet{Zeller15SkewNMixReg}  introduced the  scale mixtures of skew-normal distributions for robust mixture regressions. However, unlike our proposed NNMoE models, the regression mixture models of \citet{Wei2012}, \citet{Bai2012}, \citet{Song2014}, \citet{Zeller15SkewNMixReg} do not consider conditional mixing proportions, that is, mixing proportions depending on some input variables, as in the case of mixture of experts, which we investigate here. In addition, the approaches of \citet{Wei2012}, \citet{Bai2012} and \citet{Song2014} do not consider both the problem of robustness to outliers and the one to deal with possibly asymmetric data. 
%
Indeed, here we consider the mixture of experts framework for 
non-linear regression problems and model-based clustering of regression data, and we attempt to overcome the limitations of the NMoE model for dealing with asymmetric, heavy-tailed data and which may contain outliers.
We investigate the use of the skew-normal, $t$ and skew $t$ distributions for the experts, rather than the commonly used normal distribution.
First, the skew-normal mixture of experts (SNMoE) is proposed to accommodate data with possible asymmetric behavior. For heavy tailed or possibly noisy data, that is, data with   atypical observations, we first propose the $t$-mixture of experts model (TMoE) to handle the issues regarding namely the sensitivity of the NMoE to outliers. Finally, we propose the skew-$t$ mixture of experts model (STMoE)  which allows for accommodation of both skewness and heavy tails in the data and which is also robust to outliers. 
These models correspond to extensions of the unconditional mixture of skew-normal \citep{Lin07univSkewNMixture},  $t$ \citep{Mclachlan98robustTmixture,Wei2012}, and skew $t$ \citep{Lin07univSkewtMixture} models, to the  mixture of experts (MoE) framework, where the mixture means are regression functions and the mixing proportions are covariate-varying. 
For the models inference, we develop dedicated expectation-maximization (EM)  
and expectation conditional maximization (ECM) algorithms 
to estimate the parameters of the proposed models by monotonically maximizing the observed data log-likelihood.
The EM algorithms are indeed very popular and successful estimation algorithms for mixture models in general and for mixture of experts in particular.
Moreover, the EM algorithm for MoE has been shown by \citet{NgM-EM-IRLS-04} to be monotonically maximizing the MoE likelihood. The authors have showed that the EM (with IRLS in this case) algorithm has stable convergence and  the log-likelihood is monotonically increasing when a learning rate smaller than one is adopted for the IRLS procedure within the M-step of the EM algorithm.
They have further proposed an expectation conditional maximization (ECM) algorithm to train MoE, which also has desirable numerical properties. 
The MoE has also been considered in the Bayesian framework, for example one can cite the Bayesian MoE 
\cite{Waterhouse96bayesianMoE,Waterhouse1997}
and the Bayesian hierarchical MoE \citet{Bishop_BayesianMoE}. 
Beyond the Bayesian parametric framework, the MoE models have also been investigated within the Bayesian non-parametric framework.
We cite for example the  Bayesian non-parametric MoE model 
\citep{Rasmussen01infiniteMoE}
and the Bayesian non-parametric hierarchical MoE approach of \citet{ShiMT05_Hierarchical_GPR} using Gaussian Processes experts for regression.
For further models on mixture of experts for regression, the reader can be referred to for example the book of \citet{ShiGPR_Book2011}. 
In this paper, we investigate semi-parametric models under the maximum likelihood estimation framework.

The remainder of this paper is organized as follows. In Section \ref{sec: MoE} we briefly recall the MoE framework,  
the NMoE model and its maximum-likelihood estimation via EM. 
In Section \ref{sec: SNMoE}, we present the SNMoE model and 
in Section \ref{sec: MLE for the SNMoE} we present its inference technique using the ECM algorithm.
Then, in Section \ref{sec: TMoE} we present the TMoE model 
and derive its parameter estimation technique using the EM algorithm in
Section \ref{sec: MLE for the TMoE}. 
Then, in Section \ref{sec: STMoE}, we present the STMoE model and in  Section \ref{sec: MLE for the STMoE} the parameter estimation technique using the ECM algorithm.
In Section \ref{sec: Model selection for the NNMoE}, we also show how the model selection can be performed for these NNMoE models. We then investigate in Section \ref{sec: Prediction using the NNMoE} the use of the proposed models for fitting non-linear regression functions as well for prediction on future data. 
We also show in Section \ref{sec: MBC using the NNMoE} how the models can be used in a model-based clustering prospective.
In Section \ref{sec: Experimental study}, we perform experiments to assess the proposed models. Finally, in Section \ref{sec: Conclusion}, conclusions are drawn and a future work

\section{Mixture of experts for continuous data}
\label{sec: MoE}

Mixture of experts \citep{jacobsME,jordanHME} are used in a variety of contexts including regression, classification and clustering. 
Here we consider the MoE framework for fitting (non-linear) regression functions and clustering of univariate continuous data . 
The aim of regression is to explore the relationship of an observed random variable $Y$ given a covariate vector $\bsX \in \R^p$ via conditional density functions for $Y|\bsX = \bsx$ of the form $f (y|\bsx)$, rather than only exploring the unconditional distribution of $Y$. 
For their reach modeling flexibility, mixture models \citep{McLachlanFMM} has took much attention for non-linear regression problems and we  distinguish in particular mixture of regressions and mixture of experts for regression analysis.
The univariate mixture of regressions model assumes that the observed pairs of data $(\bsx,y)$ where $y \in \R$ is the response for some covariate $\bsx \in \R^p$, are generated from $K$ regression functions and are governed by a hidden categorical random variable $Z$ indicating from which component each observation is generated.
Thus, the mixture of regressions model decomposes the nonlinear regression model density $f (y|\bsx)$ into a convex weighted sum of $K$ regression component models $f_k(y|\bsx)$ and can be defined as follows: %
\begin{eqnarray}
f(y|\bsx;\bsvPsi) &=& \sum_{k=1}^K \pi_k f_k(y|\bsx; \bsvPsi_k)
\label{eq: mixture of regressions}
\end{eqnarray}where the $\pi_k$'s are defined by $\pi_k = \Pro(Z = k)$ and represent the non-negative mixing proportions that sum to 1. The model parameter vector is given by $\bsvPsi = (\pi_1,\ldots,\pi_{K-1},\bsvPsi^T_1,\ldots,\bsvPsi^T_K)^T$,  $\bsvPsi_k$ being the parameter vector of the $k$th component density.

Although similar, the mixture of experts \citep{jacobsME} differ from regression mixture models in many aspects. One of the main differences is that the MoE model consists in a fully conditional mixture while in the regression mixture, only the component densities are conditional. Indeed, the mixing proportions are constant for the regression mixture, while in the MoE, they are modeled as a function of the inputs, generally modeled by logistic or a softmax function. 

\subsection{The mixture of experts (MoE) model}
Mixture of experts (MoE) for regression analysis \citep{jacobsME,jordanHME} extend the model (\ref{eq: mixture of regressions}) by modeling the mixing proportions as function of some covariates $\bsr \in \R^q$. 
The mixing proportions, known as the gating functions in the context of MoE, are modeled by the multinomial logistic model and are defined by:
\begin{eqnarray}
\Pro(Z=k|\bsr;\bsalpha) &=&\frac{\exp{(\bsalpha_k^T\bsr)}}{\sum_{\ell=1}^K\exp{(\bsalpha_{\ell}^T \bsr)}}\nonumber \\
&=&\pi_{k}(\bsr;\bsalpha)
\label{eq: multinomial logistic}
\end{eqnarray}where $\bsr \in \R^q$ is a covariate vector, $\bsalpha_{k}$ is the $q$-dimensional coefficients vector associated with $\bsr$ and $\bsalpha = (\bsalpha^T_1,\ldots,\bsalpha^T_{K-1})^T$ is the parameter vector of the logistic model, with $\bsalpha_K$ being the null vector. 
Thus, the MoE model consists in a fully conditional mixture model where both the mixing proportions (the gating functions) and the component densities (the experts) are conditional on some covariate variables (respectively $\bsr$ and $\bsx$). 
%
The use of mixtures with mixing proportions defined through a logistic regression model has also been studied by \citet{HuangElAll2015} for penalized model-based clustering of spatial data by using a mixture of offset-normal shape factor analyzers (MOSFA). 

\subsection{The normal mixture of experts (NMoE) model and its maximum likelihood estimation}
\label{sec: MLE for the NMoE}
In the case of mixture of experts for regression, it is usually assumed that the experts are normal, that is, follow a normal distribution. A $K$-component normal mixture of experts (NMoE)  ($K>1$) has the following formulation:
\begin{eqnarray}
f(y|\bsr,\bsx;\bsvPsi) &=& \sum_{k=1}^K \pi_k(\bsr;\bsalpha) \text{N}\left(y; \mu(\bsx;\bsbeta_k), \sigma_k^2\right)
\label{eq: normal MoE}
\end{eqnarray}which involves, in the semi-parametric case, component means defined as parametric (non-)linear regression functions $\mu(\bsx;\bsbeta_k)$. 
 
The NMoE model parameters are estimated by maximizing the observed data log-likelihood by using the EM algorithm \citep{dlr, jacobsME, jordanHME, jordan_and_xu_1995, NgM-EM-IRLS-04, McLachlanEM2008}. 
%
Suppose we observe an i.i.d sample of $n$ observations $(y_1,\ldots,y_n)$ with their respective associated covariates $(\bsx_1,\ldots,\bsx_n)$ and $(\bsr_1,\ldots,\bsx_r)$.  Then under the MoE model, the observed data log-likelihood for the parameter vector $\bsvPsi$ is given by:
\begin{equation}
\log L(\bsvPsi)  = \sum_{i=1}^n  \log  \sum_{k=1}^K \pi_k(\bsr_i;\bsalpha) \text{N}\left(y_i; \mu(\bsx;\bsbeta_k), \sigma_k^2\right).
\label{eq: log-lik normal MoE}
\end{equation}
The E-Step at the $m$th iteration of the EM algorithm for the NMoE model requires the calculation of the following posterior probability that the observation $(y_i, \bsx_i,\bsr_i)$ belongs to expert $k$, given a parameter estimation $\bsvPsi^{(m)}$:
\begin{eqnarray}
\tau_{ik}^{(m)} = \Pro(Z_i=k|y_{i},\bsx_i,\bsr_i;\bsvPsi^{(m)})
 =  \frac{\pi_k(\bsr;\bsalpha^{(m)}) \text{N}\left(y_i;\mu_k(\bsx_i; \beta^{(m)}_k),{\sigma^2_k}^{(m)}\right)}{f(y_i;\bsvPsi^{(m)})}.
\label{eq: posterior prob NMoE}
\end{eqnarray}
Then, the M-step calculates the parameter update $\bsvPsi^{(m+1)}$ by maximizing the well-known $Q$-function, that is the expected complete-data log-likelihood:
\begin{equation}
\bsvPsi^{(m+1)} =  \arg \max_{\bsvPsi \in \bsOmega} Q(\bsvPsi;\bsvPsi^{(m)})
\label{eq: arg max Q}
\end{equation}where $\bsOmega$ is the parameter space.  
For example, in the case of  normal mixture of linear experts (NMoLE) where each expert's mean has the flowing linear form: 
\begin{equation}
\mu(\bsx;\bsbeta_k) = \bsbeta_k^T \bsx, 
\label{eq: linear regression mean}
\end{equation}where $\bsbeta_k \in \R^p$ is the  vector of regression coefficients of component $k$, the updates for each of the expert component parameters consist in analytically solving a weighted Gaussian linear regression problem and are given by:
\begin{eqnarray}
\bsbeta_k^{(m+1)}  &=& \Big[\sum_{i=1}^{n}\tau^{(m)}_{ik}  \bsx_i\bsx^T_i \Big]^{-1} \sum_{i=1}^{n} 
 \tau^{(q)}_{ik}  y_i \bsx_i,
\label{eq: beta_k update for NMoE}\\
{\sigma^2_{k}}^{(m+1)} &= &
\frac{\sum_{i=1}^n\tau_{ik}^{(m)}\left(y_i - {\bsbeta^T_{k}}^{(m+1)}\bsx_i\right)^2}{\sum_{i=1}^n\tau_{ik}^{(m)}}\cdot
\label{eq: sigma2k update NMoE}
\end{eqnarray}
For the mixing proportions, the parameter update $\bsalpha^{(m+1)}$  cannot however be obtained in a closed form. It is calculated by Iteratively Reweighted Least Squares (IRLS) \citep{jacobsME,jordanHME,Chen1999,Green1984,chamroukhiIJCNN2009,chamroukhi_PhD_2010}.

\bigskip
However, the normal distribution is not adapted to deal with asymmetric and heavy tailed data. It is also known that the normal distribution is sensitive to outliers. In the proposal, we first propose to address the issue regarding the skewness, by proposing the skew-normal mixture of experts (SNMoE). Then, we propose a robust fitting of the MoE, which is adapted to heavy-tailed data, by using the $t$ distribution, that is, the $t$ mixture of experts (TMoE). Finally, the proposed skew-$t$ mixture of experts (STMoE) allows for simultaneously accommodating asymmetry and heavy tails in the data and is also robust to outliers.

\section{The skew-normal mixture of experts (SNMoE) model}
\label{sec: SNMoE}

The skew-normal mixture of experts (SNMoE) model uses the skew-normal distribution as density for the expert components. We first recal the skew-normal distribution and describe its stochastic and hierarchical presentation, to then integrate them into the proposed SNMoE model.

\subsection{The skew-normal distribution}
\label{ssec: Skew-normal distribution}
As introduced by \citep{Azzalini1985,Azzalini1986}, a random variable $Y$ follows a univariate skew-normal distribution with location parameter $\mu \in \R$, scale parameter $\sigma^2\in (0,\infty)$ and skewness parameter $\lambda \in \R$ if it has the density
\begin{eqnarray}
f(y;\mu,\sigma^2,\lambda) &=& \frac{2}{\sigma} \phi(\frac{y-\mu}{\sigma}) 
 \Phi \left(\lambda (\frac{y-\mu}{\sigma})\right)
\label{eq: Skew-normal density}
\end{eqnarray}
where $\phi(.)$ and $\Phi(.)$ denote, respectively, the probability density function (pdf) and the cumulative distribution function (cdf) of the standard normal distribution. It can be seen from (\ref{eq: Skew-normal density}) that when $\lambda = 0$, the skew-normal  reduces to the normal distribution.
As presented by \citet{Azzalini1986, Henze1986}, if 
\begin{equation}
Y = \mu + \delta |U| + \sqrt{1 - \delta^2} E
\label{eq: stochastic representation skew-normal}
\end{equation}
where $\delta= \frac{\lambda}{\sqrt{1 +\lambda^2}}$, $U$ and $E$ are independent random variables following the normal distribution $\text{N}(0,\sigma^2)$, then $Y$ follows the  skew-normal distribution with pdf $\text{SN}(\mu,\sigma^2,\lambda)$ given by  (\ref{eq: Skew-normal density}). In the above,  $|U|$ denotes the magnitude of $U$. 
This stochastic representation of the skew-normal distribution leads to the following hierarchical representation 
in an incomplete data framework,  as presented in \citet{Lin07univSkewNMixture}: 
\begin{equation}
\begin{tabular}{lll}
$Y|u$ & $\sim$ & $\text{N}\left(\mu + \delta |u|, (1-\delta^2)\sigma^2\right),$\\
$U$ & $\sim$ & $\text{N}(0,\sigma^2)$.
\end{tabular} 
\label{eq: hierarchical representation skew-normal}
\end{equation}
This hierarchical representation greatly facilitates the inference for the model, namely in the skew-normal mixture model.  
Introduced by
 \citet{Lin07univSkewNMixture}, a $K$-component skew-normal mixture model is given by:
\begin{eqnarray}
f(y;\bsvPsi) &=& \sum_{k=1}^K \pi_k ~ \text{SN}(y; \mu_k, \sigma_k^2, \lambda_k)
\label{eq: Skew-normal mixture}
\end{eqnarray}where 
the mixture components have a skew-normal density $\text{SN}(.; .,.,.)$ given by (\ref{eq: Skew-normal density}). For the skew-normal mixture, the mixing proportions and the means of the mixture components are assumed to be constant. 

In the following section, we present the skew-normal mixture of experts (SNMoE) which extends the skew-normal mixture model to the case of mixture of experts framework, by considering conditional distributions for both the mixing proportions and the means of the mixture components.

\subsection{The skew-normal mixture of experts (SNMoE)}
The proposed skew-normal MoE (SNMoE) is a $K$-component MoE model with skew-normal experts. It is defined as follows. 
Let $\text{SN}(\mu,\sigma^2,\lambda)$ denotes a skew-normal distribution with location parameter $\mu$, scale parameter $\sigma$ and skewness parameter $\lambda$. A $K$-component  SNMoE is then defined by:
\begin{eqnarray}
f(y|\bsr,\bsx;\bsvPsi) &=& \sum_{k=1}^K \pi_k(\bsr;\bsalpha)  \text{SN}\!\left(y; \mu(\bsx;\bsbeta_k), \sigma_k^2, \lambda_k\right).
\label{eq: SNMoE}
\end{eqnarray}In the SNMoE model, each expert component $k$ has indeed a skew-normal distribution, whose density is defined by 
 (\ref{eq: Skew-normal density}).  
The parameter vector of the model is $\bsvPsi = (\bsalpha^T_1,\ldots,\bsalpha^T_{K-1},\bsvPsi^T_1,\ldots,\bsvPsi^T_K)^T$ with $\bsvPsi_k = (\bsbeta^T_k,\sigma^2_k,\lambda_k)^T$ the parameter vector for the $k$th skewed-normal expert component.
It is obvious to see that if the skewness parameter $\lambda_k = 0$ for each $k$, the SNMoE model  (\ref{eq: SNMoE}) reduces to the NMoE model (\ref{eq: normal MoE}).
Before going on the model inference, we first present its stochastic and hierarchical representations, which will serve to derive the ECM algorithm for maximum likelihood parameter estimation. 
The SNMoE model is characterized as follows.

\subsubsection{Stochastic representation of the SNMoE}
By using the stochastic representation (\ref{eq: stochastic representation skew-normal}) of the skew-normal distribution, the stochastic representation for the skew-normal mixture of experts (SNMoE) is as follows. Let $U$ and $E$ be independent univariate random variables following the standard normal distribution $\text{N}(0,1)$ with pdf $\phi(.)$.  
Given some covariates $\bsx_i$ and $\bsr_i$, a random variable $Y_i$ is said to follow the  SNMoE model (\ref{eq: SNMoE}) if it has the following representation:
\begin{equation}
Y_i =  \mu(\bsx_i;\bsbeta_{z_i}) + \delta_{z_i} \sigma_{z_i} |U_i| + \sqrt{1 - \delta_{z_i}^2} \, \sigma_{z_i} E_i.
\label{eq: stochastic representation SNMoE}
\end{equation}In (\ref{eq: stochastic representation SNMoE}), we have $\delta_{z_i}= \frac{\lambda_{z_i}}{\sqrt{1 +\lambda_{z_i}^2}}$ where $z_i\in \{1,\ldots,K\}$ is a realization of the categorical variable $Z_i$ which follows the multinomial distribution, that is:
\begin{equation}
Z_i|\bsr_i \sim \text{Mult}\!\left(1;\pi_{1}(\bsr_i;\bsalpha),\ldots, \pi_{K}(\bsr_i;\bsalpha)\right)
\label{eq: Multinomial}
\end{equation}where each of the probabilities $\pi_{z_i}(\bsr_i;\bsalpha) = \Pro(Z_i=z_i|\bsr_i)$ is given by the logistic function (\ref{eq: multinomial logistic}). In this incomplete data framework,  $z_i$ represents the hidden label of the component generating the $i$th observation.

The stochastic representation (\ref{eq: stochastic representation SNMoE}) of the SNMoE leads to the following hierarchical representation, which, as it will be presented in Section \ref{sec: MLE for the SNMoE}, greatly facilitates the model inference. 

\subsubsection{Hierarchical representation of the SNMoE}
 By introducing the binary latent component-indicators $Z_{ik}$ such that $Z_{ik}=1$ iff $Z_i =k$, $Z_i$ being the hidden class label of the $i$th observation,  a hierarchical model for the SNMoE model can be derived from its stochastic representation (\ref{eq: stochastic representation SNMoE}) and is as follows
\begin{eqnarray}
Y_i|u_i, Z_{ik}=1, \bsx_i &\sim & \text{N}\Big(\mu(\bsx_i;\bsbeta_k) + \delta_k |u_i|, (1-\delta^2_k)\sigma^2_k\Big),\nonumber\\
U_i|Z_{ik}=1&\sim & \text{N}(0,\sigma^2_k),\label{eq: hierarchical representation SNMoE}\\
\bsZ_i|\bsr_i &\sim & \text{Mult}\left(1;\pi_1(\bsr_i;\bsalpha),\ldots,\pi_K(\bsr_i;\bsalpha) \right)\nonumber
\end{eqnarray}where $\bsZ_i = (Z_{i1},\ldots,Z_{iK})$ and $\delta_k = \frac{\lambda_k}{\sqrt{1 +\lambda^2_k}}$.

\section{Maximum likelihood estimation of the SNMoE model}
\label{sec: MLE for the SNMoE}

The unknown parameter vector $\bsvPsi$ of the SNMoE model can be estimated by maximizing the observed-data log-likelihood.
Given an observed i.i.d sample of $n$ observations $(y_1,\ldots,y_n)$ with their respective associated covariates $(\bsx_1,\ldots,\bsx_n)$ and $(\bsr_1,\ldots,\bsx_r)$, under the SNMoE model (\ref{eq: SNMoE}), the observed data log-likelihood for the parameter vector $\bsvPsi$ is given by:
\begin{equation}
\log L(\bsvPsi) =  \sum_{i=1}^n  \log  \sum_{k=1}^K \pi_k(\bsr_i;\bsalpha) \text{SN}\left(y; \mu(\bsx;\bsbeta_k), \sigma_k^2, \lambda_k\right).
\label{eq: log-lik SNMoE}
\end{equation}The maximization of this log-likelihood can not be performed in a closed form. However, in this latent data framework, the maximization can be performed  via  expectation-maximization (EM)-type algorithms \citep{McLachlanEM2008}. 
More specifically, we propose a dedicated Expectation Conditional Maximization (ECM) algorithm to monotonically maximize (\ref{eq: log-lik SNMoE}). The ECM algorithm \citep{meng_and_rubin_ECM_93} is an EM variant that mainly aims at addressing the optimization problem in the M-step of the EM algorithm. In ECM,  the  M-step is performed by several conditional maximization (CM) steps by dividing the parameter space  into sub-spaces. The parameter vector updates are then performed sequentially, one coordinate block after another in each sub-space. 
\subsection{ECM-algorithm for the SNMoE model}
\label{ssec: ECM SNMoE}

Deriving the ECM algorithm requires the definition of the complete-data log-likelihood. From the hierarchical representation (\ref{eq: hierarchical representation SNMoE}) of the SNMoE, the complete-data log-likelihood  $\bsvPsi$, where the complete-data are $\{y_i,z_i,u_i,\bsx_i,\bsr_i\}_{i=1}^n$, is given by:
{\small \begin{eqnarray}
\log L_c(\bsvPsi)  &\!\! =\!\! &  \sum_{i=1}^n \sum_{k=1}^K Z_{ik} \big[ \log\left(\Pro\left(Z_i=k|\bsr_i\right)\right) +   \log\left(f\left(u_i|Z_{ik}=1\right)\right)+ \log\left(f\left(y_i|u_i,Z_{ik}=1,\bsx_i\right)\right)\big] \nonumber \\
& = & \log L_{c}(\bsalpha) + \sum_{k=1}^K \log L_{c}(\bsvPsi_k),
\label{eq: complete log-likelihood SNMoE}
\end{eqnarray}}
with
{\small \begin{eqnarray*} 
\log L_{c}(\bsalpha)  &\!\! = \!\!  &\sum_{i=1}^{n} \sum_{k=1}^K Z_{ik} \log \pi_k(\bsr_i;\bsalpha),
\label{eq: L1c SNMoE}\\ 
\log  L_{c}(\bsvPsi_k)  &\!\!  =\!\!  & \sum_{i=1}^{n} Z_{ik}\Big[ - \log (2 \pi) -  \log (\sigma^2_k) - \frac{1}{2} \log (1 - \delta^2_k)
 - \frac{d^2_{ik}}{2(1 - \delta^2_k)}  + \frac{\delta_k ~ d_{ik} ~ u_i}{(1 - \delta^2_k)\sigma_k} - \frac{u_i^2}{2(1 - \delta^2_k)\sigma^2_k}\Big],
\label{eq: L2c SNMoE}
\end{eqnarray*}}where $d_{ik} = \frac{y_i-\mu(\bsx_i;\bsbeta_k)}{\sigma_k}$ denotes the Mahalanobis  distance between $y_i$ and the  $k$th expert's mean (with $\sigma_k$ as the standard deviation). 
Then, the proposed ECM algorithm for the SNMoE model performs as follows. It starts with an initial parameter vector $\bsvPsi^{(0)}$ and alternates between the E- and CM- steps until a convergence criterion is satisfied.

\subsubsection{E-Step}
\label{sssec: E-Step SNMoE}

The E-Step of the ECM algorithm for the SNMoE calculates the $Q$-function, that is the conditional expectation of the complete-data log-likelihood (\ref{eq: complete log-likelihood SNMoE}),  given the observed data $\{(y_i,\bsx_i,\bsr_i)\}_{i=1}^n$ and a current parameter estimation $\bsvPsi^{(m)}$, $m$ being the current iteration:
{\begin{eqnarray}
Q(\bsvPsi;\bsvPsi^{(m)})&=& \E\big[\log L_c(\bsvPsi)|\{y_i,\bsx_i,\bsr_i\}_{i=1}^n;\bsvPsi^{(m)}\big].
\label{eq: Q-function definition SNMoE} 
\end{eqnarray}
From (\ref{eq: complete log-likelihood SNMoE}), it follows that the $Q$-function is given by: 
\begin{equation}
Q(\bsvPsi;\bsvPsi^{(m)})=Q_{1}(\bsalpha;\bsvPsi^{(m)})+\sum_{k=1}^K Q_{2}(\bsvPsi_k;\bsvPsi^{(m)}),
\label{eq: Q-function decomposition SNMoE}
\end{equation}
with
\begin{eqnarray}
Q_{1}(\bsalpha;\bsvPsi^{(m)}) &=& \sum_{i=1}^{n}\sum_{k=1}^K \tau^{(m)}_{ik}\log \pi_{k}(\bsr_i;\bsalpha),
\label{eq: Q-alpha SNMoE}\\
Q_{2}(\bsvPsi_k;\bsvPsi^{(m)}) &=& \sum_{i=1}^{n} \tau^{(m)}_{ik}\Bigg[ -  \log (2 \pi) - \log (\sigma^2_k) - \frac{1}{2} \log (1 - \delta^2_k)
\nonumber \\
& & 
+ \frac{\delta_k ~ d_{ik} ~ e_{1,ik}^{(m)}}{(1 - \delta^2_k)\sigma_k}  - \frac{e_{2,ik}^{(m)}}{2(1 - \delta^2_k)\sigma^2_k} - \frac{d^2_{ik}}{2(1 - \delta^2_k)}\Bigg]
\label{eq: Q-Psik SNMoE}
\end{eqnarray}for $k=1,\ldots,K$, where  
the required conditional expectations are given by:
\begin{eqnarray*}
\tau_{ik}^{(m)} &=& \E_{{\bsvPsi^{(m)}}}\left[Z_{ik}|y_i,\bsx_i,\bsr_i\right], \label{eq: E1 E-setp SNMoE} \\
e_{1,ik}^{(m)} &=&\E_{{\bsvPsi^{(m)}}}\left[U_{i}|Z_{ik}=1,y_i,\bsx_i,\bsr_i\right], \label{eq: E2 E-setp SNMoE}\\
e_{2,ik}^{(m)} &=& \E_{{\bsvPsi^{(m)}}}\left[U^2_{i}|Z_{ik}=1,y_i,\bsx_i,\bsr_i\right].
\label{eq: E3 E-setp SNMoE}
\end{eqnarray*}The $\tau_{ik}^{(m)}$'s represent the posterior distribution of the hidden class labels $Z_i$ and correspond to the posterior memberships of the observed data. They are given by:
\begin{eqnarray}
\tau_{ik}^{(m)} &=& \frac{\pi_k(\bsr;\bsalpha^{(m)}) \text{SN}\left(y_i;\mu(\bsx_i;\bsbeta_k^{(m)}),{\sigma^2_k}^{(m)},\lambda^{(m)}_k\right)}{f(y_i;\bsvPsi^{(m)})}\cdot
\label{eq: posterior prob SNMoE}
\end{eqnarray}
The conditional expectations $e_{1,ik}^{(m)}$ and $e_{2,ik}^{(m)}$ correspond to  the posterior distribution of the hidden variables $U_i$ and $U_i^2$, respectively. 
From the hierarchical representation (\ref{eq: hierarchical representation SNMoE}), as shown by \citet{Lin07univSkewNMixture} in the case of the skew-normal mixture model, by Bayes' theorem, the posterior distribution of $U_i$ is the following half normal:
\begin{equation}
U_i|Z_{ik}=1,y_i,\bsx_i,\bsr_i \sim HN_{[0,\infty)}\left(\mu_{u_{ik}}, \sigma^2_{u_{k}}\right) 
\nonumber
\label{eq: posterior of U in SNMoE}
\end{equation}where the posterior mean and variance in this case of SNMoE are respectively given by:
\begin{equation}
\mu_{u_{ik}} = \delta_k(y_i-\mu(\bsx_i;\bsbeta_k)) \quad \text{and} ~ \sigma^2_{u_{k}}= (1-\delta^2_k)\sigma^2_k.
\nonumber
\label{eq: posterior mean and variance of U in SNMoE}
\end{equation}Then the two conditional expectations  of $U_i$ and $U_i^2$ are respectively given by:
\begin{eqnarray}
e_{1,ik}^{(m)} &=& 
{\mu_{u_{ik}}}^{(m)} +{\sigma_{u_{k}}}^{(m)}  \frac{\phi\left(\lambda^{(m)}_k {d_{ik}}^{(m)}\right)}{\Phi\left(\lambda^{(m)}_k {d_{ik}}^{(m)}\right)},
\label{eq: posterior of u SNMoE}\\
e_{2,ik}^{(m)} &=& 
{\mu^2_{u_{ik}}}^{(m)} + {\sigma^2_{u_{k}}}^{(m)} + {\mu_{u_{ik}}}^{(m)} {\sigma_{u_{k}}}^{(m)}\frac{\phi\left(\lambda^{(m)}_k {d_{ik}}^{(m)}\right)}{\Phi\left(\lambda^{(m)}_k {d_{ik}}^{(m)}\right)}\cdot
\label{eq: posterior of u2 SNMoE}
\end{eqnarray} 
From (\ref{eq: Q-function decomposition SNMoE}), (\ref{eq: Q-alpha SNMoE}), and (\ref{eq: Q-Psik SNMoE}), it can be seen that the $Q$-function is calculated by analytically calculating the conditional expectations (\ref{eq: posterior prob SNMoE}),   (\ref{eq: posterior of u SNMoE})  and (\ref{eq: posterior of u2 SNMoE}).

\subsubsection{M-Step}
\label{sssec: M-Step SNMoE}

Then, the M-step calculates the parameter vector $\bsvPsi^{(m+1)}$ as in (\ref{eq: arg max Q}), that is by maximizing the $Q$-function (\ref{eq: Q-function decomposition SNMoE}) with respect to $\bsvPsi$. This can be performed by separately maximizing $Q_{1}(\bsalpha;\bsvPsi^{(m)})$ with respect to $\bsalpha$ and, for each component $k$ $(k=1,\ldots,K)$, the function $Q(\bsvPsi_k;\bsvPsi^{(m)})$ with respect to $\bsvPsi_k$ where $\bsvPsi_k = (\bsbeta^T_k,\sigma^2_k,\lambda_k)^T$.
We adopt the ECM extension of the EM algorithm. The M-step in this case consists of four conditional- maximization (CM)-steps, corresponding to the decomposition of the parameter vector $\bsvPsi$ into four sub-vectors 
$\bsvPsi = (\bsalpha,\bsbeta,\bssigma,\bslambda)^T$.
Thus, this leads to the following CM steps.
\paragraph{CM-Step 1}Calculate $\bsalpha^{(m+1)}$ by maximizing $Q_{1}(\bsalpha;\bsvPsi^{(m)})$:
\begin{equation}
\bsalpha^{(m+1)} =  \arg \max_{\bsalpha} Q_{1}(\bsalpha;\bsvPsi^{(m)}).
\label{eq: arg max_alpha Q-function SNMoE}
\end{equation}
Contrarily to the case of the standard skew-normal mixture model and  skew-normal regression mixture model, this maximization in the case of the proposed SNMoE does not exist in closed form. It is performed iteratively 
by Iteratively Reweighted Least Squares (IRLS).

\paragraph{The Iteratively Reweighted Least Squares (IRLS) algorithm:}
\label{par: IRLS M-Step SNMoE}
The IRLS algorithm is used to maximize $Q_{1}(\bsalpha,\bsvPsi^{(m)})$ given by (\ref{eq: Q-alpha SNMoE}) with respect to the parameter $\bsalpha$ in the M step at each iteration $m$ of the ECM algorithm.  
The IRLS is a Newton-Raphson algorithm, which consists in starting with a vector $\bsalpha^{(0)}$, and, at the $l+1$ iteration, updating the estimation of $\bsalpha$ as follows:
\begin{equation}
\bsalpha^{(l+1)}=\bsalpha^{(l)}-\Big[\frac{\partial^2 Q_{1}(\bsalpha,\bsvPsi^{(m)})}{\partial \bsalpha \partial \bsalpha^T}\Big]^{-1}_{\bsalpha=\bsalpha^{(l)}} \frac{\partial Q_{1}(\bsalpha,\bsvPsi^{(m)})}{\partial \bsalpha}\Big|_{\bsalpha=\bsalpha^{(l)}}
\label{eq: IRLS update}
\end{equation}
where $\frac{\partial^2 Q_{1}(\bsalpha,\bsvPsi^{(m)})}{\partial \bsalpha \partial \bsalpha^T}$ and $\frac{\partial Q_{1}(\bsalpha,\bsvPsi^{(m)})}{\partial \bsalpha}$ are respectively the Hessian matrix and the gradient vector of $Q_{1}(\bsalpha,\bsPsi^{(m)})$. At each IRLS iteration the Hessian and the gradient are evaluated at $\bsalpha = \bsalpha^{(l)}$ and are  computed similarly as in \citet{chamroukhi_et_al_neurocomputing2010}\citet{chamroukhi_et_al_NN2009}.
%
The parameter update $\bsalpha^{(m+1)}$  is taken at convergence of the IRLS algorithm (\ref{eq: IRLS update}).  
Then,  for  $k=1\ldots,K$,

\paragraph{CM-Step 2} Calculate $\bsbeta_k^{(m+1)}$ by maximizing $Q_{2}(\bsvPsi_k;\bsvPsi^{(m)})$ given by (\ref{eq: Q-Psik SNMoE}) w.r.t $\bsbeta_k$. 
Here we focus on the common linear case for the experts where each expert-component mean function is the one of a linear regression model and has the form (\ref{eq: linear regression mean}). It can be easily shown that the maximization problem for 
the resulting  skew-normal mixture of linear of experts (SNMoLE)   can be solved analytically and has the following solution:
\begin{eqnarray}
\bsbeta_k^{(m+1)}  &=& \Big[\sum_{i=1}^{n}\tau^{(m)}_{ik}  \bsx_i\bsx^T_i \Big]^{-1} \sum_{i=1}^{n} 
 \tau^{(q)}_{ik} \left(y_i -  \delta_{k}^{(m)}  e^{(m)}_{1,ik} \right)\bsx_i.
\label{eq: beta_k update for SNMoE}
\end{eqnarray}
\paragraph{CM-Step 3:} Calculate ${\sigma^2_{k}}^{(m+1)}$ by maximizing $Q_{2}(\bsvPsi_k;\bsvPsi^{(m)})$ given by (\ref{eq: Q-Psik SNMoE}) w.r.t $\sigma^2_{k}$. Similarly to the update of $\bsbeta_k$, the analytic solution of this problem is given by: 
\begin{eqnarray}
{\sigma^2_{k}}^{(m+1)} &\!\!\!\! =\!\!\!\! &
\frac{\sum_{i=1}^n\tau_{ik}^{(m)} \left[\left(y_i - {\bsbeta^T_{k}}^{(m+1)}\bsx_i\right)^2
 - 2 \delta_{k}^{(m+1)} e^{(m)}_{1,ik} (y_i - {\bsbeta^T_{k}}^{(m+1)}\bsx_i) 
+ e^{(m)}_{2,ik}\right]}
{2 \left( 
1 - {\delta^2_{k}}^{(m)} 
\right)\sum_{i=1}^n\tau_{ik}^{(m)}}\cdot
\label{eq: sigma2k update SNMoE}
\end{eqnarray}

\paragraph{CM-Step 4}Calculate $\lambda_{k}^{(m+1)}$ by maximizing $Q_{2}(\bsvPsi_k;\bsvPsi^{(m)})$ given by (\ref{eq: Q-Psik SNMoE}) w.r.t $\lambda_{k}$,  with $\bsbeta_k$ and $\sigma^2_{k}$ fixed at $\bsbeta_{k}^{(m+1)}$ and ${\sigma^2_{k}}^{(m+1)}$, respectively. This consists in solving the following equation for $\delta_k$ to obtain $\delta_{k}^{(m+1)}$ $(k=1,\ldots,K)$ as the solution of:
\begin{eqnarray}
{\sigma^2_{k}}^{(m+1)} \delta_{k} (1-\delta^2_{k}) \sum_{i=1}^n\tau_{ik}^{(m)}
+
(1+\delta^2_{k}) \sum_{i=1}^n\tau_{ik}^{(m)} (y_i - {\bsbeta^T_{k}}^{(m+1)}\bsx_i) ~ e^{(m)}_{1,ik} \nonumber \\
-  
\delta_{k} \sum_{i=1}^n\tau_{ik}^{(m)} \Big[ e^{(m)}_{2,ik} 
+ 
\left(y_i - {\bsbeta^T_{k}}^{(m+1)}\bsx_i\right)^2\Big]=0\cdot
\label{eq: deltak update SNMoE}
\end{eqnarray}Then, given  the update $\delta^{(m+1)}_k$, 
the update of the skewness parameter $\lambda_k$ is  calculated  as $\lambda^{(m+1)}_k = \frac{\delta^{(m+1)}_k}{\sqrt{1 - {\delta^2_k}^{(m+1)}}}$. 

It is obvious to see that when the skewness parameter $\lambda_k = \delta_k =0$ for all $k$, the parameter updates for the SNMoE corresponds to those of the NMoE. Hence, compared to the standard NMoE, the SNMoE model is characterized by an additional flexibility feature, that is the one  
 to be handle possibly skewed data. 
However, while the SNMoE model is tailored to model the skewness in the data, it may be not adapted to handle data containing groups or a group with heavy-tailed distribution. 
The NMoE and the SNMoE may thus be affected by outliers.
In the next section, we address the problem of sensitivity of normal mixture of experts to outliers and heavy tails. We first propose a robust  mixture of experts modeling by using the $t$ distribution.

\section{The $t$ mixture of experts (TMoE) model}
\label{sec: TMoE}
The proposed $t$ mixture of experts (TMoE) model is based on the $t$ distribution, which is known as a robust generalization of the normal distribution.  The $t$ distribution is recalled in the following section. We also described  its stochastic and hierarchical representations, which will be used to derive those of the TMoE model.

\subsection{The $t$ distribution}
 
The use of the $t$ distribution for mixture components has been shown to be more robust than the normal distribution to handle outliers in the data and accommodate data  with heavy tailed distribution. This has been shown in terms of density modeling and cluster analysis for multivariate data \citep{Mclachlan98robustTmixture,Peel2000robusTtmixture} as well as  for univariate data \citep{Lin07univSkewtMixture}.
The $t$-distribution with location parameter $\mu \in \R$, scale parameter $\sigma^2\in (0,\infty)$ and degrees of freedom $\nu \in (0,\infty)$ has the probability density function
\begin{equation}
  f(y;\mu, \sigma^2, \nu) = \frac{\Gamma(\frac{\nu+1}{2})} {\sqrt{\nu\pi}\,\Gamma(\frac{\nu}{2})} \left(1+\frac{d_y^2}{\nu} \right)^{-\frac{\nu+1}{2}},\!
  \label{eq: t density}
\end{equation}where $d_y = \frac{y-\mu}{\sigma}$ denotes the Mahalanobis distance between $y$ and $\mu$ ($\sigma$ being the scale parameter), and $\Gamma$ is the gamma function given by $\Gamma(x) = \int_0^\infty x^{t-1} e^{-x}\, d x$.
%
%
The $t$ distribution can be characterized as follows. Let $E$ be an univariate random variable with a standard normal distribution with pdf given by $\phi(.)$. Then, let $W$ be a random variable independent of $E$ and following the gamma distribution, that is $W \sim \text{gamma}(\frac{\nu}{2},\frac{\nu}{2})$ where the density function of the gamma distribution is given by $f (x;a,b) = \{b^a x^{a-1}/\Gamma(a)\} \exp(-bx)\Indicatrice_{(0,\infty)}(x); \quad (a,b)>0$ and the indicator function $\Indicatrice_{(0,\infty)}(x)=1$ for $x > 0$ and is zero elsewhere.
Then, a random variable $Y$ having the following representation:
\begin{equation}
Y = \mu + \sigma \frac{E}{\sqrt{W}}
\label{eq: stochastic representation t}
\end{equation}
follows the  $t$ distribution $t_{\nu}(\mu,\sigma^2, \nu)$ with pdf given by (\ref{eq: t density}). 
As given in  \citet{LiuAndRubin95} for the multivariate case, a hierarchical representation of the $t$ distribution in this univariate case can be expressed from the stochastic representation (\ref{eq: stochastic representation t}) as: 
\begin{equation}
\begin{tabular}{lll}
$Y_i|w_i$ & $\sim$ & $\text{N}\left(\mu, \frac{\sigma^2}{w_i}\right)$\\
$W_i$ & $\sim$ & $\text{gamma}\left(\frac{\nu}{2},\frac{\nu}{2}\right)$.
\end{tabular} 
\label{eq: hierarchical representation t}
\end{equation} 

 \subsection{The $t$ mixture of experts (TMoE) model}
The proposed   $t$ mixture of experts (TMoE) model extends the $t$ mixture model to the MoE framework.
%
The mixture of $t$ distributions have been first proposed by \citet{Mclachlan98robustTmixture,Peel2000robusTtmixture}
 for multivariate data. For the univariate case, a $K$-component $t$ mixture model takes the following form
\begin{eqnarray}
f(y;\bsvPsi) &=& \sum_{k=1}^K \pi_k ~ t_{\nu_k}(y; \mu_k, \sigma_k^2, \nu_k)
\label{eq: t mixture}
\end{eqnarray}where each of the mixture components has a $t$ density given by (\ref{eq: t density}).
\citet{Wei2012} considered the $t$-mixture model for the regression context on univariate data where the means $\mu_k$ in (\ref{eq: t mixture}) are (linear) regression functions of the form $\mu(\bsx;\bsbeta_k)$. However, this model do not explicitly model the mixing proportions as function the inputs; they are assumed to be constant.

The proposed $t$ mixture of experts (TMoE) is MoE model with $t$-distributed experts and is defined as follows.  
Let $t_{\nu}(\mu,\sigma^2,\nu)$ denotes a $t$ distribution with location parameter $\mu$, scale parameter $\sigma$ and degrees of freedom $\nu$, whose density is given by (\ref{eq: t density}). A $K$-component TMoE model is then defined by:
\begin{eqnarray}
f(y|\bsr,\bsx;\bsvPsi) &=& \sum_{k=1}^K \pi_k(\bsr;\bsalpha) ~ t_{\nu_k}\left(y; \mu(\bsx;\bsbeta_k), \sigma_k^2, \nu_k\right).
\label{eq: TMoE}
\end{eqnarray}The parameter vector of the TMoE model is given by $\bsvPsi = (\bsalpha^T_1,\ldots,\bsalpha^T_{K-1},\bsvPsi^T_1,\ldots,\bsvPsi^T_K)^T$ where $\bsvPsi_k = (\bsbeta^T_k,\sigma^2_k,\nu_k)^T$ is the parameter vector for the $k$th $t$ expert component which has a $t$ distribution. 
One can see that when the robustness parameter $\nu_k \rightarrow \infty$ for each $k$, the TMoE model  (\ref{eq: TMoE}) approaches the NMoE model (\ref{eq: normal MoE}).

In the following section, we present the stochastic and hierarchical characterizations of the proposed TMoE model and then derive the model maximum likelihood inference procedure.
\subsubsection{Stochastic representation of the TMoE}
\label{ssec: stochastic TMoE}
By using the stochastic representation (\ref{eq: stochastic representation t}) of the $t$ distribution, the stochastic representation for the $t$ mixture of experts (TMoE) is as follows. Let $E$ be a univariate random variable following the standard normal distribution $E \sim \phi(.)$.  
Suppose that, conditional on the hidden variable $Z_i=z_i$, a random variable $W_{i}$ is distributed as $\text{gamma}(\frac{\nu_{z_i}}{2},\frac{\nu_{z_i}}{2})$. Then, given the covariates $(\bsx_i, \bsr_i)$, a random variable $Y_i$ is said to follow the TMoE model (\ref{eq: TMoE}) if it has the following representation:
\begin{equation}
Y_i =  \mu(\bsx_i;\bsbeta_{z_i}) +  \sigma_{z_i}  \frac{E_i}{\sqrt{W_{z_i}}},
\label{eq: stochastic representation TMoE}
\end{equation}where the categorical variable $Z_i$ conditional on the covariate $\bsr_i$ follows the multinomial distribution as in (\ref{eq: Multinomial}).\\
Similarly to the case of the previously presented SNMoE model, the stochastic representation (\ref{eq: stochastic representation TMoE}) leads to the following hierarchical representation of the TMoE, which facilitates the model inference  as it will be presented in Section \ref{sec: MLE for the TMoE}.

\subsubsection{Hierarchical representation of the TMoE}
\label{ssec: hierarchical rep TMoE}
From (\ref{eq: stochastic representation t}) and (\ref{eq: stochastic representation TMoE}), following the hierarchical representation of the mixture of multivariate $t$-distributions (see for example  \citet{Mclachlan98robustTmixture}), the hierarchical representation of the TMoE model is written as:
\begin{eqnarray}
Y_i| w_i, Z_{ik}=1, \bsx_i &\sim& \text{N}\left(\mu(\bsx_i;\bsbeta_k), \frac{\sigma^2_k}{w_i}\right), \nonumber \\
W_i|Z_{ik}=1&\sim& \text{gamma}\left(\frac{\nu_k}{2},\frac{\nu_k}{2}\right)\\
\bsZ_i|\bsr_i &\sim & \text{Mult}\left(1;\pi_1(\bsr_i;\bsalpha),\ldots,\pi_K(\bsr_i;\bsalpha) \right).\nonumber 
\label{eq: hierarchical representation TMoE}
\end{eqnarray}

\section{Maximum likelihood estimation of the TMoE model}
\label{sec: MLE for the TMoE} 
Given an i.i.d sample of $n$ observations,  
the unknown parameter vector $\bsvPsi$ can be estimated by maximizing the observed-data log-likelihood,  which, under the TMoE model, is given by:
\begin{equation}
\log L(\bsvPsi) =  \sum_{i=1}^n  \log  \sum_{k=1}^K \pi_k(\bsr_i;\bsalpha) t{\nu_k}\left(y; \mu(\bsx;\bsbeta_k), \sigma_k^2, \nu_k\right).
\label{eq: log-lik TMoE}
\end{equation}To perform this maximization, we first use the EM algorithm and then described an ECM extension \citep{meng_and_rubin_ECM_93}    as in \citet{LiuAndRubin95} for a single $t$ distribution and as in \citet{Mclachlan98robustTmixture, Peel2000robusTtmixture} for mixture of $t$-distributions.

\subsection{The EM algorithm for the TMoE model} 
\label{ssec: EM TMoE}
To maximize the log-likelihood function (\ref{eq: log-lik TMoE}), the EM algorithm for the TMoE model starts with an initial parameter vector $\bsvPsi^{(0)}$ and alternates between the E- and M- steps until convergence. The E-step computes the expected completed data log-likelihood (the $Q$-function) and the M-Step maximize it.
 From the hierarchical representation of the TMoE (\ref{eq: hierarchical representation TMoE}), the complete data consist of the responses $(y_1,\ldots,y_n)$ and their corresponding covariates $(\bsx_1,\ldots,\bsx_n)$ and $(\bsr_1,\ldots,\bsr_n)$, as well as the latent variables $(w_1,\ldots,w_n)$ and the latent labels $(z_1,\ldots,z_n)$. Thus, the complete-data log-likelihood of $\bsvPsi$  is given by:
{\small \begin{eqnarray}
\log L_c(\bsvPsi)  &=&  \sum_{i=1}^n \sum_{k=1}^K Z_{ik} \big[ \log\left(\Pro\left(Z_i=k|\bsr_i\right)\right) +   \log\left(f\left(w_i|Z_{ik}=1\right)\right)+ \log\left(f\left(y_i|u_i,Z_{ik}=1,\bsx_i\right)\right)\big] \nonumber \\
& = & \log L_{1c}(\bsalpha) + \sum_{k=1}^K \big[\log L_{2c}(\bsvPsi_k) + \log L_{3c}(\nu_k)\big],
\label{eq: complete log-likelihood TMoE}
\end{eqnarray}}where 
{\small \begin{eqnarray}
\log L_{1c}(\bsalpha) &= & \sum_{i=1}^{n} \sum_{k=1}^K Z_{ik} \log \pi_k(\bsr_i;\bsalpha), \label{eq: L1c TMoE}\\
\log L_{1c}(\bsvPsi_k) &=& \sum_{i=1}^{n} Z_{ik} \Big[- \frac{1}{2}\log (2 \pi) -  \frac{1}{2}\log (\sigma^2_k) -  \frac{1}{2} w_i d^2_{ik}\Big], \label{eq: L2c TMoE}\\
\log L_{3c}(\nu_k) &=& \sum_{i=1}^{n} Z_{ik}  \Big[ - \log \Gamma \left(\frac{\nu_k}{2}\right) + \left(\frac{\nu_k}{2}\right) \log \left(\frac{\nu_k}{2}\right)    +  \left(\frac{\nu_k}{2}-1\right)  \log (w_i)
 - \left(\frac{\nu_k}{2}\right) w_i\Big]. \label{eq: L3c TMoE}
\end{eqnarray}}
\subsection{E-Step}
\label{ssec: E-Step TMoE}
The E-Step of the EM algorithm for the TMoE calculates the $Q$-function, that is the conditional expectation of the complete-data log-likelihood (\ref{eq: complete log-likelihood STMoE}),  given the observed data and a current parameter estimation $\bsvPsi^{(m)}$, $m$ being the current iteration.
It can be seen from (\ref{eq: L1c TMoE}), (\ref{eq: L2c TMoE}) and (\ref{eq: L3c TMoE}) that computing the $Q$-function requires the following conditional expectations:
\begin{eqnarray*}
\tau_{ik}^{(m)} &=& 
\E_{{\bsvPsi^{(m)}}}\left[Z_{ik}|y_i,\bsx_i,\bsr_i \right], \label{eq: E[Zik|yi] defintion TMoE}\\
w_{ik}^{(m)} &=& \E_{{\bsvPsi^{(m)}}}\left[W_{i}|y_i, Z_{ik}=1,\bsx_i,\bsr_i\right], \label{eq: E[Wi|yi,Zik] definition TMoE}\\
e_{1,ik}^{(m)} &=& \E_{{\bsvPsi^{(m)}}}\left[\log(W_{i})|y_i, Z_{ik}=1,\bsx_i,\bsr_i\right]\cdot \label{eq: E[log Wi|yi,Zik] definition TMoE}
\end{eqnarray*}
It follows that the $Q$-function is given by
\begin{equation}
Q(\bsvPsi;\bsvPsi^{(m)})=Q_{1}(\bsalpha;\bsvPsi^{(m)})+\sum_{k=1}^K \left[Q_{2}(\bsvPsi_k,\bsvPsi^{(m)}) + Q_{3}(\nu_k,\bsvPsi^{(m)})\right],
\label{eq: Q-function decomposition TMoE}
\end{equation}
where
{\small \begin{eqnarray*}
Q_{1}(\bsalpha;\bsvPsi^{(m)}) &= & \sum_{i=1}^{n} \sum_{k=1}^K \tau^{(m)}_{ik} \log \pi_k(\bsr_i;\bsalpha), \label{eq: Q_alpha TMoE}\\ 
Q_{2}(\bsvPsi_k;\bsvPsi^{(m)}) &=& \sum_{i=1}^{n} \tau^{(m)}_{ik}
\Big[- \frac{1}{2}\log (2 \pi) -  \frac{1}{2}\log (\sigma^2_k) -  \frac{1}{2} ~ w^{(m)}_{ik} d^2_{ik}\Big].\label{eq: Q_Psik TMoE}\\
Q_{3}(\nu_k;\bsvPsi^{(m)}) &=& \sum_{i=1}^{n} \tau^{(m)}_{ik}  \left[- \log \Gamma \left(\frac{\nu_k}{2}\right) + \left(\frac{\nu_k}{2}\right) \log \left(\frac{\nu_k}{2}\right)    
 - \left(\frac{\nu_k}{2}\right) ~ w^{(m)}_{ik} 
  +  \left(\frac{\nu_k}{2}-1\right) e^{(m)}_{1,ik} \right]. \label{eq: Q_nuk TMoE}
\end{eqnarray*}}The required conditional expectations are given as follows. 
First, the conditional expectation (\ref{eq: E[Zik|yi] defintion TMoE}) corresponds the posterior membership probabilities and is given by:
\begin{eqnarray}
\tau_{ik}^{(m)} &=& \frac{\pi_k(\bsr;\bsalpha^{(m)}) t_{\nu_k}(y_i;\mu(\bsx_i;\bsbeta_k^{(m)}),{\sigma^2_k}^{(m)}, \nu^{(m)}_k)}{f(y_i;\bsvPsi^{(m)})}\cdot
\label{eq: posterior prob TMoE}
\end{eqnarray}
Then, it can be easily shown (see for example \citet{Mclachlan98robustTmixture}
 and \citet{Peel2000robusTtmixture} \citet{LiuAndRubin95} for details) that:
{\small \begin{eqnarray}
\!\!\!\!\!\!\!\!\!\! \E_{{\bsvPsi^{(m)}}}\left[W_{i}|y_i, Z_{ik}=1,\bsx_i,\bsr_i \right] &\!\!\!\! =\!\!\!\! & \frac{\nu^{(m)}_k+1}{\nu^{(m)}_k+{d^2_{ik}}^{(m)}} 
= w^{(m)}_{ik}, \label{eq: E[Wi|yi,Zik] expression TMoE}\\
\!\!\!\!\!\!\!\!\!\! \E_{{\bsvPsi^{(m)}}}\left[\log(W_{i})|y_i, Z_{ik}=1,\bsx_i,\bsr_i \right] &\!\!\!\! =\!\!\!\! & \log\left(w^{(m)}_{ik}\right) + \left\{\psi\left(\frac{\nu^{(m)}_k +1}{2}\right) -  \log\left(\frac{\nu^{(m)}_k + 1}{2}\right) \right\} 
= e^{(m)}_{1,ik}, \label{eq: E[log Wi|yi,Zik] expression TMoE}
\end{eqnarray}}where $\psi(x) = \left\{\partial \Gamma(x)/\partial x \right\}/\Gamma(x)$ is the Digamma function.

\subsection{M-Step}
\label{ssec: M-Step TMoE}In the M-step, as it can be seen from (\ref{eq: Q-function decomposition TMoE}), the $Q$-function can be maximized by independently maximizing $Q_{1}(\bsalpha;\bsvPsi^{(m)})$, and, for each $k$, $Q_{2}(\bsPsi_k;\bsvPsi^{(m)})$, $Q_{3}(\nu_k;\bsvPsi^{(m)})$, with respect to $\bsalpha$, $\bsPsi_k$ and $\nu_k$, respectively.
Thus, on the $(m+1)$th iteration of the M-step, the model parameters are updated as follows.

\paragraph{M-Step 1}Calculate $\bsalpha^{(m+1)}$ by maximizing $Q_{1}(\bsalpha;\bsvPsi^{(m)})$ w.r.t $\bsalpha$. 
This can be performed iteratively via IRLS (\ref{eq: IRLS update}) as for the mixture of SNMoE.

\paragraph{M-Step 2} Calculate $\bsPsi_k^{(m+1)}$ by maximizing $Q_{3}(\bsvPsi_k;\bsvPsi^{(m)})$ 
w.r.t $\bsPsi_k=(\bsbeta^T_k,\sigma^2_k)^T$. This is achieved by first maximizing $Q_{3}(\bsvPsi_k;\bsvPsi^{(m)})$  with respect to $\bsbeta_k$ and then with respect to $\sigma^2_{k}$.
For the  $t$ mixture of linear experts (TMoLE) case where the expert means are of the form  (\ref{eq: linear regression mean}), this maximization can be performed analytically and provides the following updates: 
\begin{eqnarray}
\bsbeta_k^{(m+1)}  &=& \Big[\sum_{i=1}^{n}\tau^{(m)}_{ik} w_{ik}^{(m)}  \bsx_i\bsx^T_i \Big]^{-1} \sum_{i=1}^{n} 
 \tau^{(q)}_{ik}w_{ik}^{(m)}  y_i  \bsx_i,
\label{eq: beta_k update for TMoE}
\end{eqnarray}
\begin{eqnarray}
{\sigma^2_{k}}^{(m+1)} &= &
\frac{1}{\sum_{i=1}^n\tau_{ik}^{(m)}}\sum_{i=1}^n\tau_{ik}^{(m)} w_{ik}^{(m)} \left(y_i - {\bsbeta^T_{k}}^{(m+1)}\bsx_i\right)^2.
\label{eq: sigma2k update TMoE}
\end{eqnarray}
 Here, we note that, following \citet{Kent1994multivariateT}  in the
case of ML estimation for single component $t$ distribution and \citet{Mclachlan98robustTmixture, Peel2000robusTtmixture} for mixture of multivariate $t$ distributions, the EM algorithm can be modified slightly by replacing the divisor $\sum_{i=1}^n\tau_{ik}^{(m)}$ in (\ref{eq: sigma2k update TMoE})  by $\sum_{i=1}^n\tau_{ik}^{(m)} w_{ik}^{(m)}$. The modified algorithm may converge faster than the conventional EM algorithm.

\paragraph{M-Step 3}Calculate $\nu_k^{(m+1)}$ by maximizing $Q_{3}(\nu_k;\bsvPsi^{(m)})$ 
w.r.t $\nu_{k}$. 
%
The degrees of freedom update $\nu^{(m+1)}_k$ is therefore obtained by iteratively solving the following equation for $\nu_k$:
\begin{eqnarray}
& & - \psi \left(\frac{\nu_k}{2}\right) + \log \left(\frac{\nu_k}{2}\right) + 1 +
\frac{1}{\sum_{i=1}^n\tau_{ik}^{(m)}}\sum_{i=1}^n\tau_{ik}^{(m)} \left( \log(w^{(m)}_{ik}) - w^{(m)}_{ik} \right) 
\nonumber \\
& &
 + \psi\left(\frac{\nu^{(m)}_k +1}{2}\right) -  \log\left(\frac{\nu^{(m)}_k + 1}{2}\right) = 0.
\label{eq: nuk update TMoE}
\end{eqnarray}This scalar non-linear equation can be solved  with a root finding algorithm, such
as Brent's method \citep{Brent1973}.
 
It is obvious to see that, as mentioned previously,  if the number of degrees of freedom $\nu_k$ is fixed at $\infty$ for all $k$, then the parameter updates  for the TMoE  model are exactly those of the NMoE model (since $w_{ik}$ tends to $1$ in this case). The TMoE model constitutes therefore a robust generalization of the NMoE model that is able to model data with density heaving longer tails than those of the NMoE model. 

After deriving the EM algorithm for the TMoE parameter estimation, now we described and ECM extension.

\subsection{The ECM algorithm for the TMoE model}
\label{ssec: ECM TMoE}
Following the ECM extension of the EM algorithm for a single $t$ distribution proposed by  \citet{LiuAndRubin95} and the one of the EM algorithm for the $t$-mixture model  \citep{Mclachlan98robustTmixture, Peel2000robusTtmixture}, the EM algorithm for the TMoE model can also be modified to give an ECM version by adding  an additional E-Step between the two M-steps 2 and 3. 
This additional E-step consists in taking the parameter vector $\bsvPsi$ with $\bsvPsi_k = \bsvPsi_k^{(m+1)}$ instead of $\bsvPsi_k^{(m)}$, that is
$$Q_2(\nu_k;\bsvPsi^{(m)}) = Q_2(\nu_k;\bsalpha^{(m)},\bsvPsi_k^{(m+1)},\nu_k^{(m)}).$$
Thus, the M-Step 3 in the above is replaced by a Conditional-Maximization (CM)-Step in which the degrees of freedom update (\ref{eq: nuk update TMoE}) is calculated with the conditional expectation (\ref{eq: E[Wi|yi,Zik] expression TMoE}) and (\ref{eq: E[log Wi|yi,Zik] expression TMoE}) computed with the updated parameters $\bsbeta_k^{(m+1)}$ and ${\sigma^2_{k}}^{(m+1)}$  respectively given by (\ref{eq: beta_k update for TMoE}) and (\ref{eq: sigma2k update TMoE}).

\bigskip
The SNMoE presented before allows to deal with asymmetric data. The TMoE handles the problem of heavy tailed data possibly affected by outliers. Now, we propose the skew $t$ mixture of experts (STMoE) model which attempts to simultaneously accommodate heavy tailed data with possible outliers and with asymmetric distribution.

\section{The skew $t$ mixture of experts (STMoE) model}
\label{sec: STMoE}

The proposed  skew $t$ mixture of experts (STMoE) model is a MoE model in which the expert components have a skew-$t$ density, rather than the standard normal one as in the NMoE model, or the previously presented skew-normal and $t$ ones as the SNMoE and the TMoE, respectively. The skew-$t$ distribution as well as its stochastic and hierarchical representations are recalled in the following section.
\subsection{The skew $t$ distribution}
Let us denote  by $t_{\nu}(.)$ and $T_{\nu}(.)$  respectively  the  probability  density  function  (pdf)  and  the  cumulative distribution function (cdf) of  the  $t$ distribution  with  degrees  of  freedom $\nu$.
The skew $t$ distribution, introduced by \citet{AzzaliniAndCapitanio2003}, can be characterized as follows. Let $U$ be an univariate random variable with a standard skew normal distribution $U \sim \text{SN}(0,1,\lambda)$ (which can be shortened as $U \sim \text{SN}(\lambda)$) with pdf given by (\ref{eq: Skew-normal density}). 
Then, let $W$ be an univariate random variable independent of $U$ and following the gamma distribution, that is, $W \sim \text{gamma}(\frac{\nu}{2},\frac{\nu}{2})$. 
A random variable $Y$ having the following representation:
\begin{equation}
Y = \mu + \sigma \frac{U}{\sqrt{W}}
\label{eq: stochastic representation skew-t}
\end{equation}follows the skew $t$ distribution $\text{ST}(\mu,\sigma^2,\lambda,\nu)$ with 
location parameter $\mu$, scale parameter $\sigma$, skewness parameter $\lambda$ and degrees of freedom $\nu$,   
 whose density is defined by:
\begin{equation}
f(y;\mu,\sigma^2,\lambda,\nu) = \frac{2}{\sigma} ~ t_{\nu}(d_y) ~  T_{\nu + 1} \left(\lambda ~ d_y \sqrt{\frac{\nu+1}{\nu+d^2_y}}\right)
\label{eq: skew-t density}
\end{equation} where $d_y = \frac{y-\mu}{\sigma}$.
From the hierarchical distribution of the skew-normal (\ref{eq: hierarchical representation skew-normal}), a further hierarchical representation of the stochastic representation (\ref{eq: stochastic representation skew-t}) of the skew $t$ distribution is given by:
\begin{eqnarray} 
Y_i|u_i, w_i&\sim & \text{N}\left(\mu + \delta |u_i|, \frac{1-\delta^2}{w_i}\sigma^2\right), \nonumber \\
U_i|w_i&\sim & \text{N}(0,\frac{\sigma^2}{w_i}),\\
W_i&\sim& \text{gamma}\left(\frac{\nu}{2},\frac{\nu}{2}\right). \nonumber
\label{eq: hierarchical representation skew-t}
\end{eqnarray}

\subsection{The skew $t$ mixture of experts (STMoE) model}
The skew proposed $t$ mixture of experts (STMoE) model extends the skew $t$ mixture model, which was first introduced by \citet{Lin07univSkewtMixture},  to the MoE framework. A $K$-component skew $t$ mixture model is given by
\begin{eqnarray}
f(y;\bsvPsi) &=& \sum_{k=1}^K \pi_k~f(y; \mu_k, \sigma_k^2, \lambda_k,\nu_k)
\label{eq: skew-t mixture}
\end{eqnarray}where each of the mixture components is a skew $t$ density given by (\ref{eq: skew-t density}).
In the skew-$t$ mixture model (\ref{eq: skew-t mixture}), the mixing proportions and the expert means are constant, that is, they are not function of the inputs. In the proposed STMoE, we consider skew-$t$ expert components with regression mean functions, and covariate varying mixing proportions. 
A $K$-component mixture of skew $t$ experts (STMoE) is therefore defined by:
\begin{eqnarray}
f(y|\bsr,\bsx;\bsvPsi) &=& \sum_{k=1}^K \pi_k(\bsr;\bsalpha)  \,\text{ST}(y; \mu(\bsx;\bsbeta_k), \sigma_k^2, \lambda_k,\nu_k)\cdot
\label{eq: skew-t MoE}
\end{eqnarray}The parameter vector of the STMoE model is $\bsvPsi = (\bsalpha^T_1,\ldots,\bsalpha^T_{K-1},\bsvPsi^T_1,\ldots,\bsvPsi^T_K)^T$ where $\bsvPsi_k = (\bsbeta^T_k,\sigma^2_k,\lambda_k,\nu_k)^T$ is the parameter vector for the $k$th skew $t$ expert component whose density is defined by 
\begin{equation}
f\big(y|\bsx;\mu(\bsx;\bsbeta_k),\sigma^2,\lambda,\nu\big) = \frac{2}{\sigma} ~ t_{\nu}(d_y(\bsx)) ~  T_{\nu + 1} \left(\lambda ~ d_y(\bsx) \sqrt{\frac{\nu+1}{\nu+d^2_y(\bsx)}}\right)
\label{eq: skew-t density}
\end{equation}where $d_y(\bsx) = \frac{y-\mu(\bsx;\bsbeta_k)}{\sigma}$ represents the Mahalanobis distance between  $y$ and $\mu(\bsx;\bsbeta_k)$.

\bigskip
It can be seen that, when the robustness parameter $\nu_k \rightarrow \infty$ for each $k$, the STMoE model (\ref{eq: skew-t MoE}) reduces to the SNMoE model  (\ref{eq: SNMoE}). On the other hand, if the skewness parameter $\lambda_k = 0$ for each $k$, the STMoE model    reduces to the TMoE model  (\ref{eq: TMoE}). Moreover, when  $\nu_k \rightarrow \infty$ and  $\lambda_k = 0$ for each $k$, it approaches the standrad NMoE model (\ref{eq: normal MoE}). This therefore makes the STMoE flexible as it generalizes the previously described models to accommodate situations with asymmetry, heavy tails, and outliers.

\subsection{Stochastic representation of the STMoE model}
The skew $t$ mixture of experts model is characterized as follows. 
Suppose that conditional on a  categorical variable $z_i \in \{1,\ldots,K\}$ representing the hidden label of the component generating the $i$th observation and following the multinomial distribution (\ref{eq: Multinomial}),  
a random variable  has the following representation: 
\begin{equation}
Y_i =  \mu(\bsx_i;\bsbeta_{z_i}) + \sigma_{z_i} \frac{E_i}{\sqrt{W_i}}
\label{eq: stochastic representation skew-t MoE}
\end{equation}where   $E_i$ and $W_i$ are independent univariate random variables with, respectively, a standard skew normal distribution $E_{i} \sim \text{SN}(\lambda_{z_i})$, and a Gamma distribution $W_i \sim \text{gamma}(\frac{\nu_{z_i}}{2},\frac{\nu_{z_i}}{2})$, and $\bsx_i$ and $\bsr_i$ are some given covariate variables. Then, the variable $Y_i$ is said to follow the skew $t$ mixture of experts (STMoE) defined by (\ref{eq: skew-t MoE}).

\subsection{Hierarchical representation of the STMoE model}
From the hierarchical representation (\ref{eq: hierarchical representation skew-t}) of the skew $t$ distribution, a hierarchical model for the proposed STMoE model (\ref{eq: skew-t MoE}) can be derived from its stochastic representation (\ref{eq: stochastic representation skew-t MoE}) and is as follows:
\begin{eqnarray} 
Y_i|u_i, w_i, Z_{ik}=1, \bsx_i &\sim& \text{N}\left(\mu(\bsx_i;\bsbeta_k) + \delta_k |u_i|, \frac{1-\delta^2_k}{w_i}\sigma^2_k\right), \nonumber \\
U_i|w_i, Z_{ik}=1&\sim & \text{N}\left(0,\frac{\sigma^2_k}{w_i}\right),\\
W_i|Z_{ik}=1&\sim& \text{gamma}\left(\frac{\nu_k}{2},\frac{\nu_k}{2}\right)\nonumber\\
\bsZ_i|\bsr_i &\sim & \text{Mult}\big(1;\pi_1(\bsr_i;\bsalpha),\ldots,\pi_K(\bsr_i;\bsalpha) \big).\nonumber
\label{eq: hierarchical representation skew-t MoE}
\end{eqnarray}This hierarchical representation will be used to derive the maximum likelihood estimation of the STMoE model parameters $\bsvPsi$ by using the ECM algorithm. 

\section{Maximum likelihood estimation of the STMoE model}
\label{sec: MLE for the STMoE}
The unknown parameter vector $\bsvPsi$ of the STMoE model is estimated by maximizing the following observed-data log-likelihood given an observed i.i.d sample of $n$ observations $y_i$ and their corresponding covariates $\bsx_i$ and $\bsr_i$:
\begin{equation}
\log L(\bsvPsi) = \sum_{i=1}^n  \log  \sum_{k=1}^K \pi_k(\bsr_i;\bsalpha) \text{ST}(y; \mu(\bsx_i;\bsbeta_k), \sigma_k^2, \lambda_k,\nu_k)\cdot
\label{eq: log-lik skew-t MoE}
\end{equation}
%
We perform this iteratively by a dedicated ECM algorithm. 
The  complete data consist of the observations $(y_1,\ldots,y_n)$, their corresponding covariates $(\bsx_1,\ldots,\bsx_n)$ and $(\bsr_1,\ldots,\bsr_n)$, as well as the latent variables $(u_1,\ldots,u_n)$ and $(w_1,\ldots,w_n)$ and the latent labels $(z_1,\ldots,z_n)$. Then, from the hierarchical representation of the STMoE (\ref{eq: hierarchical representation skew-t MoE}), the complete-data log-likelihood of $\bsvPsi$ is given by:
{\small \begin{eqnarray}
\log L_c(\bsvPsi)  & = &  \sum_{i=1}^n \sum_{k=1}^K Z_{ik} \Big[ \log\left(\Pro\left(Z_i=k|\bsr_i\right)\right) +   \log\left(f\left(w_i|Z_{ik}=1\right)\right) + \nonumber \\
& & \log\left(f\left(u_i|w_i, Z_{ik}=1\right)\right)+ \log\left(f\left(y_i|u_i,Z_{ik}=1,\bsx_i\right)\right)\Big] \nonumber \\
& = & \log L_{1c}(\bsalpha)  + \sum_{k=1}^K \big[\log L_{2c}(\bsvPsi_k) + \log L_{3c}(\nu_k) \big]
\label{eq: complete log-likelihood STMoE}
\end{eqnarray}}
where $\bsvPsi_k = (\bsbeta^T_k,\sigma_k^2,\lambda_k)^T$ and
{\small \begin{eqnarray*}
\log L_{1c}(\bsalpha) &= & \sum_{i=1}^{n} \sum_{k=1}^K Z_{ik} \log \pi_k(\bsr_i;\bsalpha), \\
\log L_{2c}(\bsvPsi_k) 
&=&\sum_{i=1}^{n}  Z_{ik} \Big[ -  \log (2 \pi) - \log (\sigma^2_k) - \frac{1}{2} \log (1 - \delta^2_k) 
 - \frac{w_i ~ d^2_{ik}}{2(1 - \delta^2_k)} + \frac{w_i~ u_i ~ \delta_k ~ d_{ik}}{(1 - \delta^2_k)\sigma_k}  - \frac{ w_i ~ u_i^2}{2(1 - \delta^2_k)\sigma^2_k}  
\Big],\\
\log L_{3c}(\nu_k) &=& \sum_{i=1}^{n}  Z_{ik}  \Big[ - \log \Gamma \left(\frac{\nu_k}{2}\right) + \left(\frac{\nu_k}{2}\right) \log \left(\frac{\nu_k}{2}\right)    +  \left(\frac{\nu_k}{2}\right)  \log (w_i)
 - \left(\frac{\nu_k}{2}\right) w_i\Big].
\label{eq: complete log-lik decomposition skew-t MoE}
\end{eqnarray*}}

\subsection{The ECM algorithm for the STMoE model}
The ECM algorithm for the STMoE model starts with an initial parameter vector $\bsvPsi^{(0)}$ and alternates between the E- and CM- steps until convergence.
\subsection{E-Step}
The E-Step of the CEM algorithm for the STMoE calculates the $Q$-function, that is the conditional expectation of the complete-data log-likelihood (\ref{eq: complete log-likelihood STMoE}),  given the observed data $\{y_i,\bsx_i,\bsr_i\}_{i=1}^n$ and a current parameter estimation $\bsvPsi^{(m)}$, $m$ being the current iteration.
From (\ref{eq: complete log-likelihood STMoE}), it can be seen that computing the $Q$-function requires the following conditional expectations:
\begin{eqnarray*} 
\tau_{ik}^{(m)} &=& \E_{{\bsvPsi^{(m)}}}\left[Z_{ik}|y_i,\bsx_i,\bsr_i\right],\label{eq: E[Zi|yi] definition STMoE}\\
w_{ik}^{(m)} &=&\E_{{\bsvPsi^{(m)}}}\left[W_{i}|y_i, Z_{ik}=1,\bsx_i,\bsr_i\right],\label{eq: E[Wi|yi,Zik] definition STMoE}\\
e_{1,ik}^{(m)} &=& \E_{{\bsvPsi^{(m)}}}\left[W_{i}U_{i}|y_i, Z_{ik}=1,\bsx_i,\bsr_i\right],\label{eq: E[WiUi|yi,Zik] definition STMoE}\\
e_{2,ik}^{(m)} &=& \E_{{\bsvPsi^{(m)}}}\left[W_{i}U^2_{i}|y_i, Z_{ik}=1,\bsx_i,\bsr_i\right],\label{eq: E[WiUi2|yi,Zik] definition STMoE}\\
e_{3,ik}^{(m)} &=& \E_{{\bsvPsi^{(m)}}}\left[\log(W_{i})|y_i, Z_{ik}=1,\bsx_i,\bsr_i\right]\label{eq: E[logWi|yi,Zik] definition STMoE}\cdot
\end{eqnarray*}
The $Q$-function being given by:
{\small \begin{equation}
Q(\bsvPsi;\bsvPsi^{(m)})=Q_{1}(\bsalpha;\bsvPsi^{(m)})+\sum_{k=1}^K \left[Q_{2}(\bsvPsi_k,\bsvPsi^{(m)})+ Q_{3}(\nu_k,\bsvPsi^{(m)})\right],
\label{eq: Q-function decomposition STMoE}
\end{equation}}
where 
{\small \begin{eqnarray*}
Q_{1}(\bsalpha;\bsvPsi^{(m)}) &\!\! =\!\! & \sum_{i=1}^{n} \sum_{k=1}^K \tau^{(m)}_{ik} \log \pi_k(\bsr_i;\bsalpha), \label{eq: Q_alpha STMoE}\\ 
Q_{2}(\bsvPsi_k;\bsvPsi^{(m)}) &\!\!=\!\! & \sum_{i=1}^{n} \tau^{(m)}_{ik}\Bigg[ -  \log (2 \pi) - \log (\sigma^2_k) - \frac{1}{2} \log (1 - \delta^2_k)
- \frac{ w^{(m)}_{ik} ~ d^2_{ik}}{2(1 - \delta^2_k)} + \frac{\delta_k ~ d_{ik} ~ e_{1,ik}^{(m)}}{(1 - \delta^2_k)\sigma_k}  - \frac{e_{2,ik}^{(m)}}{2(1 - \delta^2_k)\sigma^2_k} \Bigg],\label{eq: Q_Psik STMoE}\\
Q_{3}(\nu_k;\bsvPsi^{(m)}) &\!\! =\!\! & \sum_{i=1}^{n} \tau^{(m)}_{ik}  \left[- \log \Gamma \left(\frac{\nu_k}{2}\right) + \left(\frac{\nu_k}{2}\right) \log \left(\frac{\nu_k}{2}\right)    
 - \left(\frac{\nu_k}{2}\right) ~ w^{(m)}_{ik} 
  +  \left(\frac{\nu_k}{2}\right) e^{(m)}_{3,ik} \right]. \label{eq: Q_nuk STMoE}
\end{eqnarray*}}
Following the expressions of these conditional expectations given namely in the case of the standard skew $t$ mixture  model  \citep{Lin07univSkewtMixture},  the conditional expectations for the case of the proposed STMoE model can be expressed similarly as:
{\small \begin{eqnarray}
\tau_{ik}^{(m)} &=& \frac{\pi_k(\bsr;\bsalpha^{(m)}) ~\text{ST}\Big(y_i;\mu(\bsx_i;\bsbeta_k^{(m)}),\sigma^{2(m)}_k,\lambda^{(m)}_k, \nu^{(m)}_k\Big)}{f(y_i;\bsvPsi^{(m)})}, \label{eq: posterior prob STMoE}\\ 
w_{ik}^{(m)} &=& \left(\frac{\nu^{(m)}_k+1}{\nu^{(m)}_k+{d^2_{ik}}^{(m)}}\right)
\times 
\frac{T_{\nu^{(m)}_k+3} \left(M^{(m)}_{ik}\sqrt{\frac{\nu^{(m)}_k+3}{\nu^{(m)}_k + 1}}\right)}
{T_{\nu^{(m)}_k+1} \left(M^{(m)}_{ik}\right)},
\label{eq: E[Wi|yi,Zik] expression STMoE}
\end{eqnarray}}where
$M^{(m)}_{ik} = \lambda^{(m)}_k ~ d^{(m)}_{ik} \sqrt{\frac{\nu^{(m)}_k +1}{\nu^{(m)}_k+{d^2_{ik}}^{(m)}}}$,
\begin{eqnarray}
\!\!\!\! e_{1,ik}^{(m)} &\!\!\!\!  = \!\!\!\!  & 
\delta_{k}^{(m)} \left(y_i -\mu_k(\bsx_i;\bsbeta^{(m)})\right) w_{ik}^{(m)} 
+ \Bigg[\frac{\sqrt{1 -{\delta^2_{k}}^{(m)}}}{\pi f(y_i;\bsvPsi^{(m)})} \left( \frac{{d^2_{ik}}^{(m)}}{\nu^{(m)}_{k} (1 - {\delta^2_k}^{(m)})} +1 \right)^{-(\frac{\nu^{(m)}_{k}}{2}+1)}
\Bigg],\\
\label{eq: E[WiUi|yi,Zik] expression STMoE}
\!\!\!\! e_{2,ik}^{(m)} &\!\!\!\!  =\!\!\!\!  & 
{\delta^2_{k}}^{(m)} \left(y_i -\mu_k(\bsx_i;\bsbeta^{(m)})\right)^2 w_{ik}^{(m)} 
+  \Bigg[\left(1-{\delta^2_{k}}^{(m)}\right) {\sigma^2_k}^{(m)} \nonumber \\
&\!\!\!\! \!\!\!\! &
+ \frac{\delta_{k}^{(m)} \left(y_i -\mu_k(\bsx_i;\bsbeta^{(m)})\right) \sqrt{1 -{\delta^2_{k}}^{(m)}}}{\pi f(y_i;\bsvPsi^{(m)})}
\times \left( \frac{{d^2_{ik}}^{(m)}}{\nu^{(m)}_{k} (1 - {\delta^2_k}^{(m)})} +1 \right)^{-(\frac{\nu^{(m)}_{k}}{2}+1)}
\Bigg],\\
\label{eq: E[WiUi2|yi,Zik] expression STMoE}
%
\!\!\!\! e_{3,ik}^{(m)} &\!\!\!\! =\!\!\!\! & w^{(m)}_{ik} - \log\left(\frac{\nu^{(m)}_k + {d^2_{ik}}^{(m)}}{2}\right) - \left(\frac{\nu^{(m)}_k +1}{\nu^{(m)}_k + {d^2_{ik}}^{(m)}}\right) 
+ \psi\left(\frac{\nu^{(m)}_k +1}{2}\right)\nonumber \\
&\!\!\!\! \!\!\!\! & + \frac{\lambda^{(m)}_{k} d_{ik}^{(m)}\left({d^2_{ik}}^{(m)}  - 1\right)}
{\sqrt{\left(\nu^{(m)}_k+ 1\right)\left(\nu^{(m)}_k + {d^2_{ik}}^{(m)}\right)^3}}
\times 
\frac{t_{\nu^{(m)}_k+1} \left(M^{(m)}_{ik}\right)}
{T_{\nu^{(m)}_k+1} \left(M^{(m)}_{ik}\right)}\cdot
\label{eq: E[logWi|yi,Zik] expression STMoE}
\end{eqnarray}} 
We note that, for (\ref{eq: E[logWi|yi,Zik] expression STMoE}), we adopted a one-step-late (OSL) approach to compute the conditional expectation $e_{3,ik}^{(m)} $ 
as described in \citet{LeeAndMcLachlan14-skewtmix}, by setting the integral part in the expression of the corresponding conditional expectation given in \citep{Lin07univSkewtMixture} to zero, rather than using a Monte Carlo approximation.
%
%
We also mention that, for the multivariate skew $t$ mixture models, recently \citet{LeeAndMcLachlan15-CFUST} presented a series-based truncation approach, which exploits an exact representation of this conditional expectation and which can also be used in place of (\ref{eq: E[logWi|yi,Zik] expression STMoE}).

\subsection{M-Step}
\label{ssec: M-Step ECM STMoE}
The M-step maximizes the $Q$-function (\ref{eq: Q-function decomposition STMoE})  with respect to $\bsvPsi$ and provides the parameter vector update $\bsvPsi^{(m+1)}$. 
From (\ref{eq: Q-function decomposition STMoE}), it can be seen that the maximization of $Q$ 
can be performed by separately maximizing $Q_{1}$ with respect to the parameters $\bsalpha$ of the mixing proportions, and for each expert $k$ $(k=1,\ldots,K)$, $Q_{2}$ with respect to $(\bsbeta^T_k,\sigma^2_k)^T$ and $\lambda_k$, and $Q_{3}$ with respect to $\nu_k$. The maximization of $Q_2$ and $Q_3$ is carried out by conditional maximization (CM) steps by updating $(\bsbeta_k,\sigma_k^2)$ and then updating $(\lambda,\nu_k)$ with the given updated parameters.
This leads to the following CM steps.
On the $(m+1)$th iteration of the M-step, the STMoE model parameters are updated as follows.
\paragraph{CM-Step 1}Calculate the parameter $\bsalpha^{(m+1)}$ maximizing the function $Q_{1}(\bsalpha;\bsvPsi^{(m)})$ given by (\ref{eq: Q_alpha TMoE}) by using IRLS (\ref{eq: IRLS update}).
Then,  for  $k=1\ldots,K$,

\paragraph{CM-Step 2} Calculate $(\bsbeta_k^{T(m+1)},{\sigma^2_k}^{(m+1)})^T$ by maximizing $Q_{2}(\bsvPsi_k;\bsvPsi^{(m)})$ w.r.t $(\bsbeta^T_k,\sigma^2_k)^T$. 
For the  $t$ mixture of linear experts (TMoLE) case, where the expert means are linear regressors, that is, of the form  (\ref{eq: linear regression mean}), this maximization can be performed in a closed form and provides the following updates:
{\small \begin{eqnarray} 
\!\!\!\! \bsbeta_k^{(m+1)}  &\!\! =\!\! & \Big[\sum_{i=1}^{n}\tau^{(q)}_{ik} w_{ik}^{(m)} \bsx_i\bsx^T_i \Big]^{-1} \sum_{i=1}^{n} 
\tau^{(q)}_{ik}\left(w_{ik}^{(m)}y_i -  \bse^{(m)}_{1,ik} \delta_{k}^{(m+1)} \right)\bsx_i,
\label{eq: beta_k update for STMoE} \\
\!\!\!\!  {\sigma^2_{k}}^{(m+1)} &\!\! =\!\! &
\frac{\sum_{i=1}^n\tau_{ik}^{(m)} \Big[w_{ik}^{(m)}\left(\bsy_i - {\bsbeta^T_{k}}^{(m+1)}\bsx_i\right)^2
 - 2 \delta_{k}^{(m+1)} \bse^{(m)}_{1,ik} (y_i - {\bsbeta^T_{k}}^{(m+1)}\bsx_i) 
+\bse^{(m)}_{2,ik}\Big]}
{2 \left( 1 - {\delta^2_{k}}^{(m)} \right)\sum_{i=1}^n\tau_{ik}^{(m)}}\cdot
\label{eq: sigma2k update STMoE}
\end{eqnarray}}
\paragraph{CM-Step 3} The skewness parameters $\lambda_k$ are updated by maximizing $Q_{2}(\bsvPsi_k;\bsvPsi^{(m)})$ w.r.t $\lambda_k$, with $\bsbeta_k$ and $\sigma^2_k$ fixed at  the update $\bsbeta_k^{(m+1)}$ and ${\sigma^2_{k}}^{(m+1)}$, respectively.
It can be easily shown that the maximization  to obtain $\delta_{k}^{(m+1)}$ $(k=1,\ldots,K)$ consists in solving the following equation in $\delta_k$:
{\small \begin{eqnarray} 
%
\delta_{k} (1-\delta^2_{k}) \sum_{i=1}^n\tau_{ik}^{(m)} + (1+\delta^2_{k}) \sum_{i=1}^n\tau_{ik}^{(m)} \frac{d^{(m+1)}_{ik} e^{(m)}_{1,ik}}{\sigma_{k}^{(m+1)}} -  \delta_{k} \sum_{i=1}^n\tau_{ik}^{(m)} \Big[w_{ik}^{(m)}  {d^2_{ik}}^{(m+1)}+ \frac{e^{(m)}_{2,ik}}{{\sigma^2_{k}}^{(m+1)}}  \Big]=0 \cdot
\label{eq: deltak update STMoE}
\end{eqnarray}}

\paragraph{CM-Step 4} Similarly, the degrees of freedom $\nu_k$ are updated by maximizing $Q_{3}(\nu_k;\bsvPsi^{(m)})$ w.r.t $\nu_k$ with $\bsbeta_k$ and $\sigma^2_k$ fixed at  $\bsbeta_k^{(m+1)}$ and ${\sigma^2_{k}}^{(m+1)}$, respectively.
An update $\nu^{(m+1)}_k$ is calculated as solution of the following equation in $\nu_k$:
\begin{equation}
- \psi\left(\frac{\nu_k}{2}\right) + \log\left(\frac{\nu_k}{2}\right) +1 + 
\frac{\sum_{i=1}^n\tau_{ik}^{(m)} \left(\bse^{(m)}_{3,ik} - w^{(m)}_{ik}\right)}{\sum_{i=1}^n\tau_{ik}^{(m)}} = 0.
\label{eq: nu update STMoE}
\end{equation}
%
The two scalar non-linear equations (\ref{eq: deltak update STMoE}) and (\ref{eq: nu update STMoE})  can be solved  similarly as in the TMoE model, that is with a root finding algorithm, such as Brent's method \citep{Brent1973}.

\bigskip
As mentioned before, one can see that, when the robustness parameter $\nu_k \rightarrow \infty$ for all the components, the parameter updates for the  STMoE model correspond  to those of the SNMoE model. 
On the other hand,  when the skewness parameters $\lambda_k = 0$, the STMoE parameter updates correspond  to those of the TMoE model. Finally,  when  both the degrees of freedom $\nu_k \rightarrow \infty$ and  the skewness $\lambda_k = 0$, we obtain the parameter updates of the standard NMoE model. The STMoe therefore provides a more general  framework for inferring flexible MoE models. 

\section{Prediction using the NNMoE}
\label{sec: Prediction using the NNMoE}
The goal in 
regression 
is to be able to make predictions for the response variable(s) given some new value of the predictor variable(s) on the basis of a model trained on a set of training data. 
In regression analysis using mixture of experts, the aim is therefore to predict the response $y$ given new values of the predictors $(\bsx,\bsr)$, on the basis of a MoE model characterized by a parameter vector $\hat \bsvPsi$ inferred from a set of training data, here, by maximum likelihood via EM. 
%
These predictions can be expressed in terms of the predictive distribution of $y$, which is obtained by substituting the maximum likelihood parameter $\hat\bsvPsi$ into (\ref{eq: mixture of regressions})-(\ref{eq: multinomial logistic}) to give:
\begin{equation*}
f(y|\bsx,\bsr;\hat \bsvPsi) = \sum_{k=1}^K \pi_{k}(\bsr;\hat \bsalpha)  f_k(y|\bsx; \hat \bsvPsi_k).
\label{eq: predictive MoE}
\end{equation*}Using $f$, we might then predict $y$ for a given set of $\bsx$'s and $\bsr$'s as the expected value under $f$, 
that is by calculating the prediction $\hat y = \E_{{\it \hat\bsvPsi}}(Y|\bsr,\bsx)$. We thus need to compute the expectation of the mixture of experts model.
It is easy to show (see for example Section 1.2.4 in \citet{sylvia_fruhwirth_book_2006}) that the mean and the variance of a mixture of experts distribution of the form (\ref{eq: predictive MoE}) are respectively given by
{\small
\begin{eqnarray}
\E_{{\it \hat\bsvPsi}}(Y|\bsr,\bsx) &=& \sum_{k=1}^K \pi_k(\bsr;\hat \bsalpha_{n}) \E_{{\it \hat\bsvPsi}}(Y|Z=k,\bsx),
\label{eq: mean of MoE}\\
\V_{{\it \hat\bsvPsi}}(Y|\bsr,\bsx) &=& \sum_{k=1}^K \pi_k(\bsr;\hat \bsalpha_{n})
\big[\left(\E_{{\it \hat\bsvPsi}}(Y|Z=k,\bsx)\right)^2  + \V_{{\it \hat\bsvPsi}}(Y|Z=k,\bsx) \big] - \big[\E_{{\it \hat\bsvPsi}}(Y|\bsr,\bsx)\big]^2,
\label{eq: variance of MoE}
\end{eqnarray}}where $\E_{{\it \hat\bsvPsi}}(Y|Z=k,\bsx)$ and $\V_{{\it \hat\bsvPsi}}(Y|Z=k,\bsx)$  are respectively the component-specific (expert) means and variances. 
 The mean and the variance for the MoE models described here are given as follows.

\paragraph{NMoE}For the NMoE model, the normal expert means and variances are respectively given by
$\E_{{\it \hat\bsvPsi}}(Y|Z=k,\bsx) = \hat \bsbeta^T_{k} \bsx$
and 
$\V_{{\it \hat\bsvPsi}}(Y|Z=k,\bsx) = \hat \sigma^2_{k}$. Then, from (\ref{eq: mean of MoE}) it follows that the mean of the NMoE is given by
\begin{eqnarray}\E_{{\it \hat\bsvPsi}}(Y|\bsr,\bsx) &=& \sum_{k=1}^K \pi_k(\bsr;\hat \bsalpha_{n}) \hat \bsbeta^T_{k} \bsx.
\label{eq: mean of NMoE}
\end{eqnarray}
%
Then, the expected value for each of the three proposed MoE models is given as follows. 

\paragraph{SNMoE}From the mean and the variance of the skew-normal distribution, which can be calculated as in \citet{Genton2001} (for the multivariate case), and which are given in Lemma 1 in \citet{Lin07univSkewNMixture} for this scalar case, the expert means and variances for the  SNMoE model are calculated similarly and respectively given by:
$$\E_{{\it \hat\bsvPsi}}(Y|Z=k,\bsx) = \hat \bsbeta^T_{k} \bsx + \sqrt{\frac{2}{\pi}} ~ \hat \delta_{k} ~  \hat \sigma_{k}$$
and  
$$\V_{{\it \hat\bsvPsi}}(Y|Z=k,\bsx) = \left(1 - \frac{2}{\pi} \hat \delta^2_{k}\right)  \hat \sigma^2_{k}$$ where  $\hat \delta_{k}= \frac{\hat \lambda_{k}}{\sqrt{1 +\hat \lambda^2_{k}}}$. Then, from (\ref{eq: mean of MoE}) it follows that  the mean of the SNMoE model is given by:
\begin{eqnarray}
\E_{{\it \hat\bsvPsi}}(Y|\bsr,\bsx) &=& \sum_{k=1}^K \pi_k(\bsr;\hat \bsalpha)\Big(\hat \bsbeta^T_k \bsx + \sqrt{\frac{2}{\pi}} \hat \delta_{k} \hat \sigma_{k}\Big)\cdot
\label{eq: mean of SNMoE}
\end{eqnarray}

\paragraph{TMoE} For the TMoE model, by using the expressions of the mean and the variance of the $t$ distribution, it follows that for the TMoE model, for $\hat \nu_{k}>1$, the expert means are given by
$\E_{{\it \hat\bsvPsi}}(Y|Z=k,\bsx) = \hat \bsbeta^T_{k} \bsx$
and, for $\hat \nu_{k}>2$, the expert variances are given by $\V_{{\it \hat\bsvPsi}}(Y|Z=k,\bsx) = \frac{\hat \nu_{k}}{\hat \nu_{k} - 2} ~ \hat \sigma^2_{k}$. Then, from (\ref{eq: mean of MoE}), the mean  of the TMoE model is therefore given by:
\begin{eqnarray}
\E_{{\it \hat\bsvPsi}}(Y|\bsr,\bsx) &=& \sum_{k=1}^K \pi_k(\bsr;\hat \bsalpha) \hat \bsbeta^T_k \bsx.
\label{eq: mean of TMoE}
\end{eqnarray}

\paragraph{STMoE}The mean and the variance for a skew $t$ random variable, for this scalar case, can be easily computed as in Section 4.2 in \citet{AzzaliniAndCapitanio2003} for a non-zero location parameter. Thus, for the STMoE model, the expert means for $\hat \nu_{k}>1$, are given by 
$$\E_{{\it \hat\bsvPsi}}(Y|Z=k,\bsx) = \hat \bsbeta^T_{k} \bsx + \hat \sigma_{k} ~ \hat \delta_{k} ~ \xi(\hat \nu_{k})$$
and the expert variances for $\hat \nu_{k}>2$ are given by 
$$\V_{{\it \hat\bsvPsi}}(Y|Z=k,\bsx) = \left(\frac{\hat \nu_{k}}{\hat \nu_{k} - 2}  - \hat \delta^2_{k} ~ \xi^2(\hat \nu_{k})\right) \hat \sigma^2_{k},$$
where $\xi(\hat \nu_{k}) = \sqrt{\frac{\hat \nu_{k}}{\pi}} \frac{\Gamma\left(\frac{\hat \nu_{k}}{2} - \frac{1}{2}\right)}{\Gamma \left(\frac{\hat \nu_{k}}{2}\right)}$. Then, following (\ref{eq: mean of MoE}), the mean of the STMoE is thus given by:
\begin{eqnarray}
\E_{{\it \hat\bsvPsi}}(Y|\bsr,\bsx) &=& \sum_{k=1}^K \pi_k(\bsr;\hat \bsalpha)\Big(\hat \bsbeta^T_{k} \bsx + \hat \sigma_{k} ~ \hat \delta_{k} ~ \xi(\hat \nu_{k})\Big).
\label{eq: mean of STMoE}
\end{eqnarray}

\bigskip
Finally, the variance for each MoE model is obtained by using (\ref{eq: variance of MoE}) with the specified expert  mean and variance calculated in the above.

\section{Model-based clustering using the NNMoE} 
\label{sec: MBC using the NNMoE}

The MoE models can also be used for a model-based clustering perspective  to provide a partition of the regression data into $K$ clusters. 
Model-based clustering using the NNMoE consists in assuming that the observed data $\{\bsx_i,\bsr_i,y_i\}_{i=1}^n$ are  generated from a $K$ component mixture of, respectively,  skew-normal, $t$ or skew $t$ experts, with parameter vector $\bsvPsi$. The mixture components can be  interpreted as clusters and hence each cluster can be associated with a mixture component. 
The problem of clustering therefore becomes the one of estimating the MoE parameters $\bsvPsi$, which is performed here by using dedicated EM algorithms. 
Once the parameters are estimated, the provided posterior membership probabilities $\tau_{ik}$ 
 represent a fuzzy partition of the data. These posterior memberships are given by, (\ref{eq: posterior prob SNMoE}), (\ref{eq: posterior prob TMoE}), (\ref{eq: posterior prob STMoE}), for, respectively the SNMoE, the TMoE, and the STMoE.  
A hard partition of the data can then be obtained from the posterior memberships by applying the MAP rule, that is, by maximizing the posterior cluster probabilities to assign each observation to a cluster: 
\begin{eqnarray}
\hat{z}_i = \arg \max_{k=1}^K \hat \tau_{ik}
\label{eq: MAP rule for clustering}
\end{eqnarray}where $\hat{z}_i$ represents the estimated cluster label for the $i$th observation.

\section{Model selection for the NNMoE}
\label{sec: Model selection for the NNMoE}
One of the issues in mixture model-based clustering is model selection.  
The problem of model selection for the NNMoE models presented  here in their general forms, is equivalent to the one of choosing the optimal number of experts $K$, the degree $p$ of the polynomial regression and the degree $q$ for the logistic regression. 
The optimal value  of $(K,p,q)$ can be computed by using some model
selection criteria such as the Akaike Information Criterion (AIC) \citep{AIC}, the Bayesian Information Criterion (BIC) \citep{BIC} or the Integrated Classification Likelihood criterion (ICL)  \citep{ICL}, etc. 
The AIC and BIC are are penalized observed data log-likelihood criteria which can be defined as functions to be maximized and are respectively given by:
\begin{eqnarray*}
\mbox{AIC}(K,p,q) &=& \log L(\hat{\bsPsi}) - \frac{\eta_{\bsvPsi} \log(n)}{2},\\
\mbox{BIC}(K,p,q) &=&  \log L(\hat{\bsvPsi}) - \frac{\eta_{\bsvPsi} \log(n)}{2}.
\end{eqnarray*}The ICL criterion consists in a penalized complete-data log-likelihood and can be expressed as follows:
\begin{equation*}
\ICL(K,p,q) = \log L_c(\hat{\bsvPsi}) - \frac{\eta_{\bsvPsi} \log(n)}{2}.
\end{equation*}In the above, $\log L(\hat{\bsvPsi})$ and  $\log L_c(\hat{\bsvPsi})$ are respectively the incomplete (observed) data log-likelihood and the complete data log-likelihood, obtained at convergence of the E(C)M algorithm for the corresponding mixture of experts model and $\eta_{\bsvPsi}$ is the number of free model parameters. The number of free parameters $\eta_{\bsvPsi} $ is given by $\eta_{\bsvPsi} = K(p+q+3)-q-1$ for the NMoE model,
$\eta_{\bsvPsi} = K(p+q+4)-q-1$ for both the SNMoE and the TMoE models,
and $\eta_{\bsvPsi} = K(p+q+5)-q-1$ for the  STMoE model.
 
However, note that in MoE it is common to use mixing proportions modeled as logistic transformation of linear functions of the covariates, that is the covariate vector  in (\ref{eq: multinomial logistic}) is given by $\bsr_i = (1, r_i)^T$ (corresponding to $q=2$), $r_i$ being an univariate covariate variable. This is also adopted in this work.
Moreover, for the case of linear experts, that is when the experts are linear regressors with parameter vector $\bsbeta_k$ for which the corresponding covariate vector $\bsx_i$ in  (\ref{eq: linear regression mean})  is given by $\bsx_i = (1, x_i)^T$ (corresponding to $p=2$), $r_i$ being an univariate covariate variable, the model selection reduces to choosing the number of experts $K$.
Here we mainly consider this linear case. 

\section{Experimental study}  
\label{sec: Experimental study}

This section is dedicated to the evaluation of the proposed approach on simulated data and real-world data . 
We evaluated the performance of proposed EM algorithms for the NNMoE models in terms of modeling, robustness to outliers and clustering.
The algorithms have been implemented in Matlab. 

\subsection{Initialization and stopping rules}
\label{ssec: initialization and stopping}
The parameters $\bsalpha_k$ ($k=1, \ldots, K-1$) of the mixing proportions are initialized randomly, 
including an initialization at the null vector for one run (corresponding to equal mixing proportions).
Then, the common parameters $(\bsbeta_k,\sigma^2_k)$ ($k=1,\ldots, K$) are initialized from a random partition of the data into $K$ clusters. This corresponds to fitting a normal mixture of experts where the initial values of the parameters are respectively  given by  
(\ref{eq: beta_k update for NMoE}) and (\ref{eq: sigma2k update NMoE}) with the posterior memberships $\tau_{ik}$ replaced by the hard assignments $Z_{ik}$ issued from the random partition.
For the TMoE and STMoE models, the robustness parameters $\nu_k$ ($k=1,\ldots, K$) can be initialized randomly in the range [1, 200].
For the SNMoE and STMoE, the skewness parameters $\lambda_k$ ($k=1,\ldots, K$) can be initialized by randomly initializing the parameter $\delta_k$ in  $(-1,1)$ from the relation $\lambda_k = \frac{\delta_k}{\sqrt{1 -\delta^2_k}}$.
Then, the proposed E(C)M algorithm for each model is stopped when the relative variation of the observed-data log-likelihood 
$\frac{\log L(\bsvPsi^{(m+1)})- \log L(\bsvPsi^{(m)})}{|\log L(\bsvPsi^{(m)})|}$ reaches a prefixed threshold (for example $\epsilon=10^{-6}$).  
For each model, this process is repeated 10 times and and the solution corresponding the highest log-likelihood is finally selected.

\subsection{An illustrative example}
\label{sssec: Illustation on Bishop's data set}
We first start by an illustrative example by considering a non-linear arbitrary data set which was analyzed by \citet{Bishop_BayesianMoE} and elsewhere. 
This  data set consists of $n=250$ values of input variables $x_i$ generated uniformly in $(0,1)$ and output variables $y_i$ generated as $y_i = x_i + 0.3 \sin(2\pi x_i) + \epsilon_i$, with $\epsilon_i$ drawn from a zero mean Normal distribution with standard deviation $0.05$.  To apply the MoE models, we set the covariate vectors $(\bsx_i, \bsr_i)$ to $\bsx_i = \bsr_i = (1,x_i)^T$.
We considered mixture of three linear experts as in \citet{Bishop_BayesianMoE}.

\begin{figure}[H]
   \centering 
   \begin{tabular}{cc}
   \includegraphics[width=6cm]{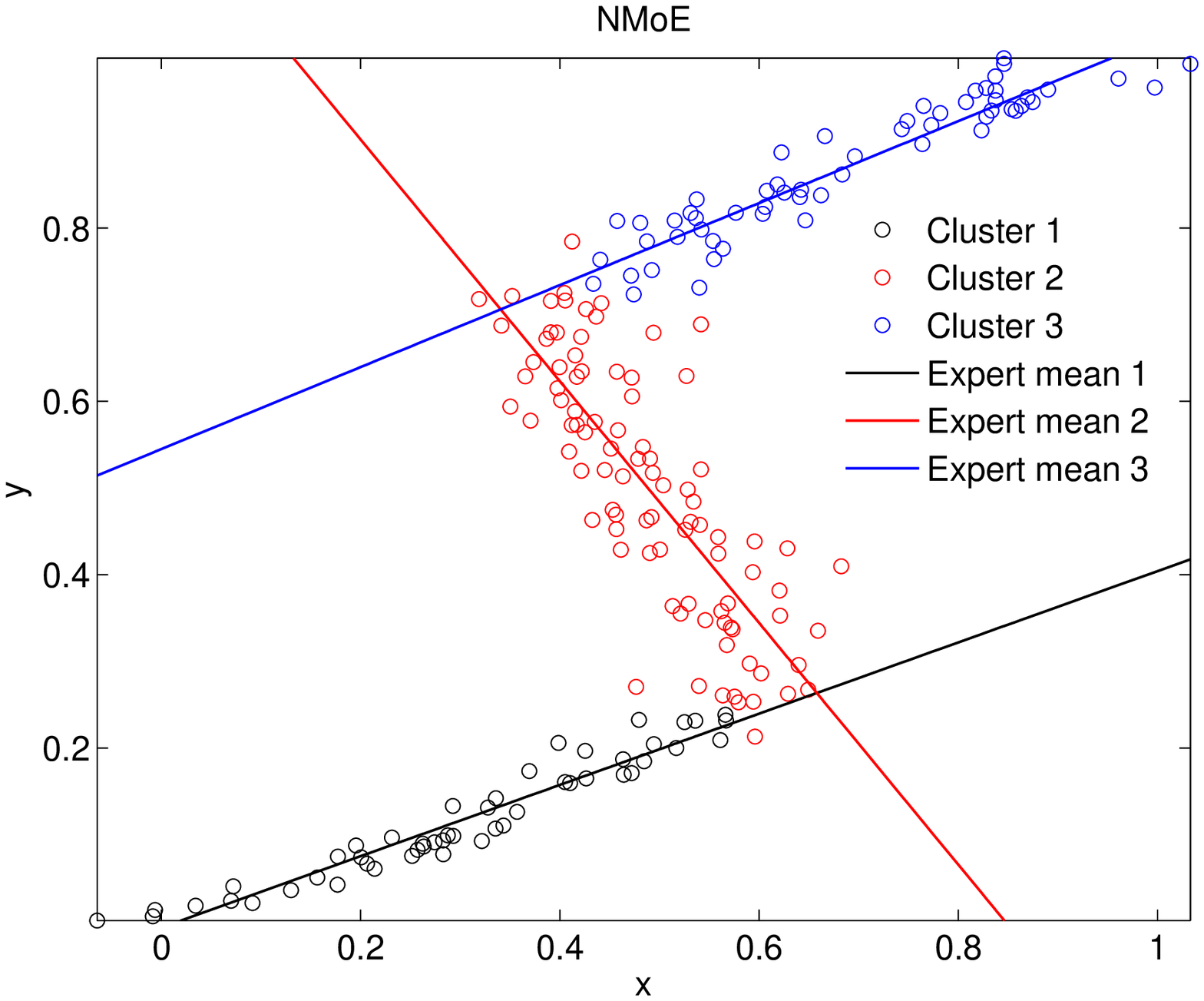} &
   \includegraphics[width=6cm]{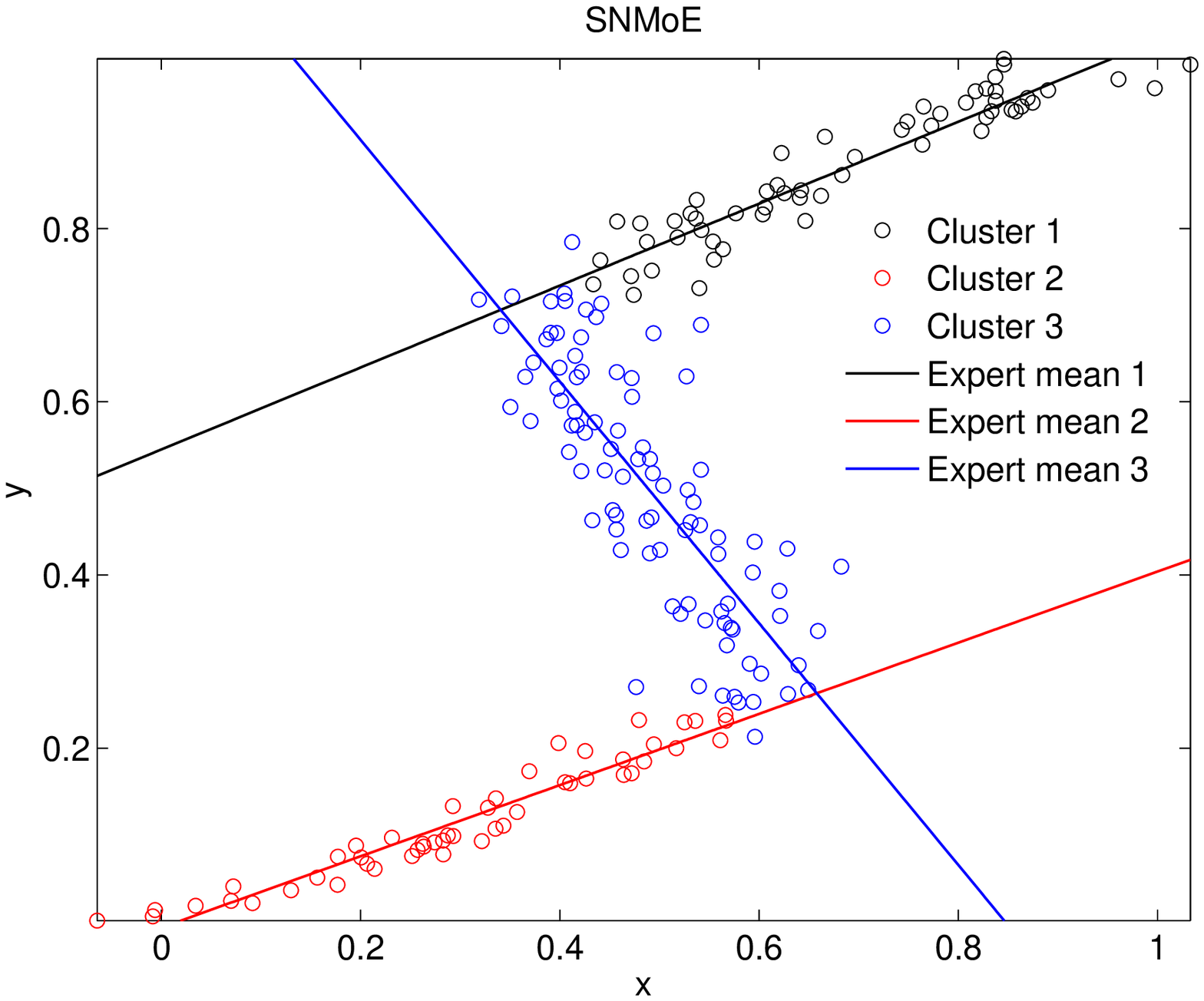}\\
   \includegraphics[width=6cm]{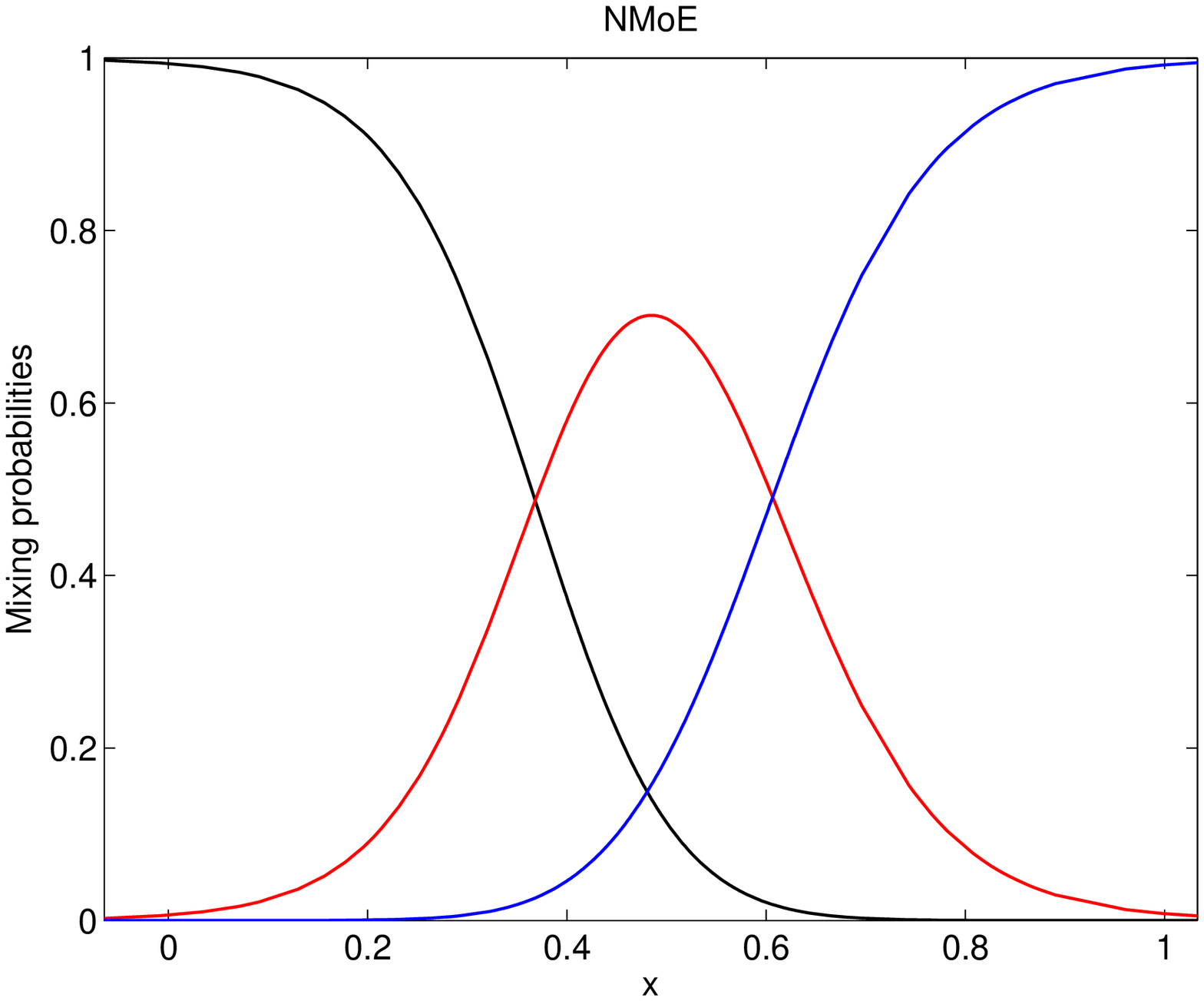} & 
   \includegraphics[width=6cm]{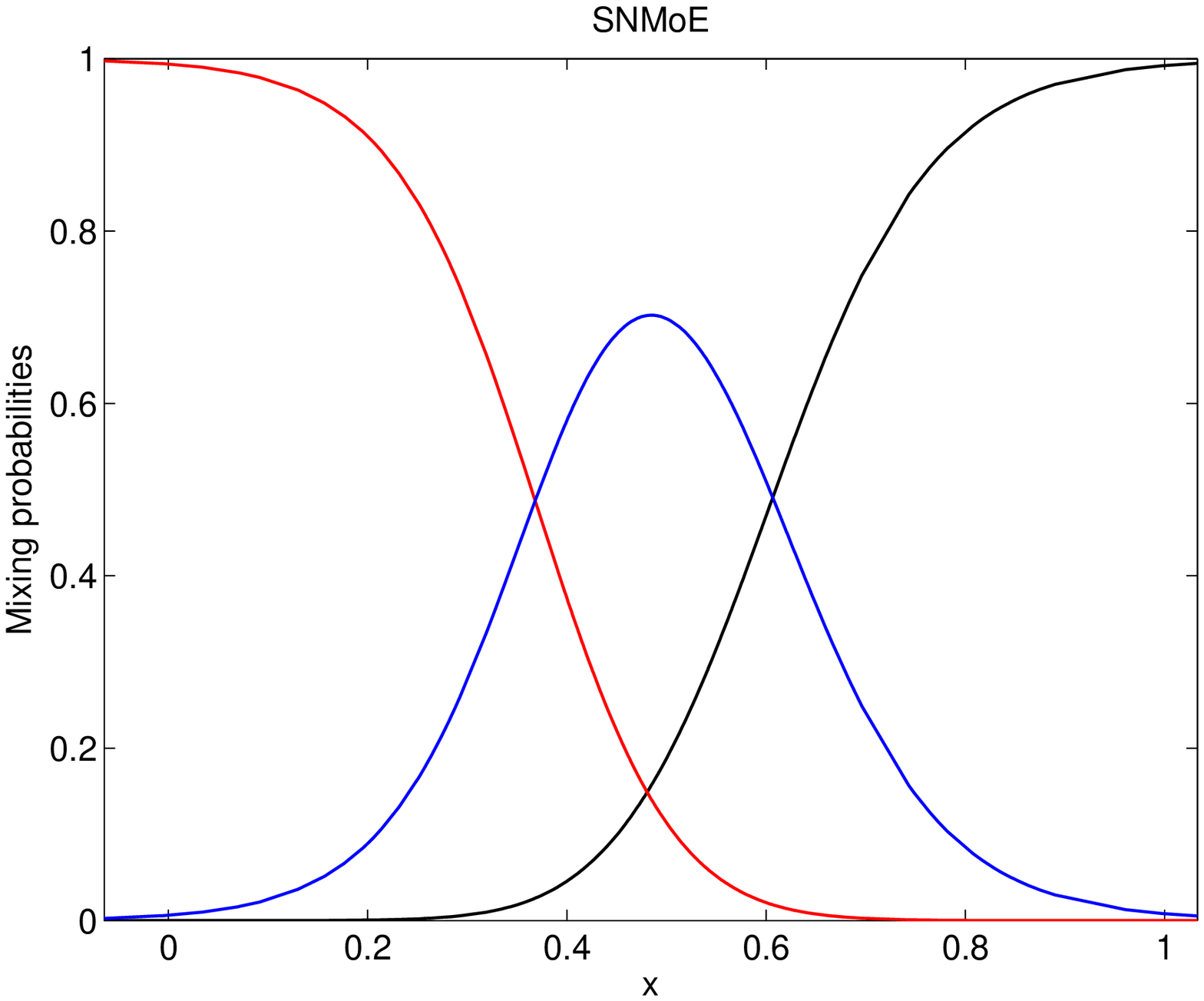}\\
   \includegraphics[width=6cm]{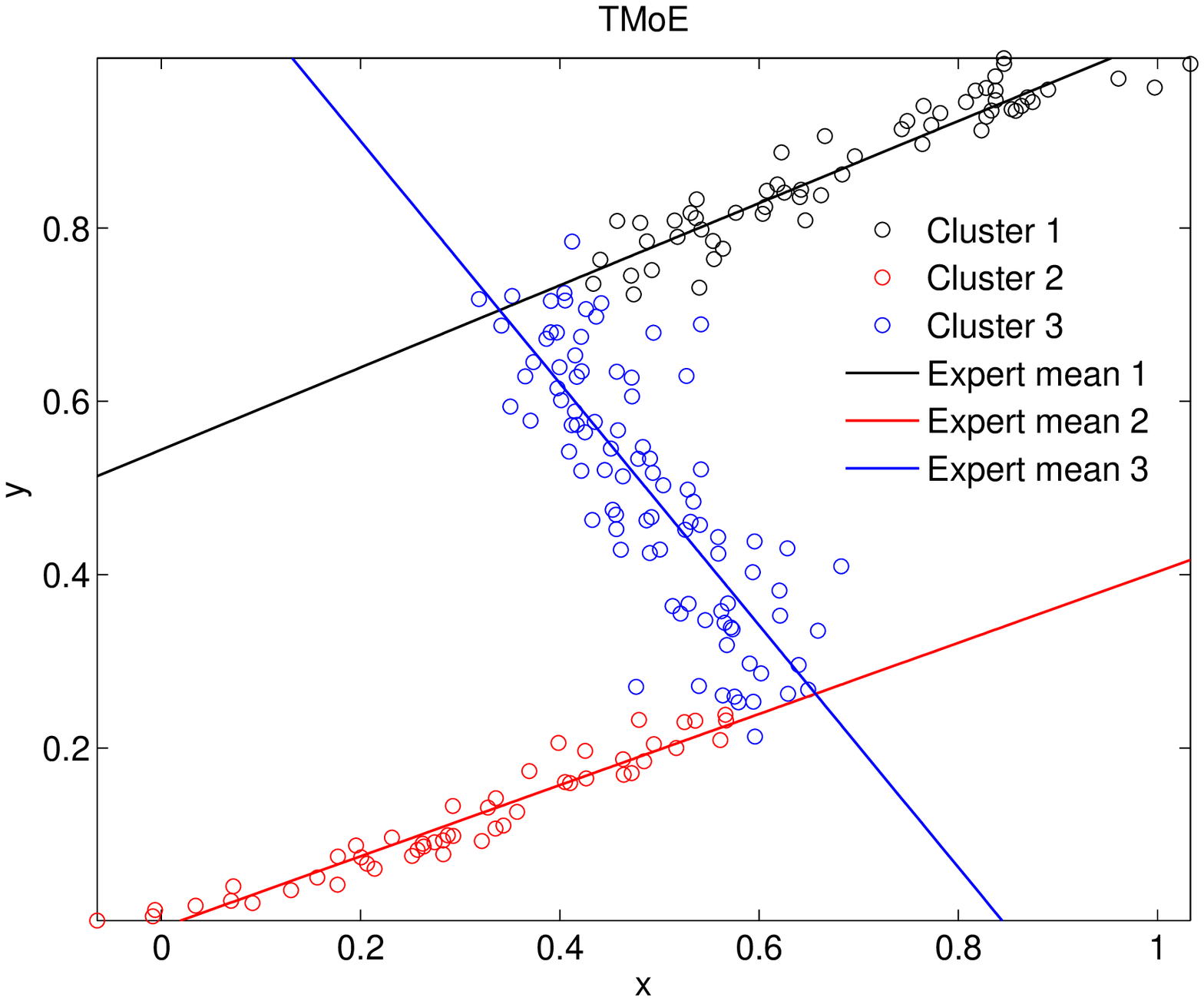} &
   \includegraphics[width=6cm]{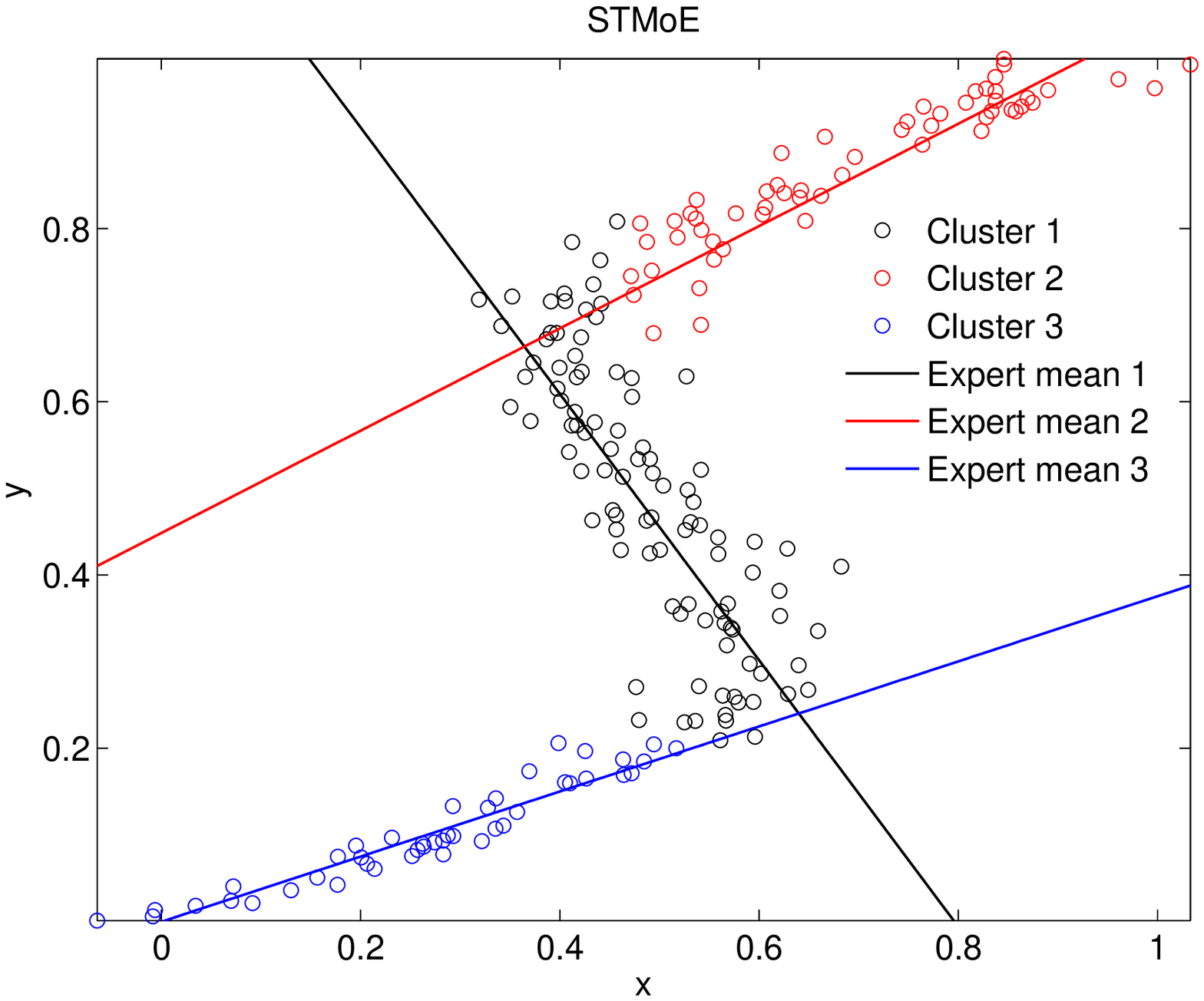}\\
   \includegraphics[width=6cm]{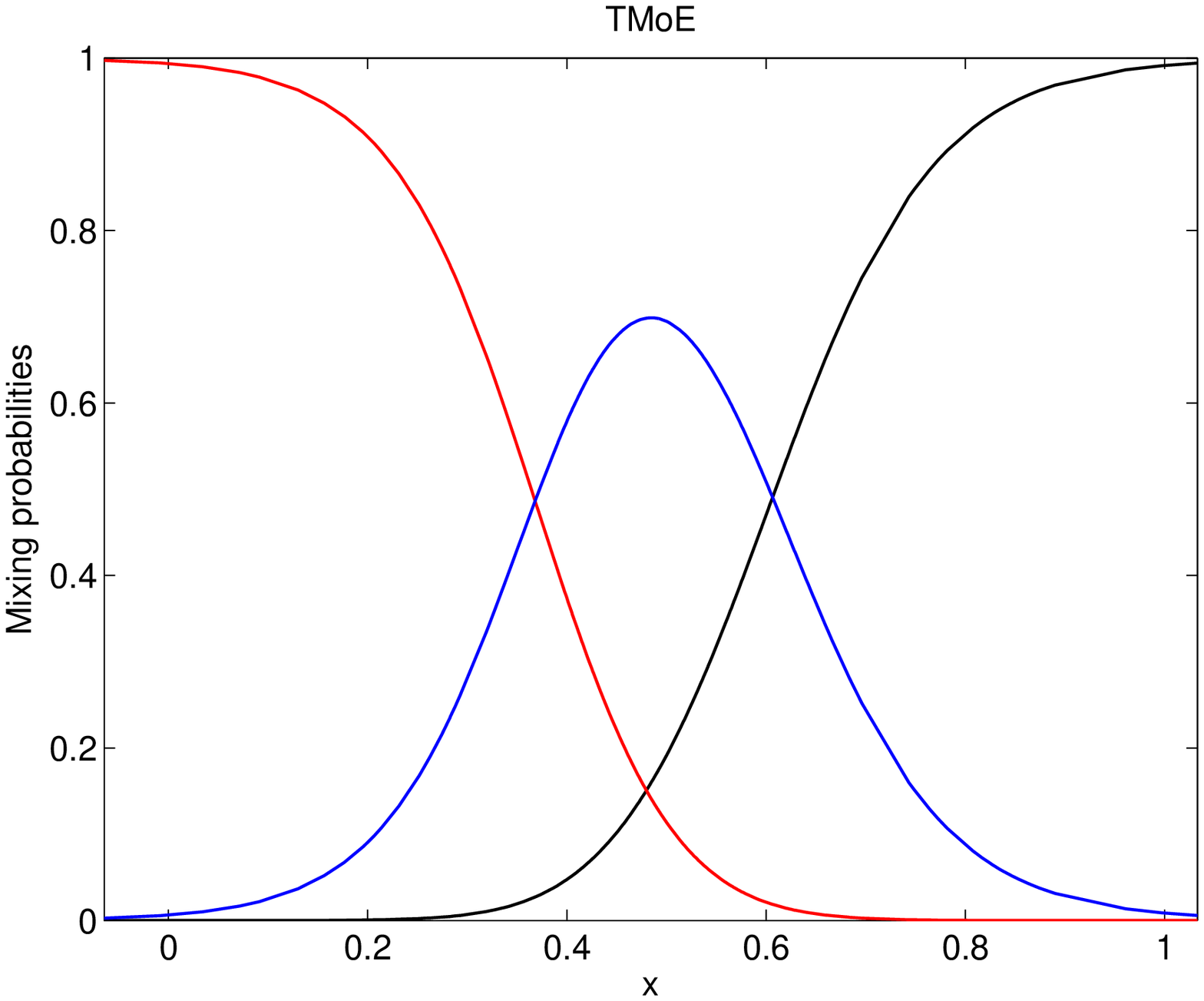} & 
   \includegraphics[width=6cm]{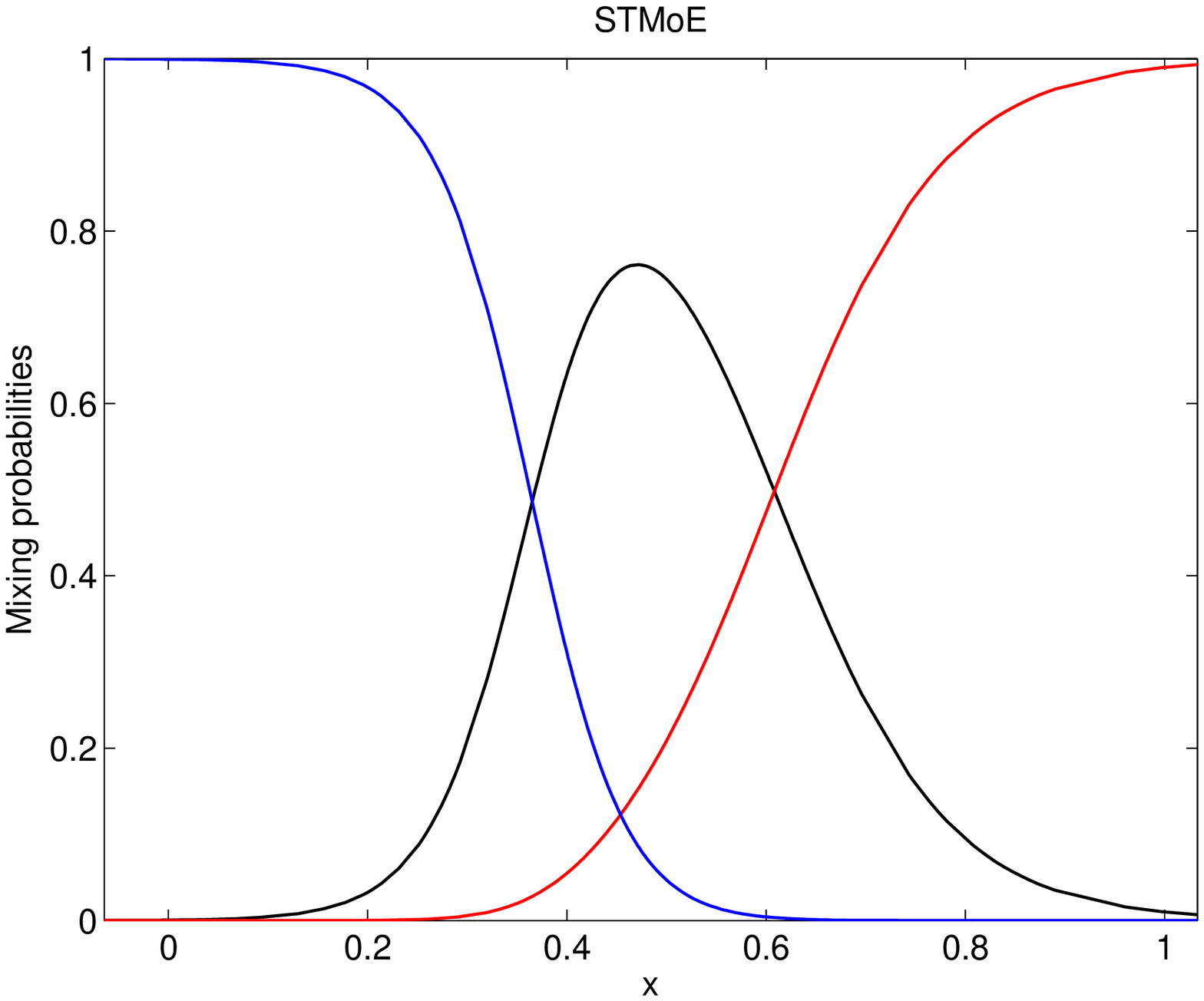}\\
   \end{tabular}
      \caption{\label{fig. Bishop_data_NMoE}Fitting the NMoE model and the three proposed non-normal mixture of experts models (SNMoE, TNMoE, STMoE) to the toy data set analyzed in \citet{Bishop_BayesianMoE}.}
\end{figure}

Figure \ref{fig. Bishop_data_NMoE} shows the expert mean functions of each of the fitted MoE models, the corresponding partitions obtained by using the Bayes' rule, and the mixing proportions as function of the inputs. One can observe that the four models are successfully applied and provide very similar results. 
The results obtained by the proposed NNMoE models are indeed close to the one obtained by the NMoE.

\subsection{Experiments on simulation data sets}
In this section we perform an experimental study on simulated data sets to apply and assess the proposed models. 
Two sets of experiments have been performed. 
The first experiment aims at observing the effect of the sample size on the estimation quality and the second one aims at observing the impact of the presence of outliers in the data on the estimation quality, that is the robustness of the models.

\subsubsection{Experiment 1}
For this first experiment on simulated data, each simulated sample consisted of $n$ observations with increasing values of the sample size $n: 50, 100, 200, 500, 1000$.  
The simulated data are generated from a two component mixture of linear experts, that is $K=2, p=q=1$.
The covariate variables $(\bsx_i, \bsr_i)$ are simulated such that $\bsx_i = \bsr_i = (1,x_i)^T$ where $x_i$ is
simulated uniformly over the interval $(-1, 1)$. 
We consider each of the four models for data generation (NNMoE, SNMoE, TMoE, STMoE), that is, given the covariates, the response $y_i|\{\bsx_i,\bsr_i;\bsvPsi\}$ is simulated according to the generative process of the models (\ref{eq: normal MoE}), (\ref{eq: SNMoE}), (\ref{eq: TMoE}), and  (\ref{eq: SNMoE}).
For each generated sample, we fit each of the four models.   
Thus, the results are reported for all the models with data generated from each of the models. We consider the mean square error (MSE) between each component of the true parameter vector and the estimated one, which is given by $\parallel \bsvPsi_j - \hat{\bsvPsi}_j\parallel^2$.
%
The squared errors
 are averaged on 100 trials. 
The used simulation parameters $\bsvPsi$ for each model are given in Table \ref{tab: simulation parameters situation 1}.
\begin{table}[H]
\centering
\small
\begin{tabular}{l | l l l l l}
\hline
& \multicolumn{5}{c}{parameters} \\
\hline
\hline
component 1 & $\bsalpha_1=(0, 10)^T$&$\bsbeta_1=(0,1)^T$ & $\sigma_1=0.1$ &$\lambda_1 = 3$   & $\nu_1 = 5$\\
component 2 & $\bsalpha_2=(0, 0)^T$&$\bsbeta_2=(0,-1)^T$  &$\sigma_2=0.1$ &$\lambda_2 = -10$ & $\nu_2 = 7$\\
 \hline
\end{tabular}
\caption{Parameter values used in simulation.}
\label{tab: simulation parameters situation 1}
\end{table}

\subsubsection{Obtained results}
 
 Tables \ref{tab. MSE for the paramters: SNMoE->SNMoE},
 \ref{tab. MSE for the paramters: TMoE->TMoE}, and \ref{tab. MSE for the paramters: STMoE->STMoE} show the obtained results in terms of the MSE for respectively  the SNMoE, the TMoE, and the STMoE.
 One can observe that, for the three proposed models, the parameter estimation error is decreasing as $n$ increases, which confirms the convergence property of the maximum likelihood estimator 
 For details on the convergence property of the MLE for mixture of experts, see for example \citep{Jiang_and_tanner_IEEEinft_99}.  
 One can also observe that the error decreases significantly  for $n\geq 500$, especially for the regression coefficients and the scale parameters.
%
%
%
%
%

{\setlength{\tabcolsep}{3pt
\begin{table}[H]
\centering
{\small
\begin{tabular}{c c c  c c c c c c c c}
\hline
param. & $\alpha_{10}$ & $\alpha_{11}$ & $\beta_{10}$ & $\beta_{11}$ & $\beta_{20}$ & $\beta_{21}$ & $\sigma_{1}$& $\sigma_{2}$ & $\lambda_{1}$ & $\lambda_{2}$ \\ 
$n$		& & & & & & & & & &\\
 \hline
 \hline
$50$     & 1.10105 &  4.1882 &  0.00916   & 0.004890   &  0.007370 & 0.00348000  & 0.001647 & 0.002234 & 3.000 & 4.999\\ 
$100$	& 0.28074 &  1.0663 &  0.008301 & 0.0006118 &  0.006360 & 0.00007904  & 0.001314 & 0.001650 & 2.999 & 5.000\\ 
$200$	& 0.03893 &  0.9343 &  0.004709 & 0.0000398 &  0.005962 & 0.00005873  & 0.001142 & 0.001552 & 2.999 & 5.000\\
$500$	& 0.02340 &  0.0908 &  0.004475 & 0.0000195 &  0.005803 & 0.00000796  & 0.001026 & 0.001521 & 3.000 & 4.999\\
$1000$	& 0.00025 &  0.0613 &  0.003912 & 0.0000012 &  0.005499 & 0.00000344  & 0.000667 & 0.001517 & 2.999 & 3.999 \\
 \hline
\end{tabular}
}
\caption{\label{tab. MSE for the paramters: SNMoE->SNMoE}MSE between each component of the estimated parameter vector of the SNMoE model and the actual one for a varying sample size $n$.}
\end{table}}
{\setlength{\tabcolsep}{3pt
\begin{table}[H]
\centering
{\small
\begin{tabular}{c c c  c c c c c c c c}
\hline
param. & $\alpha_{10}$ & $\alpha_{11}$ & $\beta_{10}$ & $\beta_{11}$ & $\beta_{20}$ & $\beta_{21}$ & $\sigma_{1}$& $\sigma_{2}$ & $\nu_{1}$ & $\nu_{2}$ \\ 
$n$		& & & & & & & & & &\\
 \hline
 \hline
$50$ 	& 1.3059 &  6.4611 & 0.0214130  & 0.0290114 & 0.0044140 & 0.0192600 &  0.0010655 & 0.0003317 & 37.956 &  11.722\\
$100$	& 1.2150 &  4.5056 & 0.0024706 &  0.0117546 & 0.0005275 & 0.0007891 &  0.0001450 & 0.0002301 & 6.1528 &  10.412\\
$200$	& 0.0341 &  3.8193 & 0.0001553 &  0.0007335 & 0.0002022 & 0.0005061 &  0.0000504 & 0.0000262 & 2.0975 &  6.3710\\
$500$	& 0.0356 &  2.2633 & 0.0000112 &  0.0000214 & 0.0001337 & 0.0002163 &  0.0000126 & 0.0000007 & 0.4859 &  5.4937\\
$1000$	& 0.0053 &  1.2510 & 0.0000018 &  0.0000258 & 0.0000005 & 0.0000427 &  0.0000126 & 0.0000004 & 0.0014 &  2.7844\\
 \hline
\end{tabular}
}
\caption{\label{tab. MSE for the paramters: TMoE->TMoE} MSE between each component of the estimated parameter vector of  the TMoE model and the actual one for a varying sample size $n$.}
\end{table}
}
{\setlength{\tabcolsep}{3pt
\begin{table}[H]
\centering
{\small
\begin{tabular}{c c c  c c c c c c c c c c}
\hline
param. & $\alpha_{10}$ & $\alpha_{11}$ & $\beta_{10}$ & $\beta_{11}$ & $\beta_{20}$ & $\beta_{21}$ & $\sigma_{1}$& $\sigma_{2}$ & $\lambda_{1}$ & $\lambda_{2}$ & $\nu_{1}$ & $\nu_{2}$ \\ 
$n$		& & & & & & & & & & & & \\
 \hline
 \hline
$50$    & 0.5256 & 5.737 & 0.000965 & 0.002440 & 0.004388 & 0.000667 & 0.000954 & 0.000608 & 3.1153 & 16.095 & 15.096 & 4.64310\\
$100$  & 0.4577 & 1.815 & 0.000847 & 0.000852 & 0.000742 & 0.000660 & 0.000844 & 0.000303 & 2.0131 & 7.844 & 5.360 & 0.26354\\
$200$  & 0.2478 & 0.785 & 0.000816 & 0.000348 & 0.000473 & 0.000556 & 0.000362 & 0.000297 & 0.7004 & 3.8470 & 3.135 & 0.16720\\ 
$500$  & 0.0316 & 0.565 & 0.000363 & 0.000091 & 0.000314 & 0.000398 & 0.000091 & 0.000061 & 0.0078 & 1.0785 & 0.2230 & 0.00860\\ 
$1000$& 0.0085 & 0.068 & 0.000261 & 0.000076 & 0.000233 & 0.000116 & 0.000026 & 0.000002 & 0.0028 & 0.5545 & 0.0494 & 0.00079\\ 
 \hline
\end{tabular}
}
\caption{\label{tab. MSE for the paramters: STMoE->STMoE} MSE between each component of the estimated parameter vector of the STMoE model and the actual one for a varying sample size $n$.}
\end{table}
} 
%
In addition to the previously showed results, we plotted in Figures 
\ref{fig. TwoClust-NMoE_NMoE},
\ref{fig. TwoClust-NMoE_SNMoE},
\ref{fig. TwoClust-NMoE_TMoE}, and 
\ref{fig. TwoClust-NMoE_STMoE} the estimated quantities provided by applying the proposed models and their true counterparts for $n=500$ for the same the data set which was generated according the NMoE model. 
The upper-left plot of each of these figures shows the estimated mean function, the estimated expert component mean functions, and the corresponding true ones.
The upper-right plot shows the estimated mean function and the estimated confidence region computed as plus and minus twice the estimated (pointwise) standard deviation of the model as presented in Section \ref{sec: Prediction using the NNMoE}, and their true counterparts. 
The middle-left plot shows the true expert component mean functions and the true partition, and the middle-right plot shows their estimated counterparts.
Finally, the bottom-left plot shows the log-likelihood profile during the EM iterations and the bottom-right plot shows  the estimated mixing probabilities.

One can clearly see that the estimations provided by each of the proposed models are very close to the true ones which  correspond to those of the NMoE model in this case. 
This provides an additional support to the fact that the proposed algorithms perform well and the corresponding proposed models are good generalizations of the normal mixture of experts (NMoE), as they clearly approach the NMoE as shown in this simulated example.

Figure \ref{fig. TwoClust-All->All-estimated models} shows the true and estimated MoE mean functions and expert mean functions by fitting the proposed NNMoE models to a simulated data set of $n=500$ observations. Each model was considered for data generation. The upper plot corresponds to the SNMoE model, the middle plot to the TMoE model and the  bottom plot to the STMoE model.
Finally, Figure \ref{fig. TwoClust-All->All-estimated partitions} shows the corresponding true and estimated partitions.
Again, one can clearly see that both the estimated models are precise. The fitted functions are close to the true ones.
In addition, one can also see that the partitions estimated by the NNMoE models are close the actual partitions.
The proposed NNMoE models can therefore be used as alternative to the NMoE model for both regression and model-based clustering. 
\begin{figure}[H]
   \centering 
   \begin{tabular}{cc}
   \includegraphics[width=7.5cm]{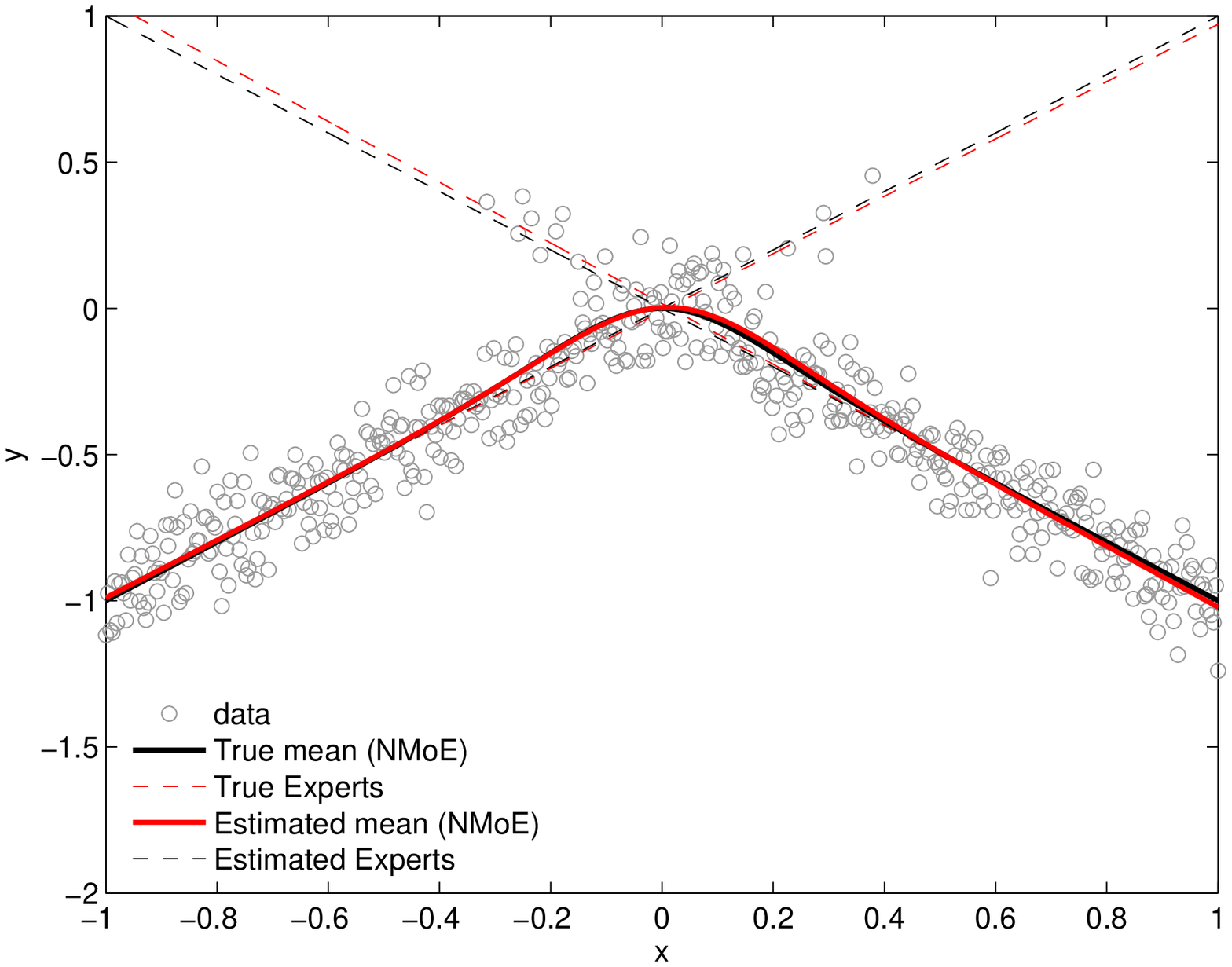}&  
   \includegraphics[width=7.5cm]{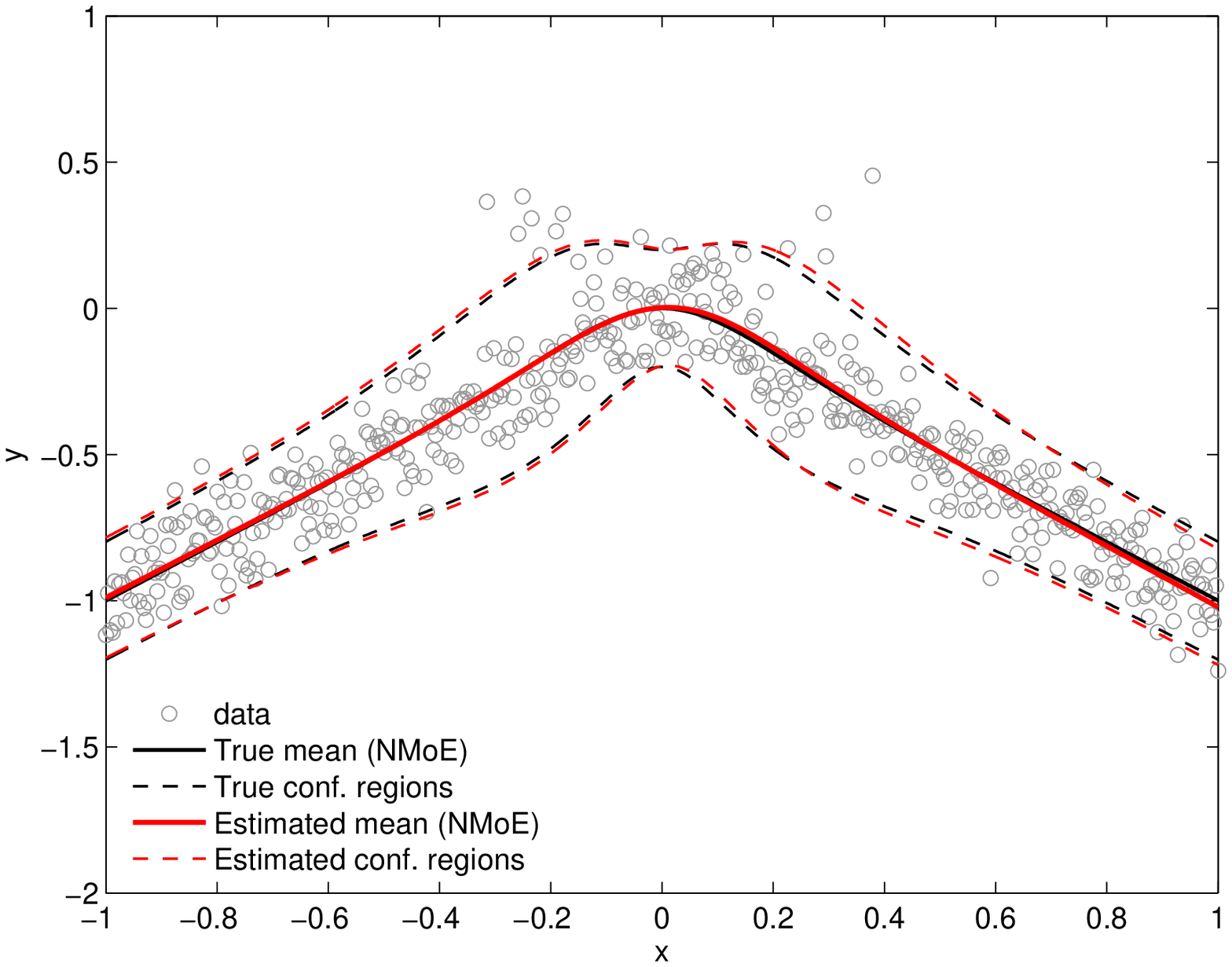}\\
\hspace{0.2cm}\includegraphics[width=7.3cm]{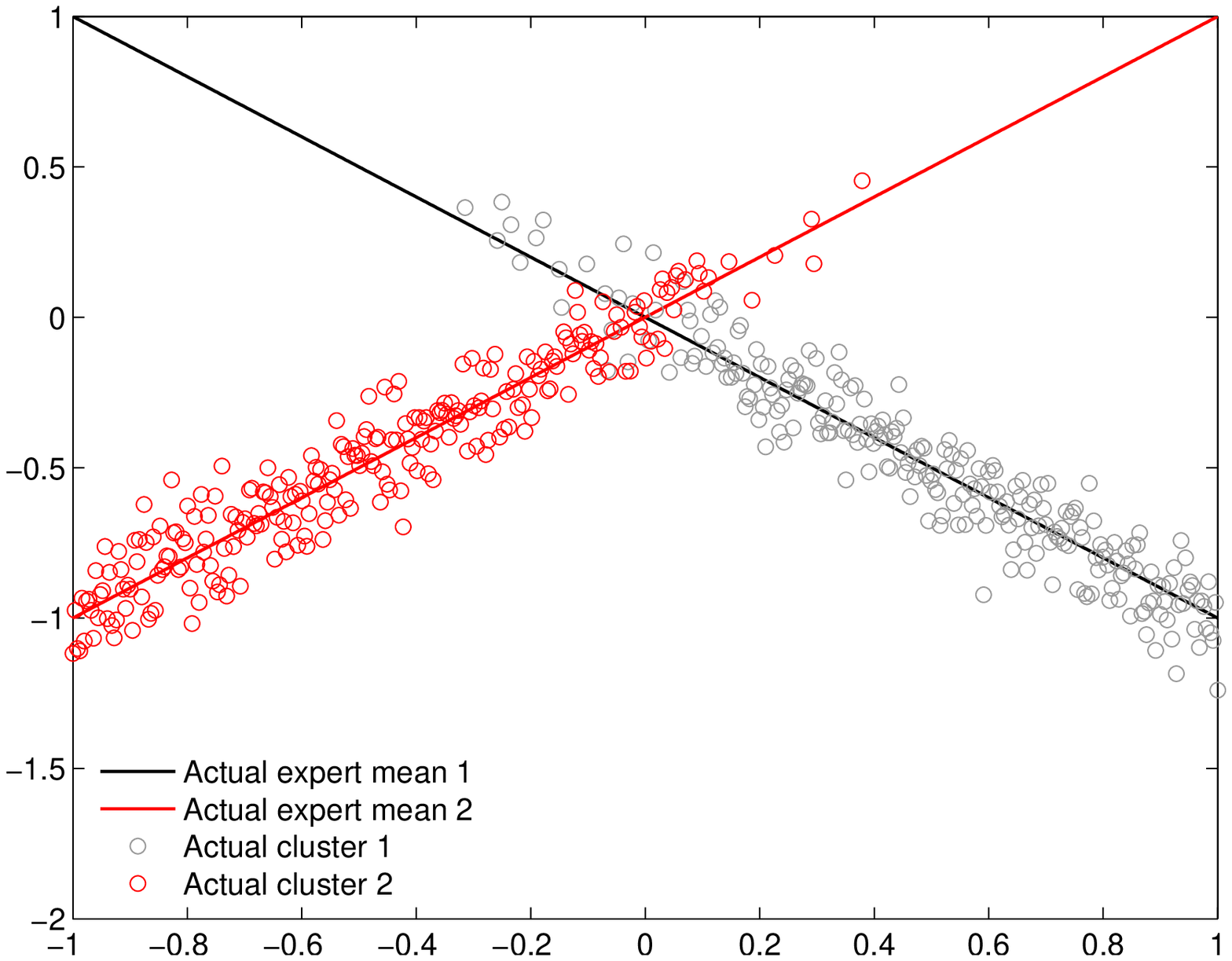}&
\hspace{0.2cm}\includegraphics[width=7.3cm]{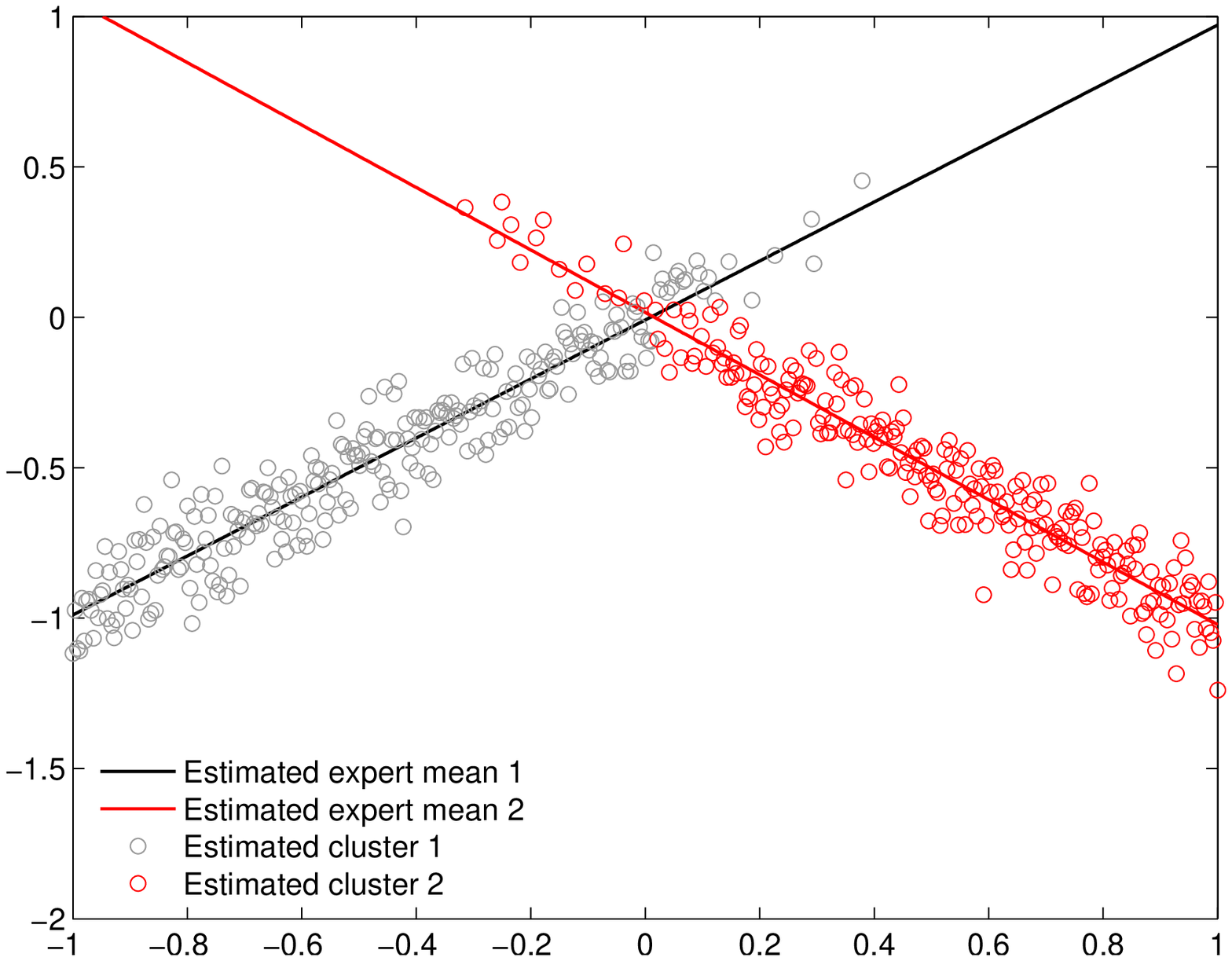}\\
   \includegraphics[width=7.5cm]{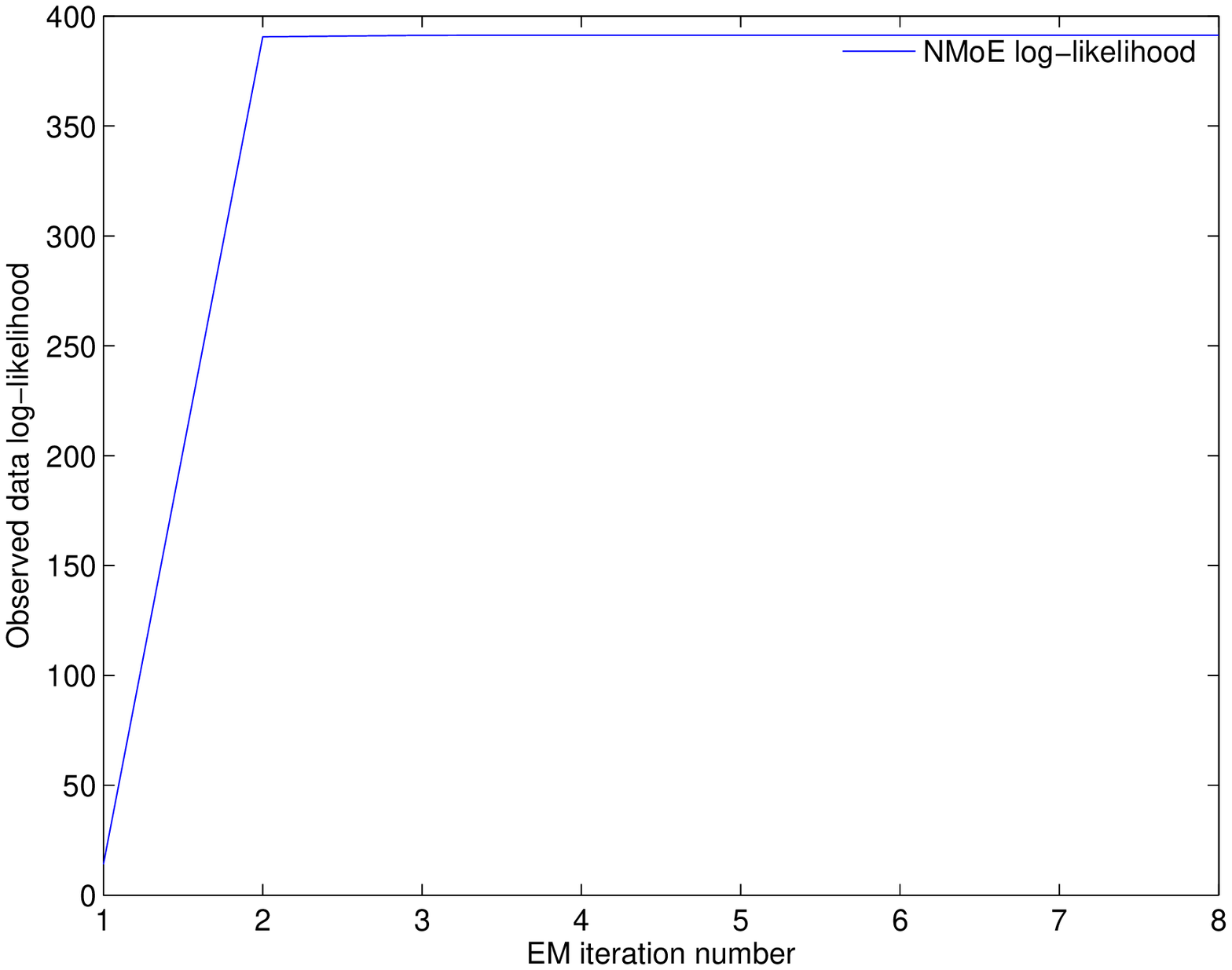} &
    \includegraphics[width=7.5cm]{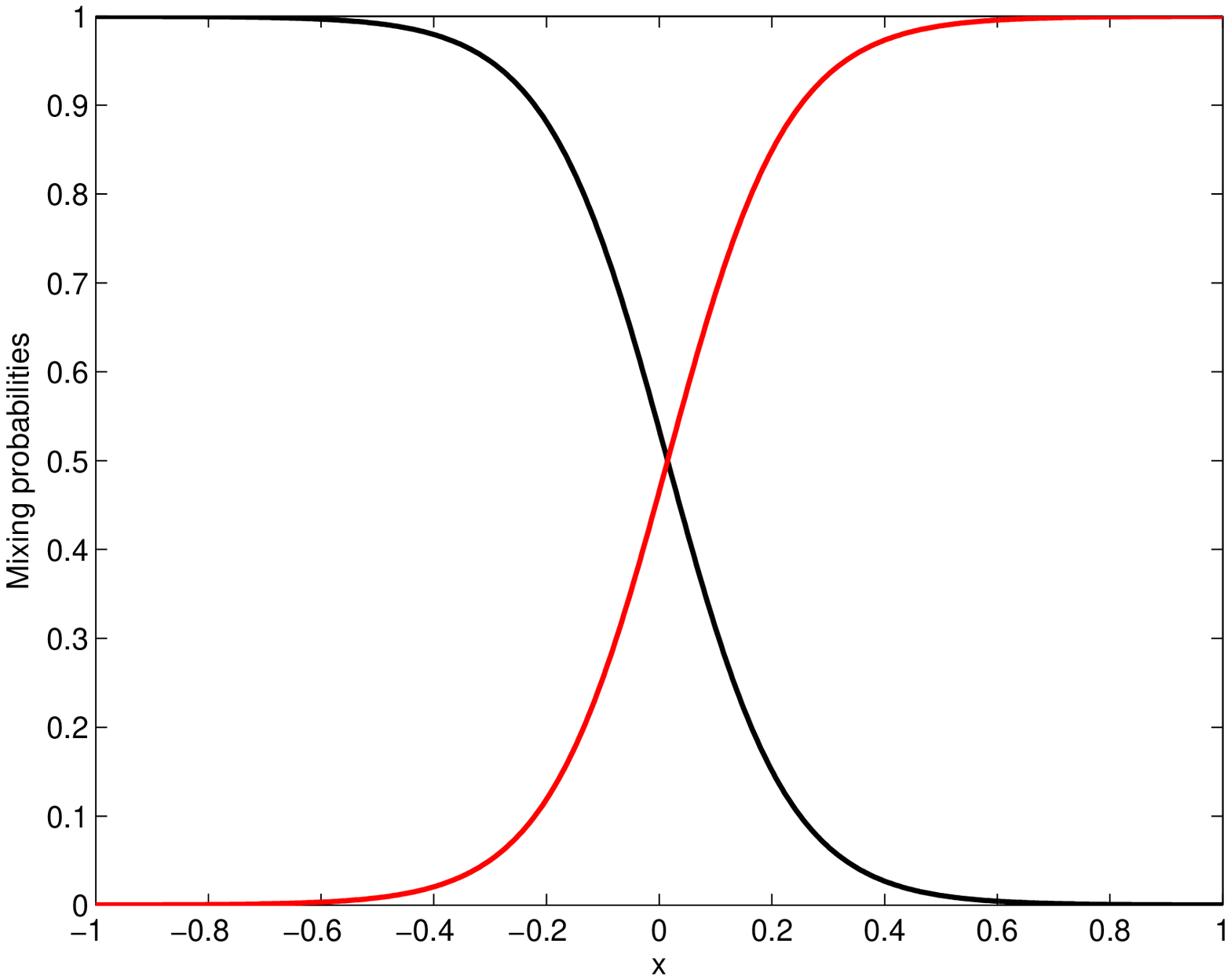}     
   \end{tabular}
      \caption{\label{fig. TwoClust-NMoE_NMoE}Fitted NMoE model to a data set generated according to the NMoE model.}
\end{figure}
\begin{figure}[H]
   \centering  
   \begin{tabular}{cc}
   \includegraphics[width=7.5cm]{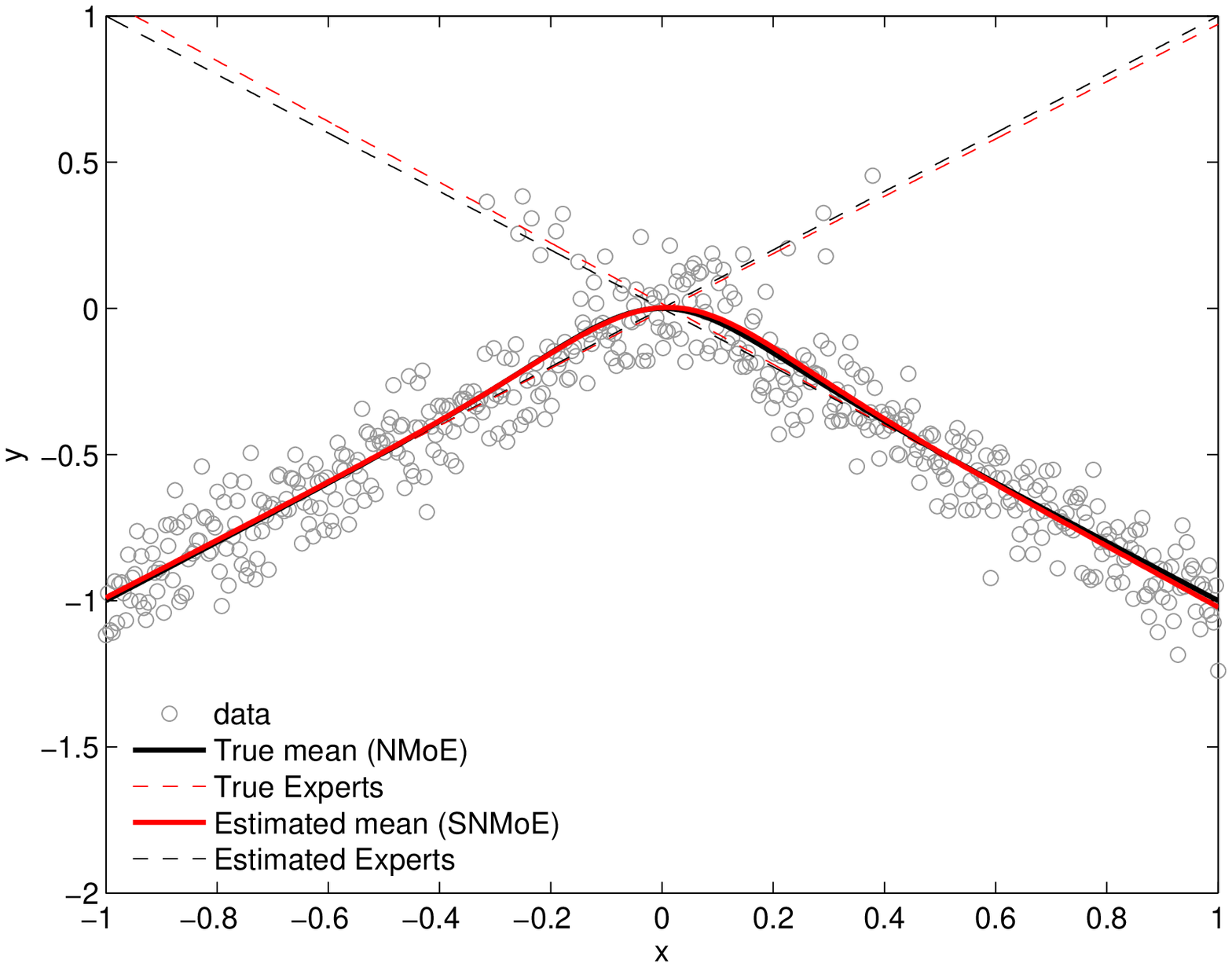}&  
   \includegraphics[width=7.5cm]{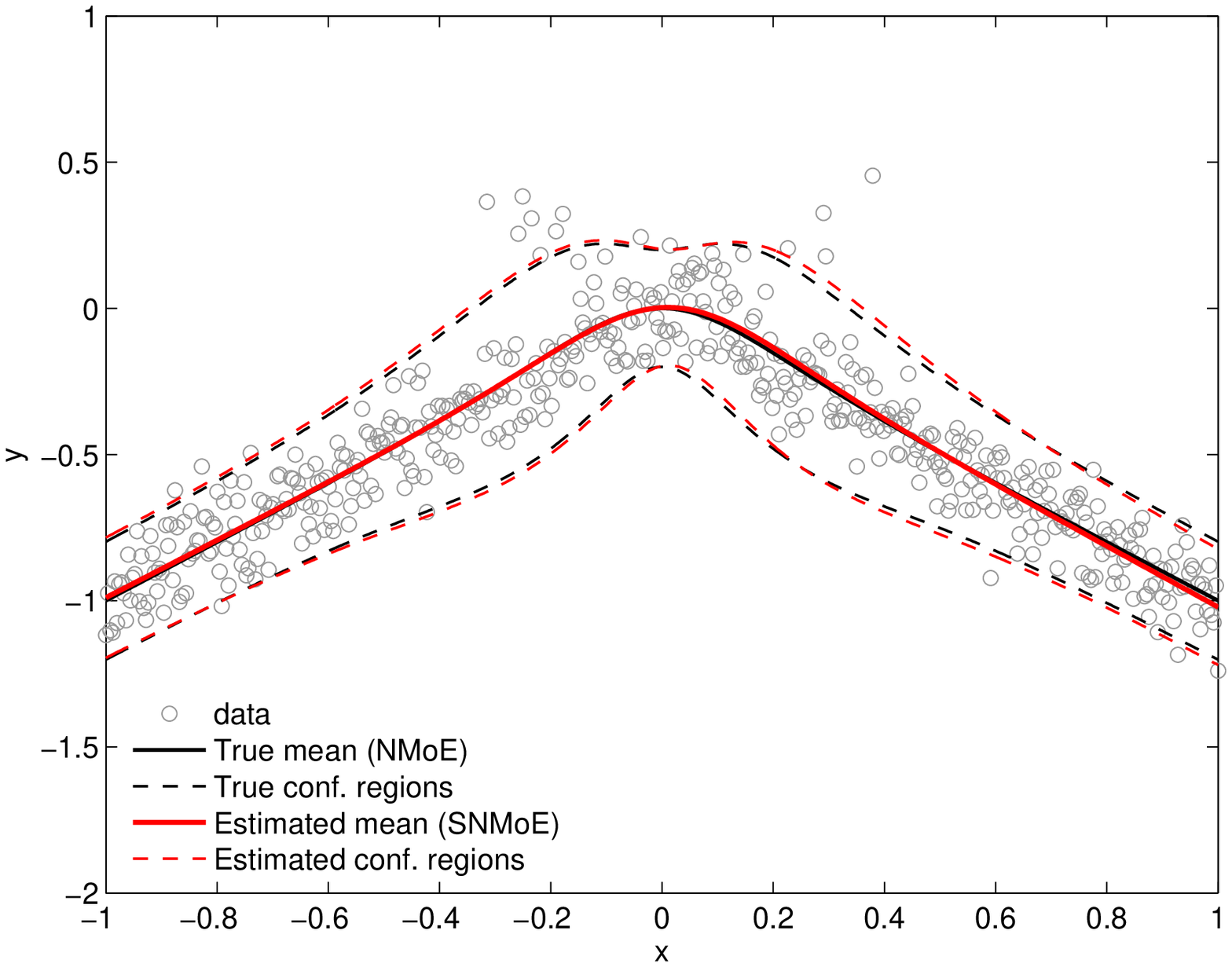}\\
   \hspace{0.2cm}\includegraphics[width=7.3cm]{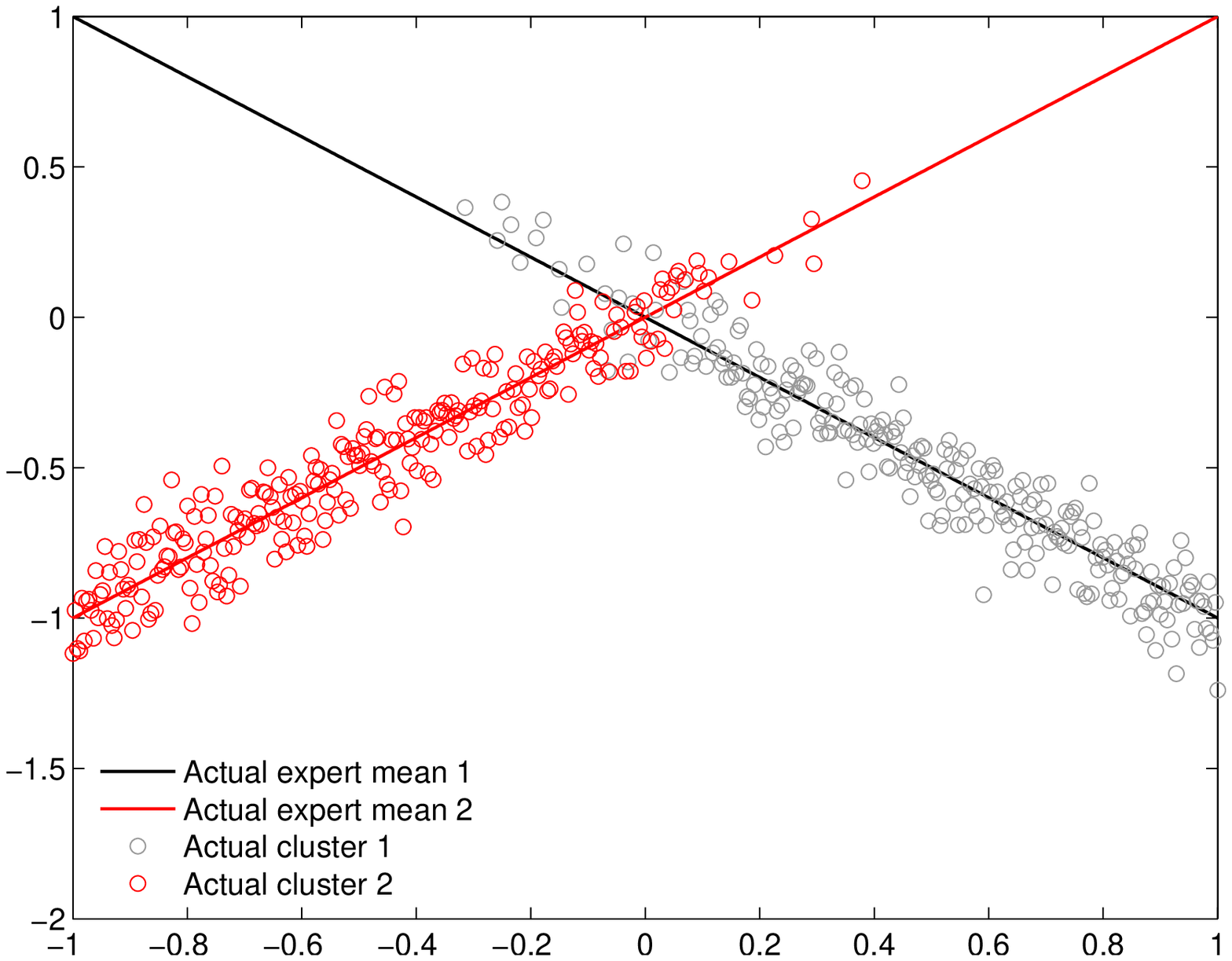}&
   \hspace{0.2cm}\includegraphics[width=7.3cm]{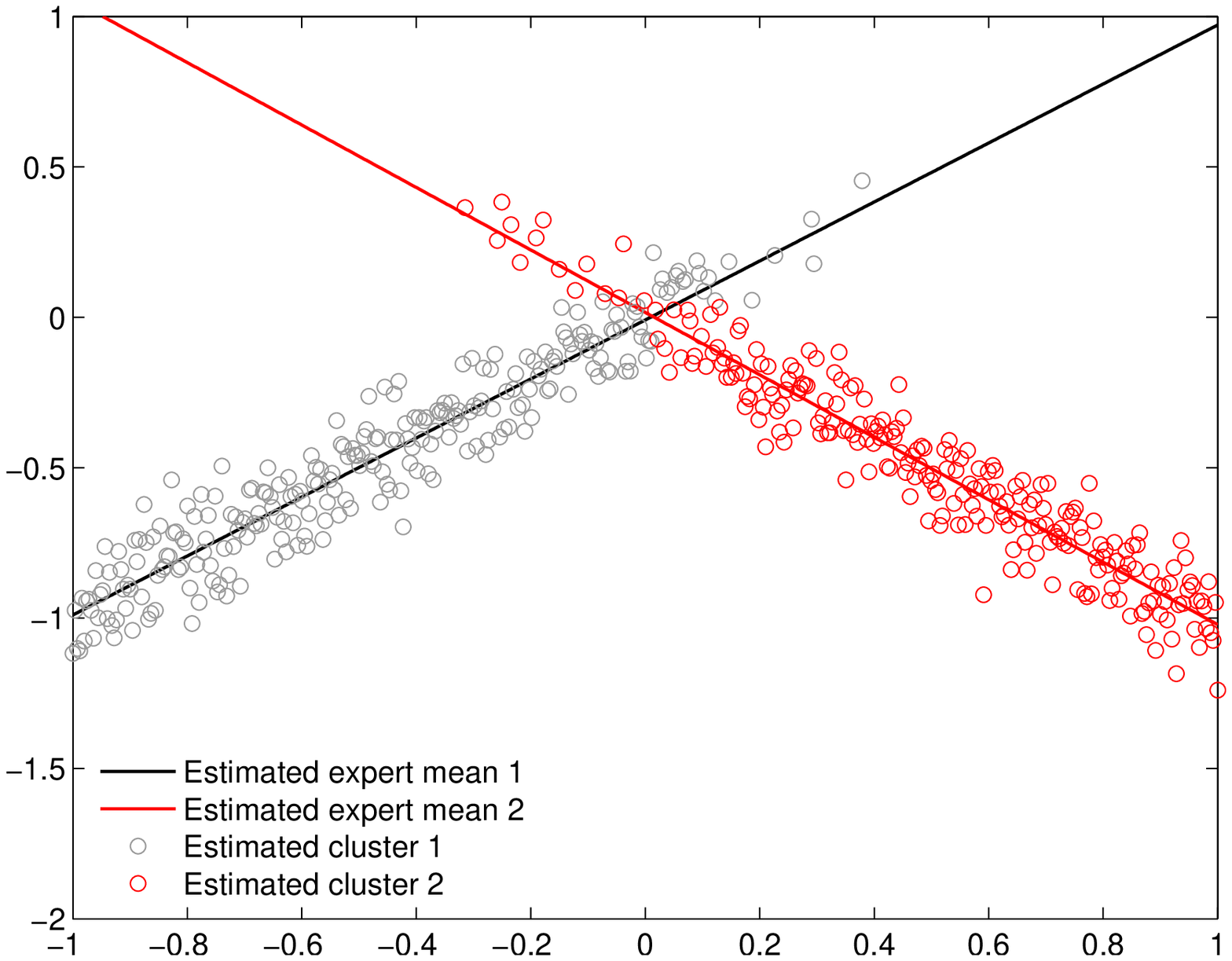}\\
   \includegraphics[width=7.5cm]{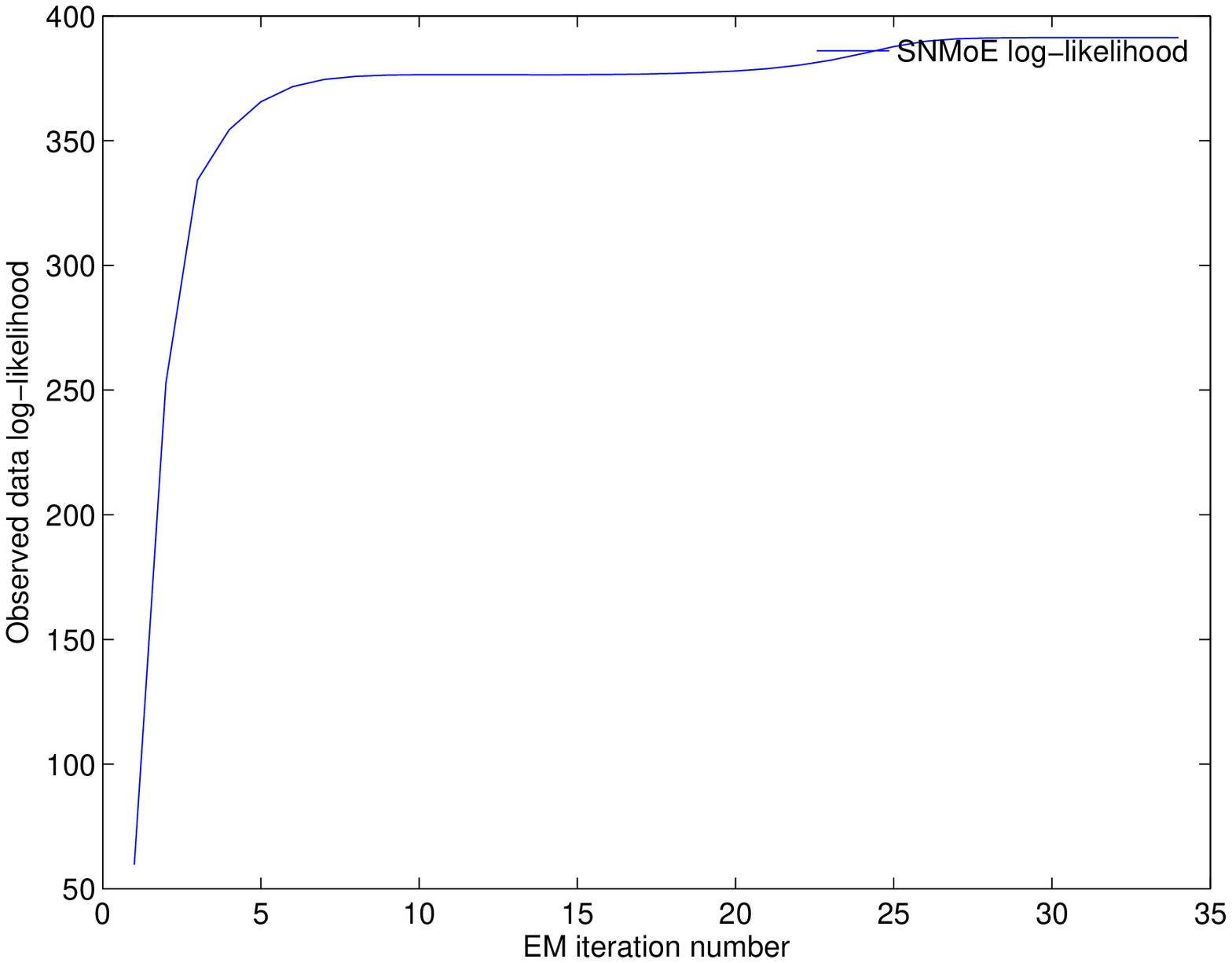} &
   \includegraphics[width=7.5cm]{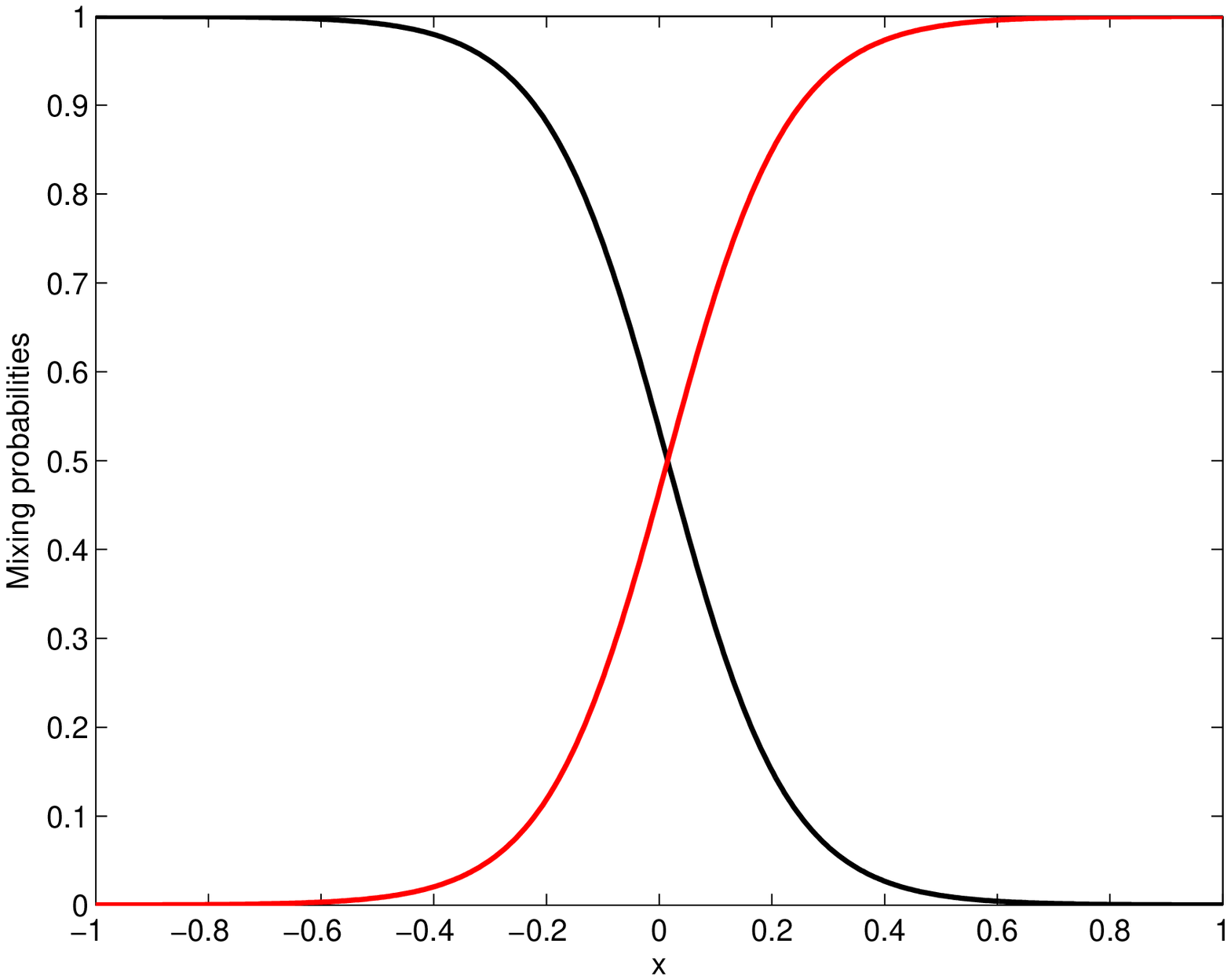}      
   \end{tabular}
      \caption{\label{fig. TwoClust-NMoE_SNMoE}Fitted SNMoE model to a data set generated according to the NMoE model.}
\end{figure}
\begin{figure}[H]
   \centering  
   \begin{tabular}{cc}
   \includegraphics[width=7.5cm]{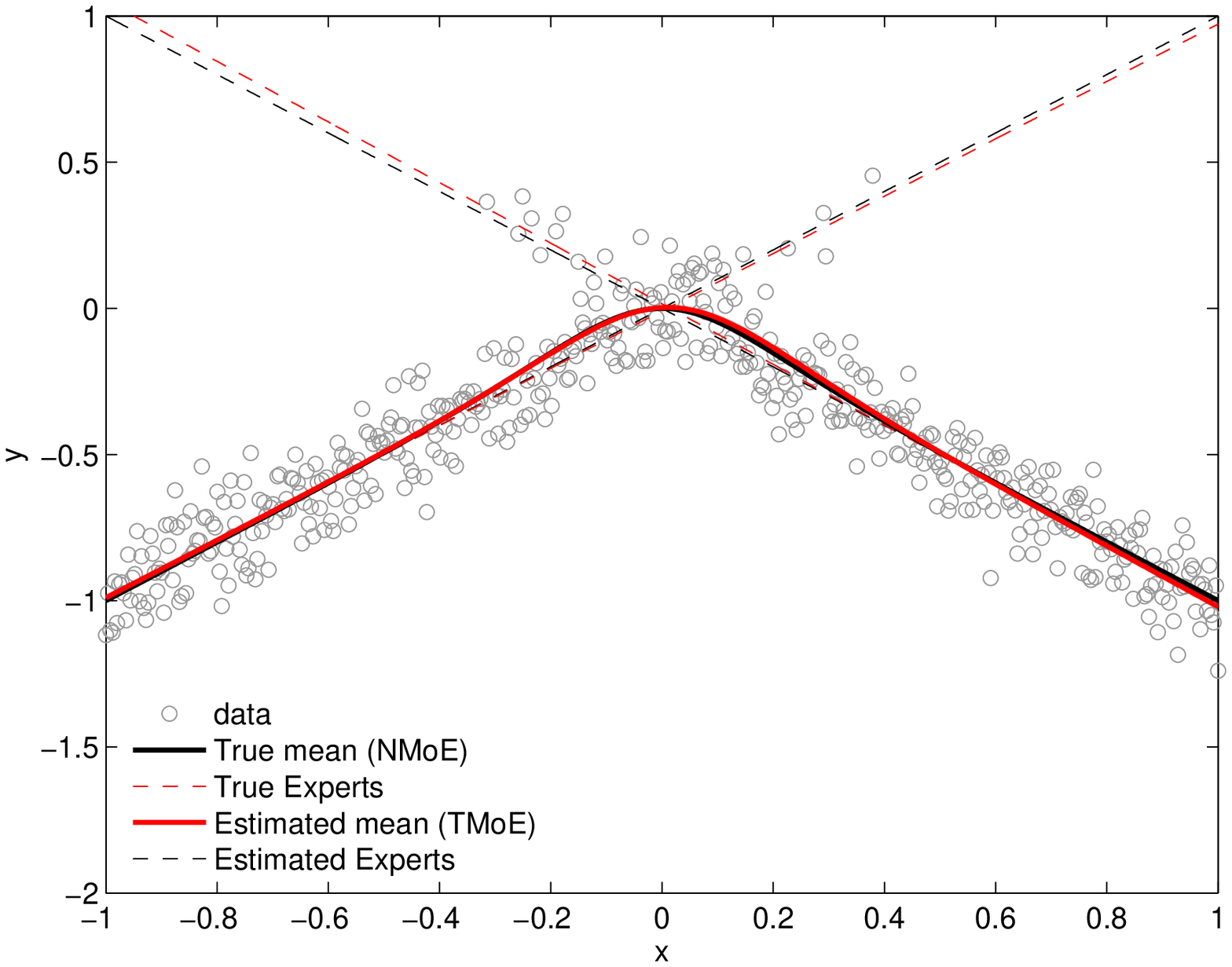}&  
   \includegraphics[width=7.5cm]{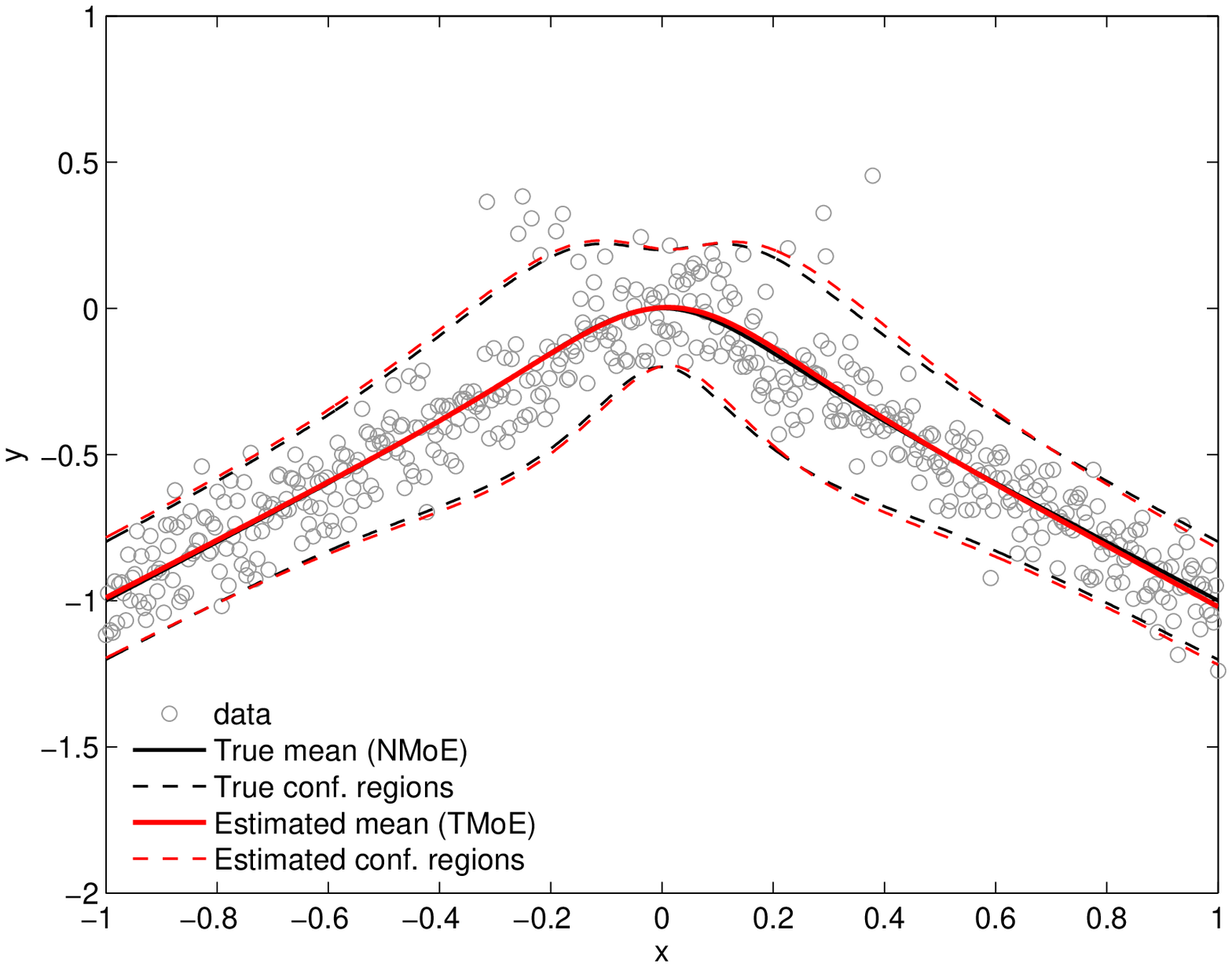}\\
\hspace{0.2cm}\includegraphics[width=7.3cm]{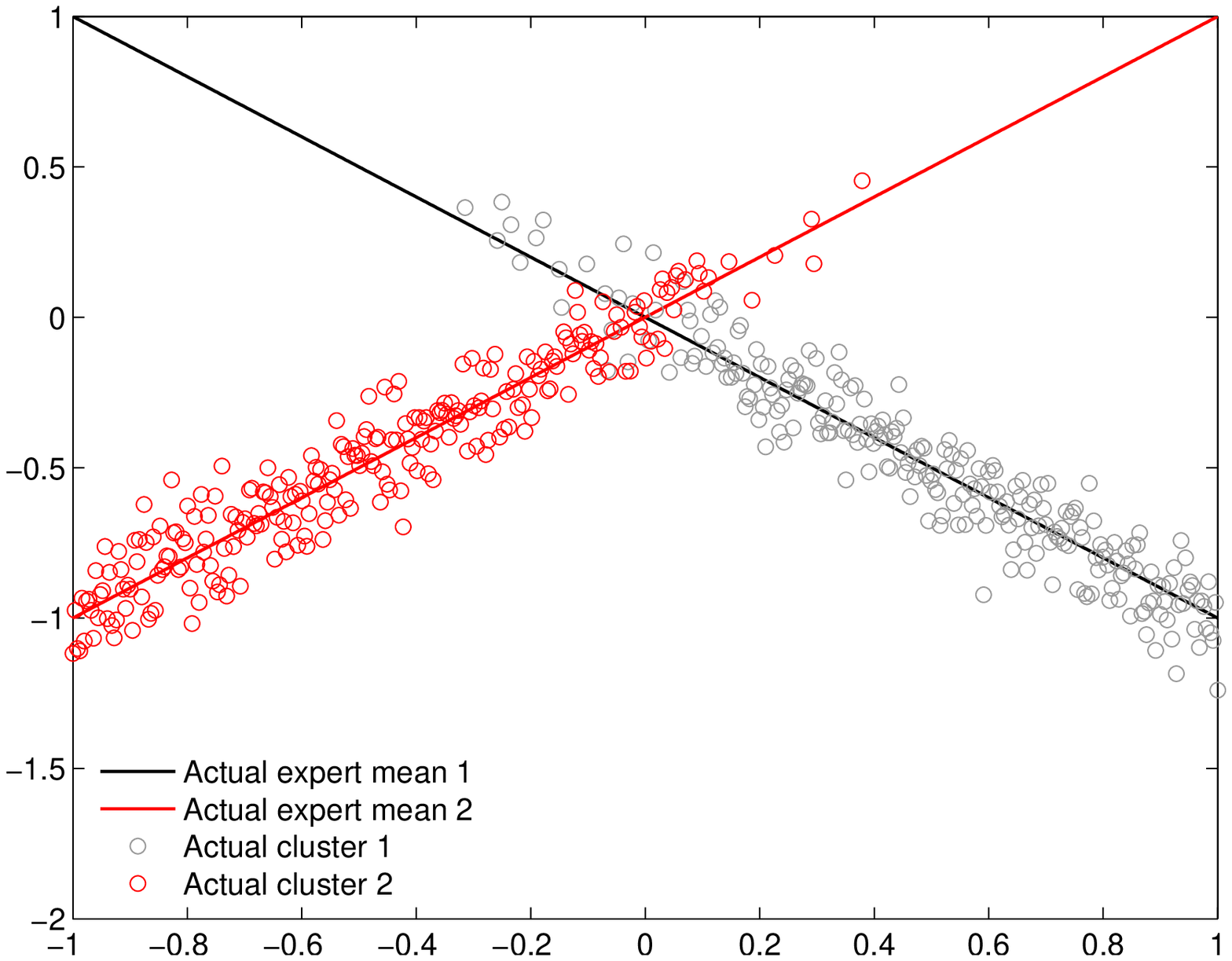}&
\hspace{0.2cm}\includegraphics[width=7.3cm]{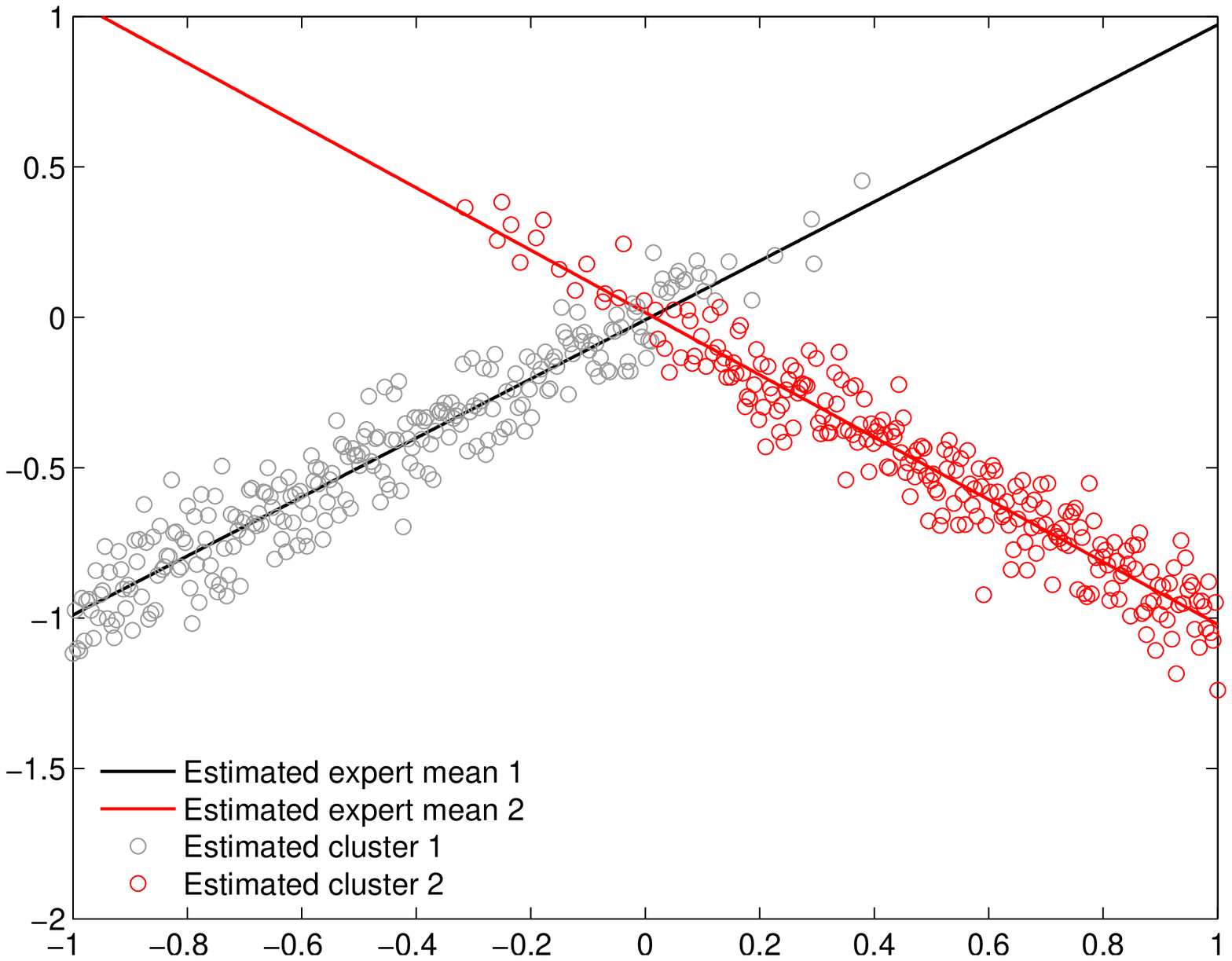}\\
   \includegraphics[width=7.5cm]{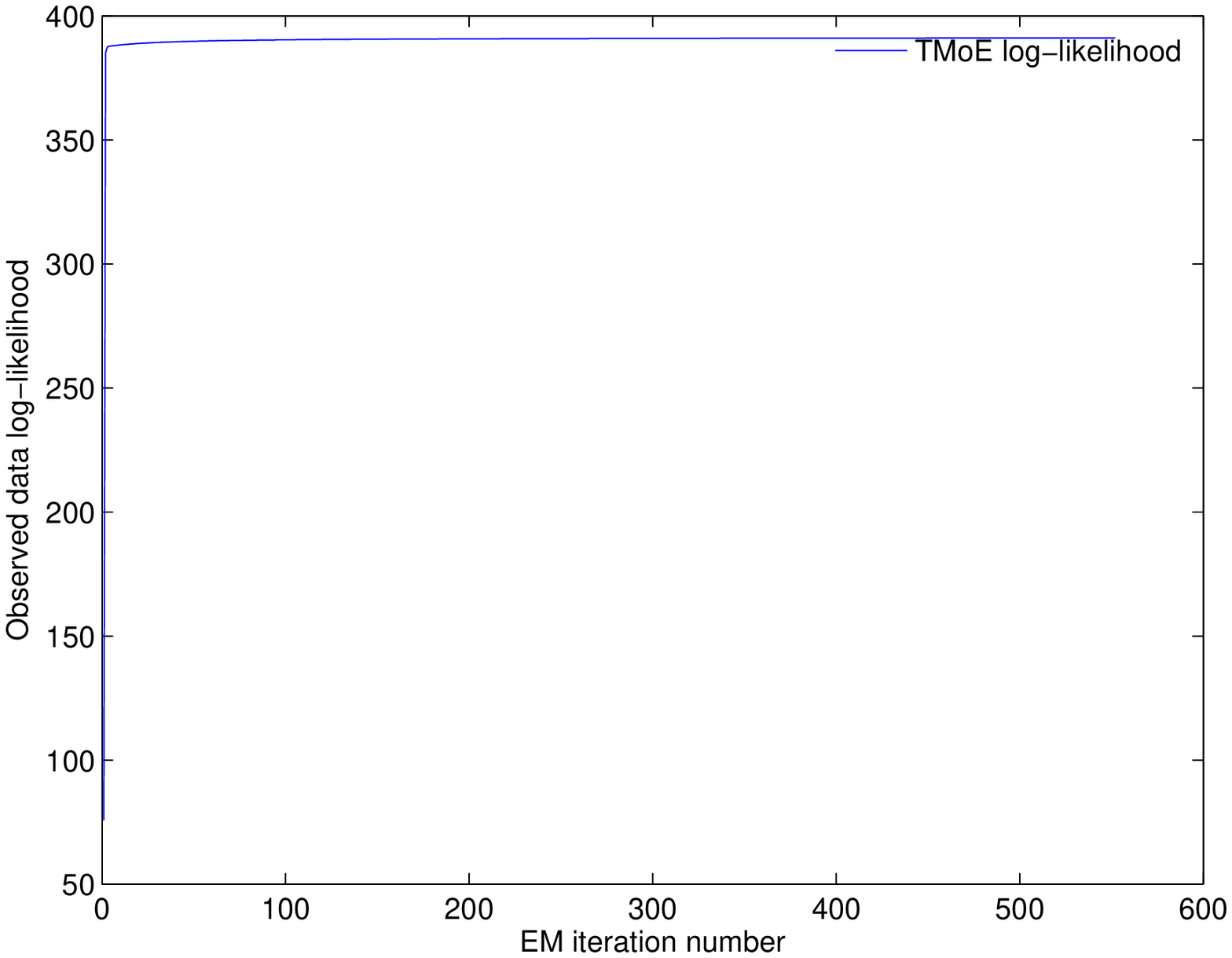} &
  \includegraphics[width=7.5cm]{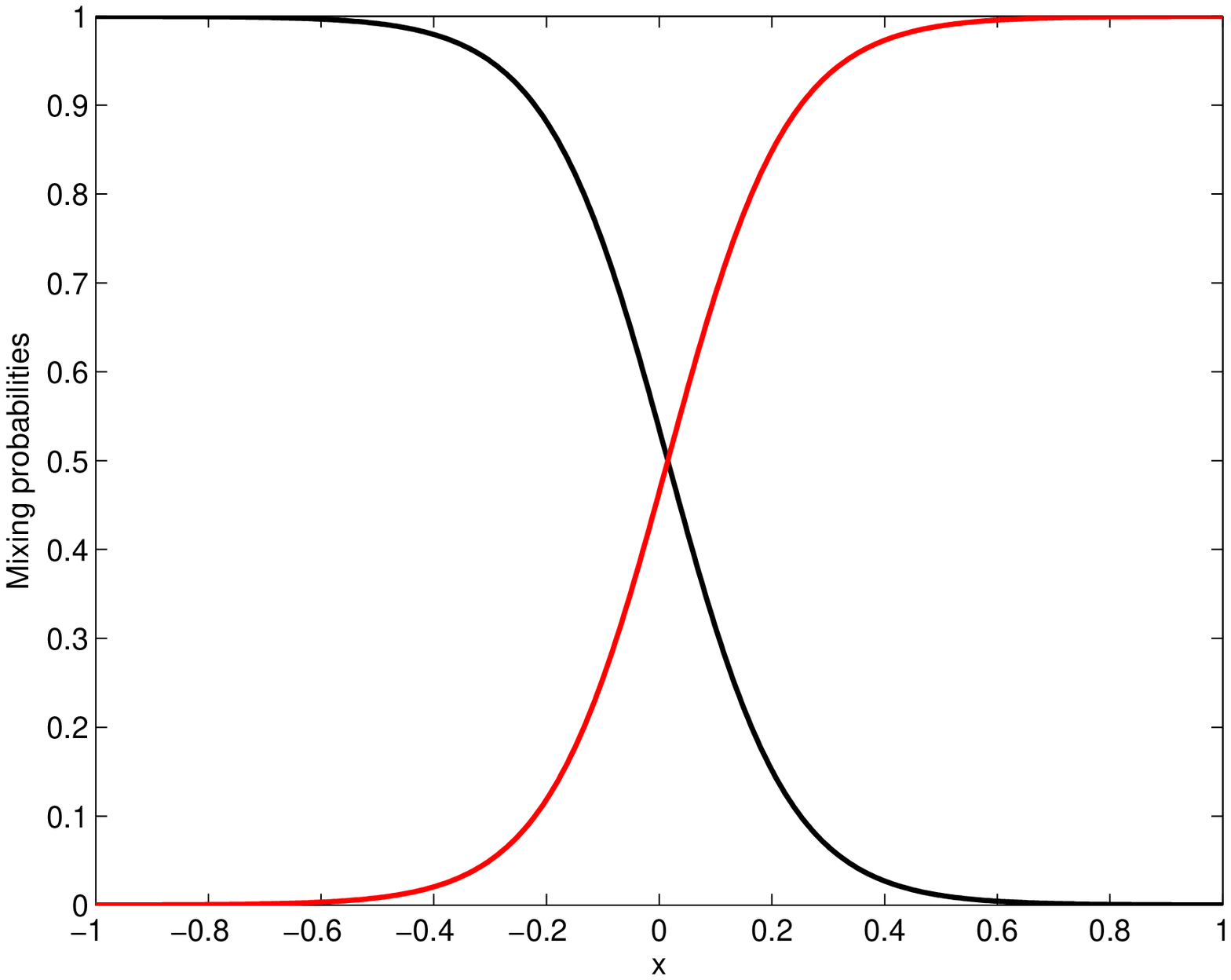}
   \end{tabular}
      \caption{\label{fig. TwoClust-NMoE_TMoE}Fitted  TMoE model to a data set generated according to the NMoE model.}
\end{figure}
\begin{figure}[H]
   \centering  
   \begin{tabular}{cc}
   \includegraphics[width=7.5cm]{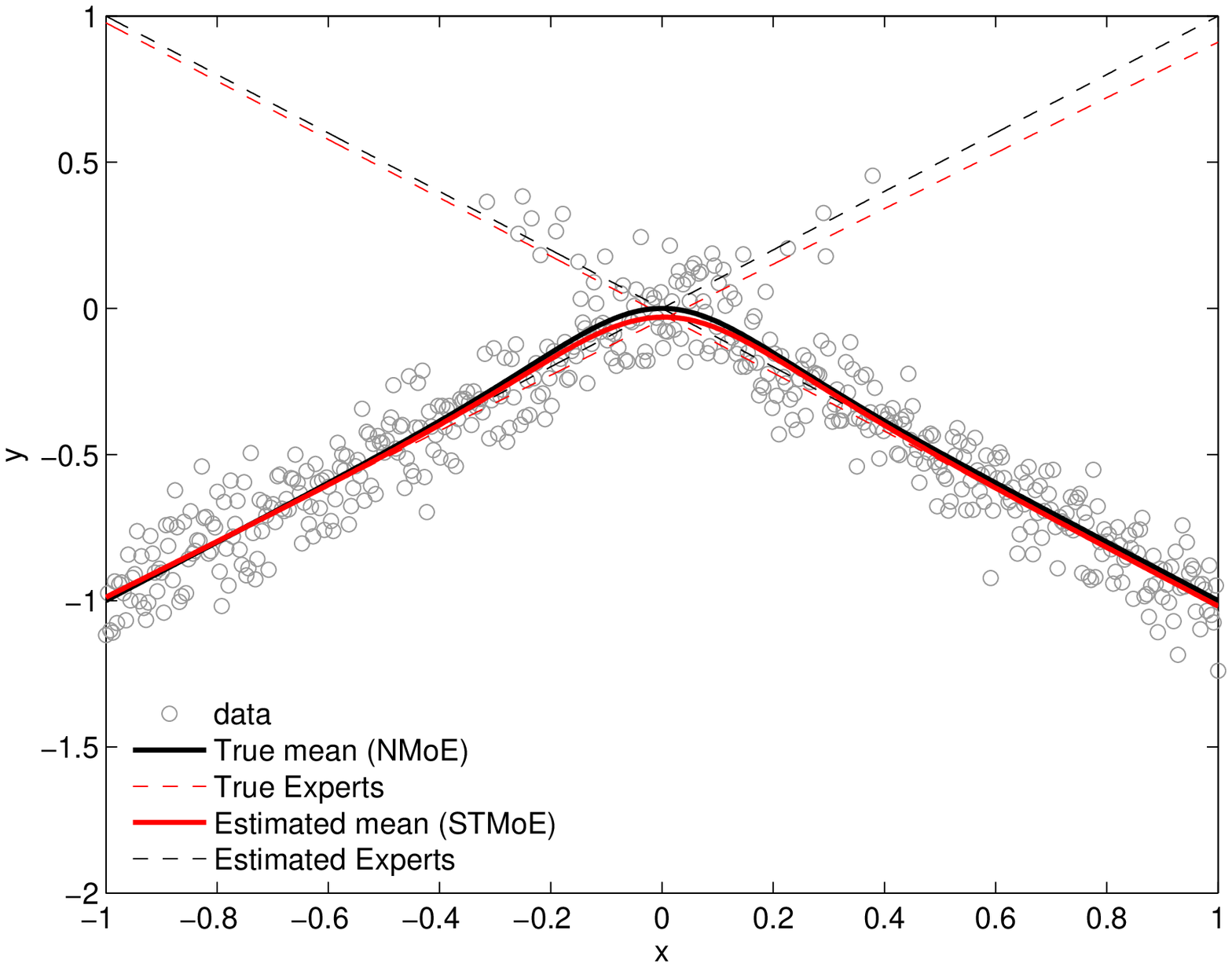}&  
   \includegraphics[width=7.5cm]{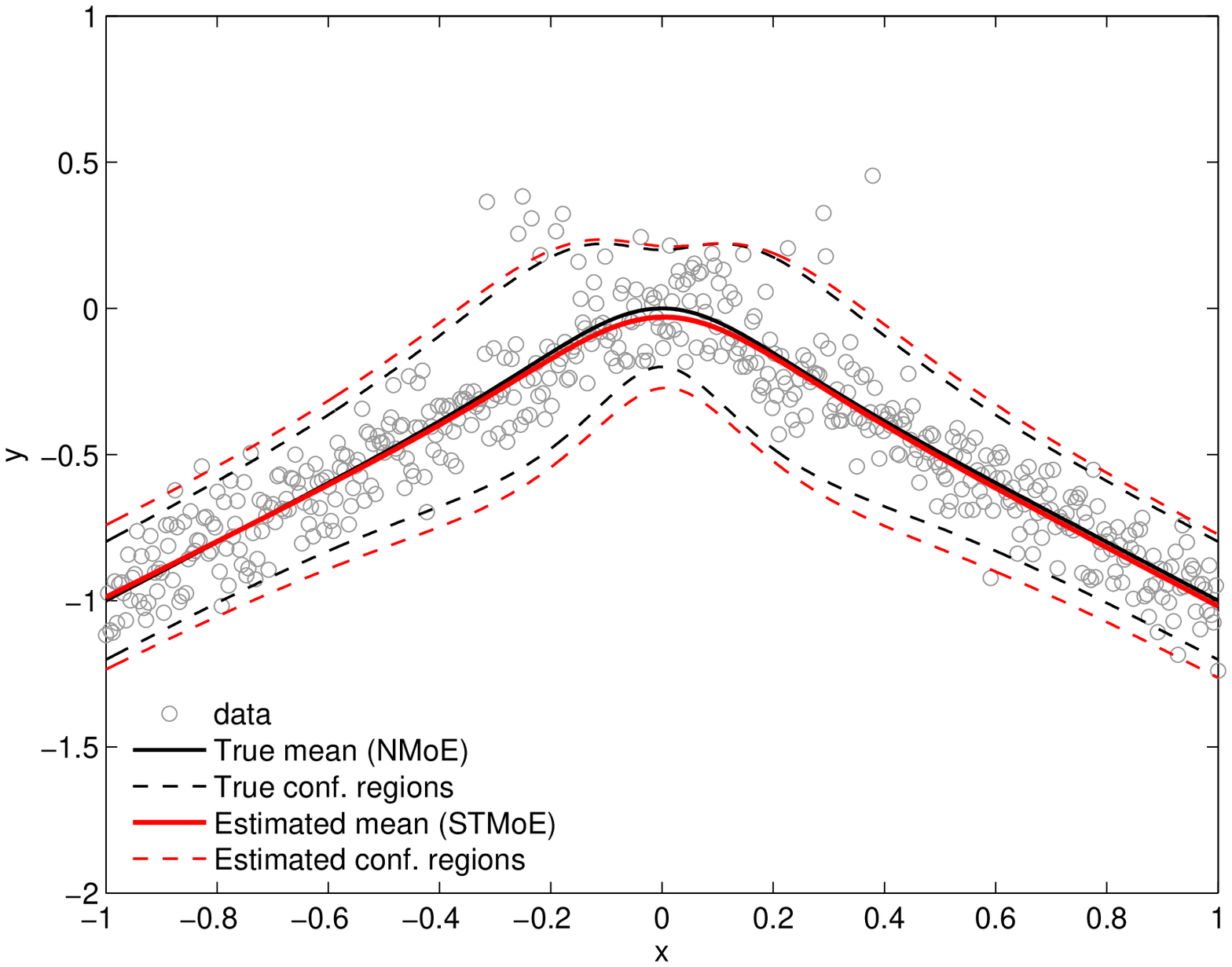}\\
  \hspace{0.2cm}\includegraphics[width=7.3cm]{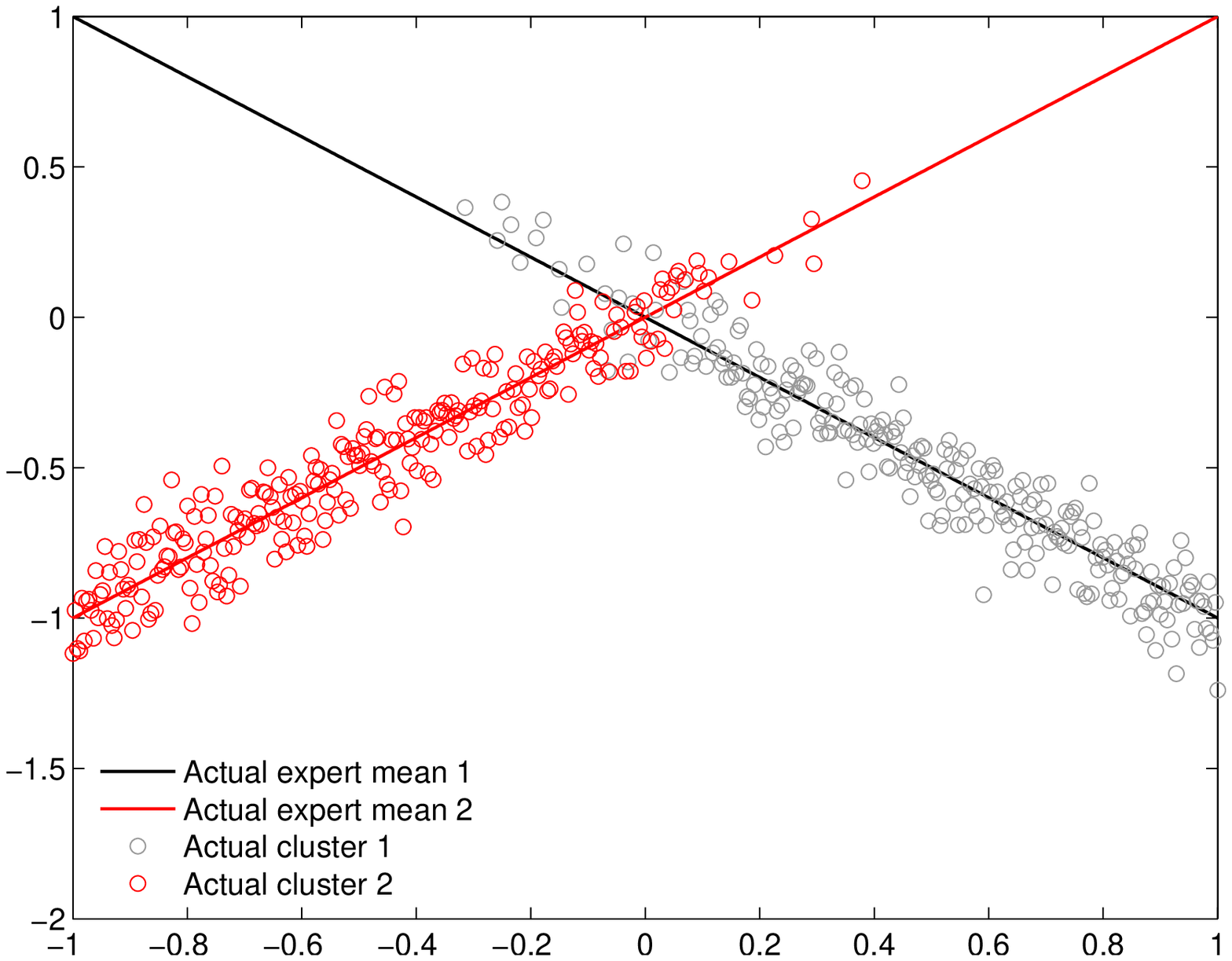}&
   \hspace{0.2cm}\includegraphics[width=7.3cm]{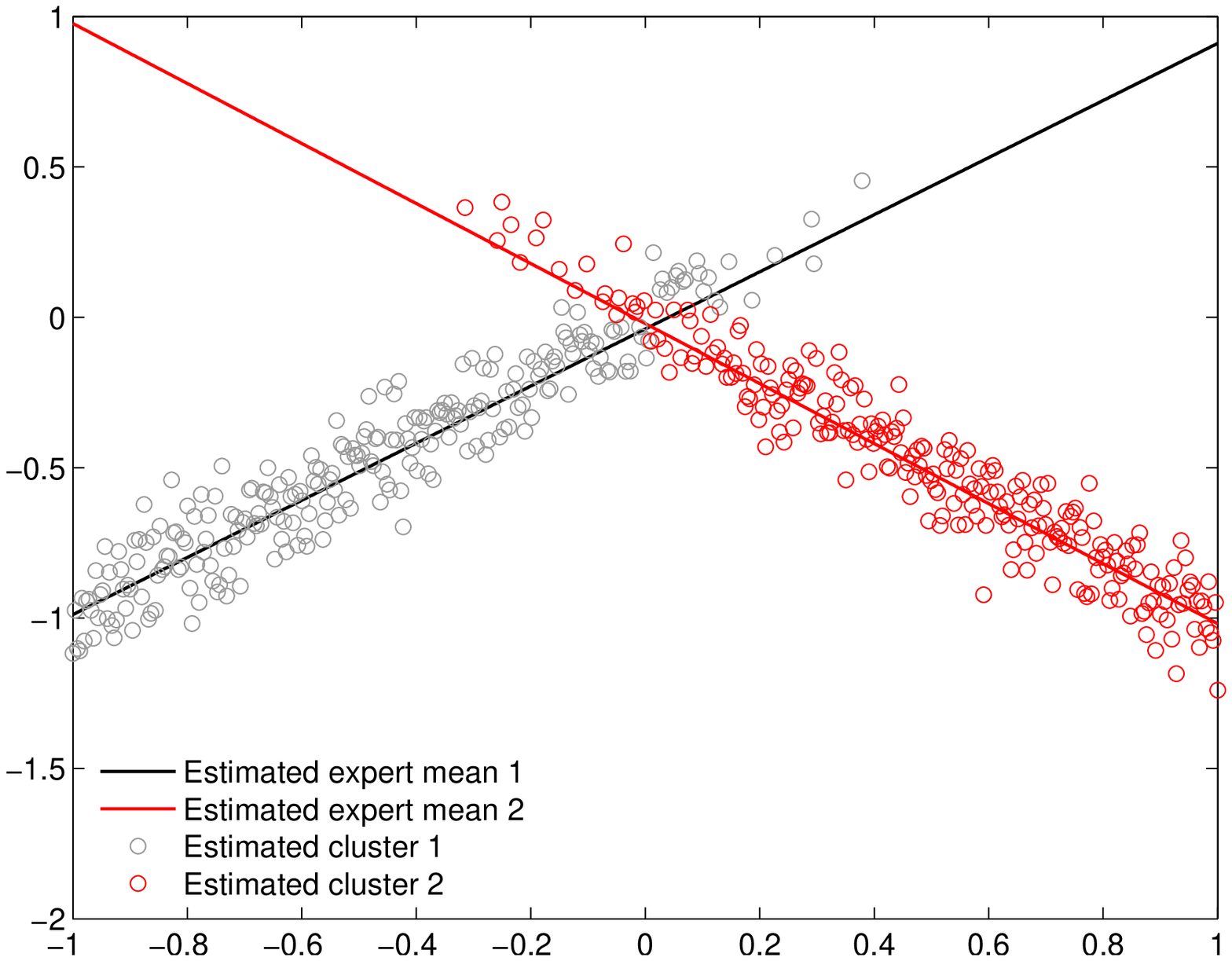}\\
   \includegraphics[width=7.5cm]{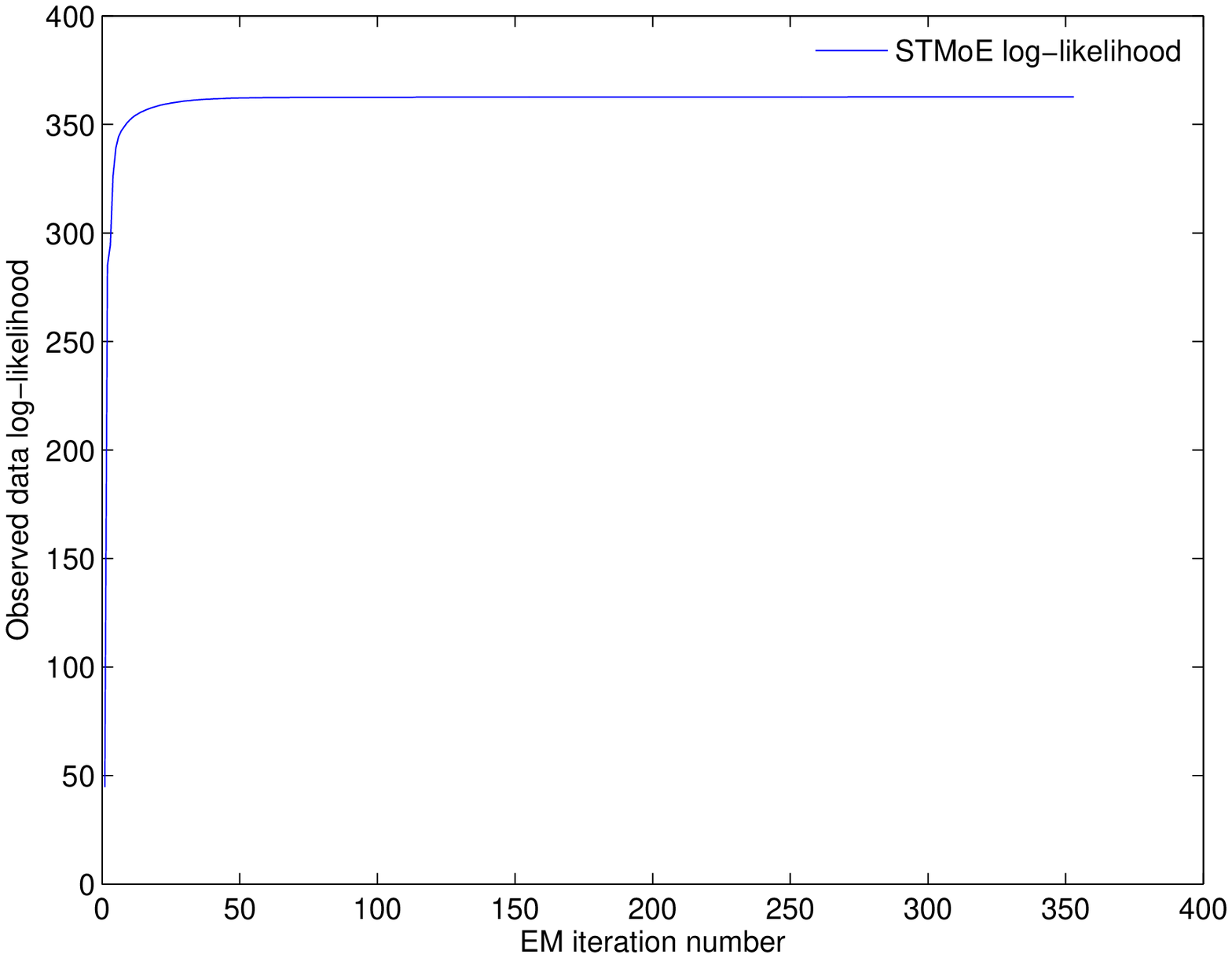} &
    \includegraphics[width=7.5cm]{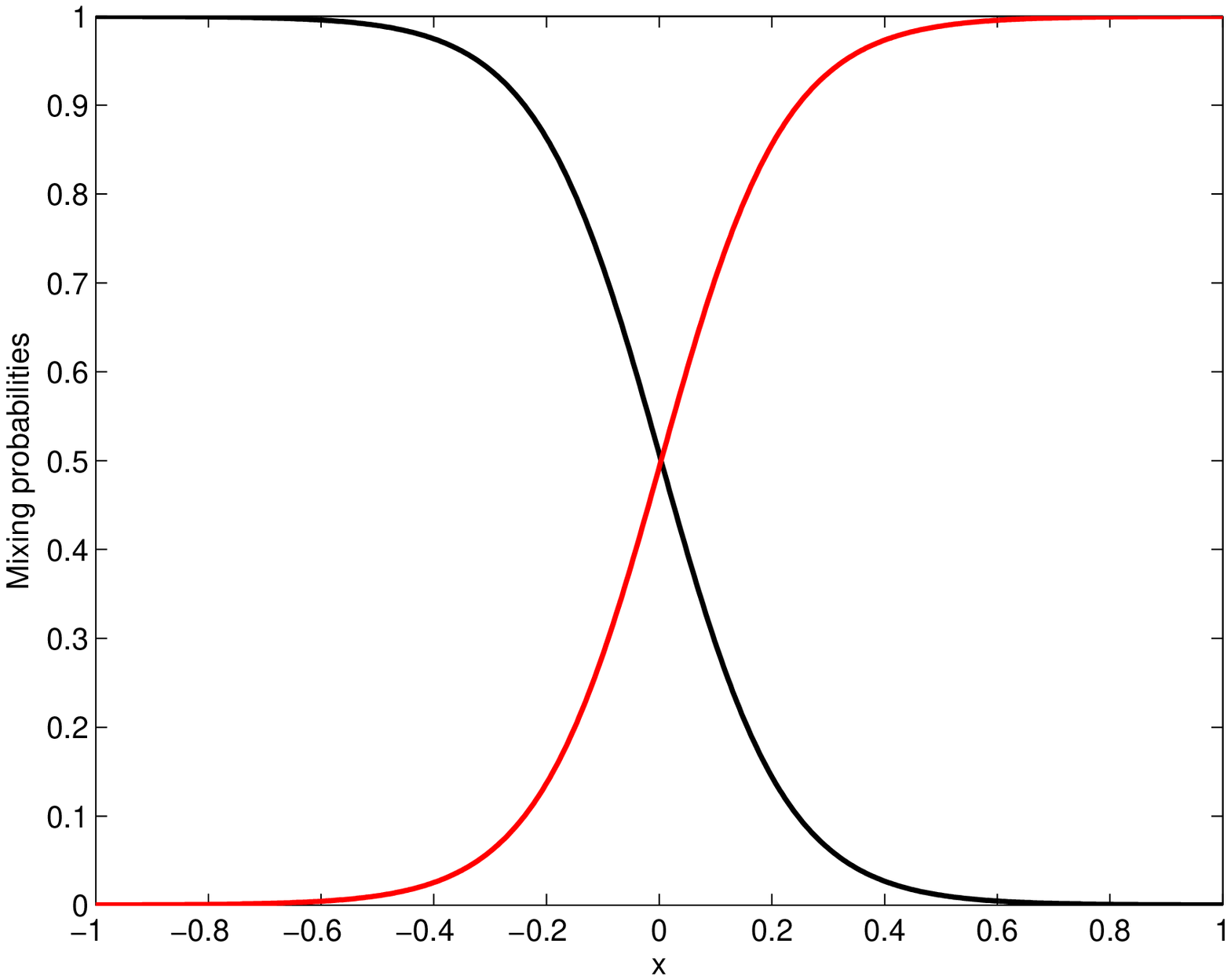}  
   \end{tabular}
      \caption{\label{fig. TwoClust-NMoE_STMoE}Fitted STMoE model to a data set generated according to the NMoE model.}
\end{figure}%
\begin{figure}[H]
   \centering  
   \begin{tabular}{cc}
  \includegraphics[width=7.5cm]{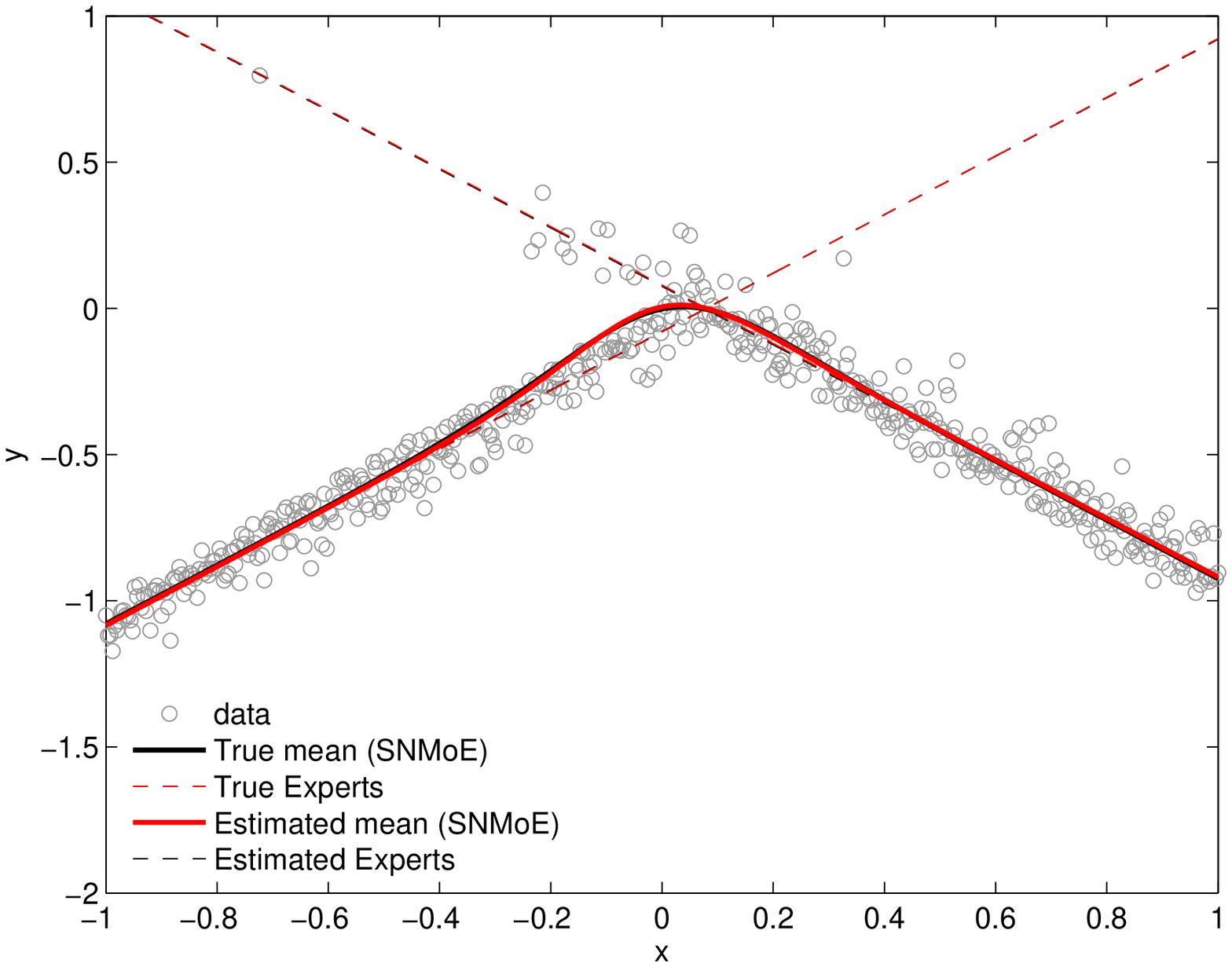}&  
   \includegraphics[width=7.5cm]{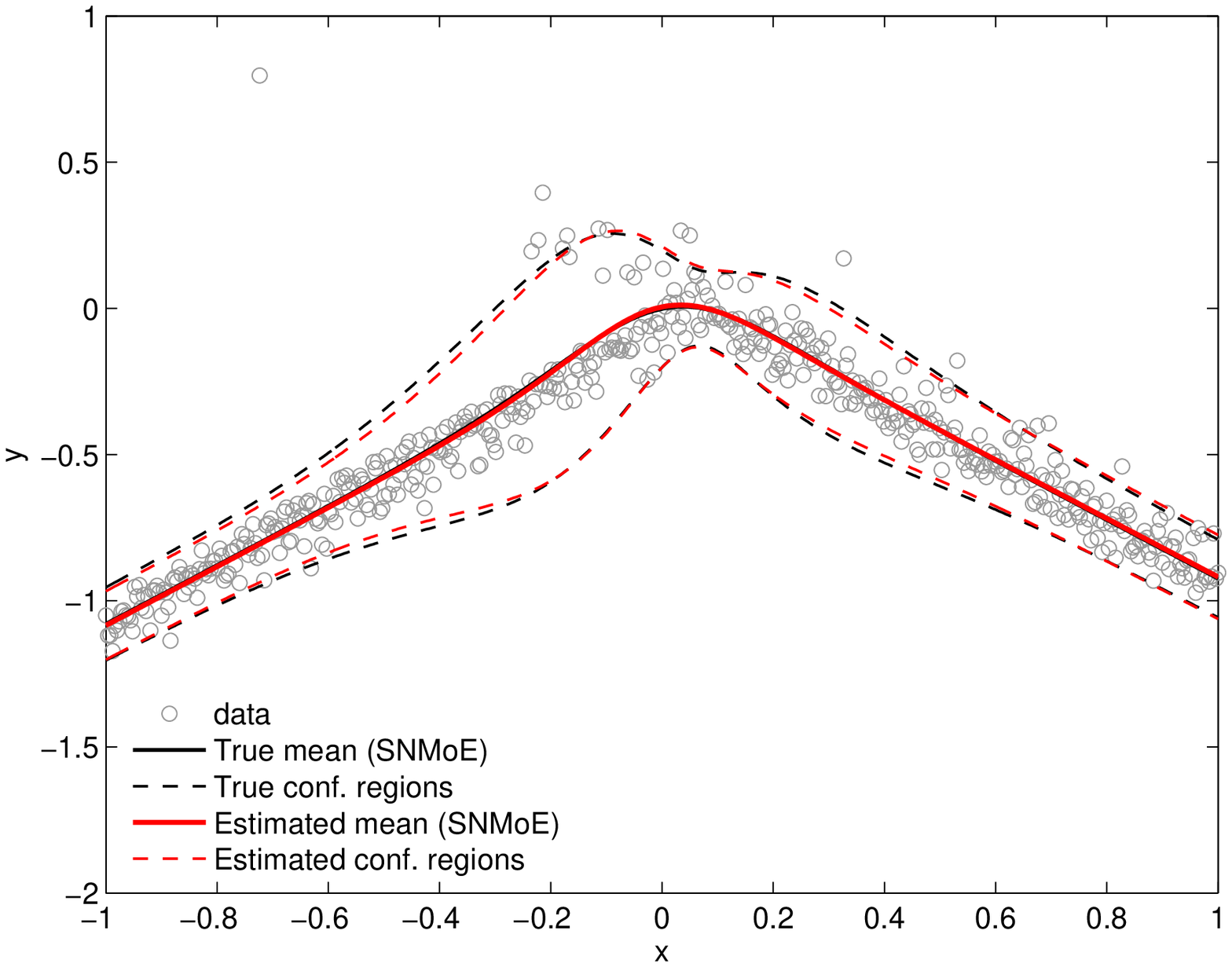}\\ 
   \includegraphics[width=7.5cm]{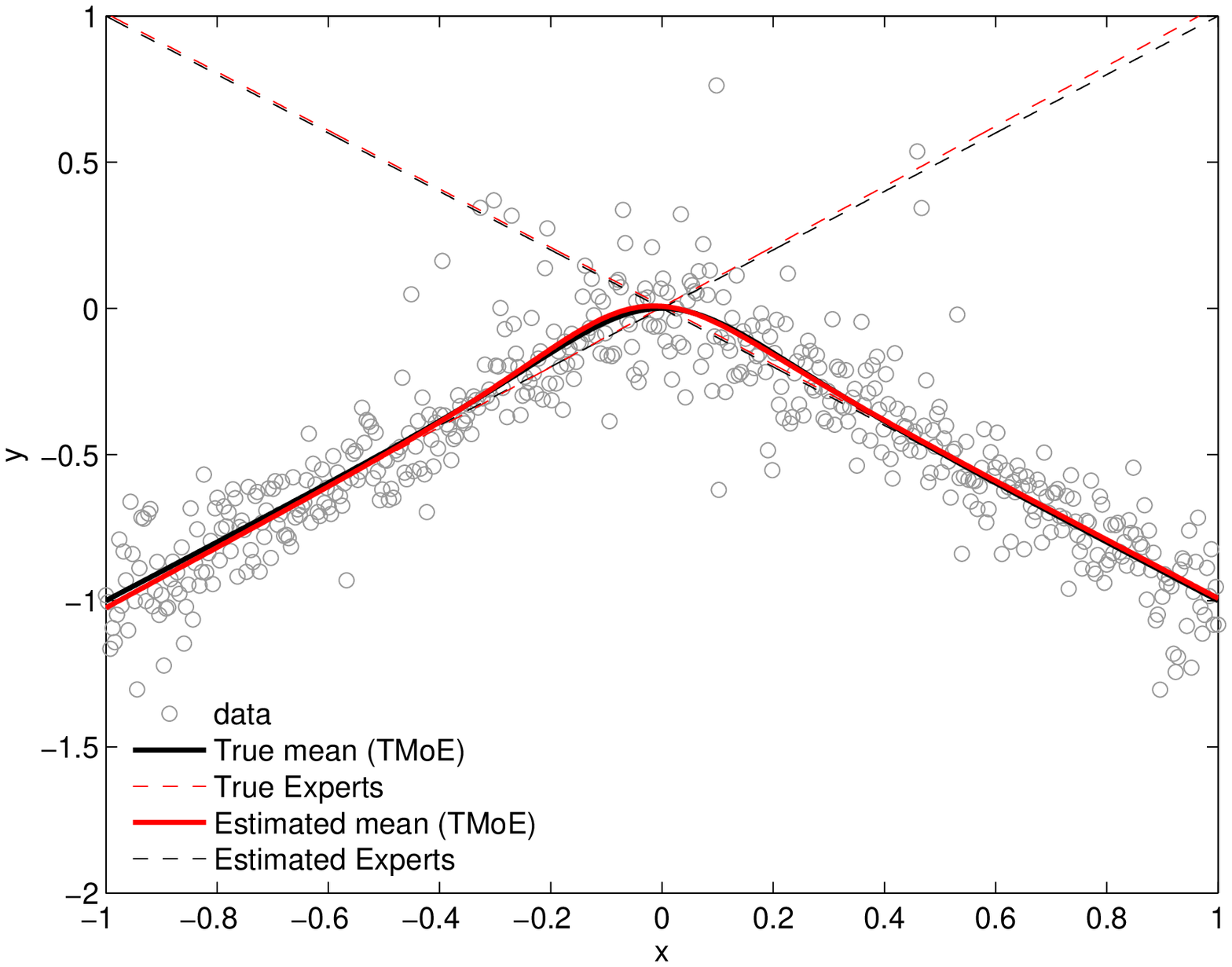}&  
   \includegraphics[width=7.5cm]{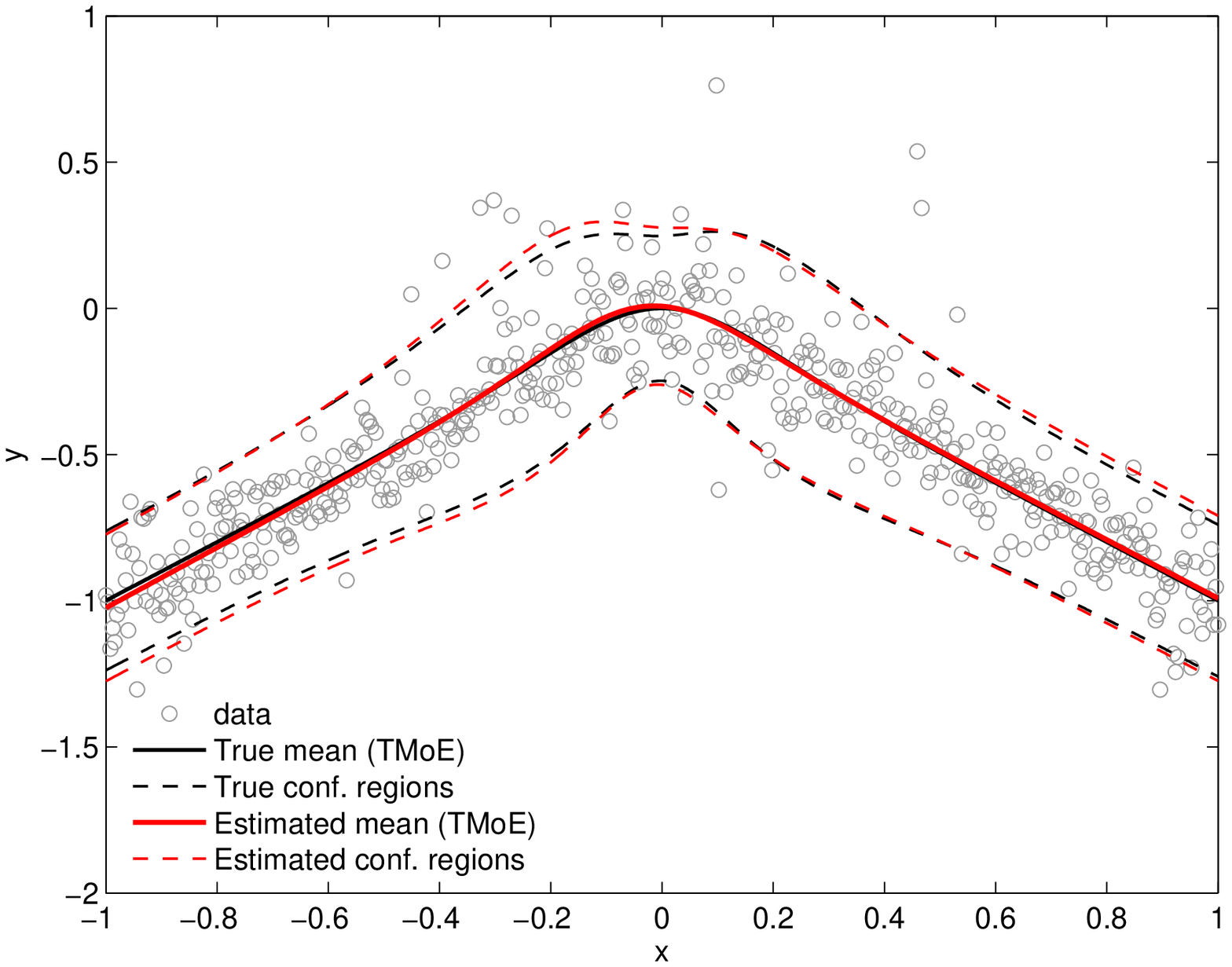}\\
   \includegraphics[width=7.5cm]{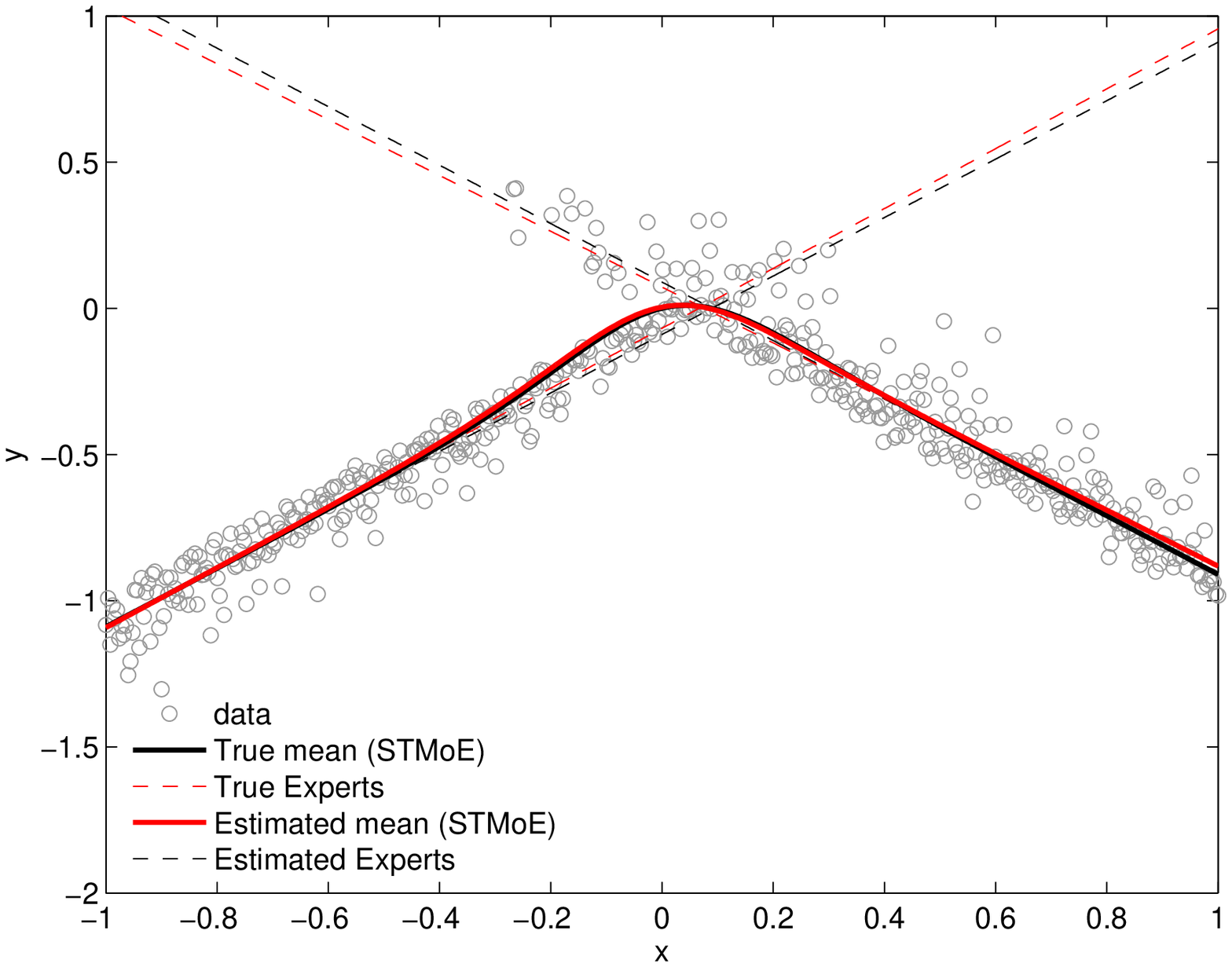}&  
   \includegraphics[width=7.5cm]{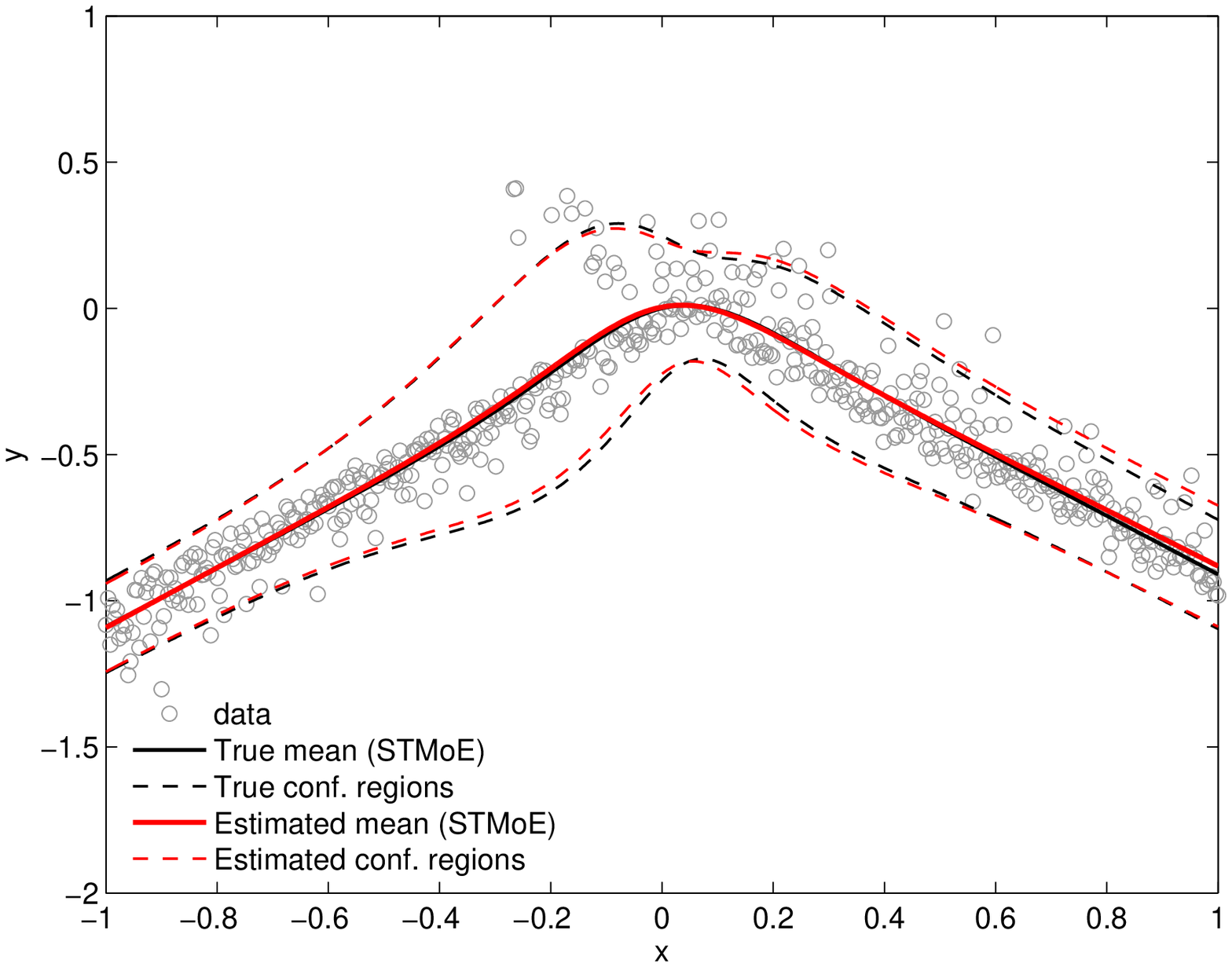}
\end{tabular}
   \caption{\label{fig. TwoClust-All->All-estimated models}The true and estimated mean function and expert mean functions by fitting the proposed NNMoE models to a simulated data set of $n=500$ observations. Up: the SNMoE model; Middle: the TMoE model; Bottom: the STMoE model.}
\end{figure}
\begin{figure}[H]
   \centering 
\begin{tabular}{cc}
  \hspace{0.2cm}\includegraphics[width=7.3cm]{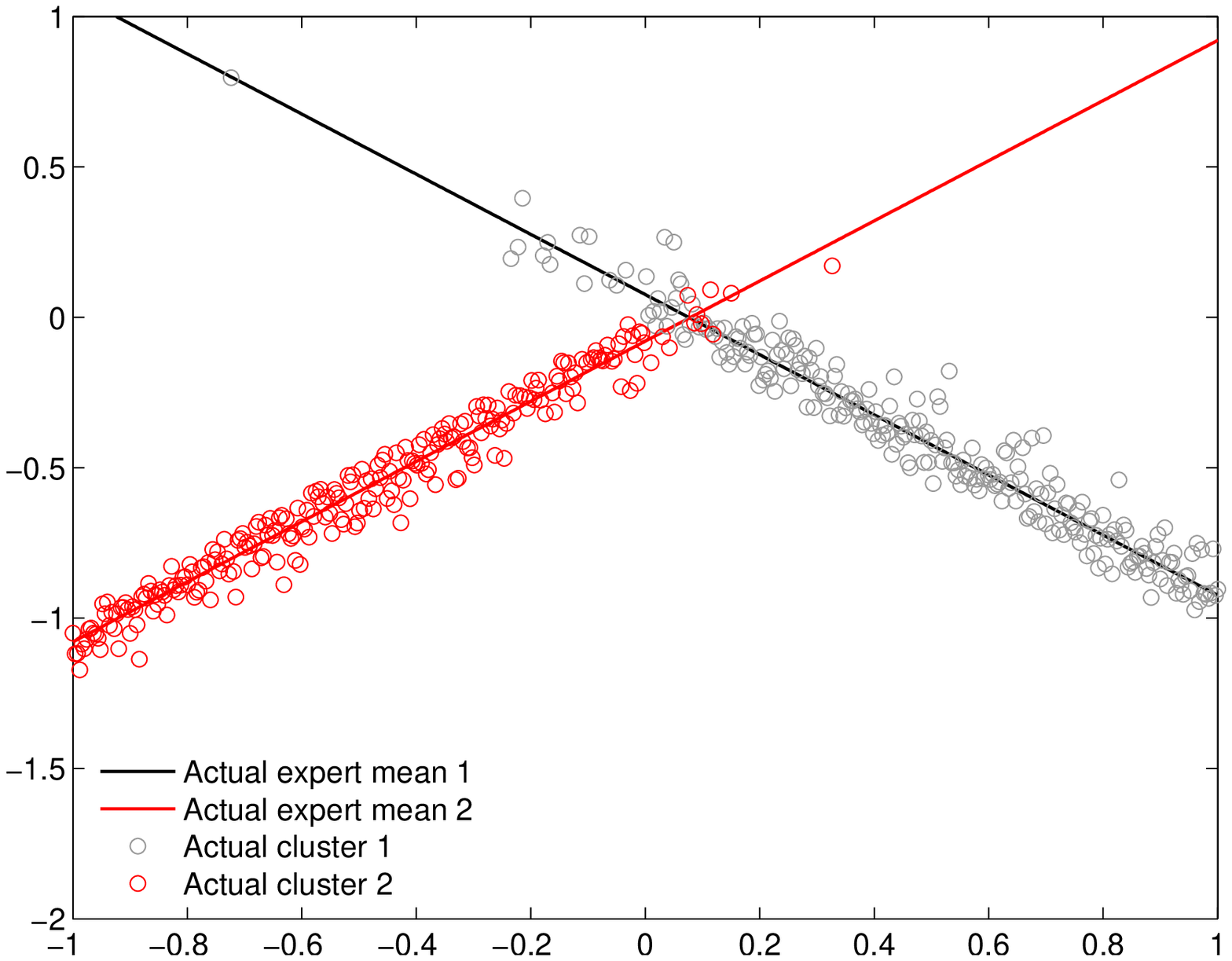}&
  \hspace{0.2cm}\includegraphics[width=7.3cm]{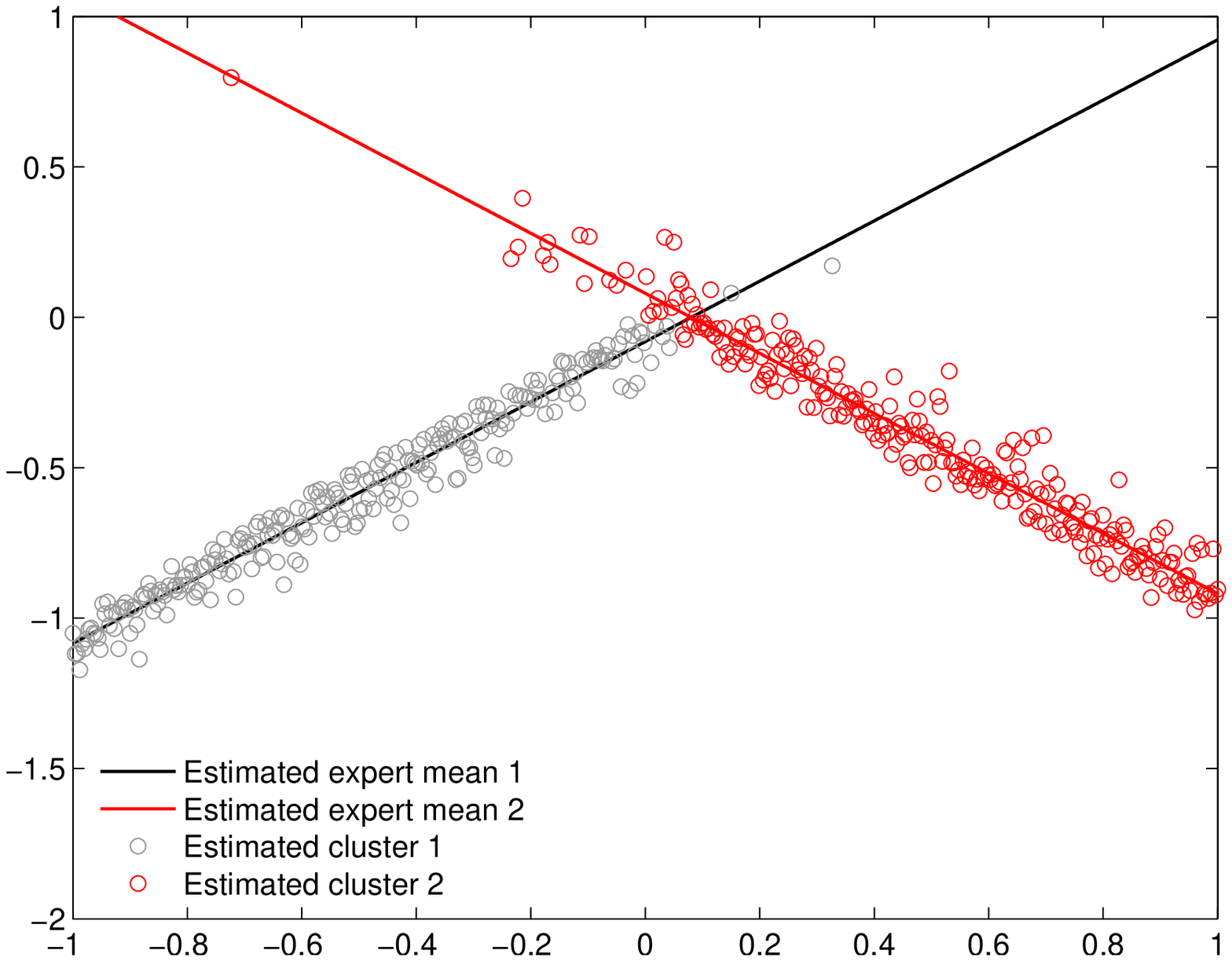}\\ 
  \hspace{0.2cm}\includegraphics[width=7.3cm]{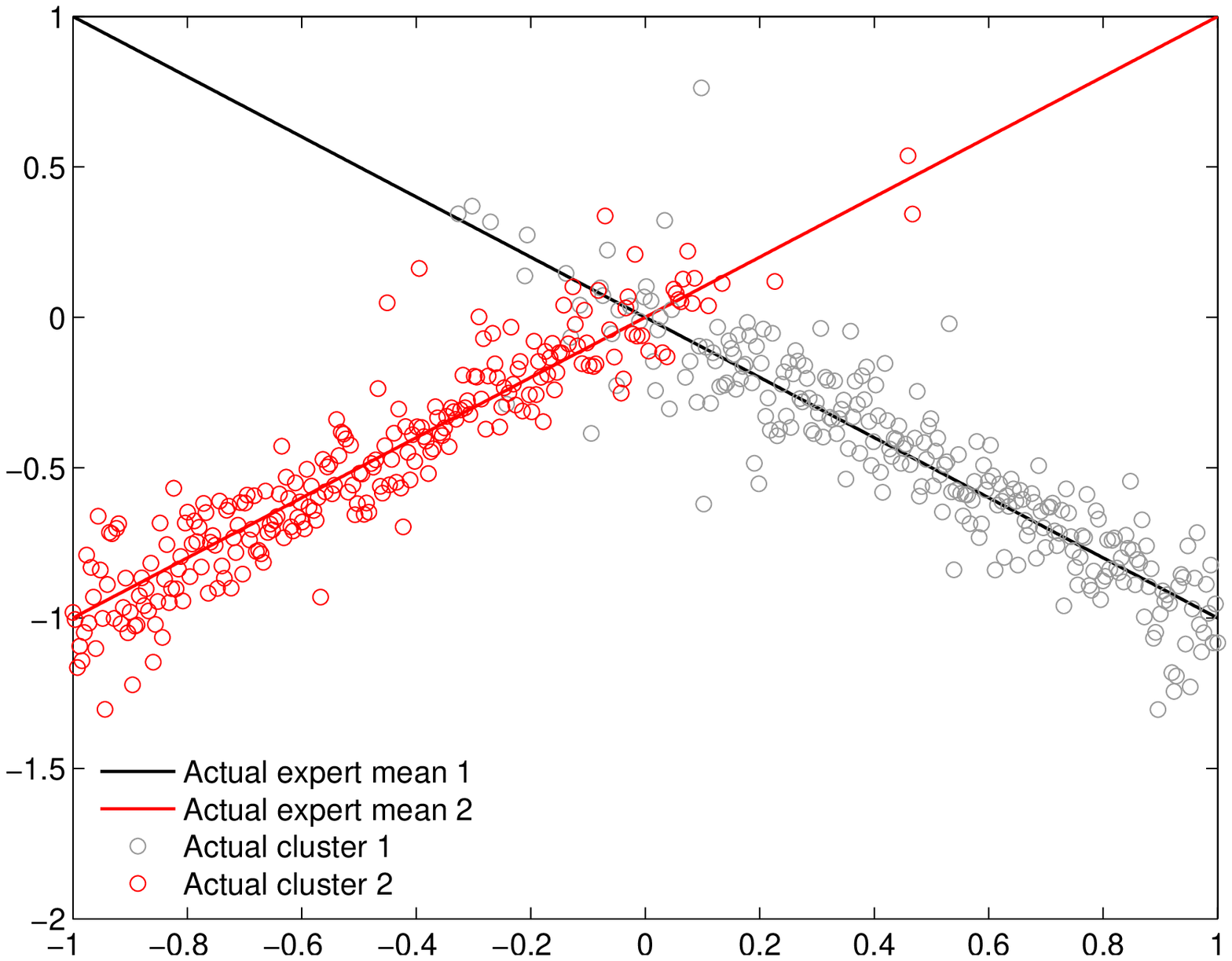}&
  \hspace{0.2cm}\includegraphics[width=7.3cm]{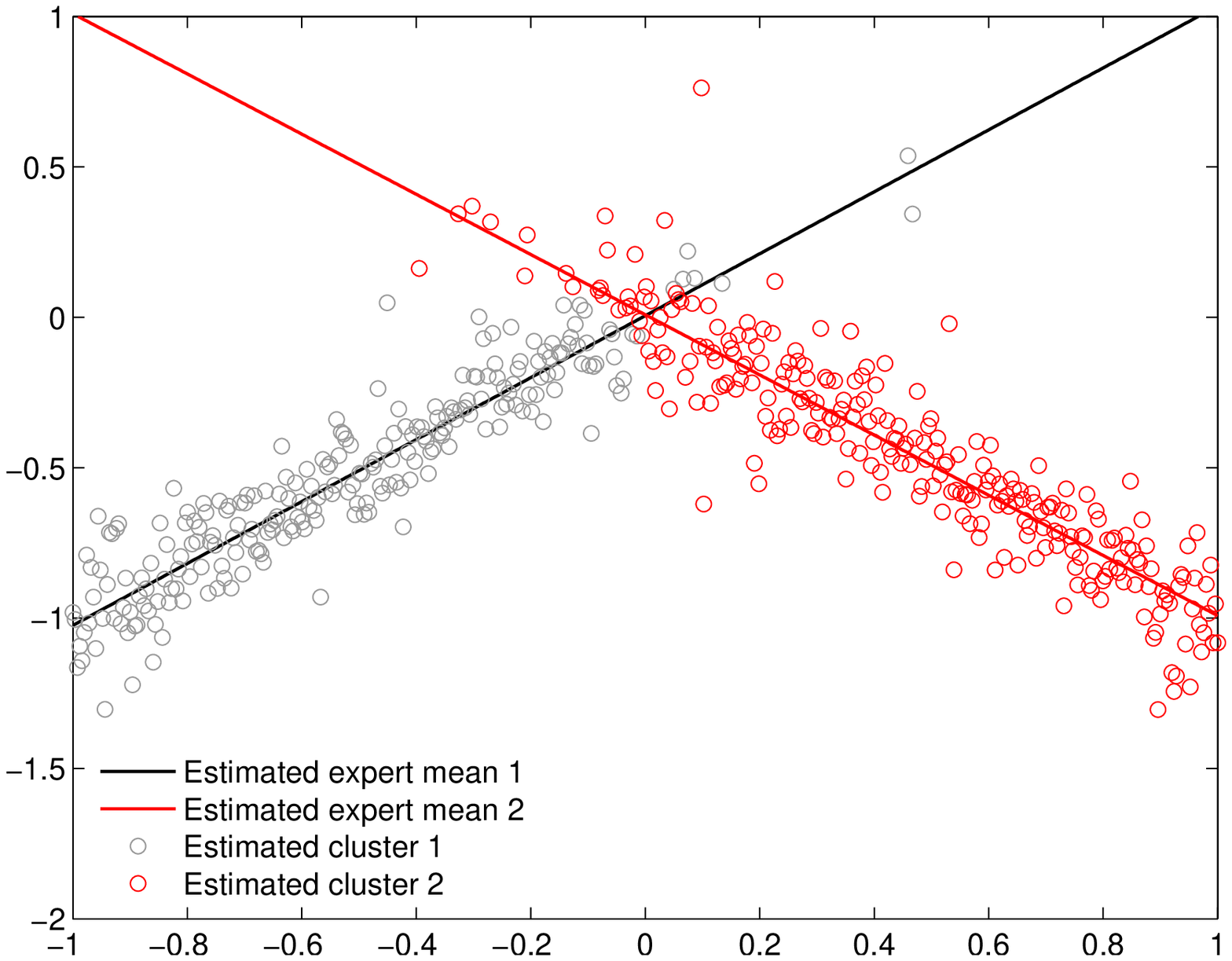}\\ 
  \hspace{0.2cm}\includegraphics[width=7.3cm]{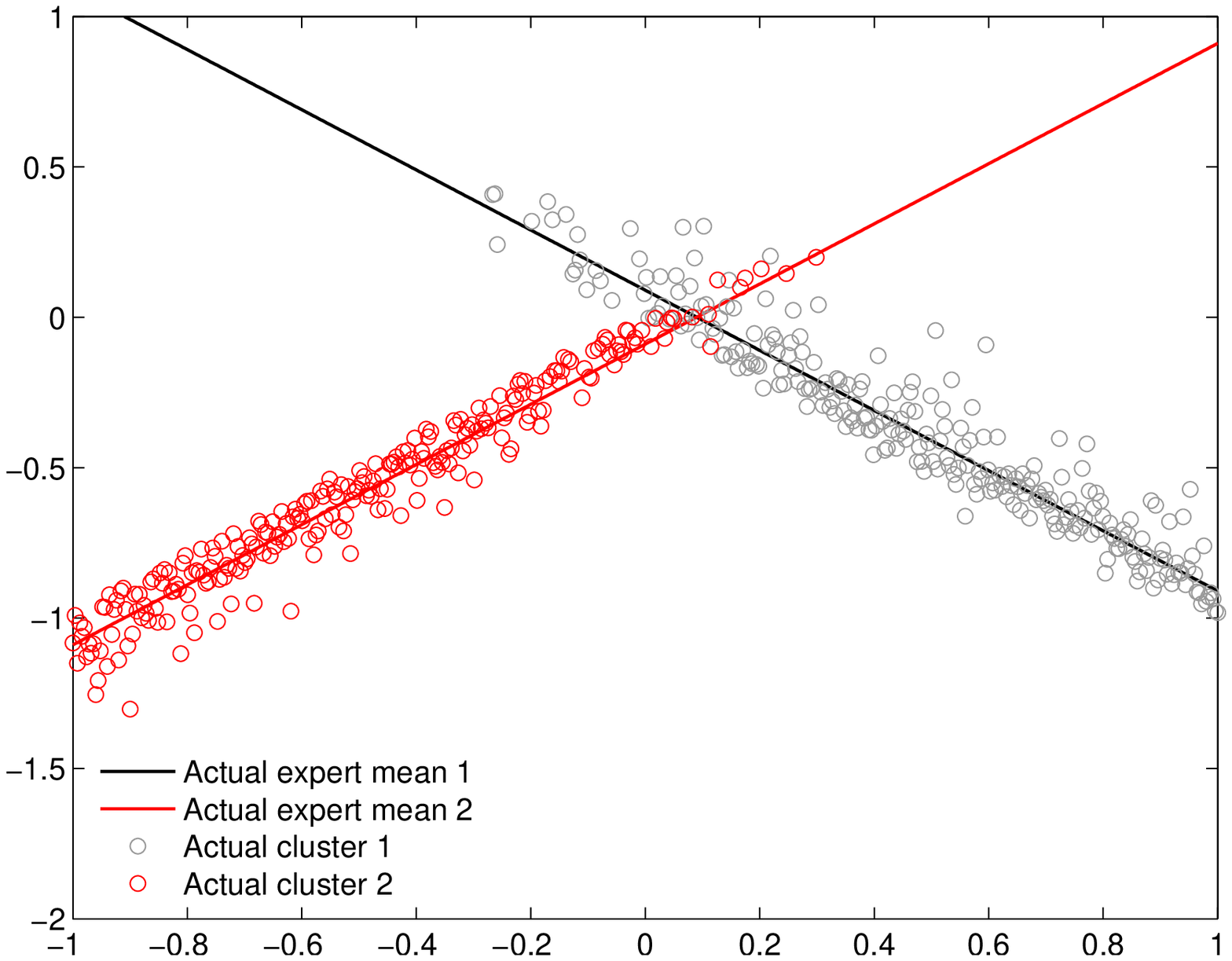}&
  \hspace{0.2cm}\includegraphics[width=7.3cm]{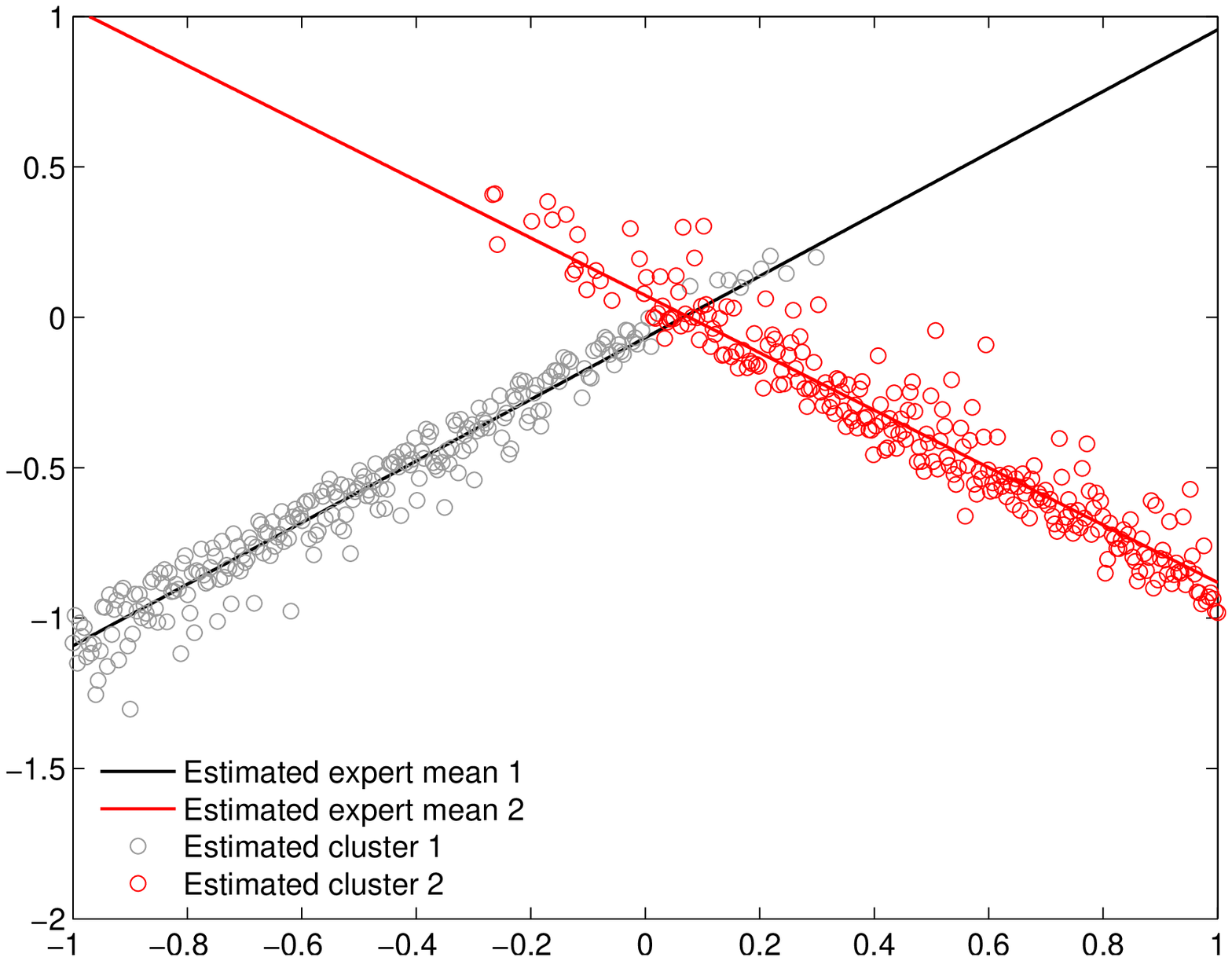} 
   \end{tabular}
   \caption{\label{fig. TwoClust-All->All-estimated partitions}The true and estimated partitions by fitting the proposed NNMoE models to the simulated data set  shown in Figure \ref{fig. TwoClust-All->All-estimated models}. Up: the SNMoE model; Middle: the TMoE model; Bottom: the STMoE model.}
\end{figure}

\subsubsection{Experiment 2}

In this experiment we examine the robustness of the proposed models to outliers  versus the standard NMoE one. 
For that, we considered each of the four models (NMoE, SNMoE, TMoE, and STMoE) for data generation. For each generated sample, each of the four models in considered for the inference. The data were generated  exactly in the same way as in Experiment 1, except for some observations which were generated with a probability $c$ from a class of outliers. We considered the same class of outliers as in \citet{Nguyen2014-MoLE}, that is the predictor $x$  is generated uniformly over the interval $(-1, 1)$ and the response $y$ is set the value $-2$. 
We apply the MoE models by setting the covariate vectors as before, that is,  $\bsx = \bsr  = (1, x)^T$. 
We considered varying probability of outliers $c = 0\%, 1\%, 2\%, 3\%, 4\%, 5\%$ and the sample size of the generated data is $n=500$. An example of simulated sample containing $5\%$ outliers is shown in Figure \ref{fig. TwoClust-Outliers-NMoE_NMoE}. 
As a criterion of evaluation of the impact of the outliers on the quality of the results, we considered the MSE between the true regression mean function and the estimated one. This MSE is calculated as
$\frac{1}{n}\sum_{i=1}^n\!\parallel \! \E_{{\it \bsvPsi}}(Y_i|\bsr_i,\bsx_i) - \E_{{\it \hat\bsvPsi}}(Y_i|\bsr_i,\bsx_i)\!\parallel^2$ where the expectations are computed as in Section \ref{sec: Prediction using the NNMoE}.
%
\subsubsection{Obtained results}

Table \ref{tab. MSE for the mean function - Noisy simulated data : All->all} shows,  for each of the fours models, the results in terms of mean squared error between the true mean function and the estimated one, for an increasing number of outliers in the data.
First, one can see that, when there is no outliers ($c=0\%$), the error of the TMoE is less than those of the other models, for the four situations, that is including the case where the data are not generated according to the TMoE model, which is somewhat surprising. This includes the case where the data aregenrated according to the NMoE model, for which the TMoE error is slightly less than the one of the NMoE model. 
Then, it can be seen that when there is outliers, the TMoE model outperforms the other models for almost all the situations, except the one in which the data are generated according to the STMoE model. When the data do not contain outliers and are generated from the STMoE, this one indeed  outperforms the NMoE and SNMoE models.
For the situation when there is no outliers and the data are generated according to the TMoE or the STMoE, these two models may provide quasi-identical results. 
 In the case of presence of outliers in data generated from the STMoE,  this one outperforms the NMoE and SNMoE models for all the situations, and outperforms the TMoE for the majority of situations, namely when the number of the outliers is more than $2\%$. 
One can also see that, for all the situations with outliers, as expected, the TMoE and STMoE models always provide the best  results. These two models are indeed much more robust to outliers compared to the normal and skew-normal ones because the expert components in these two models follow a robust distribution, that is the $t$ distribution for the TMoE, and the skew $t$ distribution for the STMoE.
The NMoE and SNMoE are  sensitive to outliers. When there is outliers, the SNMoE behavior is comparable to the one of the NMoE. 
However, when the number of outliers is increasing, it can be seen that the increase in the error of the NMoE and SNMoE model is more pronounced compared to the one of the TMoE and STMoE models. The error for both the TMoE and STMoE may indeed slightly increase, remain stable or even decrease in some situations. This supports the expected robustness of the TMoE and STMoE models.
{\setlength{\tabcolsep}{6pt
\begin{table}[H]
\centering
{\small
\begin{tabular}{l l c c c c c c }
\hline
& \hspace{1cm}$c$& $0\%$ & $1\%$ & $2\%$ & $3\%$ & $4\%$ & $5\%$\\
\multicolumn{2}{c}{Model} & & & & & &\\
 \hline
 \hline 
\multirow{4}{*}{\rotatebox[origin=c]{90}{NMoE}}
& NMoE    & 	0.0001783	& 0.001057 & 0.001241 & 0.003631 &	 0.013257 &	0.028966 \\
& SNMoE  & 	0.0001798	& 0.003479 & 0.004258 & 0.015288 &	 0.022056 &	0.028967 \\
& TMoE    & 	\underline{0.0001685}	& \underline{0.000566} & \underline{0.000464} & \underline{0.000221} &	 \underline{0.000263} &	\underline{0.000045} \\
& STMoE  & 	0.0002586	& 0.000741 & 0.000794 & 0.000696 &	 0.000697 &	0.000626 \\
 \hline 
 \hline 
\multirow{4}{*}{\rotatebox[origin=c]{90}{SNMoE}}
& NMoE    & 0.0000229 & 0.000403 & 0.004012 & 0.002793 & 0.018247 & 0.031673 \\
& SNMoE  & 0.0000228 & 0.000371 & 0.004010 & 0.002599 & 0.018247 & 0.031674 \\
& TMoE    & \underline{0.0000325} & \underline{0.000089} & \underline{0.000130} & \underline{0.000513} & \underline{0.000108} & \underline{0.000355} \\
& STMoE  & 0.0000562 & 0.000144 & 0.000022 & 0.000268 & 0.000152 & 0.001041 \\ 
 \hline 
 \hline 
\multirow{4}{*}{\rotatebox[origin=c]{90}{TMoE}}
& NMoE    & 0.0002579 & 0.0004660 & 0.002779 & 0.015692 & 0.005823 & 0.005419 \\
& SNMoE  & 0.0002587 & 0.0004659 & 0.006743 & 0.015686 & 0.005835 & 0.004813 \\
& TMoE    & \underline{0.0002529} & \underline{0.0002520} & \underline{0.000144} & \underline{0.000157} & \underline{0.000488} & \underline{0.000045}\\
& STMoE  & 0.0002473& 0.0002451 & 0.000173 & 0.000176 & 0.000214 & 0.000291 \\
 \hline 
 \hline 
\multirow{4}{*}{\rotatebox[origin=c]{90}{STMoE}} 
& NMoE    & 0.000710 & 0.0007238 & 0.001048 & 0.006066 & 0.012457 & 0.031644 \\
& SNMoE  & 0.000713 & 0.0009550 & 0.001045 & 0.006064 & 0.012456 & 0.031644 \\
& TMoE    & \underline{0.000279} & 0.0003808 & \underline{0.000371} & 0.000609 & 0.000651 & 0.000609 \\
& STMoE  & 0.000280 & \underline{0.0001865} & 0.000447 & \underline{0.000600} & \underline{0.000509} & \underline{0.000602} \\
 \hline
\end{tabular}
}
\caption{\label{tab. MSE for the mean function - Noisy simulated data : All->all}MSE between the estimated mean function and the true one for each of the four  models for a varying probability $c$ of outliers for each simulation. The first column indicates the model used for generating the data and the second one indicates the model used for inference.}
\end{table} 
}

To highlight the robustness to noise of the TMoE and STMoE models, in addition to the previously shown numerical results, 
figures 
\ref{fig. TwoClust-Outliers-NMoE_NMoE},
\ref{fig. TwoClust-Outliers-NMoE_SNMoE},
\ref{fig. TwoClust-Outliers-NMoE_TMoE}, 
and
\ref{fig. TwoClust-Outliers-NMoE_STMoE}
show an example of results obtained on the same data set by, respectively, the NMoE, the SNMoE, TMoE, and the STMoE. The data are generated by the NMoE model and contain $c=5\%$ of outliers. 

In this example, we clearly see that both the NMoE model and the SNMoE are severely affected by the outliers. They provide a rough fit especially for the second component whose estimation is affected by the outliers.
However, one can  see that both the TMoE and the STMoE model clearly provide a precise fit; the estimated mean functions and expert components are very close to the true ones. The TMoE and the STMoE are  robust to outliers, in terms of estimating the true model as well as in terms of estimating the true partition of the data (as shown in the middle plots of the data).  
 %
Notice that for the TMoE and the STMoE, the confidence regions are not shown because for this situation the estimated degrees of freedom are less than $2$ ($1.5985$ and   $1.5253$ for the TMoE, and $1.6097$    $1.5311$ for the STMoE); Hence the variance for these models in that case is not defined (see Section \ref{sec: Prediction using the NNMoE}). 
The TMoE and STMoE models provide indeed components with small degrees of freedom corresponding to highly heavy tails, which allow to handle outliers in this noisy case.
\begin{figure}[H]
   \centering 
\begin{tabular}{cc}
   \includegraphics[width=7.5cm]{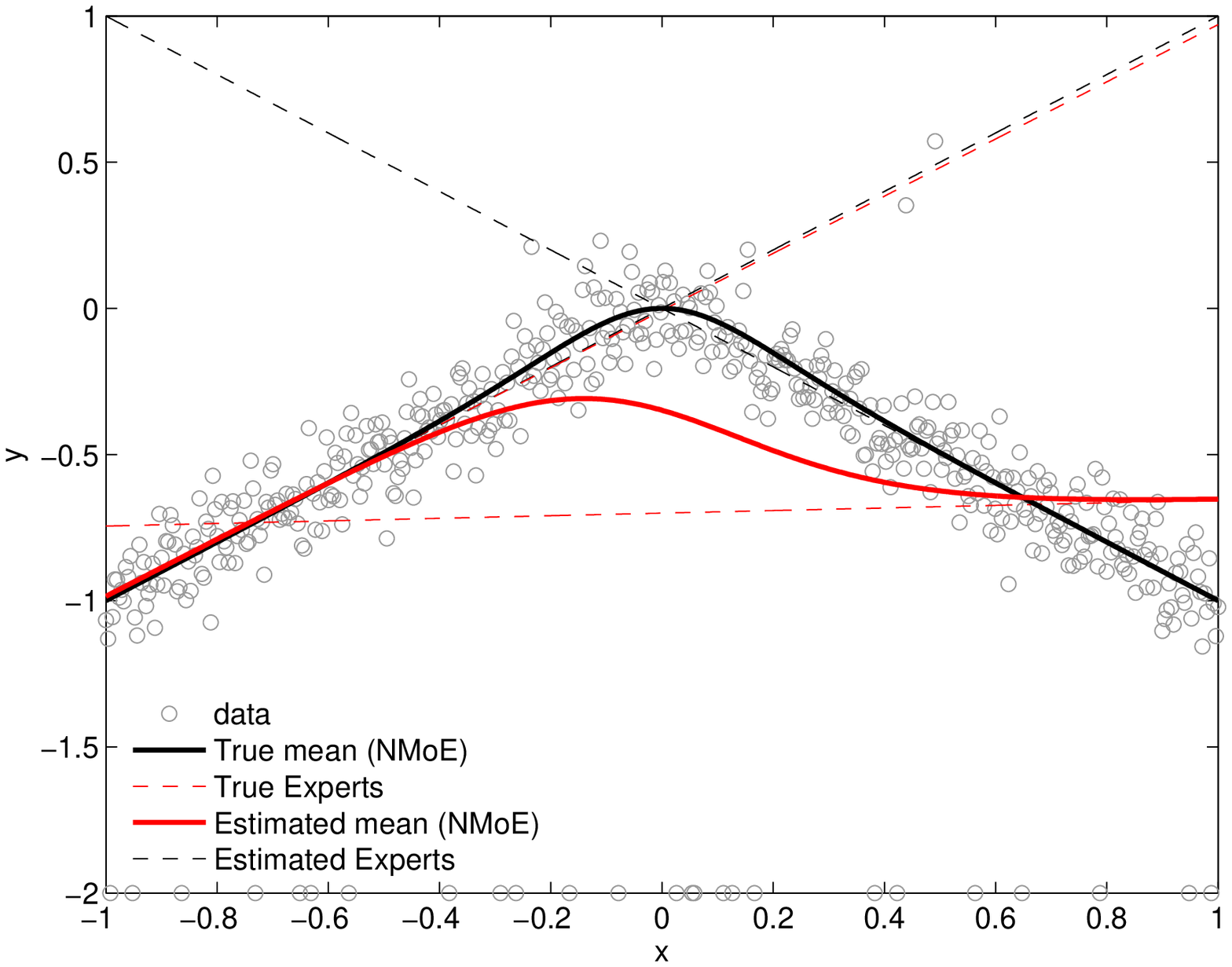}&  
   \includegraphics[width=7.5cm]{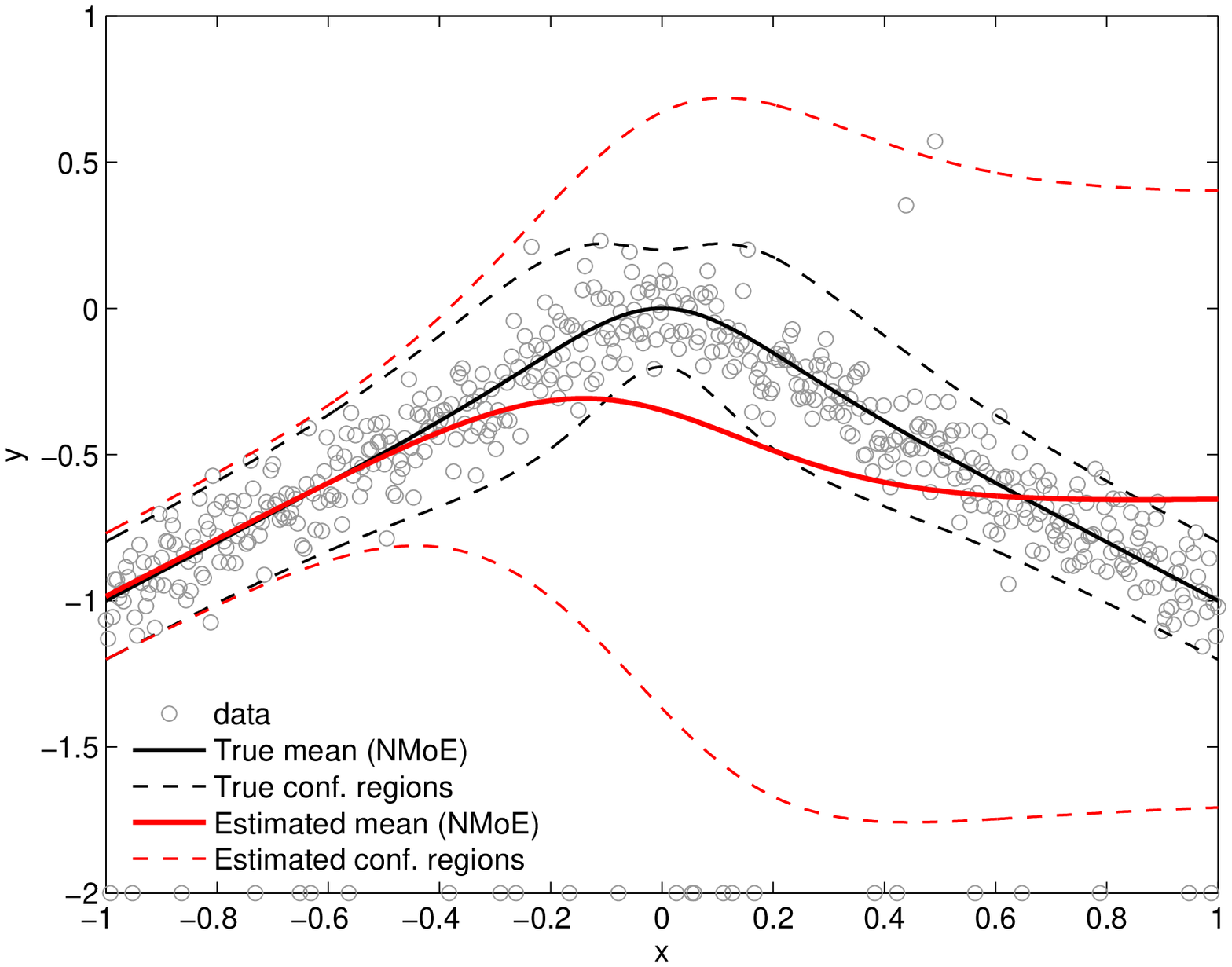}\\
   \hspace{0.3cm}\includegraphics[width=7.2cm]{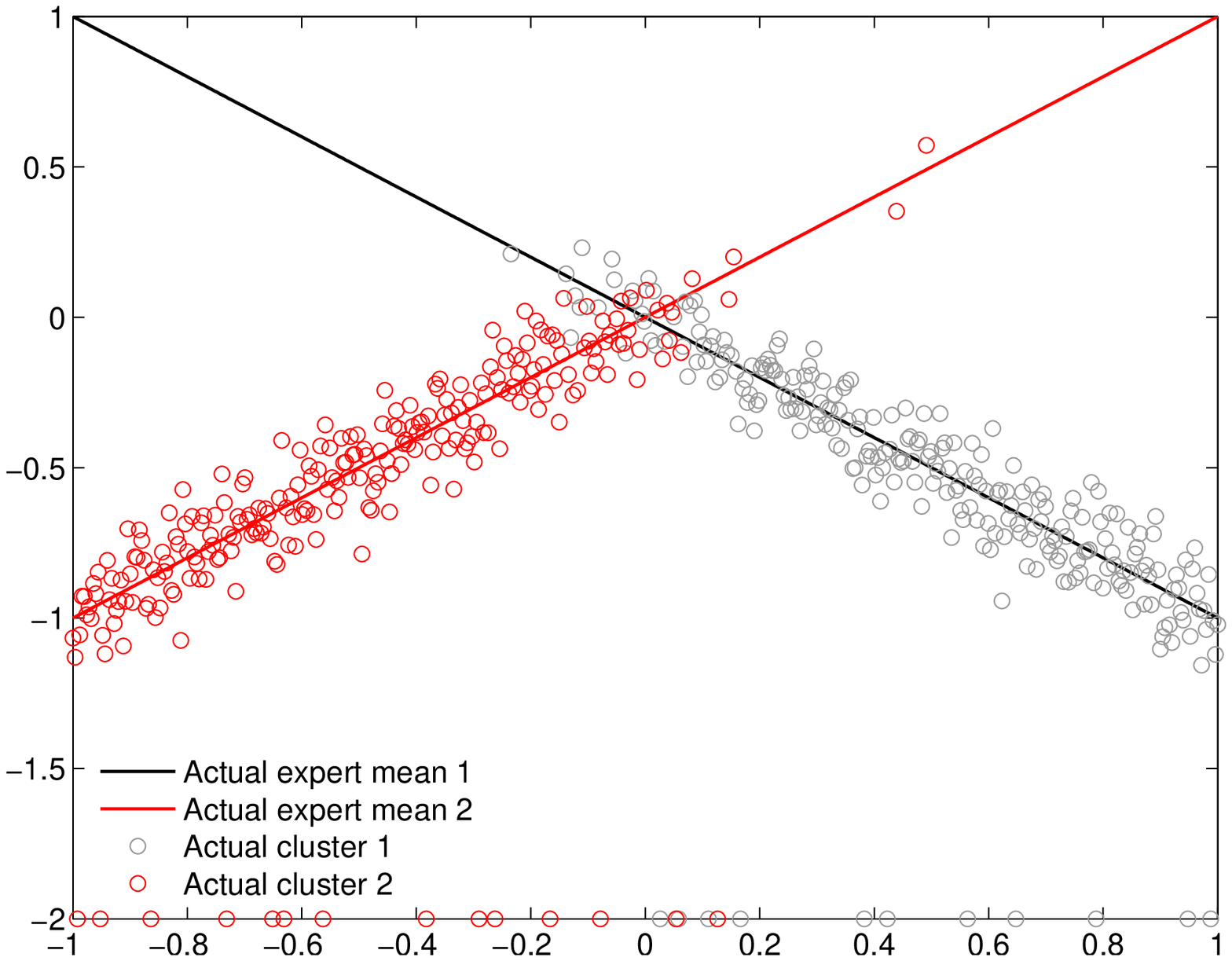}&
   \hspace{0.2cm}\includegraphics[width=7.2cm]{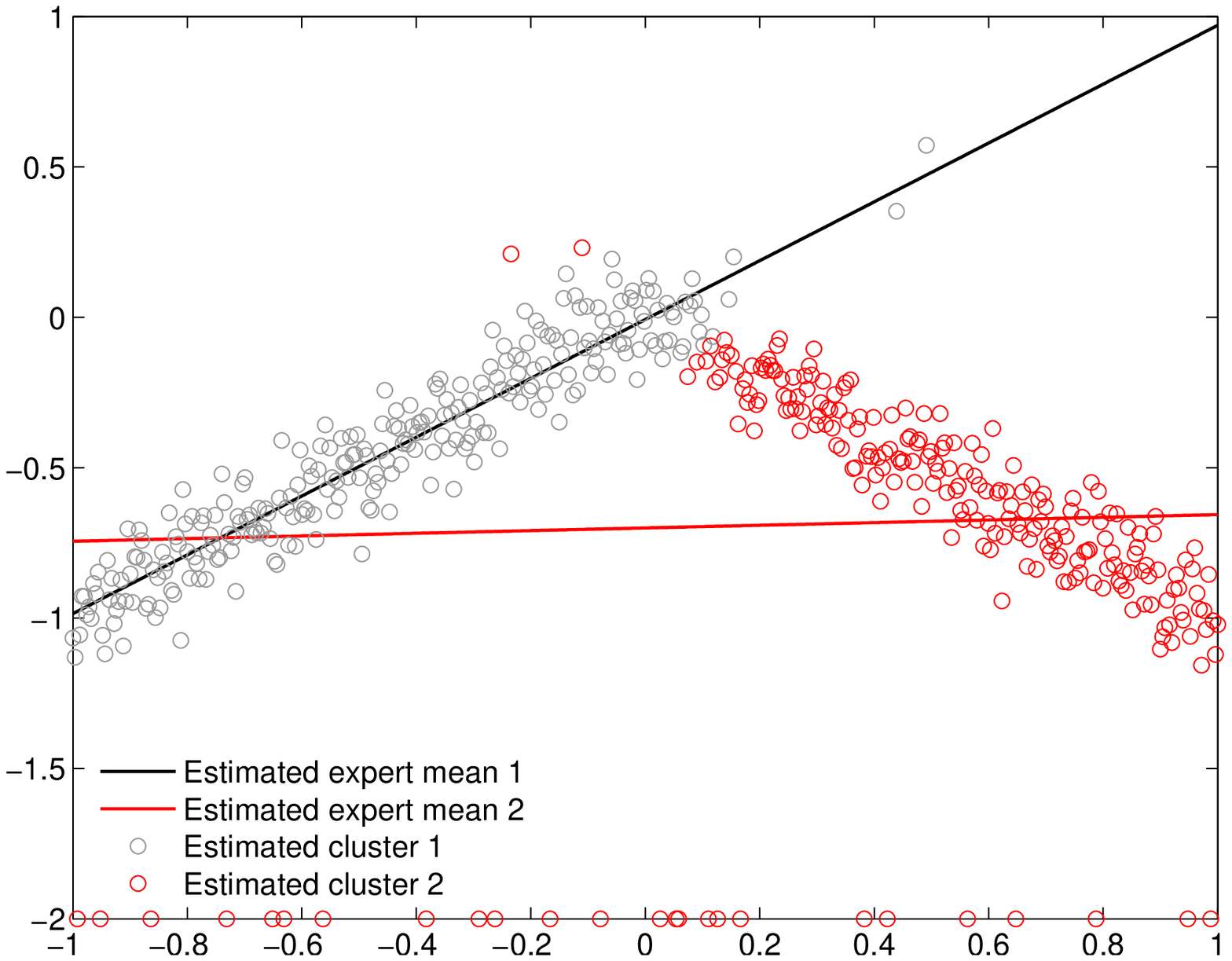} \\
   \includegraphics[width=7.5cm]{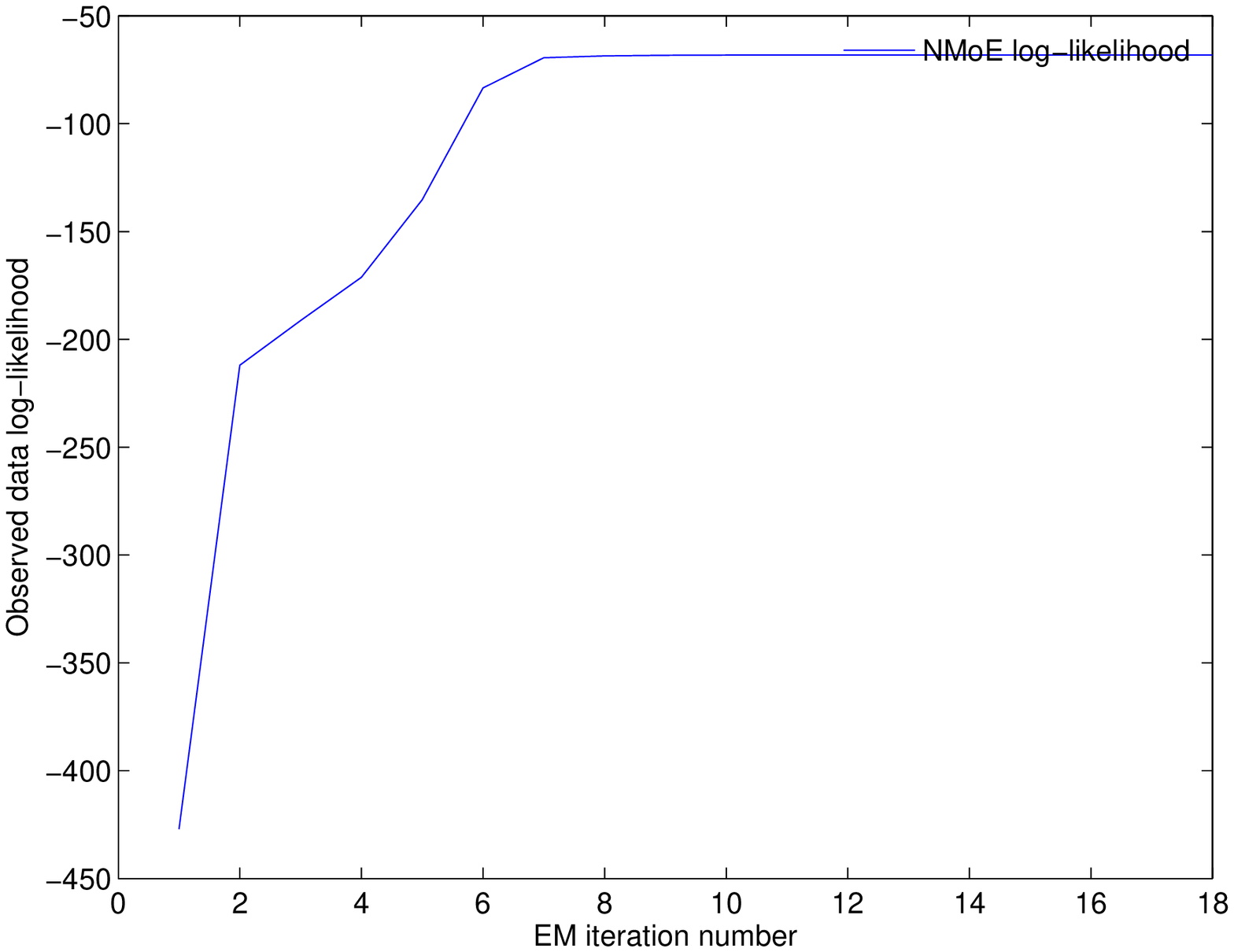} &
   \includegraphics[width=7.5cm]{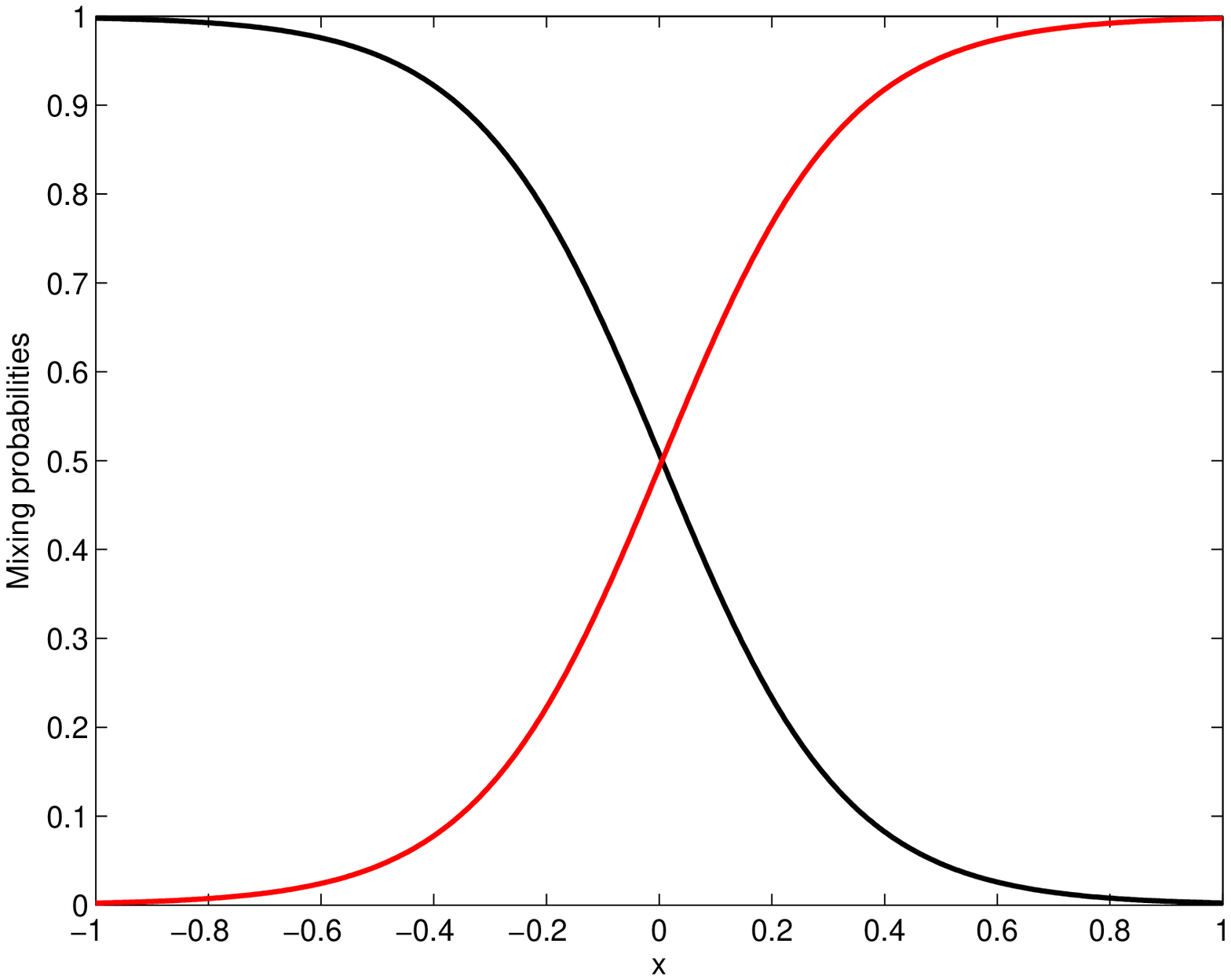}\\
   \end{tabular}
      \caption{\label{fig. TwoClust-Outliers-NMoE_NMoE}Fitted NMoE model to a data set of $n=500$ observations generated according to the NMoE model  and including $5\%$ of outliers.}
\end{figure}
\begin{figure}[H]
   \centering  
   \begin{tabular}{cc}
   \includegraphics[width=7.5cm]{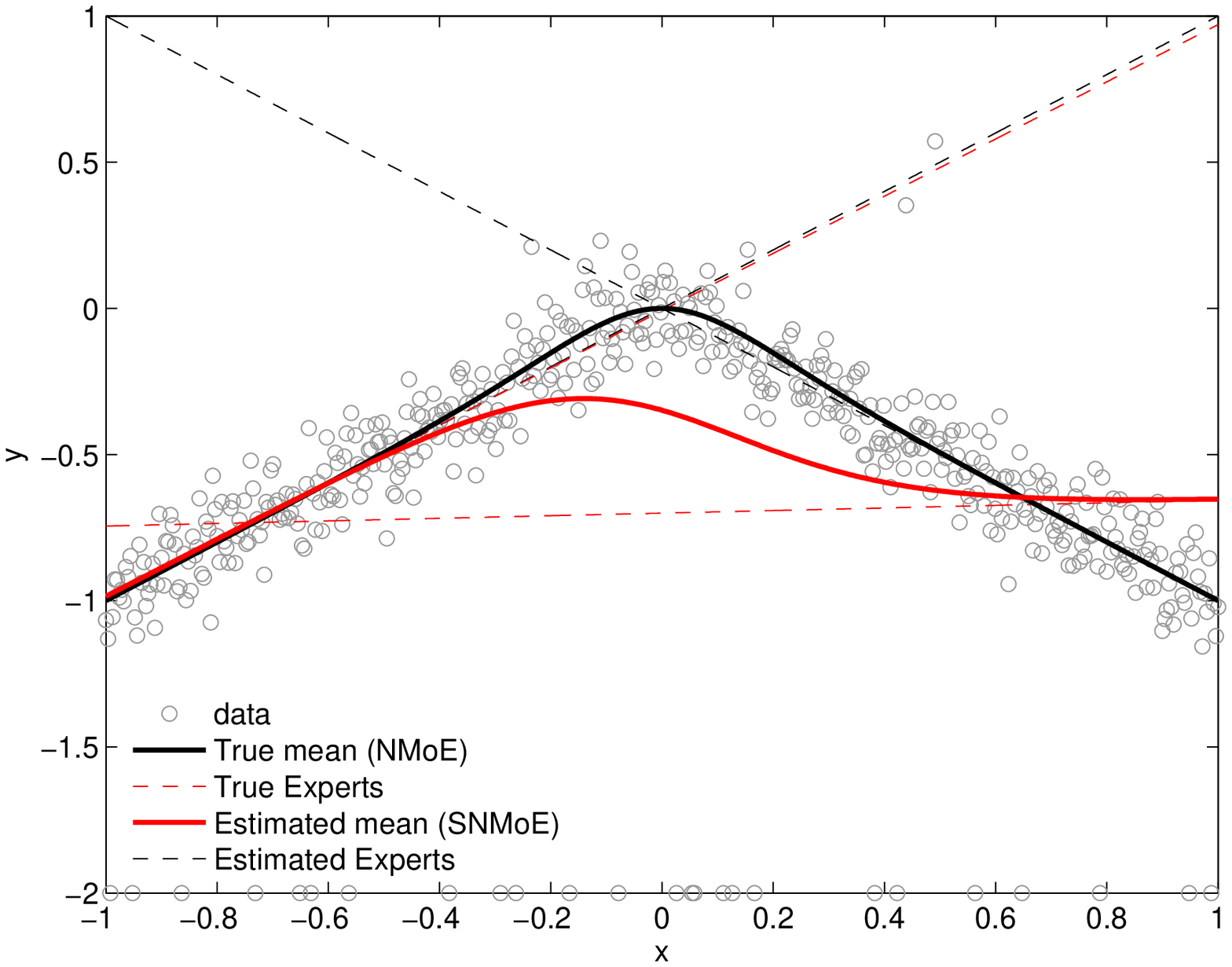}&  
   \includegraphics[width=7.5cm]{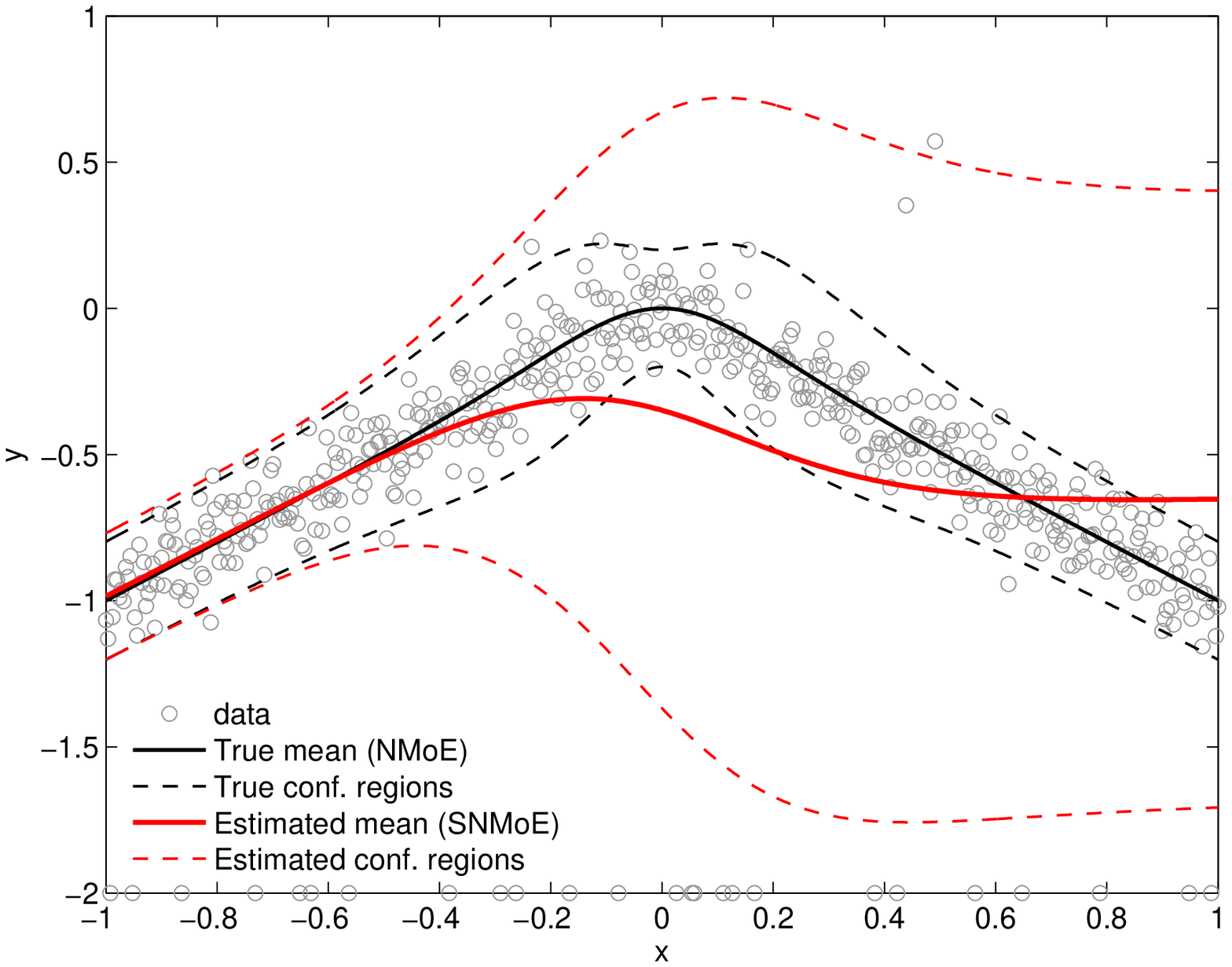}\\
   \hspace{0.3cm}\includegraphics[width=7.2cm]{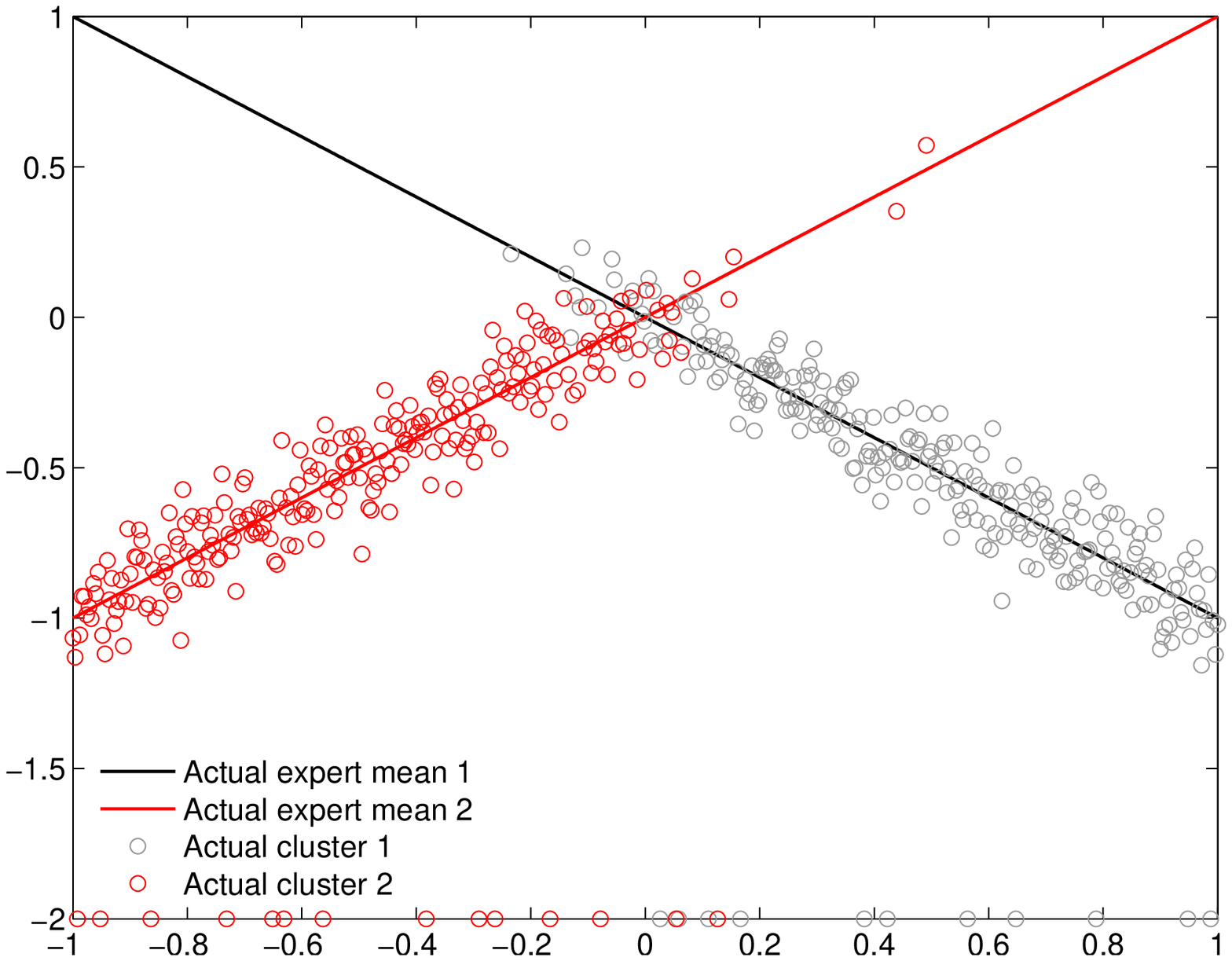}&
    \hspace{0.2cm}\includegraphics[width=7.2cm]{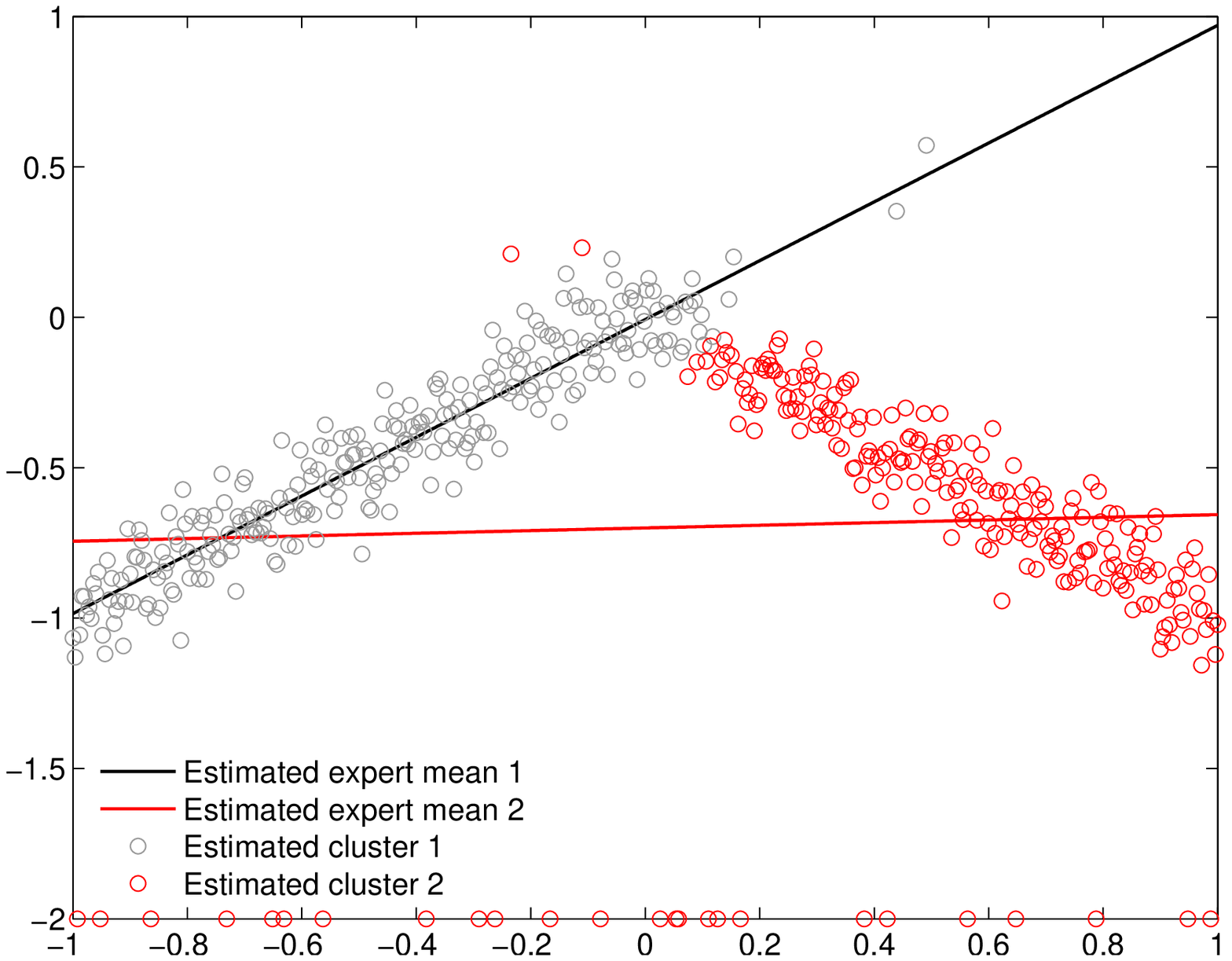} \\
   \includegraphics[width=7.5cm]{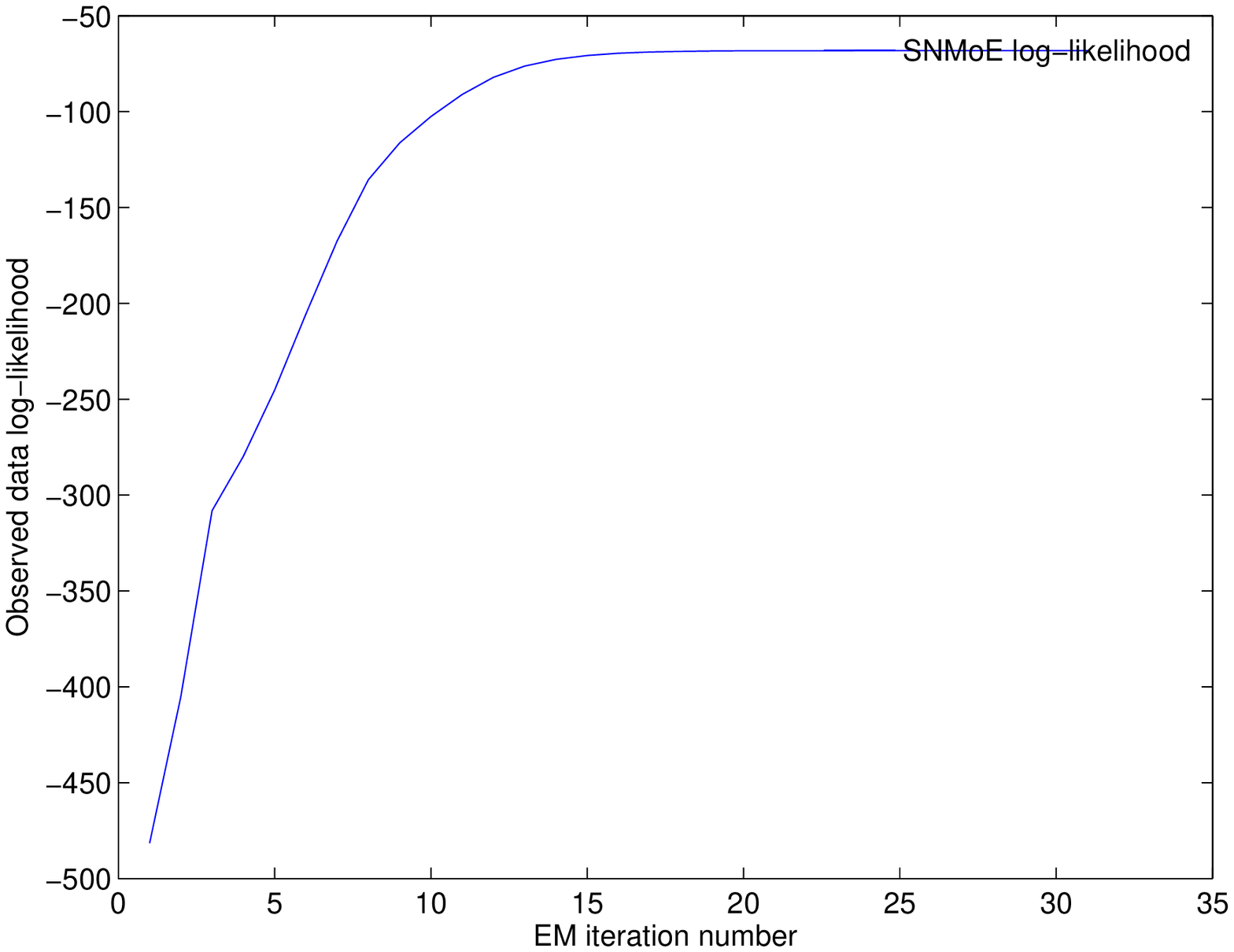}&
    \includegraphics[width=7.5cm]{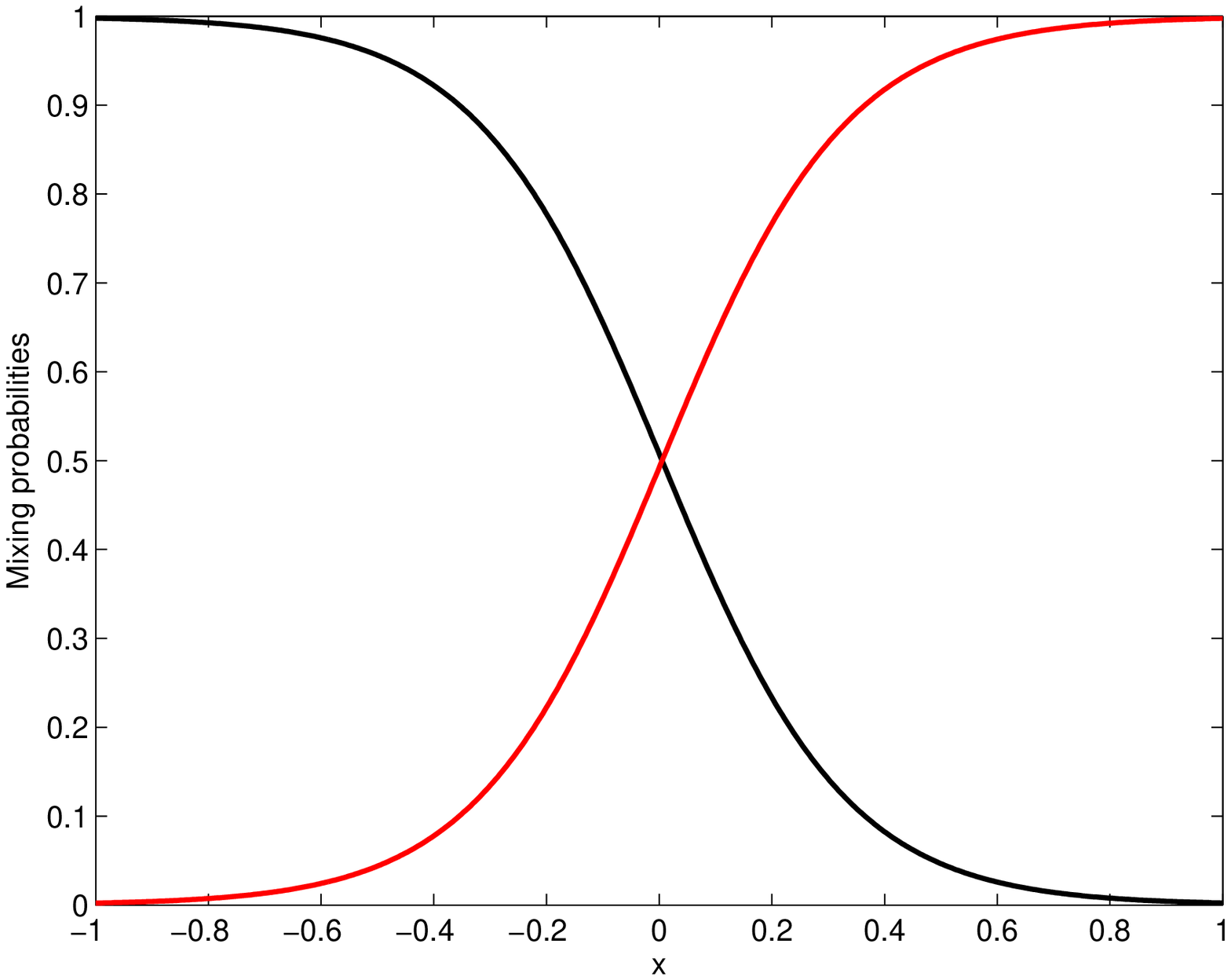}
   \end{tabular}
      \caption{\label{fig. TwoClust-Outliers-NMoE_SNMoE}Fitted SNMoE model to a data set of $n=500$ observations generated according to the NMoE model  and including $5\%$ of outliers.}
\end{figure} 
\begin{figure}[H]
   \centering  
   \begin{tabular}{cc}
   \includegraphics[width=7.5cm]{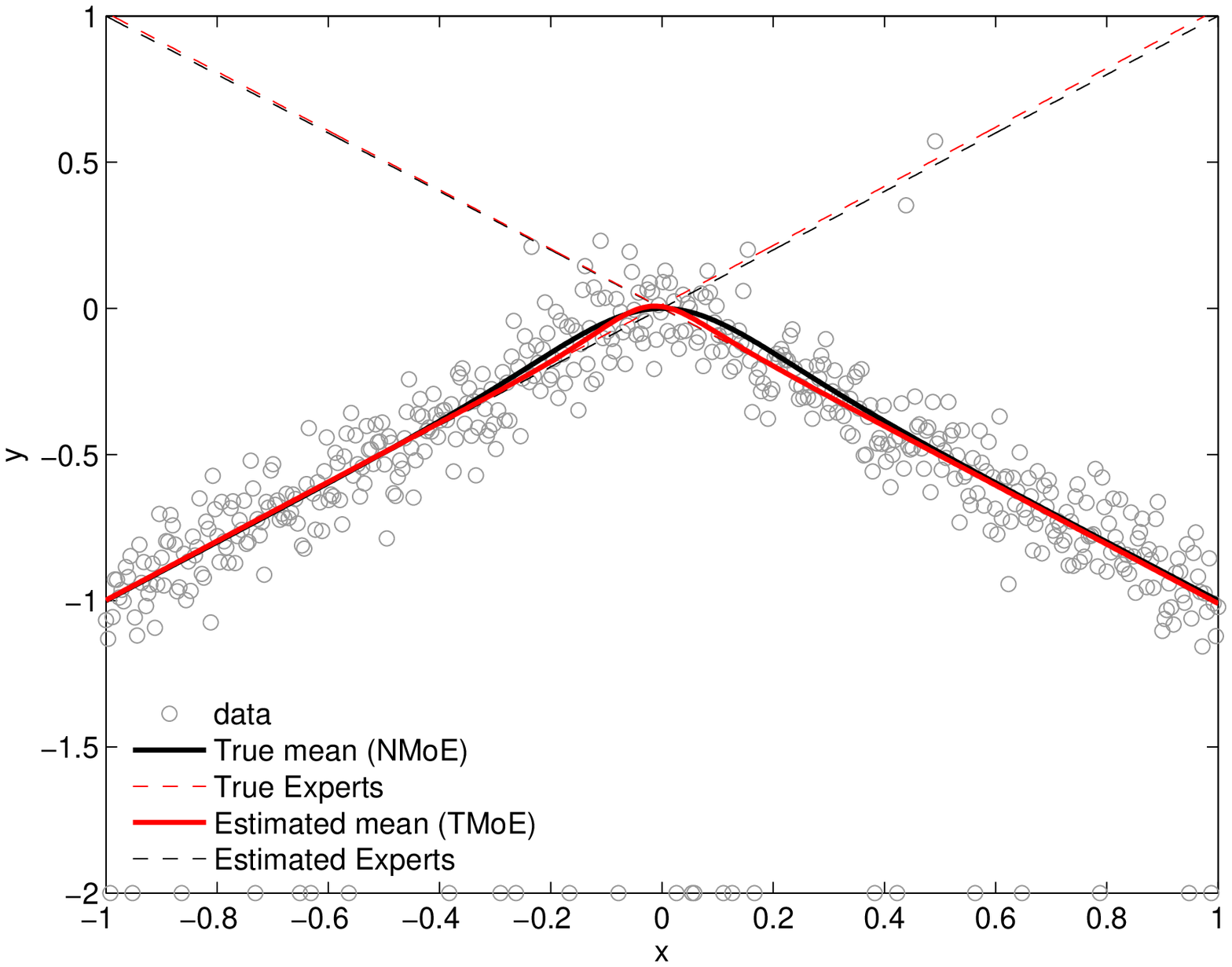}&  
   \includegraphics[width=7.5cm]{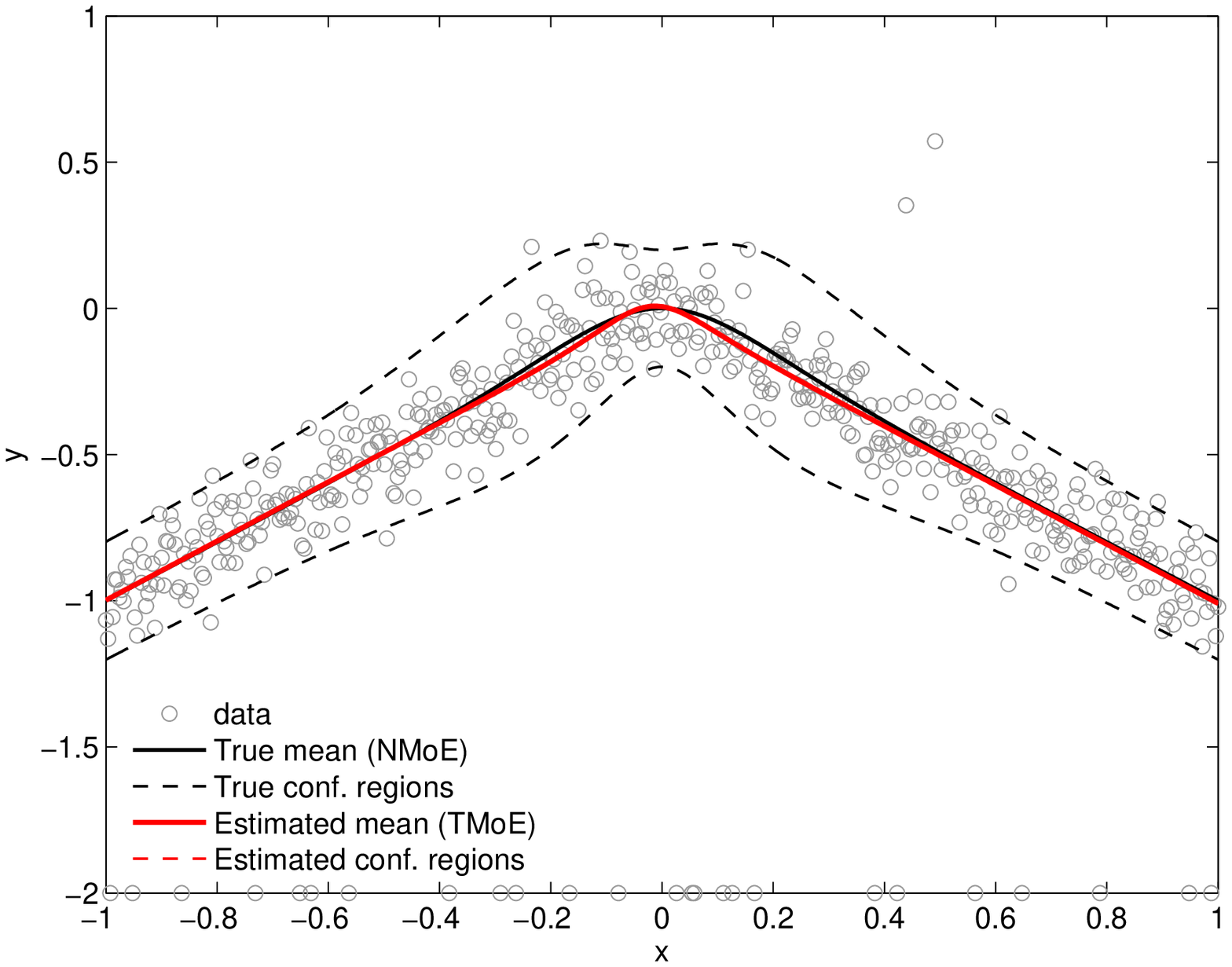}\\
   \hspace{0.3cm}\includegraphics[width=7.2cm]{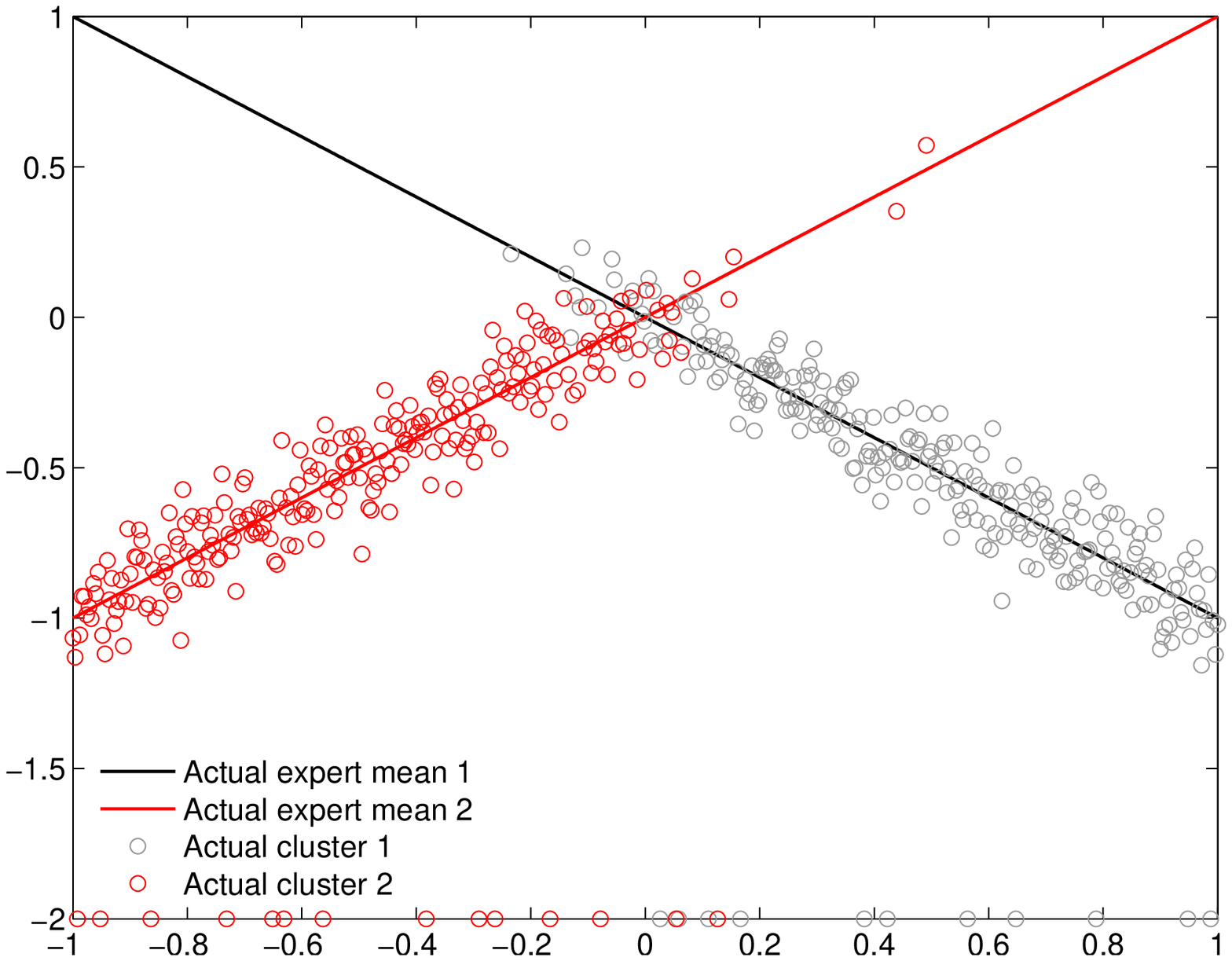}&
   \hspace{0.2cm}\includegraphics[width=7.2cm]{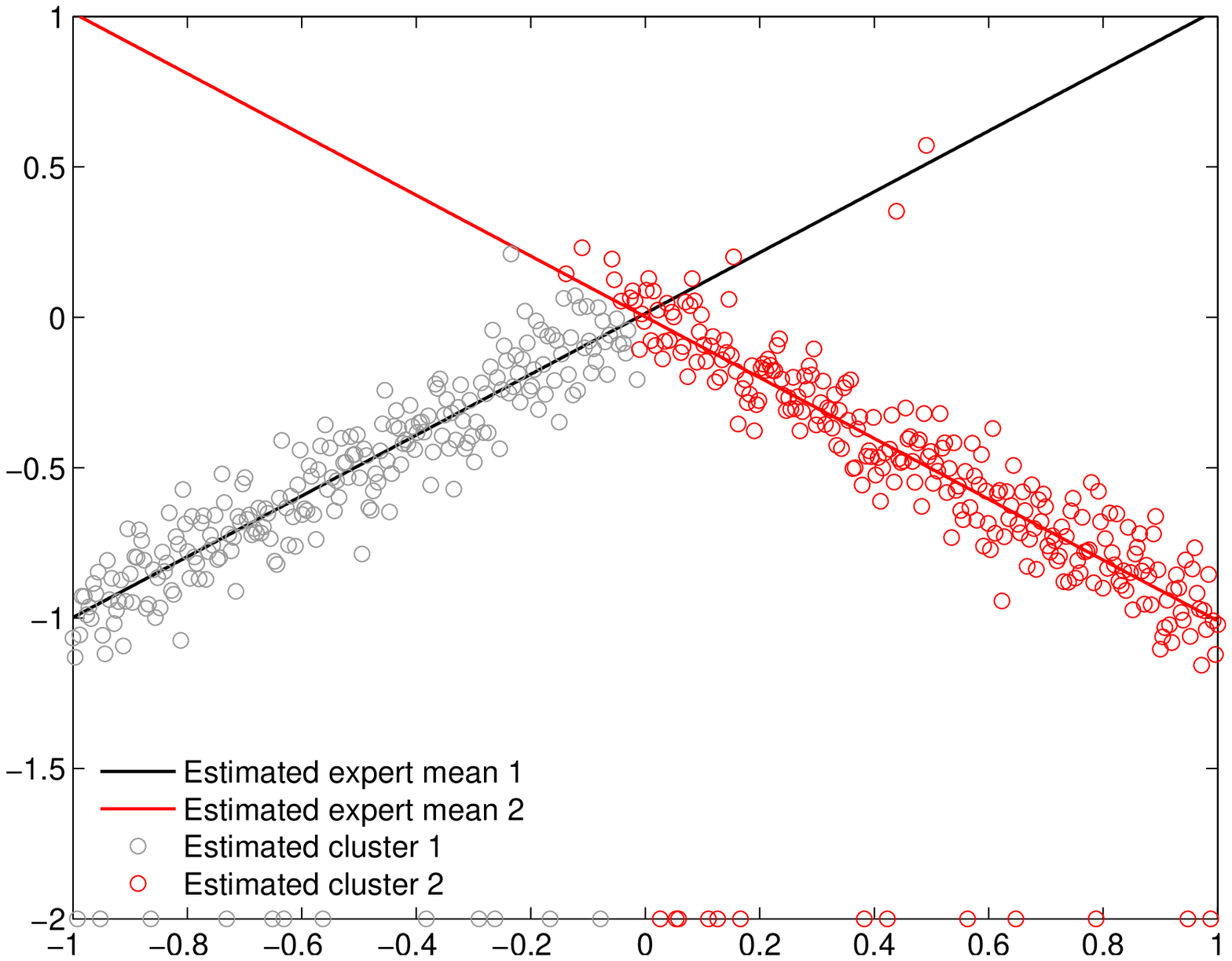}\\
   \includegraphics[width=7.5cm]{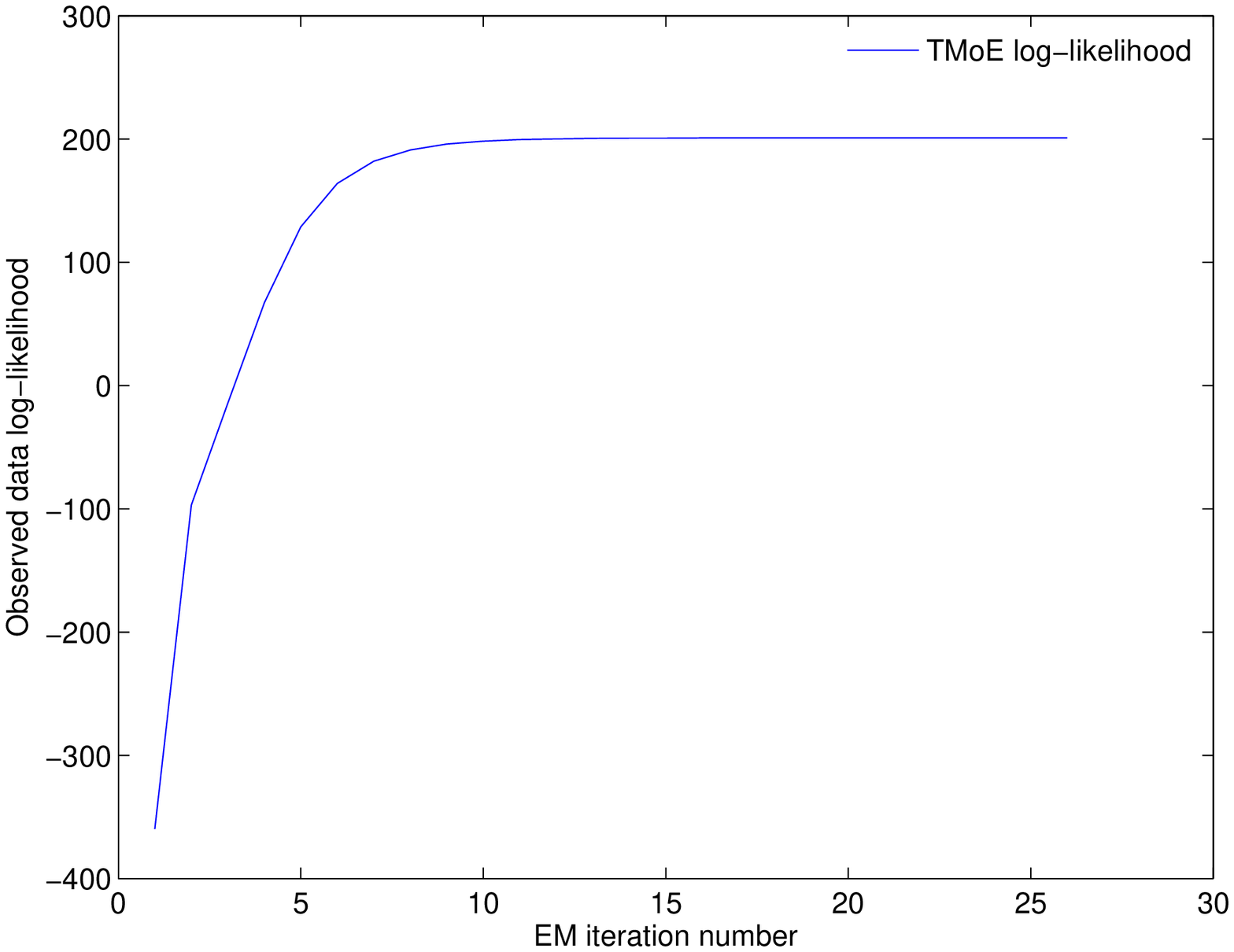} &
   \includegraphics[width=7.5cm]{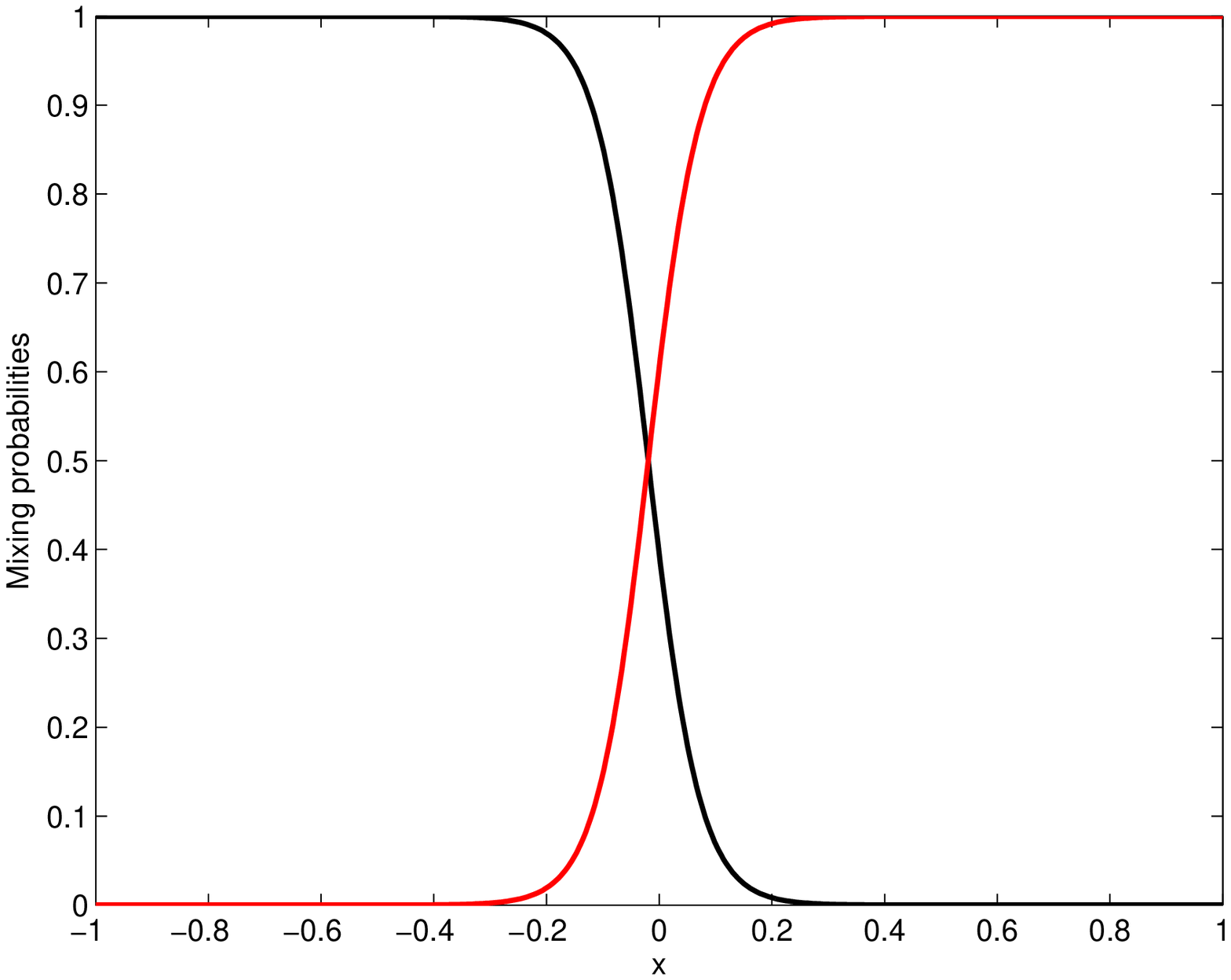}
   \end{tabular}
      \caption{\label{fig. TwoClust-Outliers-NMoE_TMoE}Fitted TMoE model to a data set of $n=500$ observations generated according to the NMoE model  and including $5\%$ of outliers.}
\end{figure}
\begin{figure}[H]
   \centering  
   \begin{tabular}{cc}
   \includegraphics[width=7.5cm]{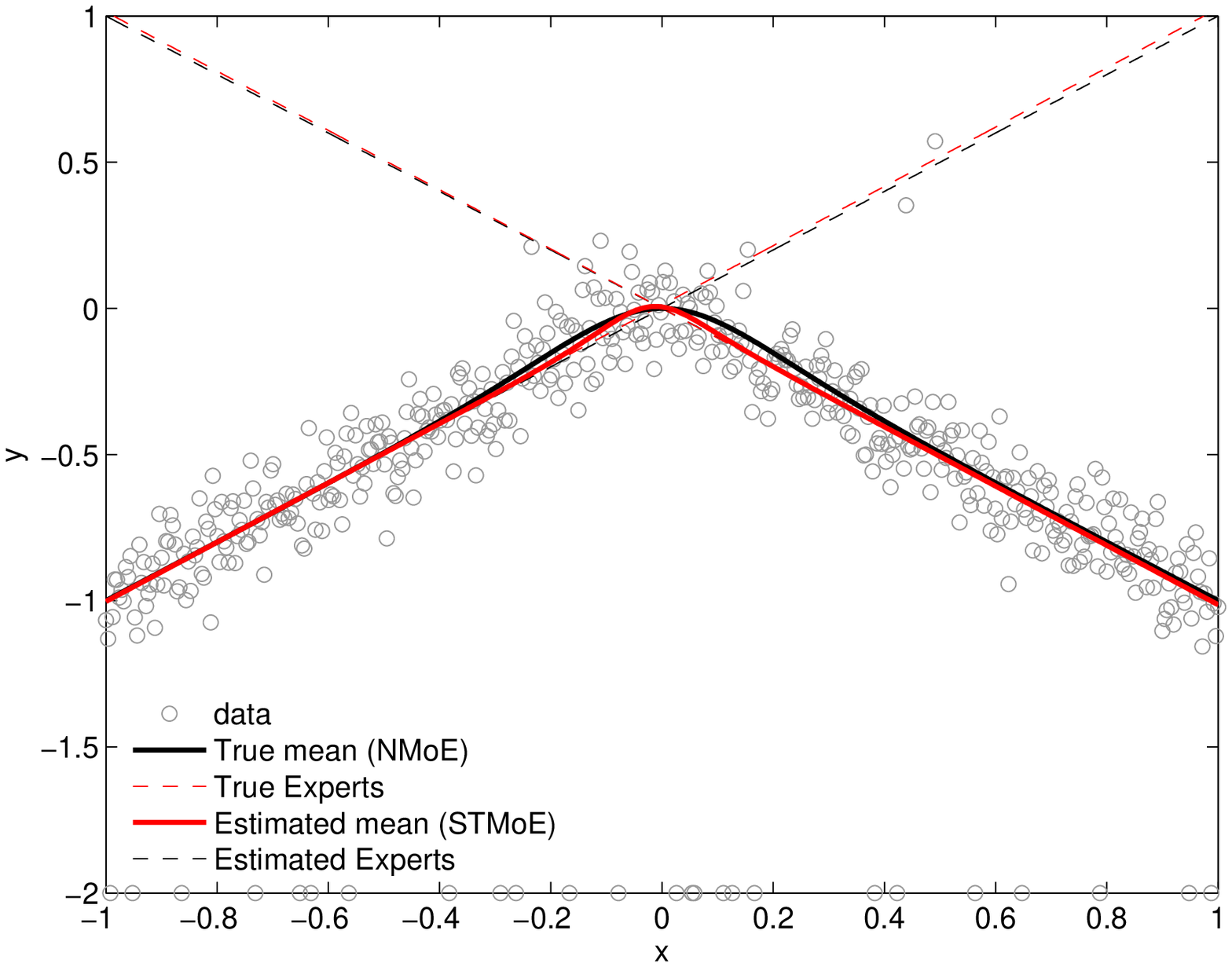} &
   \includegraphics[width=7.5cm]{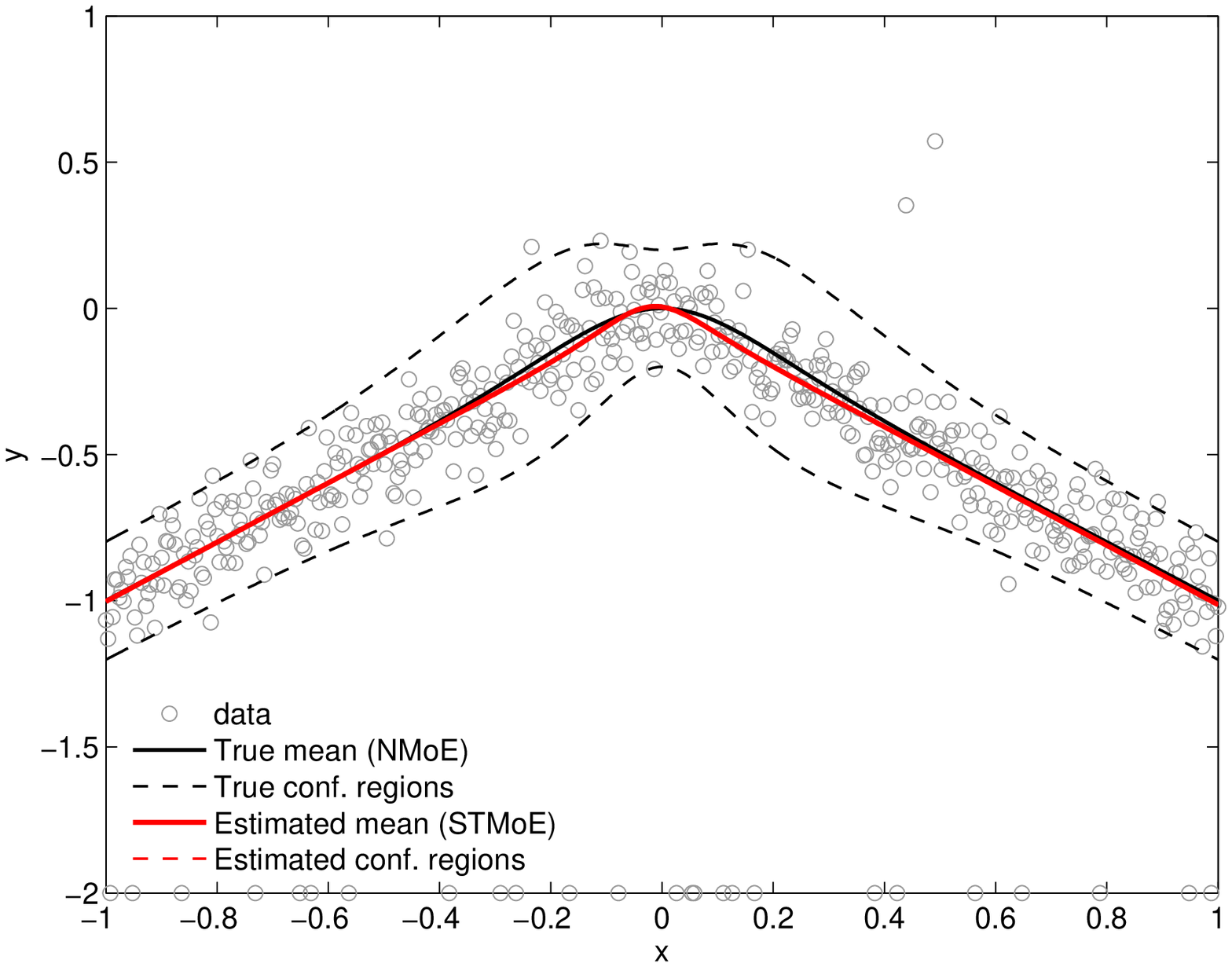}\\
   \hspace{0.3cm}\includegraphics[width=7.2cm]{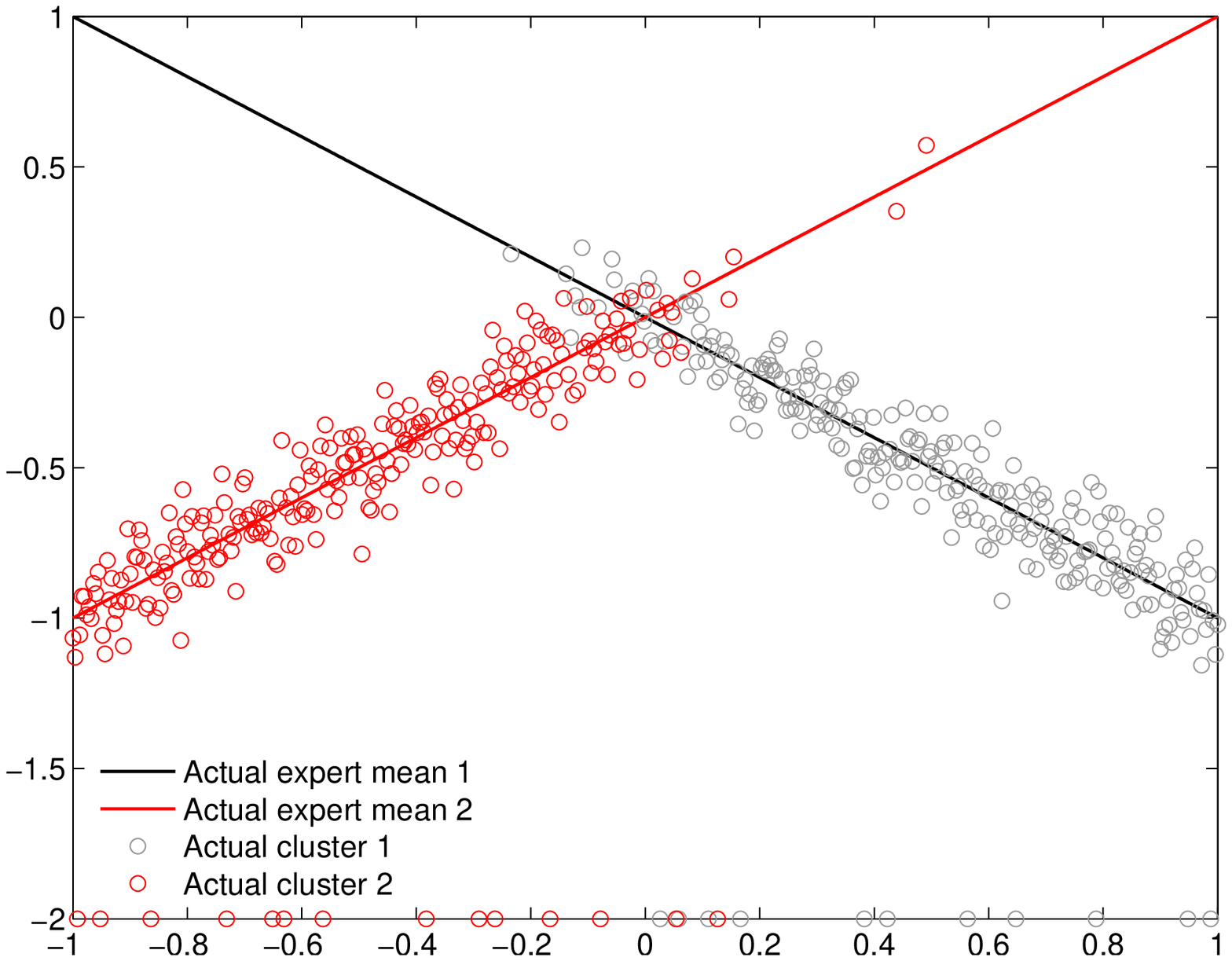}&
   \hspace{0.2cm}\includegraphics[width=7.2cm]{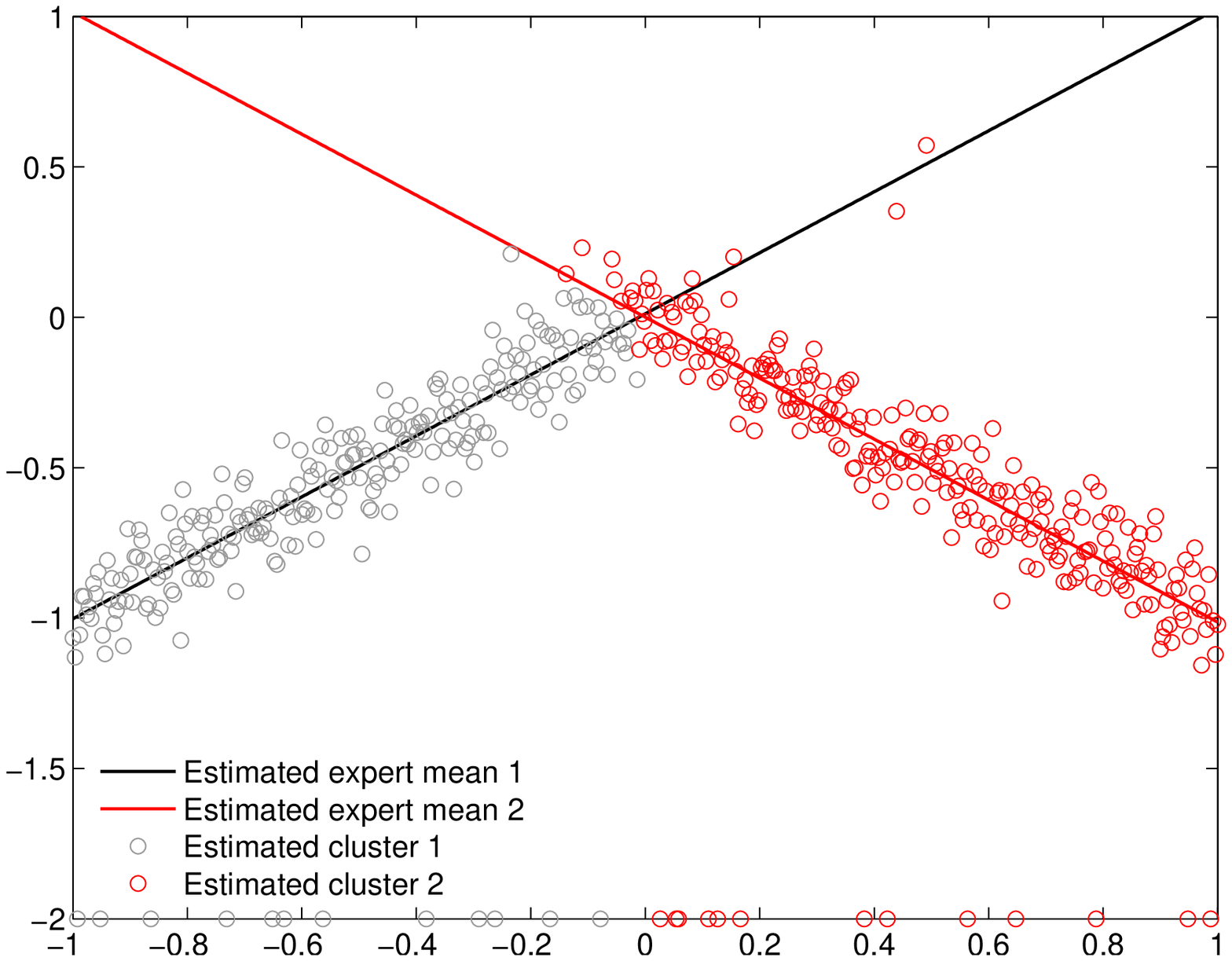}\\
   \includegraphics[width=7.5cm]{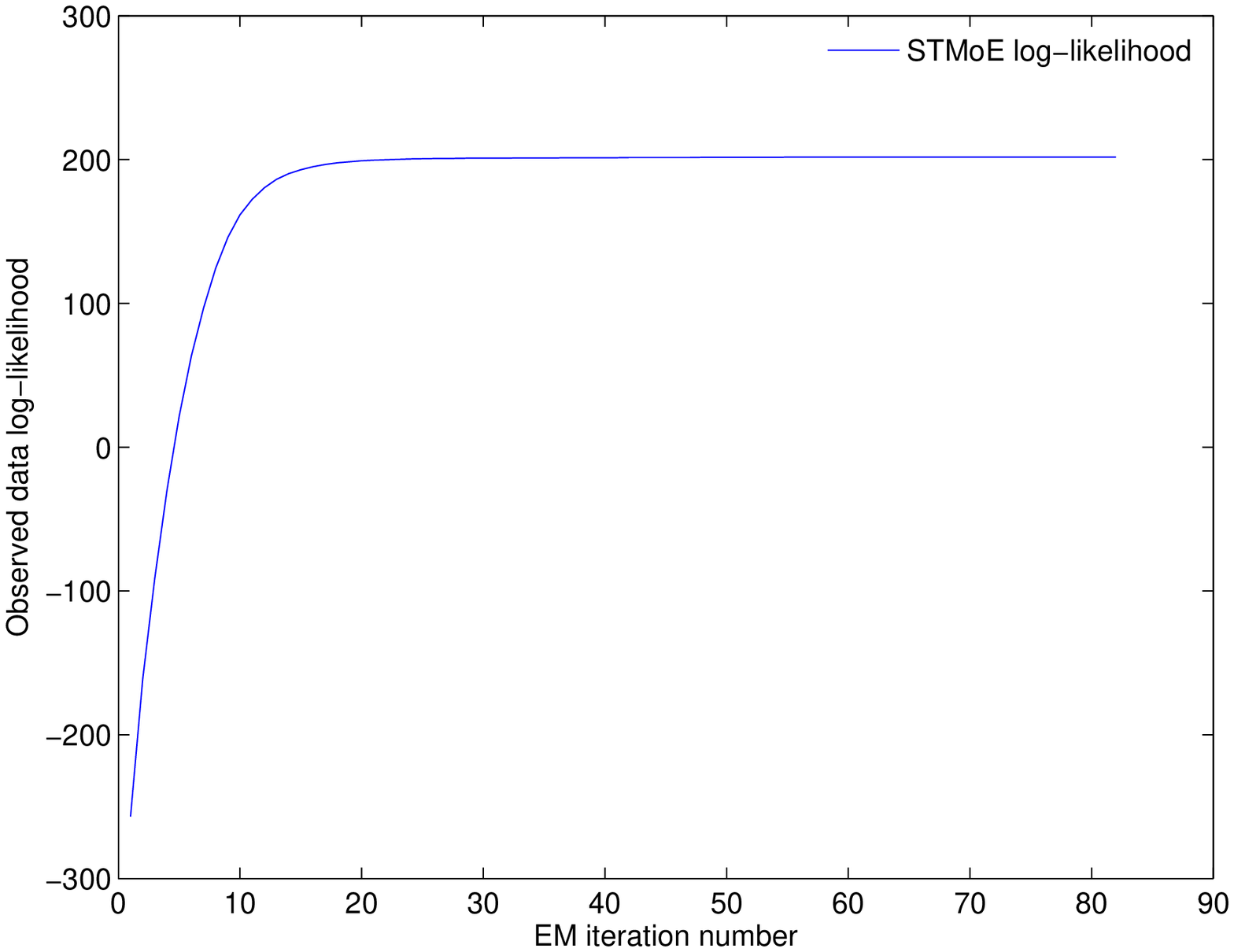} & 
   \includegraphics[width=7.5cm]{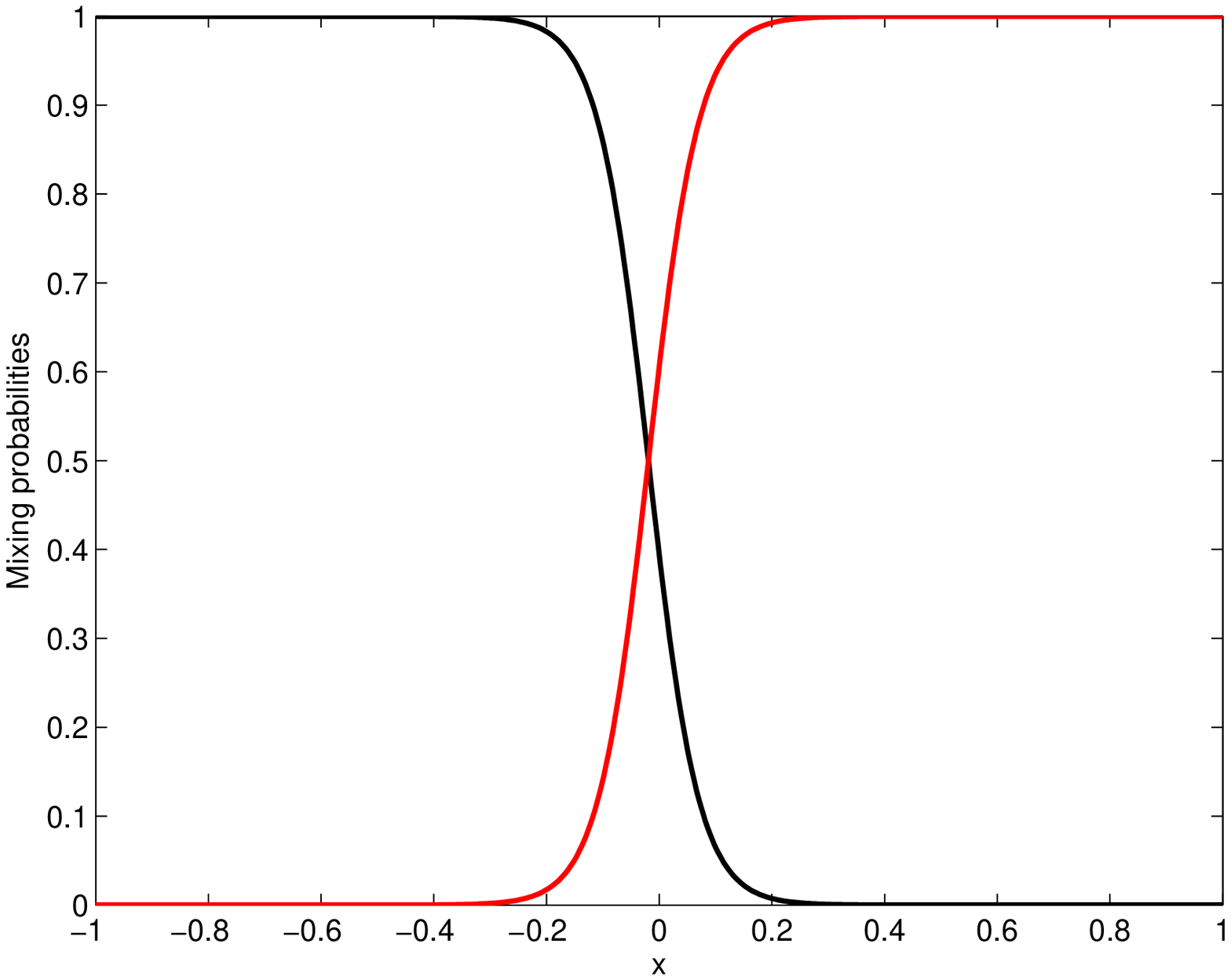}  
   \end{tabular}
      \caption{\label{fig. TwoClust-Outliers-NMoE_STMoE}Fitted STMoE model to a data set of $n=500$ observations generated according to the NMoE model  and including $5\%$ of outliers.}
\end{figure}

\subsection{Application to two real-world data sets}

In this section, we consider an application to two real-world data sets: the tone perception data set and the temperature anomalies data set shown in Figure \ref{fig. Tone and temperature anomalies data}.
\begin{figure}[H]
   \centering 
   \begin{tabular}{cc}
   \includegraphics[width=7.5cm]{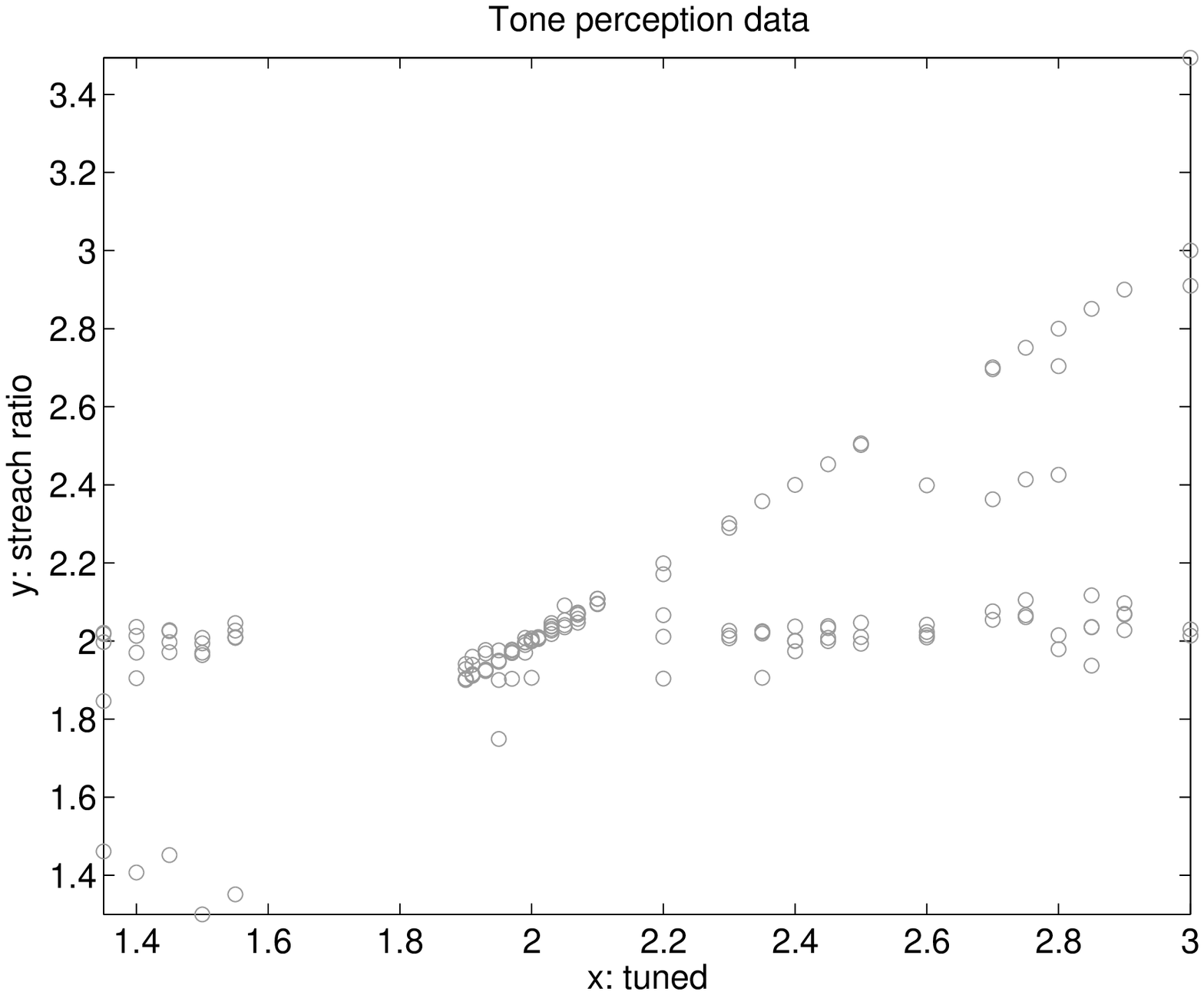} & 
   \includegraphics[width=7.5cm]{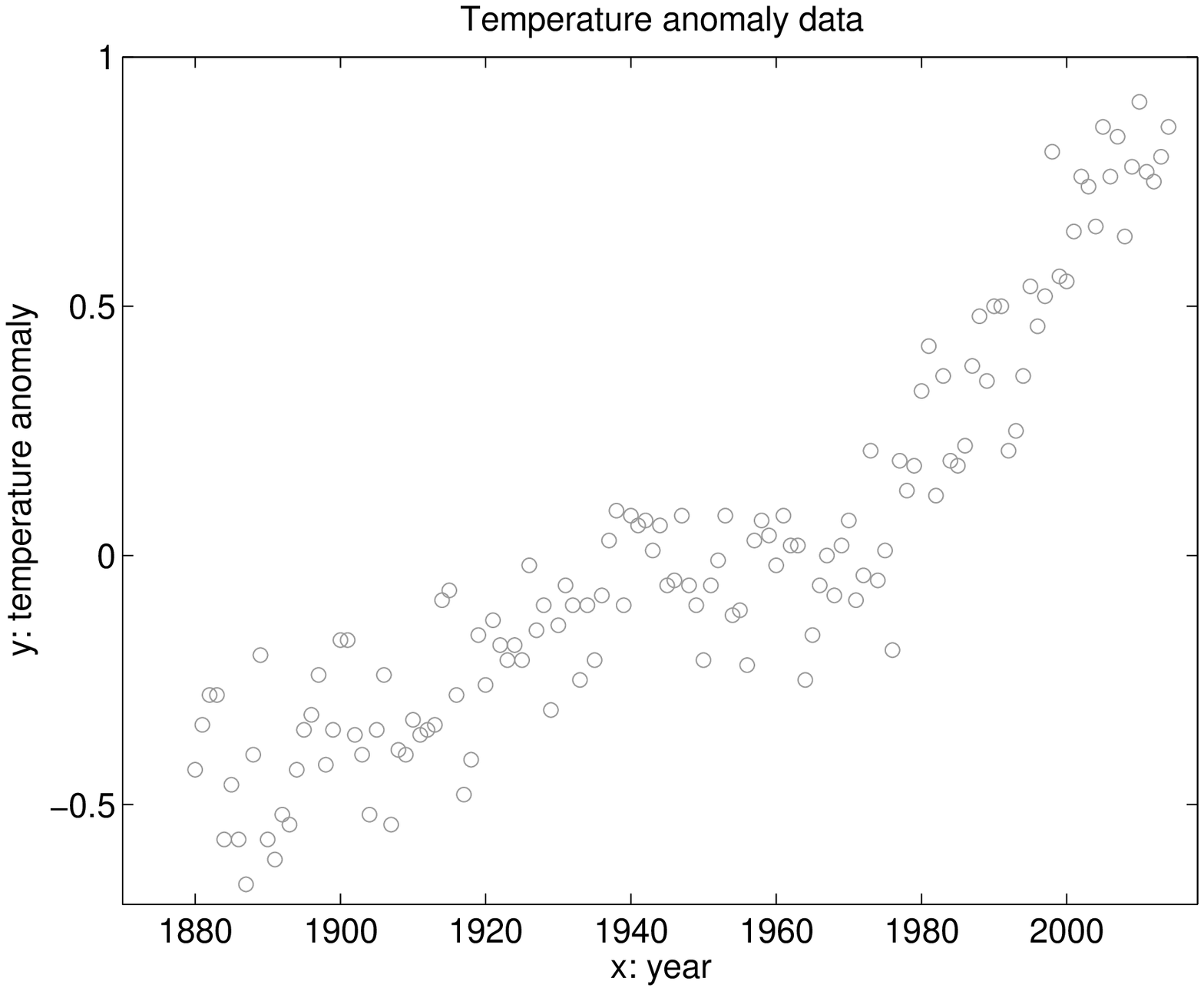} 
   \end{tabular}
      \caption{\label{fig. Tone and temperature anomalies data}Scatter plots of the tone perception data and the temperature anomalies data.}
\end{figure}

\subsubsection{Tone perception data set} The first analyzed data set is the real tone perception data set\footnote{Source: \url{http://artax.karlin.mff.cuni.cz/r-help/library/fpc/html/tonedata.html}} which goes back to \citet{Cohen1984}. It was recently studied by \citet{Bai2012} and \citet{Song2014} by using robust regression mixture models based on, respectively, the $t$ distribution and the Laplace distribution.
In the tone perception experiment, a pure fundamental tone was played to a trained musician. Electronically generated overtones were added, determined by a stretching ratio (``stretch ratio" = 2) which corresponds to the harmonic pattern usually heard in traditional definite pitched instruments. 
The musician was asked to tune an adjustable tone to the octave above the fundamental tone and a ``tuned'' measurement gives the ratio of the adjusted tone to the fundamental. 
The obtained data consists of $n=150$ pairs of ``tuned'' variables, considered here as  predictors  ($x$), and their corresponding ``strech ratio'' variables considered as responses ($y$). 
To apply the proposed MoE models, we set the response $y_i (i=1,\ldots,150)$ as the ``strech ratio'' variables and the covariates $\bsx_i = \bsr_i = (1,x_i)^T$ where $x_i$ is the ``tuned'' variable of the $i$th observation. We also follow the study in \citet{Bai2012} and \citet{Song2014} by using two mixture components. Model selection results are given later in Table \ref{tab. Model selection Tone data}.
 
Figure  \ref{fig. Original Tone data and all models} shows the scatter plots of the tone perception data and the linear expert components of the fitted NMoE model and the proposed SNMoE, TMoE, and STMoE models. 
One can observe that we obtain a good fit with all the models. 
The NMoE and SNMoE are quasi-identical, and differ very slightly from those of the TMoE and STMoE, which are very similar.
%
The two regression lines may correspond to correct tuning and tuning to the first overtone, respectively, as analyzed in \citet{Bai2012}. 
%
%
\begin{figure}[htbp]
   \centering 
   \begin{tabular}{cc}
   \includegraphics[width=7.5cm]{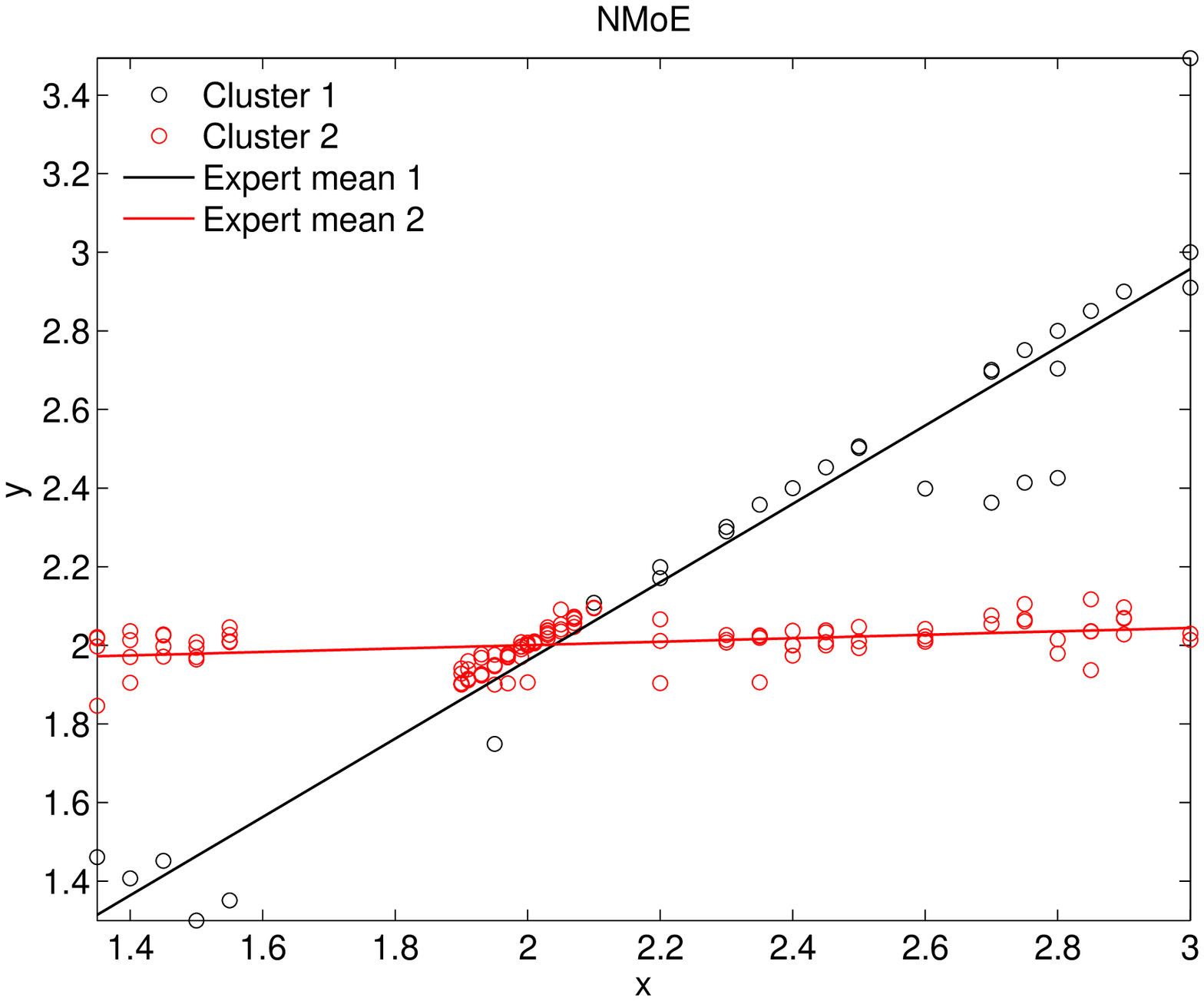} & \includegraphics[width=7.5cm]{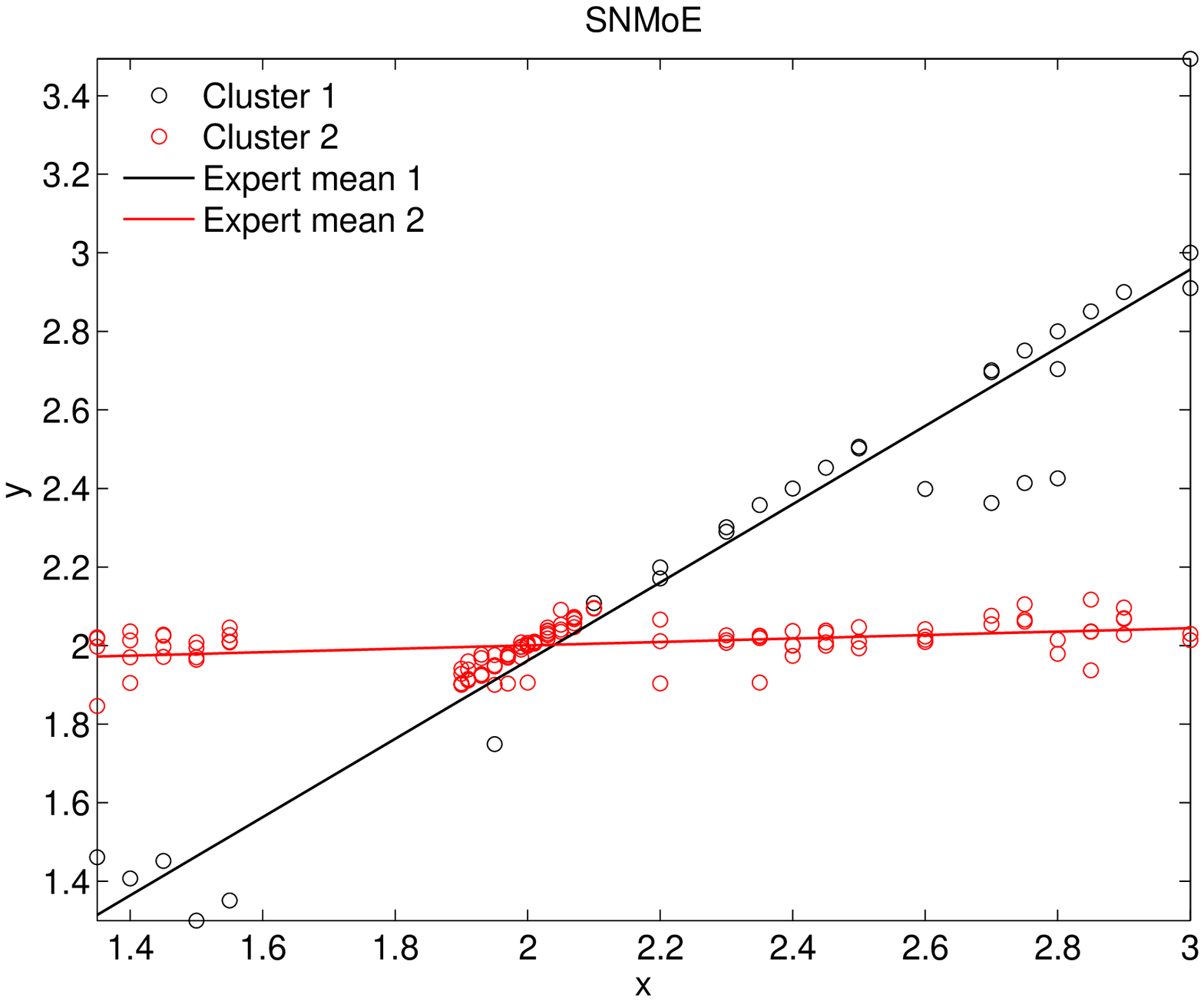}\\
\includegraphics[width=7.5cm]{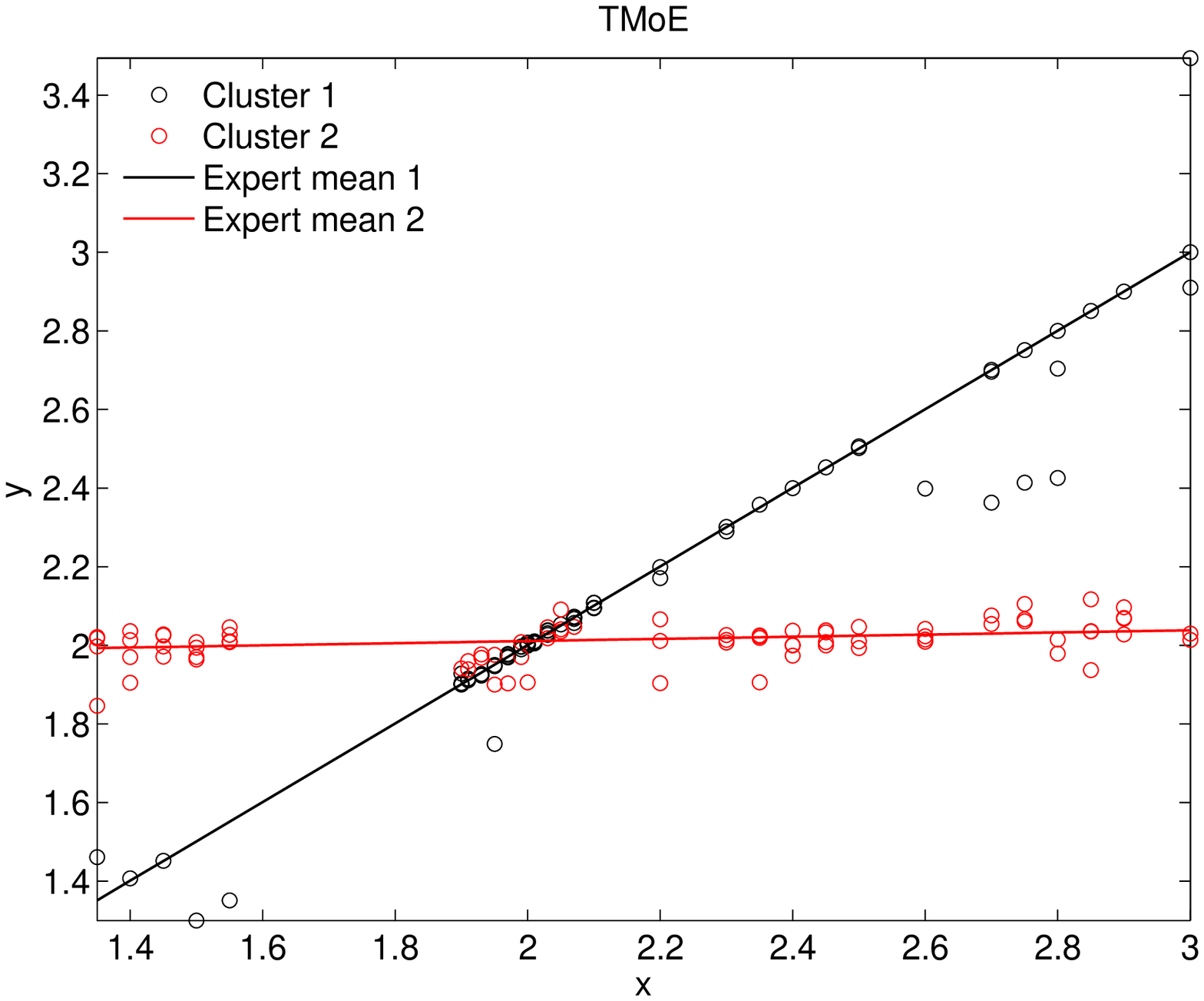} & \includegraphics[width=7.5cm]{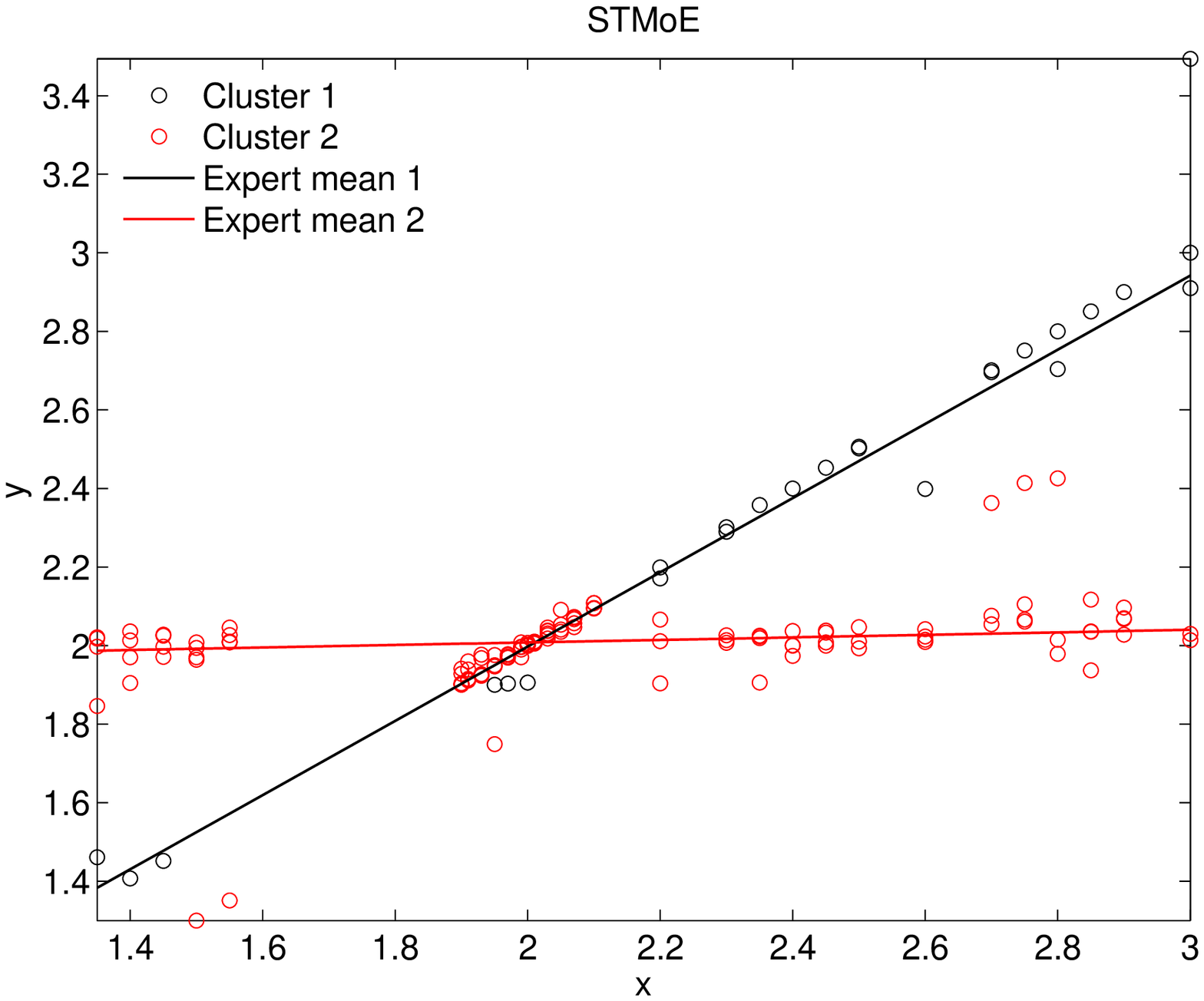}\\
   \end{tabular}
      \caption{\label{fig. Original Tone data and all models}The fitted MoLE to the original tone data set. Upper-left: NMoE model, Upper-right: SNMoE model, Bottom-left: TMoE model, Bottom-right: STMoE model. The predictor $x$ is the actual tone ratio and the response $y$ is the perceived tone ratio.}
\end{figure}
%
Figure \ref{fig. Tone data and all loglik models} shows the log-likelihood profiles for each of the four models. It can namely be seen that training the $t$ mixture of experts for this experiment may take more iterations than the normal and the skew-normal MoE models. The STMoE has indeed more parameters to estimate than the other ones. However, in terms of computing time, all the models converge in only few seconds on a personal laptop (withe 2,9 GHz processor and and 8 GB memory). 
\begin{figure}[htbp]
   \centering 
   \begin{tabular}{cc}
   \includegraphics[width=7.5cm]{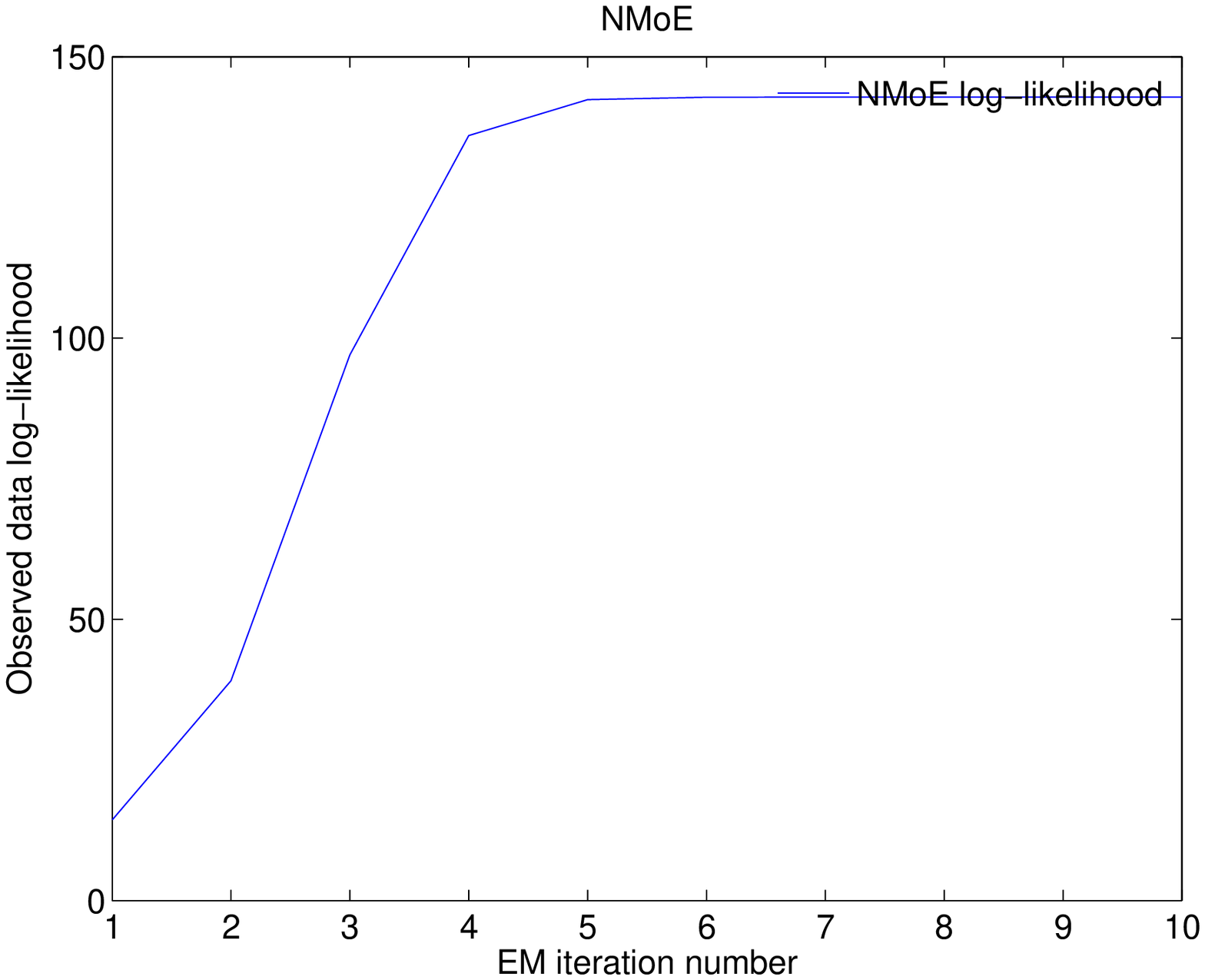} & 
   \includegraphics[width=7.5cm]{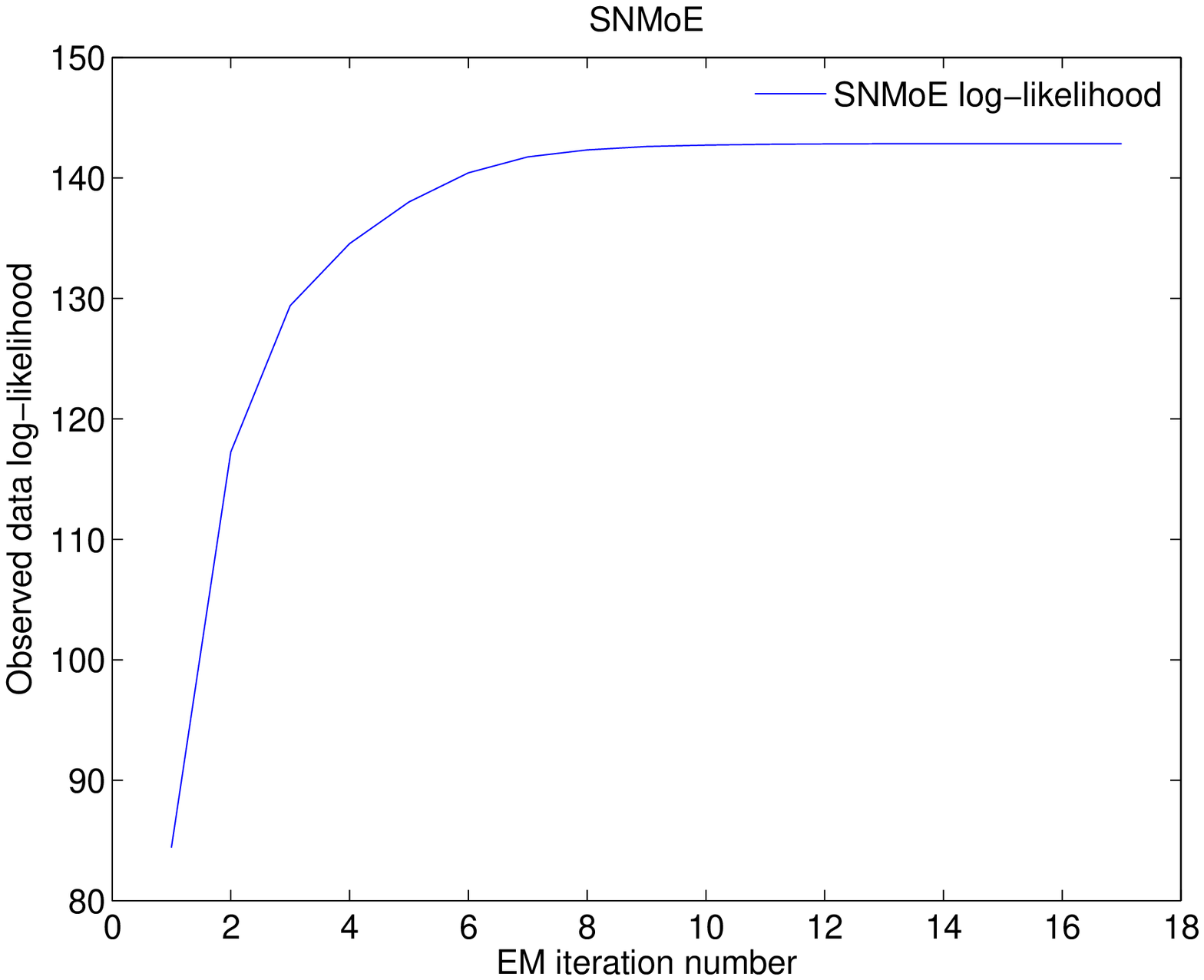}\\
   \includegraphics[width=7.5cm]{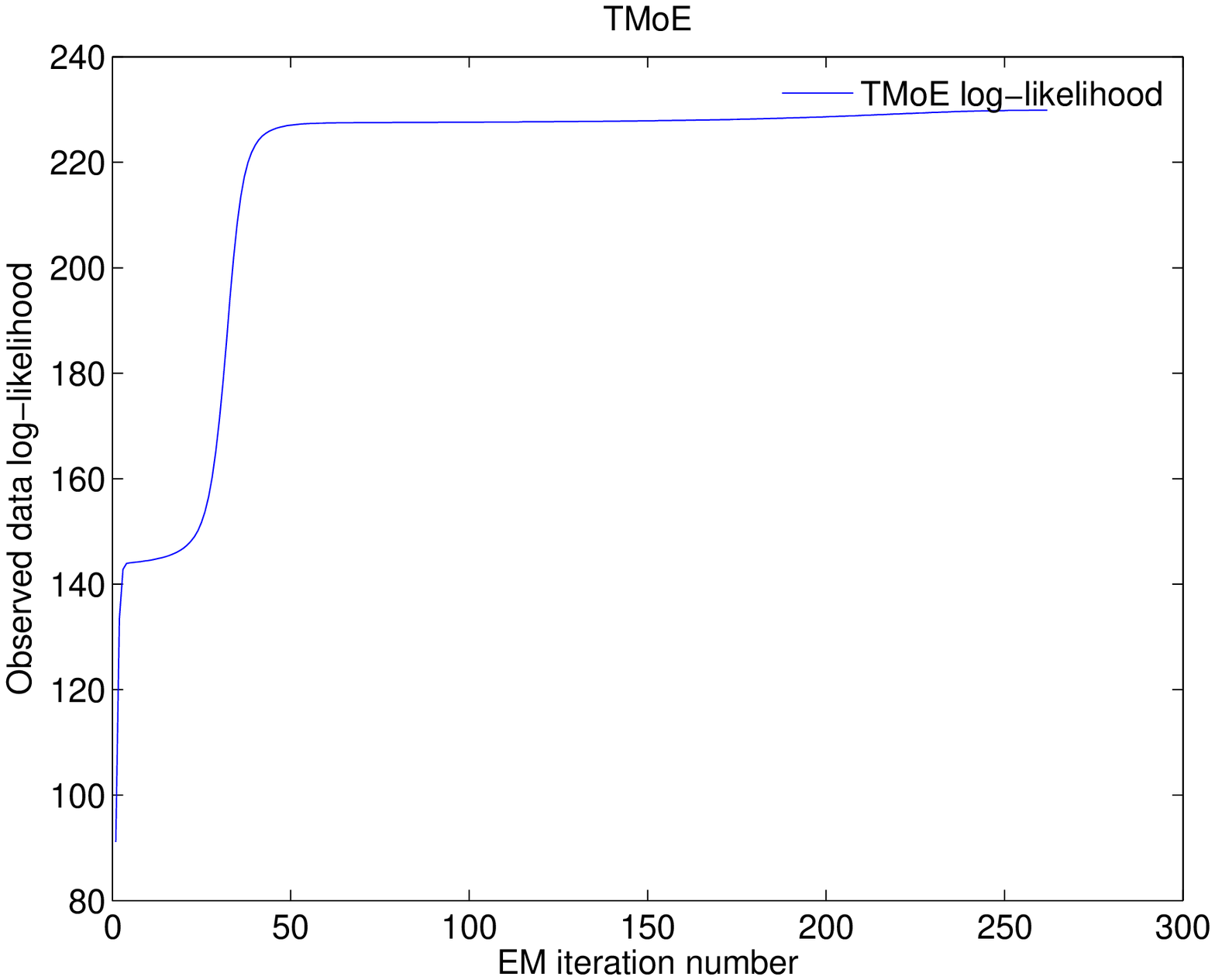} & 
   \includegraphics[width=7.5cm]{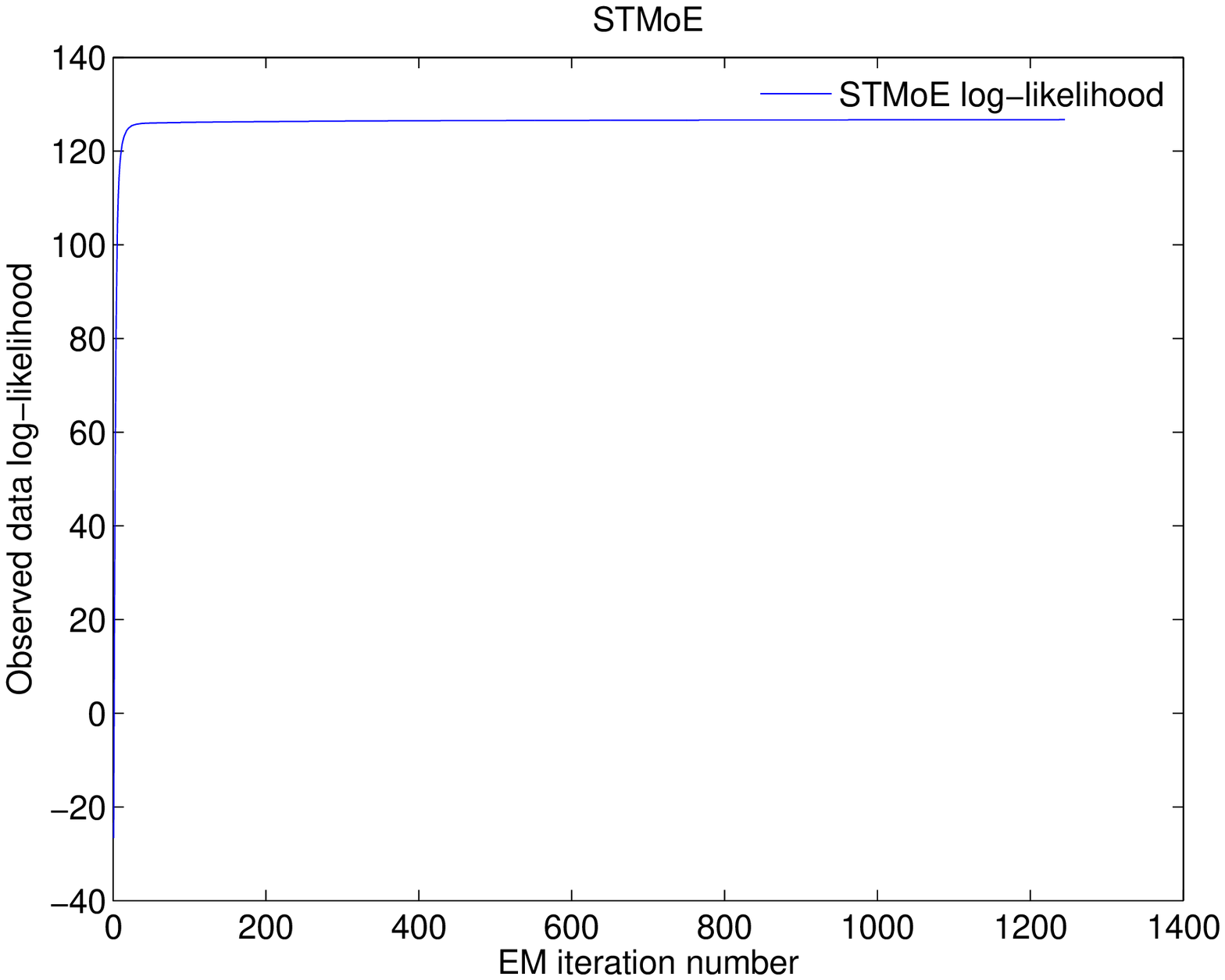}\\
   \end{tabular}
      \caption{\label{fig. Tone data  and all loglik models}The log-likelihood during the EM iterations when fitting the MoLE models to the original tone data set. Upper-left: NMoE model, Upper-right: SNMoE model, Bottom-left: TMoE model, Bottom-right: STMoE model.}
\end{figure}
The values of estimated parameters for the tone perception data set are given in Table \ref{tab. Estimated parameters for the tone perception data set}.
One can see that the regression coefficients are very similar for all the models, except for the first component of the TMoE model. This can be observed on the fit in Figure \ref{fig. Original Tone data and all models} where the first expert component for the TMoE model slightly differ from the one of the other ones. 
One can also see that the SNMoE model parameters are  identical to those of the NMoE, with a skewness close to zero.
For the STMoE model, it retrieves a skewed component and with  high degrees of freedom compared to the other component. This component may be seen as approaching the one of the SNMoE model, while the second one in approaching a $t$ distribution, that is the one of the TMoE model.
{\setlength{\tabcolsep}{3pt
\begin{table}[htbp]
\centering
{\small \begin{tabular}{l c c  c c c c c c c c c c}
\hline
param. & $\alpha_{10}$ & $\alpha_{11}$ & $\beta_{10}$ & $\beta_{11}$ & $\beta_{20}$ & $\beta_{21}$ & $\sigma_{1}$& $\sigma_{2}$ & $\lambda_{1}$ & $\lambda_{2}$ & $\nu_{1}$ & $\nu_{2}$ \\ 
model		& & & & & & & & & & & & \\
 \hline
 \hline
NMoE  	&  -2.690 &	0.796 & 	-0.029 &	0.995 & 1.913 &	 0.043 &	0.137 & 0.047 &	 - & - & - & -	\\
SNMoE	&  -2.694 &	0.797 & 	-0.029 &	0.995 & 1.913 &	 0.043 &	0.137 & 0.047 &	 5.2e-13 & -1.65e-13 & - & - \\
TMoE		&  -0.058 &	-0.070& 	 0.002  &	0.999 & 1.956 &	 0.027 &	0.002 & 0.029 &    -	    & - 	&	0.555 & 2.017	\\
 STMoE	&  -3.044  &	0.824& -0.058    &	  0.944 & 1.944 &	 0.032     &	0.200 & 0.032 &	93.386 & -0.011 &	19.070 & 1.461\\
\hline
\end{tabular}}
\caption{\label{tab. Estimated parameters for the tone perception data set}Values of the estimated MoE parameters for the original Tone perception data set.}
\end{table}
} 

We also performed a model selection procedure on this data set to choose the best number of MoE components for a number of components between 1 and 5. We used BIC, AIC, and ICL. Table  \ref{tab. Model selection Tone data} gives the obtained values of the  model selection criteria. 
One can see that for the NMoE model overestimate the number oc components.  AIC performs poorly for all the models.
BIC provides the correct number of components for the three proposed models. ICL too estimated the correct number of components for both the SNMoE and STMoE models, but hesitates between 2 (the correct number) and 3 components for the TMoE model. One can conclude that the BIC is the criterion to be suggested for the analysis. 

%
{\setlength{\tabcolsep}{2pt
\begin{table}[htbp]
\centering
{\small 
\begin{tabular}{l |ccc | ccc |  ccc | ccc}
\hline
	& \multicolumn{3}{c|}{NMoE}	&	\multicolumn{3}{c|}{SNMoE}	&	\multicolumn{3}{c|}{TMoE}  & \multicolumn{3}{c}{STMoE}\\
\cline{2-13}
K			 &   BIC 	 & 	AIC 	&    ICL		 &   BIC 	 & 	AIC 	&    ICL		 &   BIC 	 & 	AIC 	&    ICL		 &   BIC 	 & 	AIC 	&    ICL \\
\hline
\hline
	1		& 1.8662  &  	6.3821    &	1.8662	&	-0.6391 &   5.3821  &    -0.6391&71.3931 &  77.4143 &71.3931 & 69.5326  & 77.0592 &  69.5326\\
	2		&122.8050&  134.8476&  107.3840&\underline{117.7939}&  132.8471& \underline{102.4049}& \underline{204.8241}&  219.8773&  186.8415& \underline{92.4352} & 110.4990 &  \underline{82.4552}\\
	3		&118.1939&  137.7630&   76.5249&122.8725 & 146.9576 &  98.0442&199.4030  &223.4880 & 183.0389 &77.9753&  106.5764&   52.5642\\
	4		&121.7031&  148.7989&   94.4606 & 109.5917 & 142.7087 & 97.6108 & 201.8046 & \underline{234.9216}&  \underline{187.7673}& 77.7092 & 116.8474&   56.3654\\
	5		&\underline{141.6961}&\underline{176.3184} & \underline{123.6550}&107.2795 &  \underline{149.4284}&   96.6832&187.8652 & 230.0141 & 164.9629& 79.0439 & \underline{128.7194} &  67.7485\\
\hline 
\end{tabular}
\caption{\label{tab. Model selection Tone data}Choosing the number of expert components $K$ for the original tone perception data by using the information criteria BIC, AIC, and ICL. Underlined numbers indicate the highest value for each criterion.}
}
\end{table}}

Now we examine the sensitivity of the MoE models to outliers based on this real data set.  For this, we adopt the same scenario used in \citet{Bai2012} and \citet{Song2014} (the last and more difficult scenario) by adding 10 identical pairs $(0,4)$ to the original data set as outliers in the $y$-direction, considered as high leverage outliers. We apply the MoE models in the same way as before.

The upper plots in Figure \ref{fig. Tone data with outliers and all models} show that the normal and the skew-normal mixture of experts provide almost identical fits and are sensitives to outliers.
However, in both cases, compared to the normal regression mixture result in \citet{Bai2012}, and the Laplace regression mixture and the $t$ regression mixture results in \citet{Song2014}, the fitted NMoE and SNMoE model are affected less severely by the outliers 
This may be attributed to the fact that the mixing proportions here are depending on the predictors, which is not the case in these regression mixture models, namely the ones of \citet{Bai2012}, and  \citet{Song2014}.  One can also see that, even the regression mean functions are  affected severely by the outliers, the provided partitions are still reasonable and similar to those provided in the previous non-noisy case.
Then, the bottom plots in Figure \ref{fig. Tone data with outliers and all models}  clearly show that the TMoE and the STMoE provide a robust good fit. For the TMoE, the obtained fit is quasi-identical to the first one on the original data without outliers, shown in the bottom-left plot of Figure \ref{fig. Original Tone data and all models}. For the STMoE, even if the results differ very slightly compared to the case with outliers, the obtained fits for both situations (with and without outliers) are very reasonable.
Moreover, we notice that, as showed in \citet{Song2014}, for this situation with outliers, 
the $t$ mixture of regressions fails; The  fit is affected severely by the outliers.
However, for the proposed TMoE and STMoE, the ten high leverage outliers have no significant impact on the fitted experts.  
This is because here the mixing proportions depend on the inputs, which is not the case for the regression mixture model  described in \citet{Song2014}. 
%
%
%
\begin{figure}[htbp]
   \centering 
   \begin{tabular}{cc}
   \includegraphics[width=7.5cm]{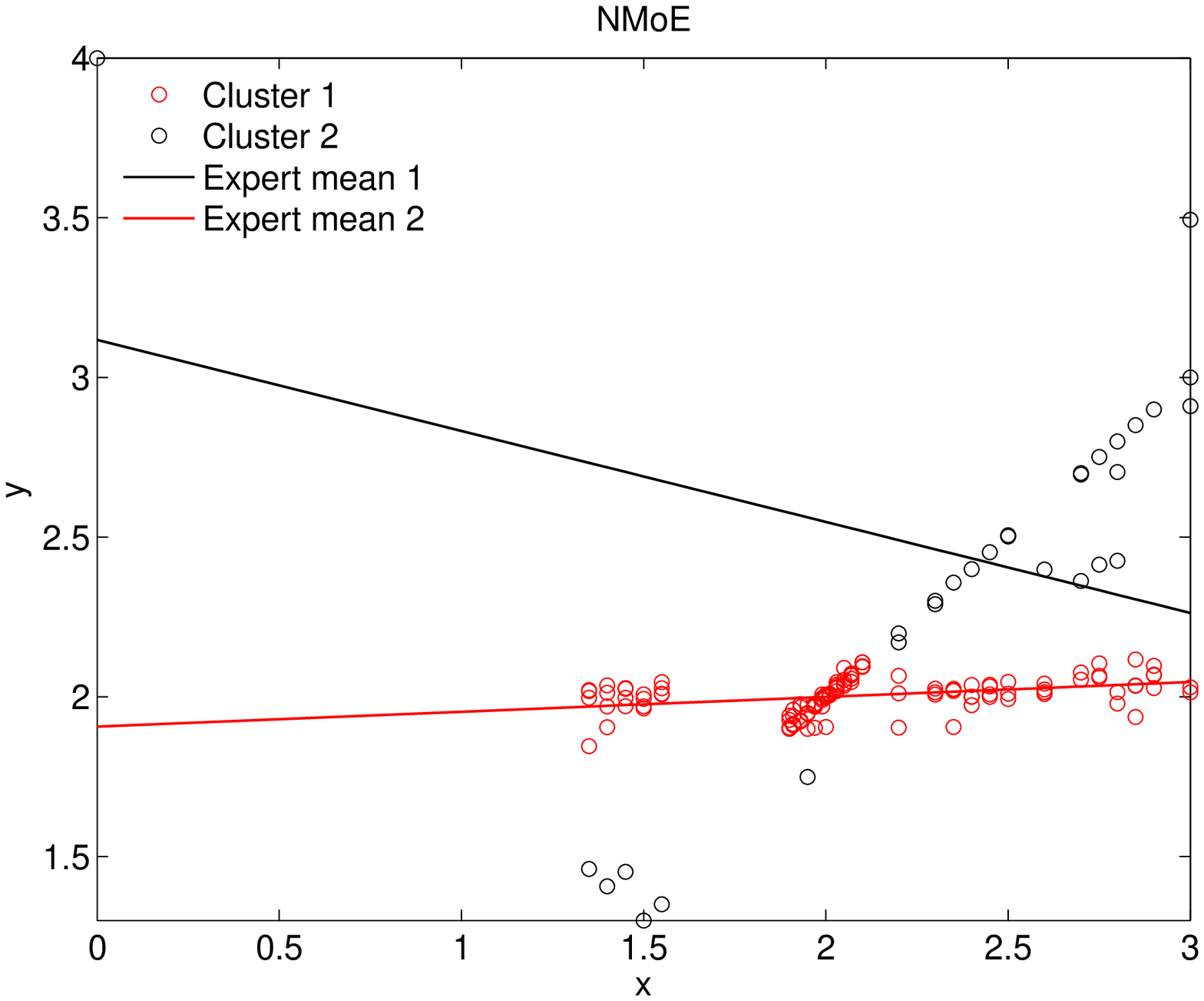} & \includegraphics[width=7.5cm]{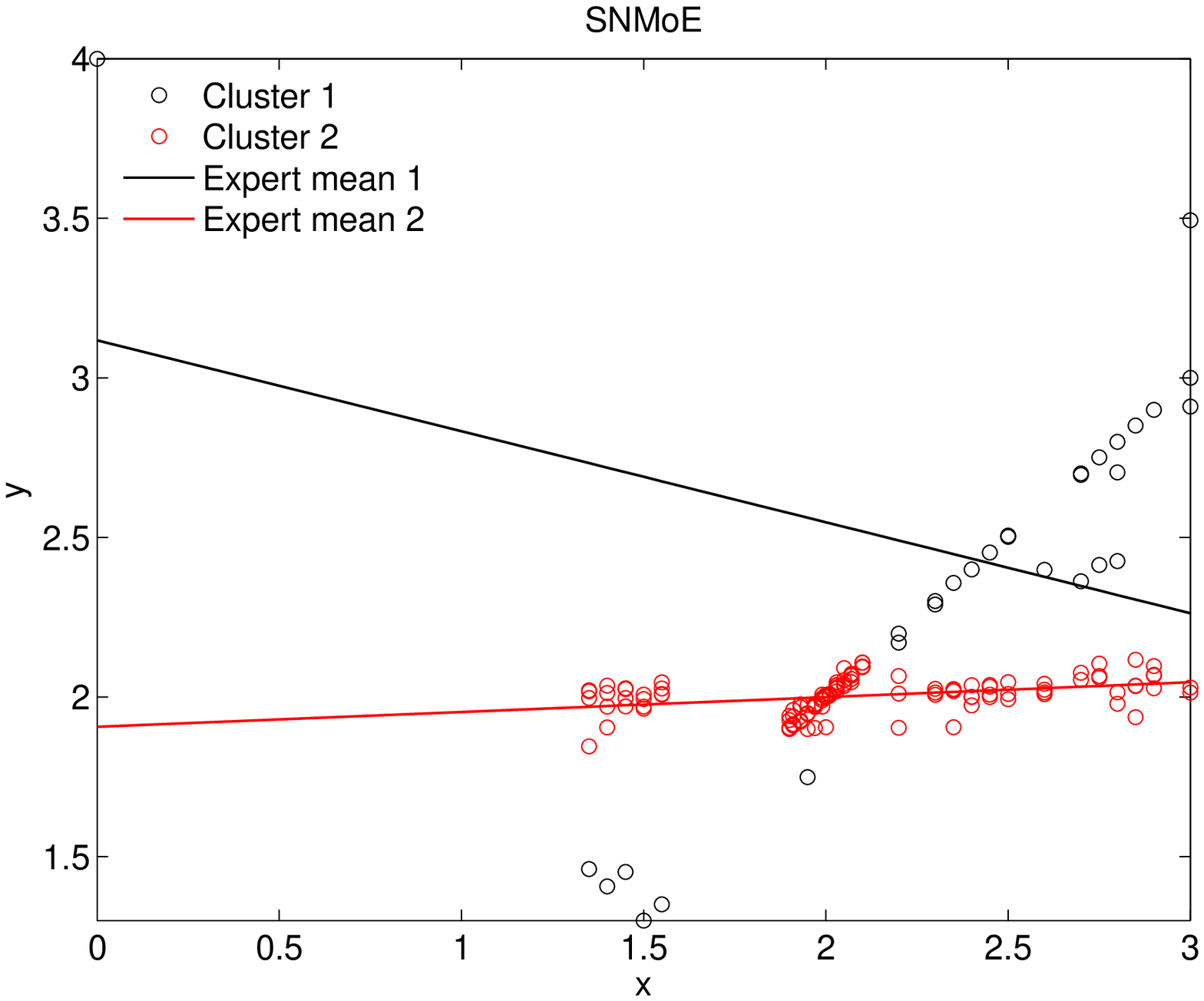}\\
   \includegraphics[width=7.5cm]{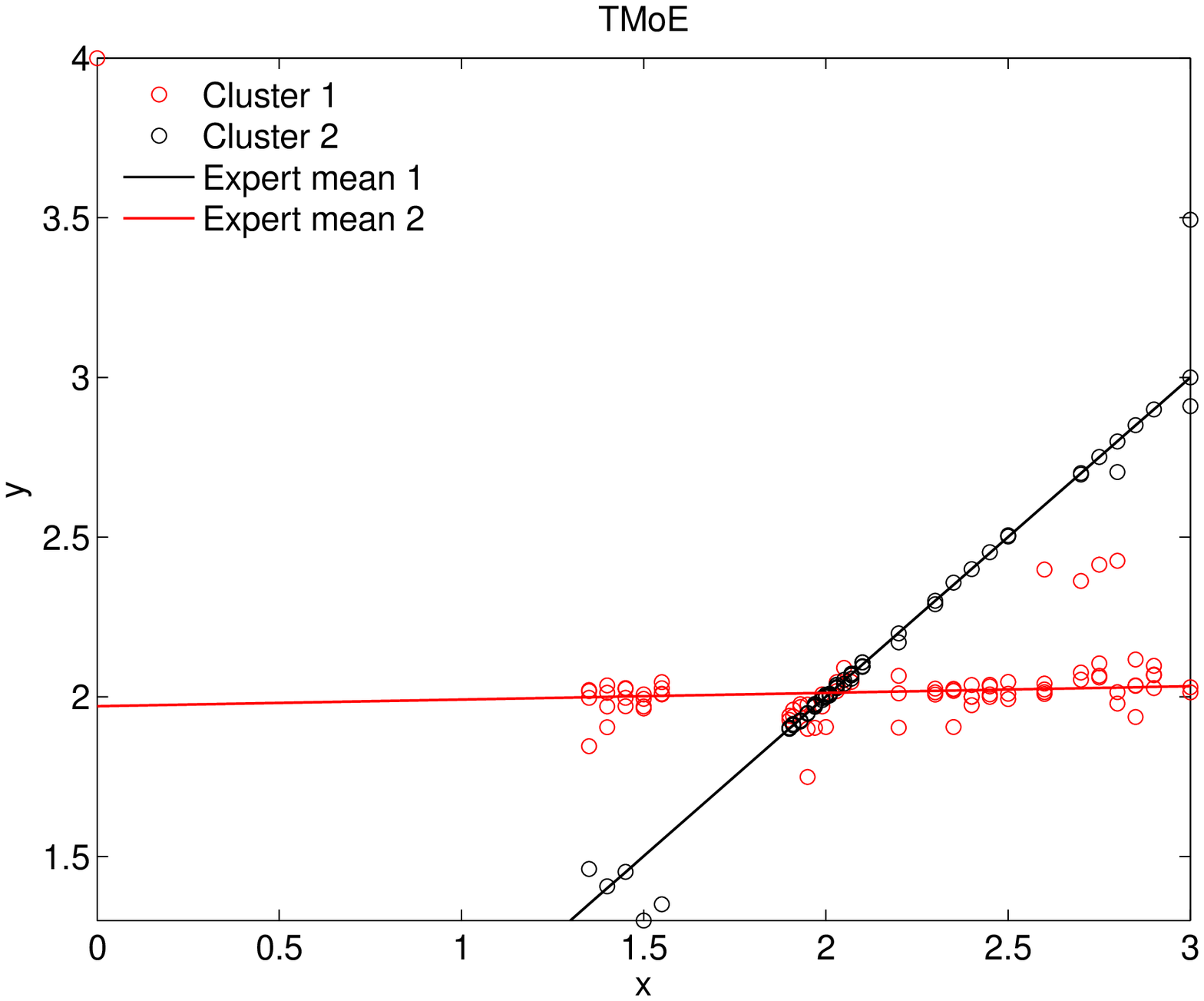} & \includegraphics[width=7.5cm]{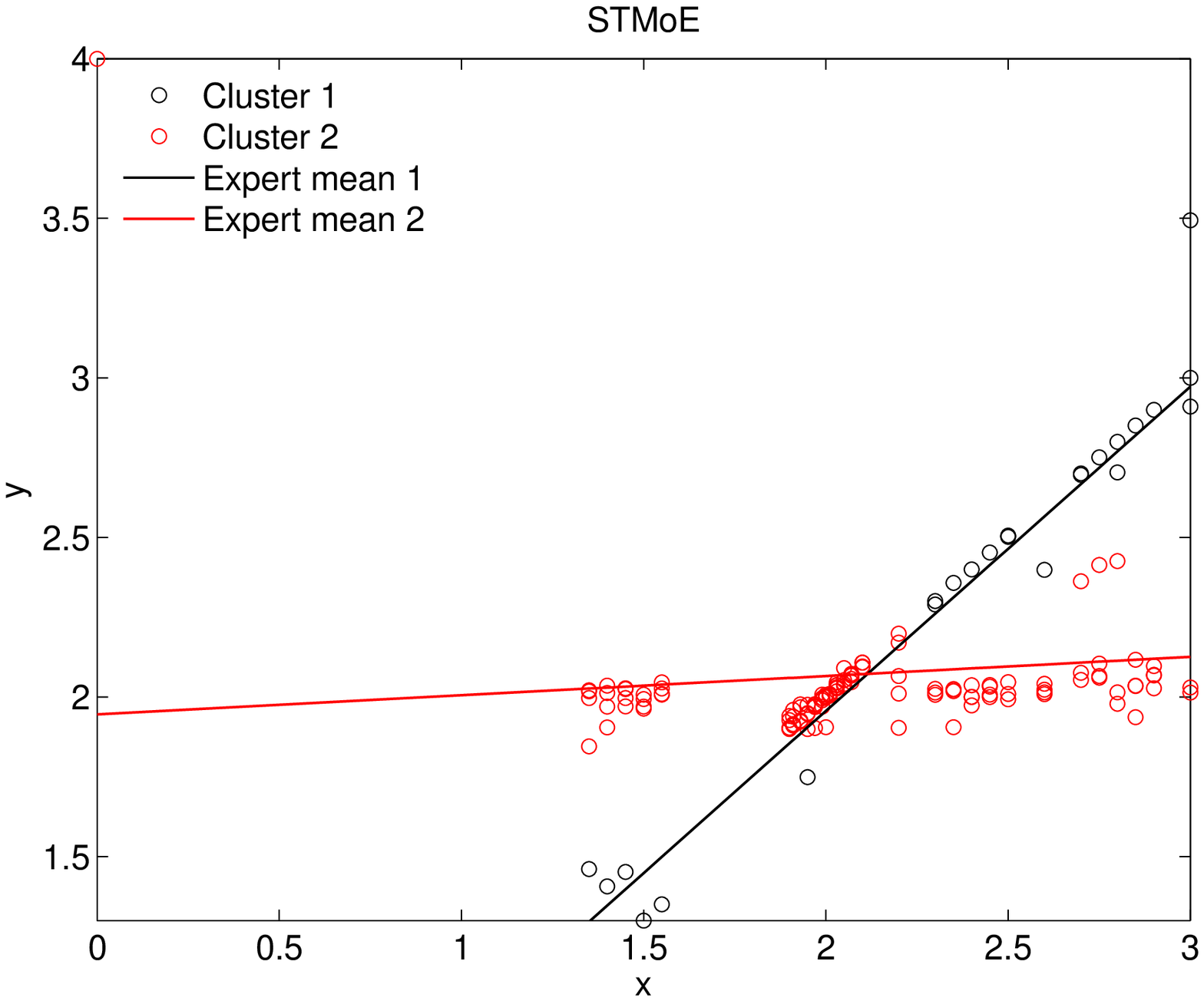}\\
   \end{tabular}
      \caption{\label{fig. Tone data with outliers and all models}Fitting MoLE to the tone data set with ten added outliers $(0,4)$. Upper-left: NMoE model, Upper-right: SNMoE model, Bottom-left: TMoE model, Bottom-right: STMoE model. The predictor $x$ is the actual tone ratio and the response $y$ is the perceived tone ratio.}
\end{figure}
Figure \ref{fig. Tone data with outliers and loglik} shows the log-likelihood profiles for each of the four models, which show a similar behavior than the one in the case without outliers.
\begin{figure}[htbp]
   \centering 
   \begin{tabular}{cc}
   \includegraphics[width=7.5cm]{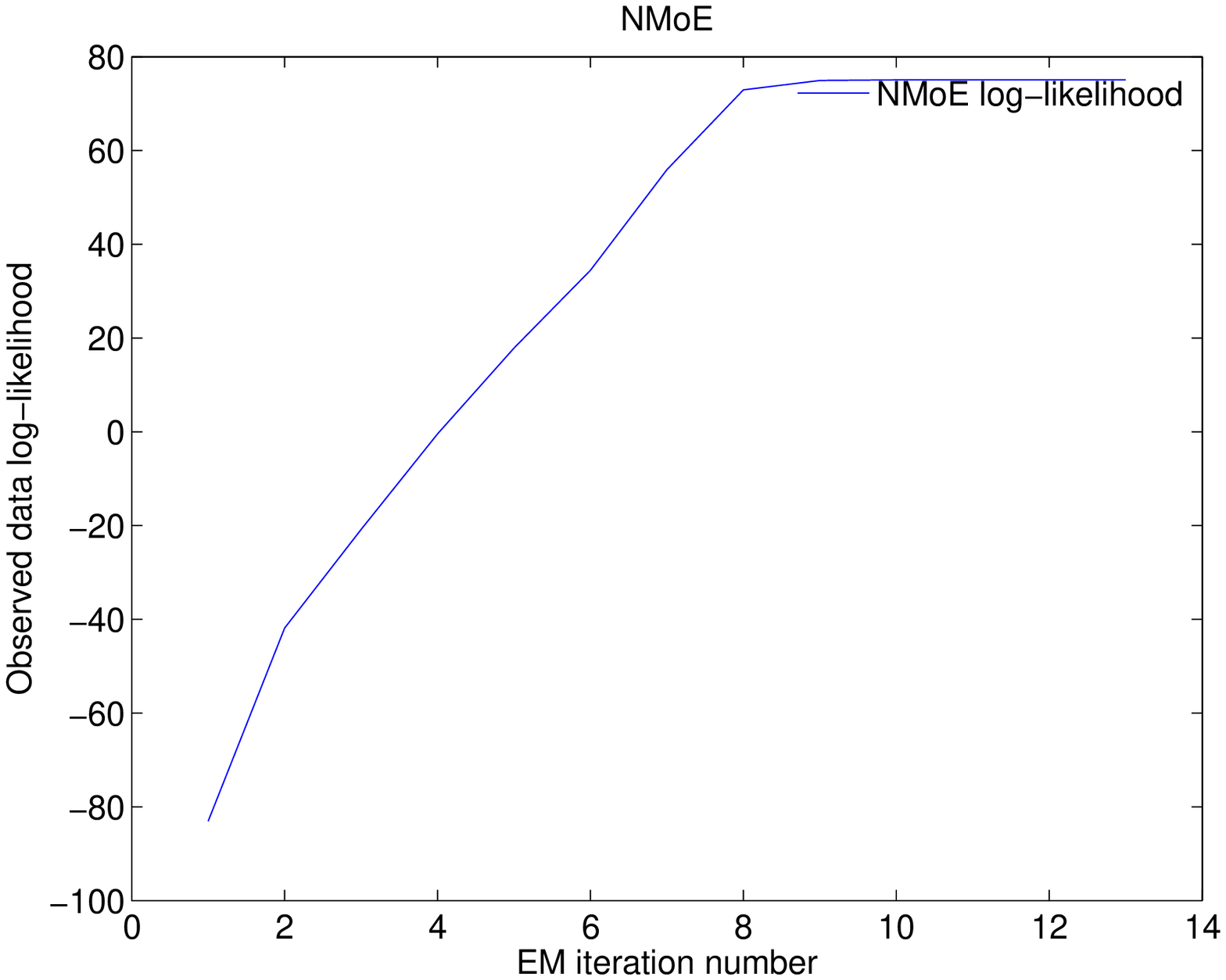} & \includegraphics[width=7.5cm]{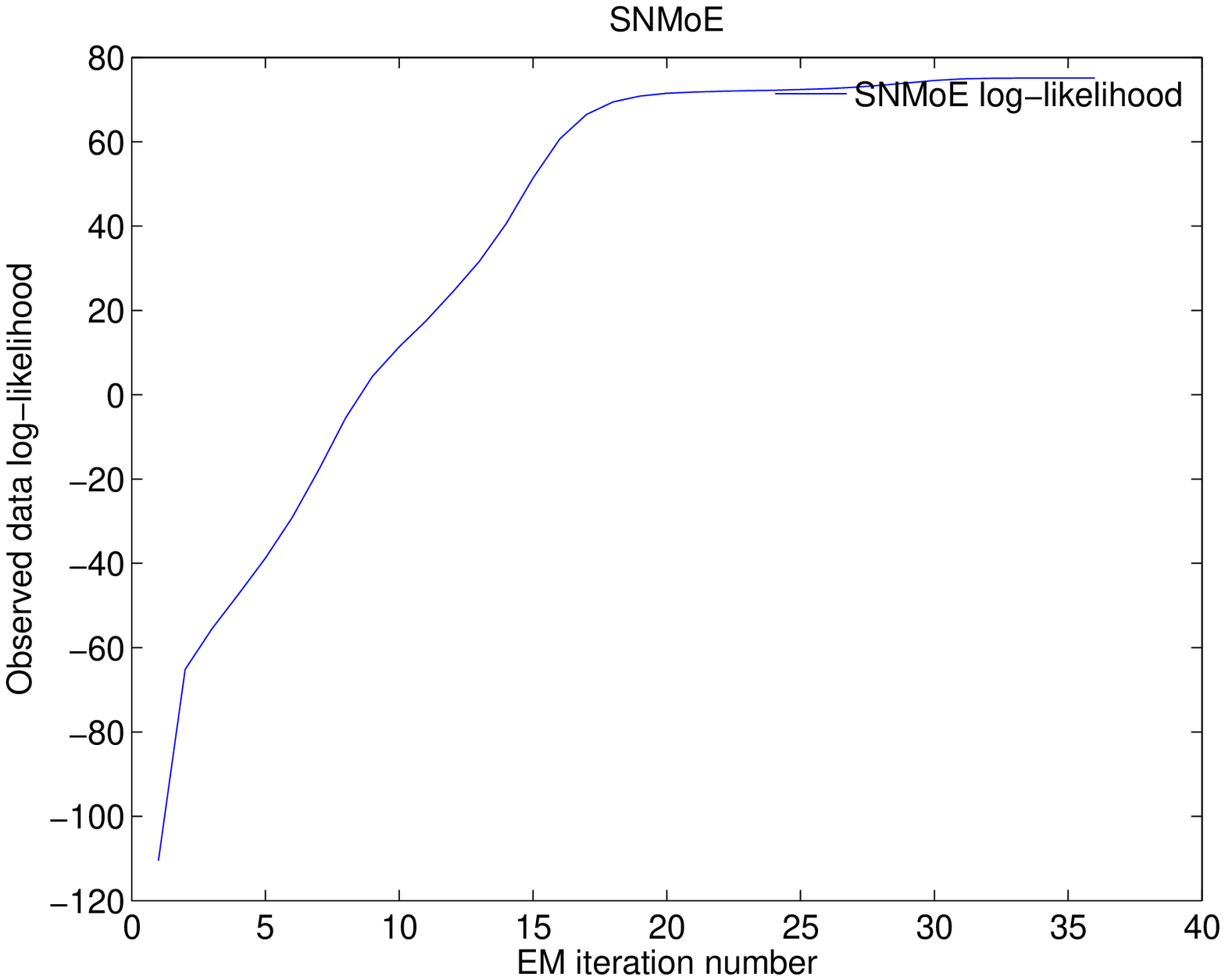}\\
   \includegraphics[width=7.5cm]{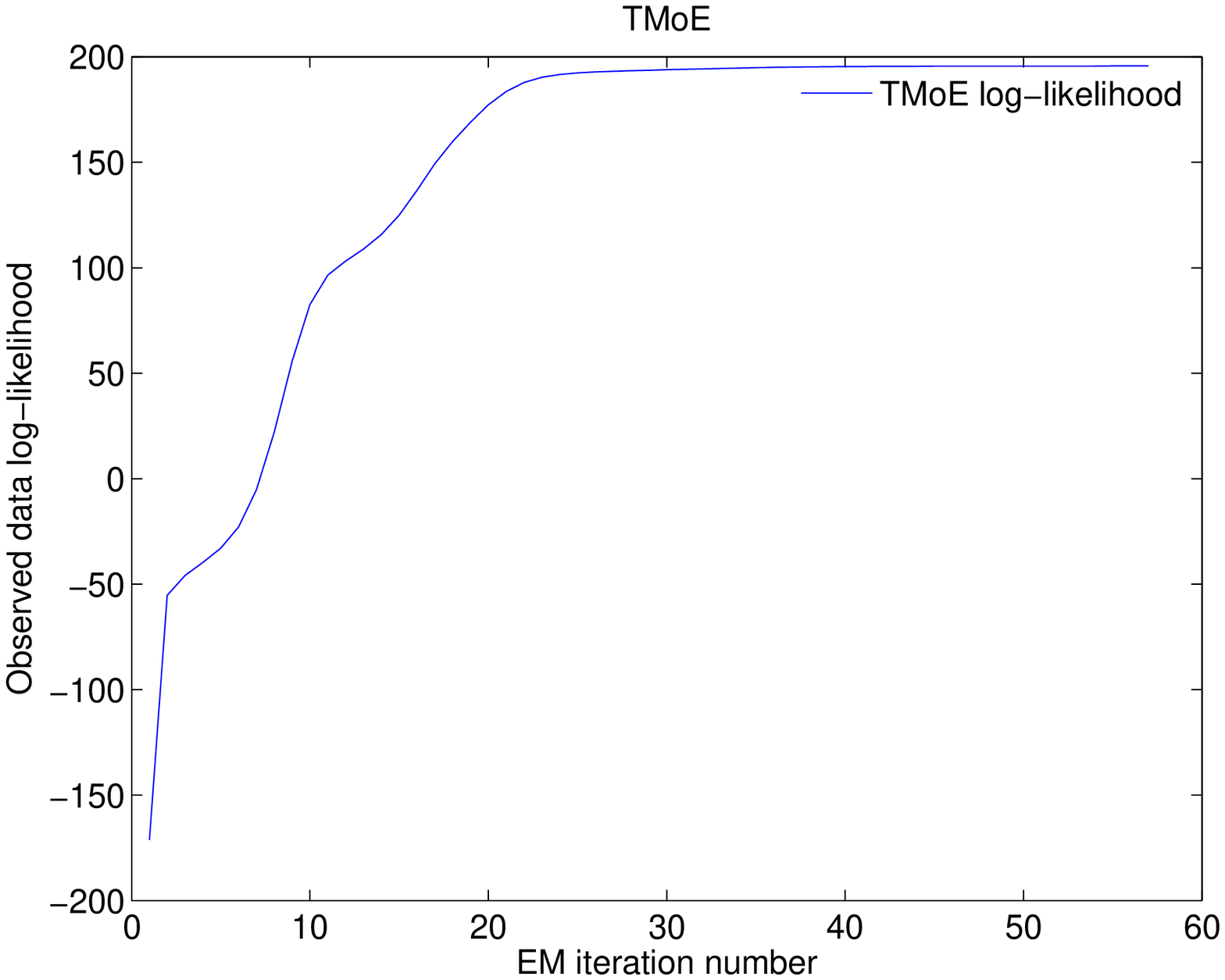} & \includegraphics[width=7.5cm]{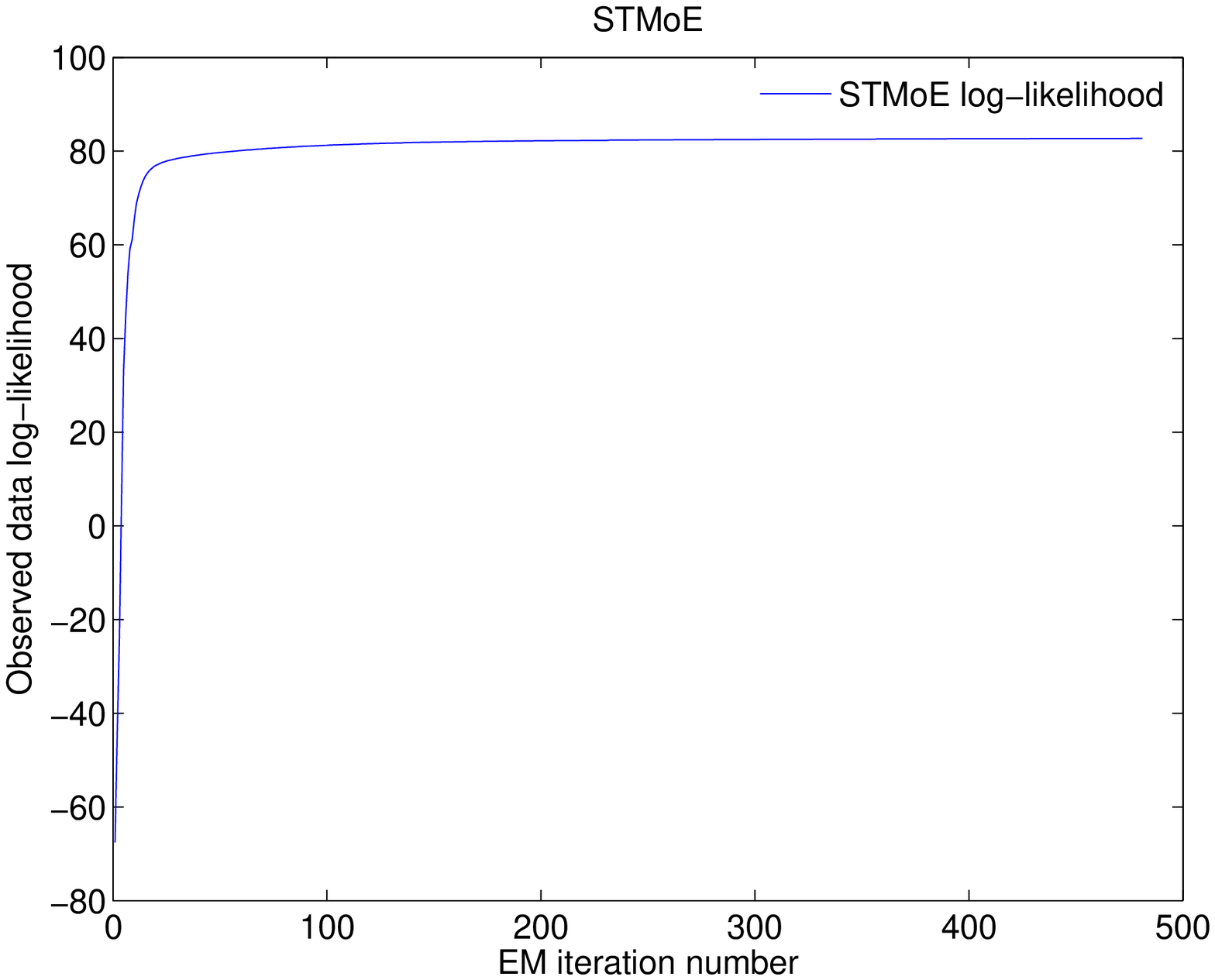}\\
   \end{tabular}
      \caption{\label{fig. Tone data with outliers and loglik}The log-likelihood during the EM iterations when fitting the MoLE models to the tone data set with ten added outliers $(0,4)$. Upper-left: NMoE model, Upper-right: SNMoE model, Bottom-left: TMoE model, Bottom-right: STMoE model.}
\end{figure}

The values of estimated MoE parameters in this case with outliers are given in Table \ref{tab. Estimated parameters for the tone perception data set with outliers}.
%
One can  see that the SNMoE model parameters are  identical to those of the NMoE, with a skewness close to zero.
The regression coefficients for the second expert component are very similar for all the models. For the first component, the TMoE model is still retrieving a more heavy tailed component. 
For the STMoE model, it retrieves a skewed normal component  while the second component in approaching a $t$ distribution with a small degrees of freedom. 
{\setlength{\tabcolsep}{3pt
\begin{table}[htbp]
\centering
{\small \begin{tabular}{l c c  c c c c c c c c c c}
\hline
param. & $\alpha_{10}$ & $\alpha_{11}$ & $\beta_{10}$ & $\beta_{11}$ & $\beta_{20}$ & $\beta_{21}$ & $\sigma_{1}$& $\sigma_{2}$ & $\lambda_{1}$ & $\lambda_{2}$ & $\nu_{1}$ & $\nu_{2}$ \\ 
model		& & & & & & & & & & & & \\
 \hline
 \hline
NMoE  	&  0.811	&  0.150	 & 3.117 & -0.285	&	1.907 & 0.046  & 0.700 & 0.050  & - & -	 & - & - \\
SNMoE	&  -0.810 &	-0.150 & 3.117  & -0.285 &	1.907 & 0.046  & 0.701 &	0.050 & 5.5e-08 &	 1e-08 &	- & - \\
TMoE		&  0.888   &	-0.236 & 0.002  & 0.999   & 1.971 & 0.020	& 0.002    & 0.024  & -	&	- & 0.682 & 0.812\\
STMoE	&  -3.004 &	0.732  & -0.246  & 1.016 & 1.808	 &	0.060 & 0.212	&	0.088 &	156.240 & 1.757	&	81.355 & 1.630\\
 \hline
\end{tabular}}
\caption{\label{tab. Estimated parameters for the tone perception data set with outliers} Values of the estimated MoE parameters  for the  tone perception data set with added outliers.}
\end{table}
}
		    
\subsubsection{Temperature anomalies data set}
In this experiment, we examine another real-world data set related to climate change analysis. 
The NASA GISS Surface Temperature (GISTEMP) analysis provides a measure of the changing global surface temperature with monthly resolution for the period since 1880, when a reasonably global distribution of meteorological stations was established. 
The GISS analysis is updated monthly, however the data presented here\footnote{source: from \citet{TemperatureAnomalyData}, \url{http://cdiac.ornl.gov/ftp/trends/temp/hansen/gl_land.txt}} are updated annually as issued from the Carbon Dioxide Information Analysis Center (CDIAC), which has served as the primary climate-change data and information analysis center of the U.S. Department of Energy since 1982.
The data consist of $n = 135$ yearly measurements of the global annual temperature anomalies (in degrees C) computed using data from land meteorological stations for the period of $1882-2012$. 
These data have been analyzed earlier by \citet{Hansen1999,Hansen2001} and recently by \citet{Nguyen2014-MoLE} by using the Laplace mixture of linear experts (LMoLE). 

To apply the proposed non-normal mixture of expert models, we consider mixtures of two experts as in \citet{Nguyen2014-MoLE}. 
This number of components is also the one provided by the model selection criteria as shown later in Table \ref{tab. Model selection temprature anomalies data}. 
Indeed, as mentioned by \citet{Nguyen2014-MoLE},  \citet{Hansen2001} found that the data could be segmented into two periods of global warming (before 1940 and after 1965), separated by a transition period where there was a slight global cooling (i.e. 1940 to 1965). Documentation of the basic analysis method is provided by \citet{Hansen1999,Hansen2001}. 
We set the response $y_i (i=1,\ldots,135)$ as the temperature anomalies and the covariates $\bsx_i = \bsr_i = (1,x_i)^T$ where $x_i$ is the year of the $i$th observation.

%
%
Figures \ref{fig. temperature anomalies data and models experts}, 
\ref{fig. temperature anomalies data and models means}, and 
\ref{fig. temperature anomalies data and models loglik}
respectively show, for each of the MoE models, the two fitted linear expert components, 
the corresponding means and confidence regions computed as plus and minus twice the estimated (pointwise) standard deviation as presented in Section \ref{sec: Prediction using the NNMoE},
and the log-likelihood profiles.
One can observe that the four models are successfully applied on the data set and provide very similar results. 
 These results are also similar to those found by \citet{Nguyen2014-MoLE} who used a Laplace mixture of linear experts.
\begin{figure}[htbp]
   \centering 
   \begin{tabular}{cc}
   \includegraphics[width=7.5cm]{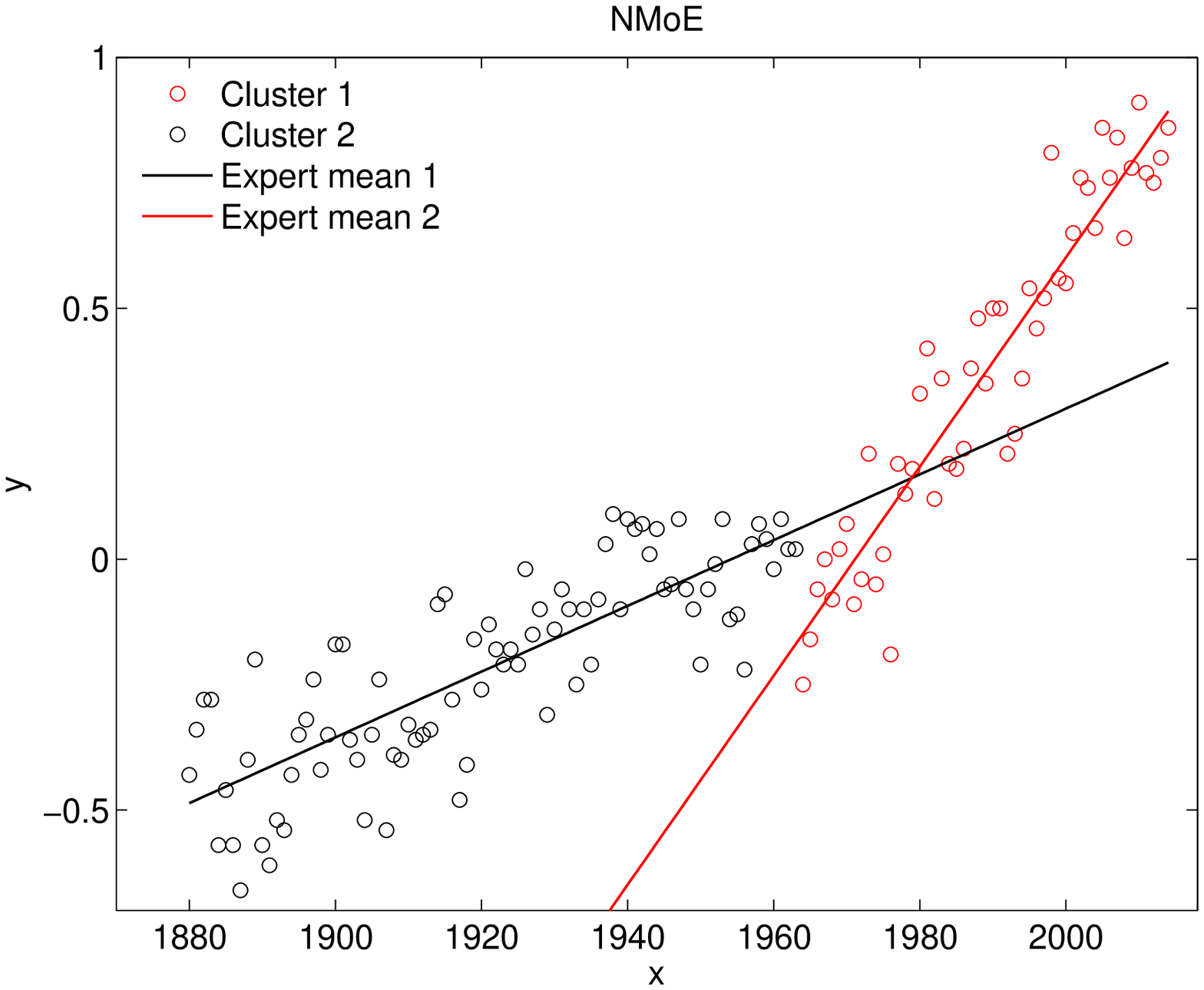} & \includegraphics[width=7.5cm]{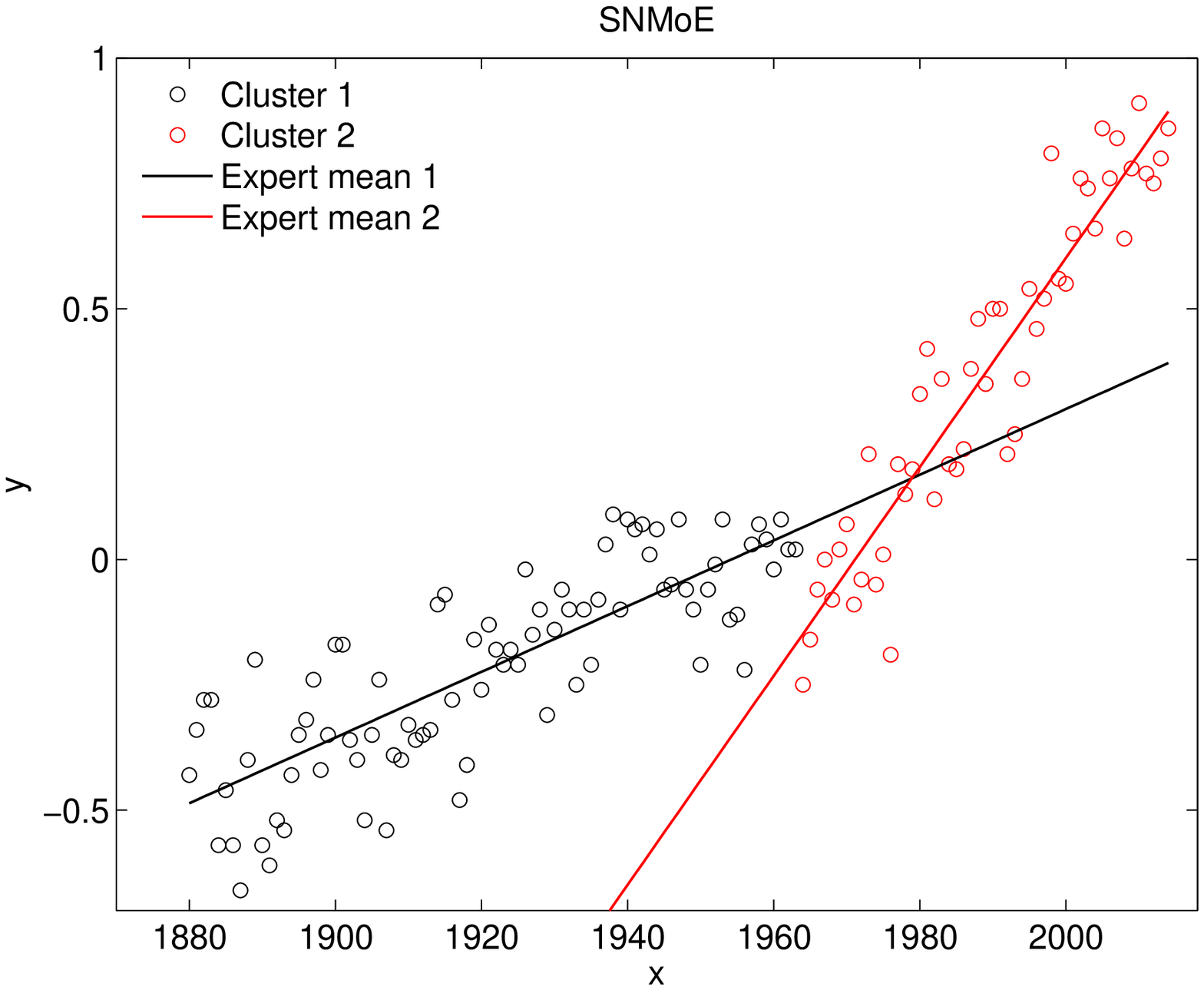}\\
\includegraphics[width=7.5cm]{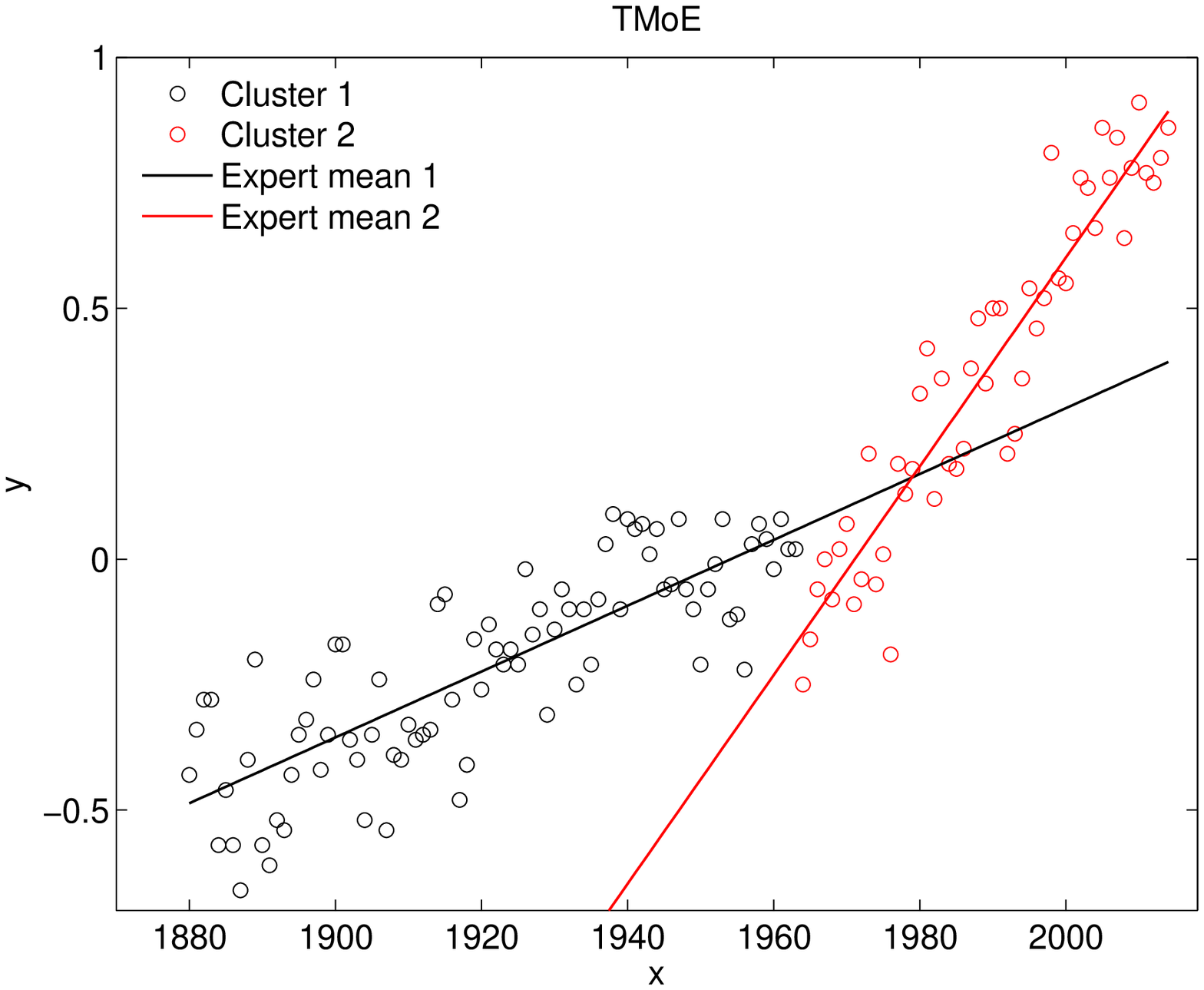} & \includegraphics[width=7.5cm]{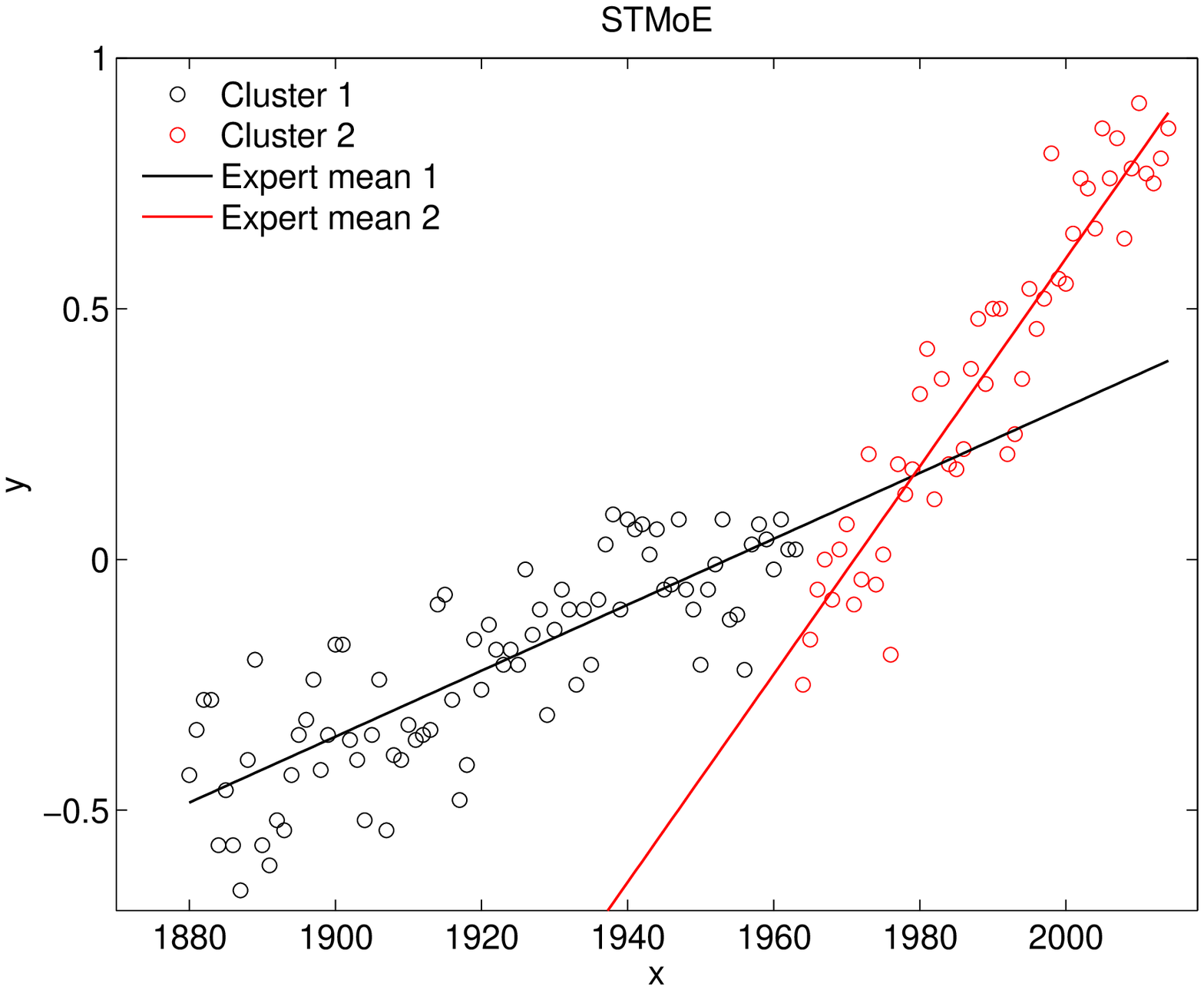}\\
   \end{tabular}
      \caption{\label{fig. temperature anomalies data and models experts}Fitting the MoLE models to the temperature anomalies data set. Upper-left: NMoE model, Upper-right: SNMoE model, Bottom-left: TMoE model, Bottom-right: STMoE model. The predictor $x$ is the year and the response $y$ is the temperature anomaly.}
\end{figure}
\begin{figure}[htbp]
   \centering 
   \begin{tabular}{cc}
   \includegraphics[width=7.5cm]{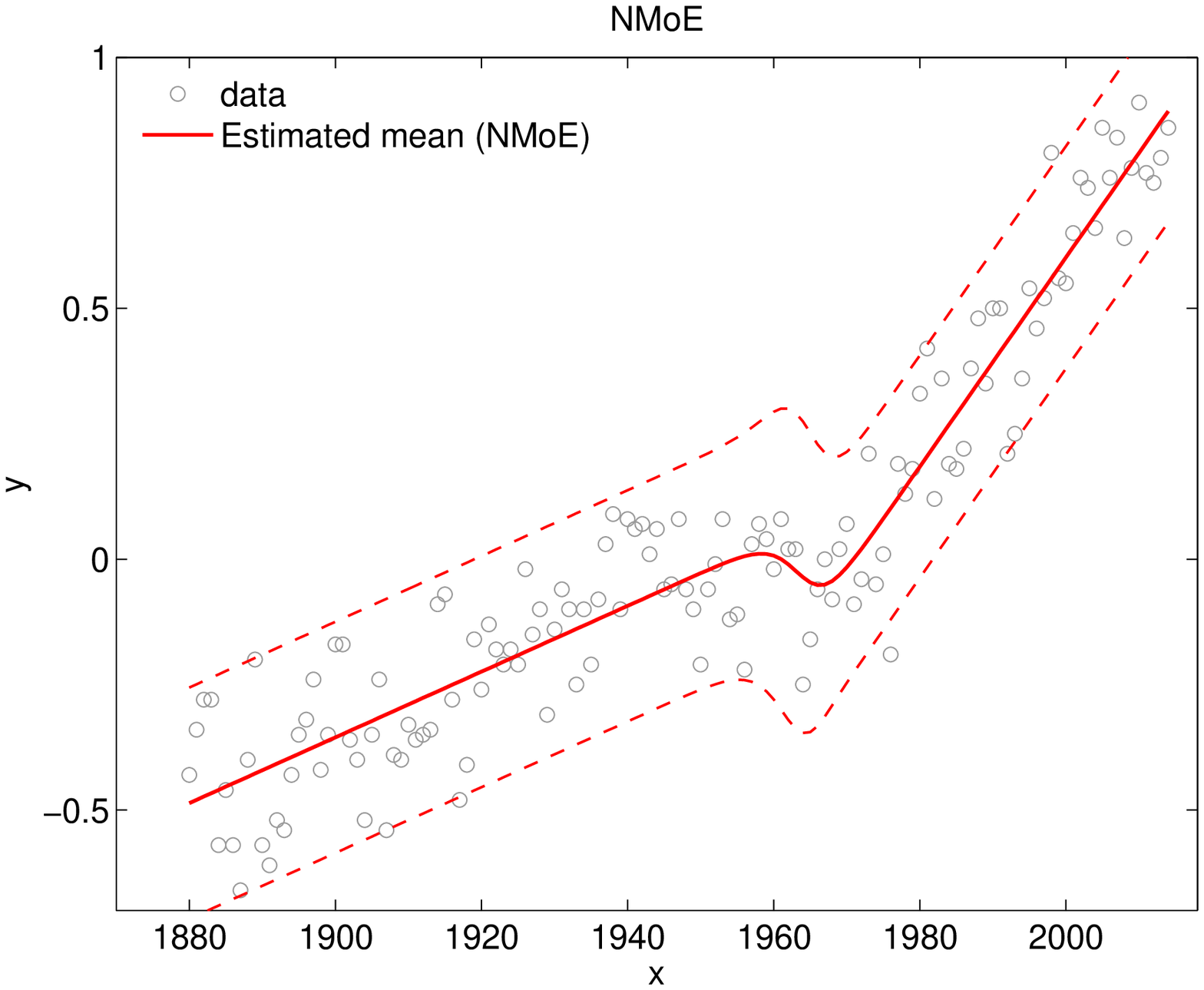} & \includegraphics[width=7.5cm]{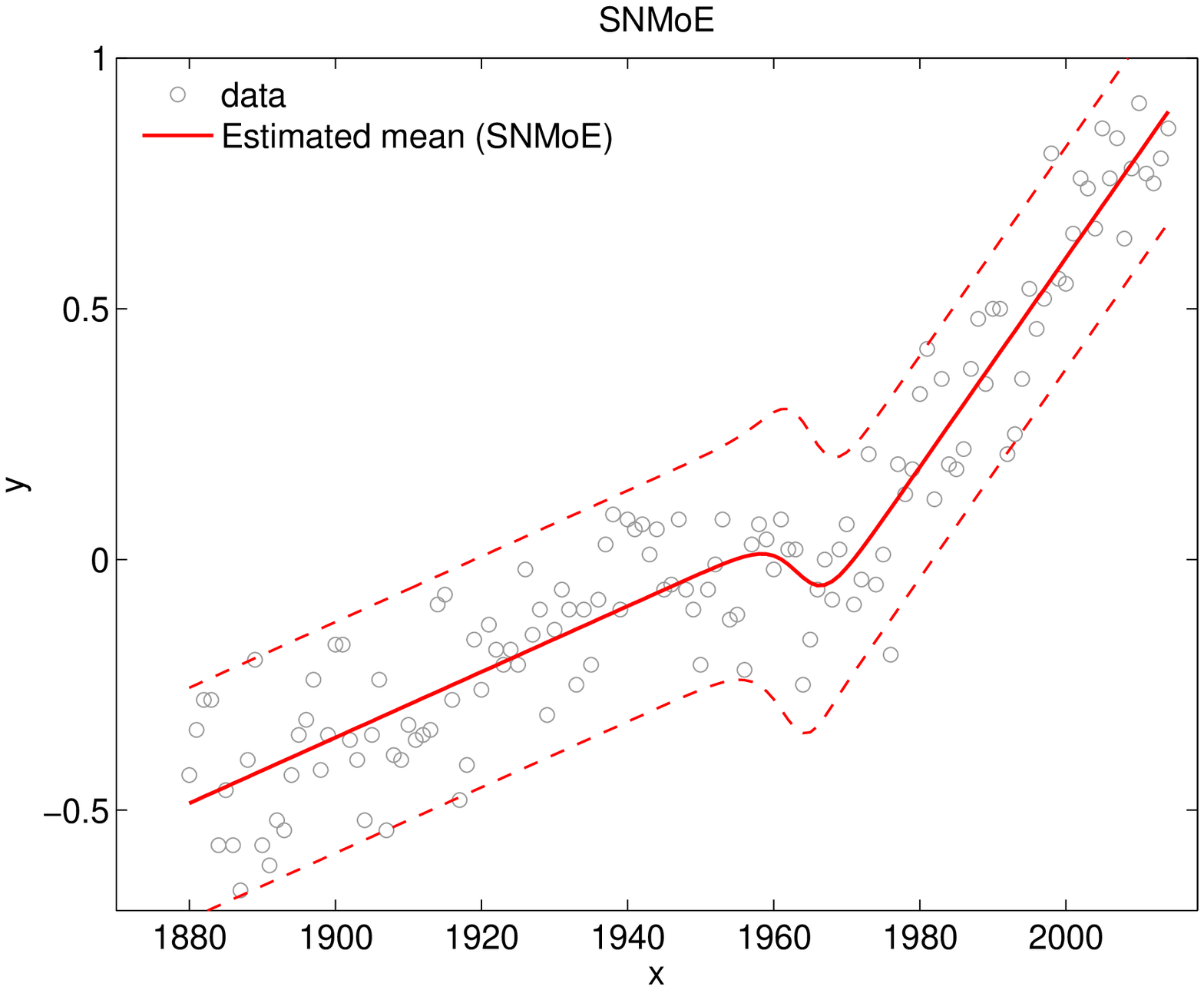}\\
\includegraphics[width=7.5cm]{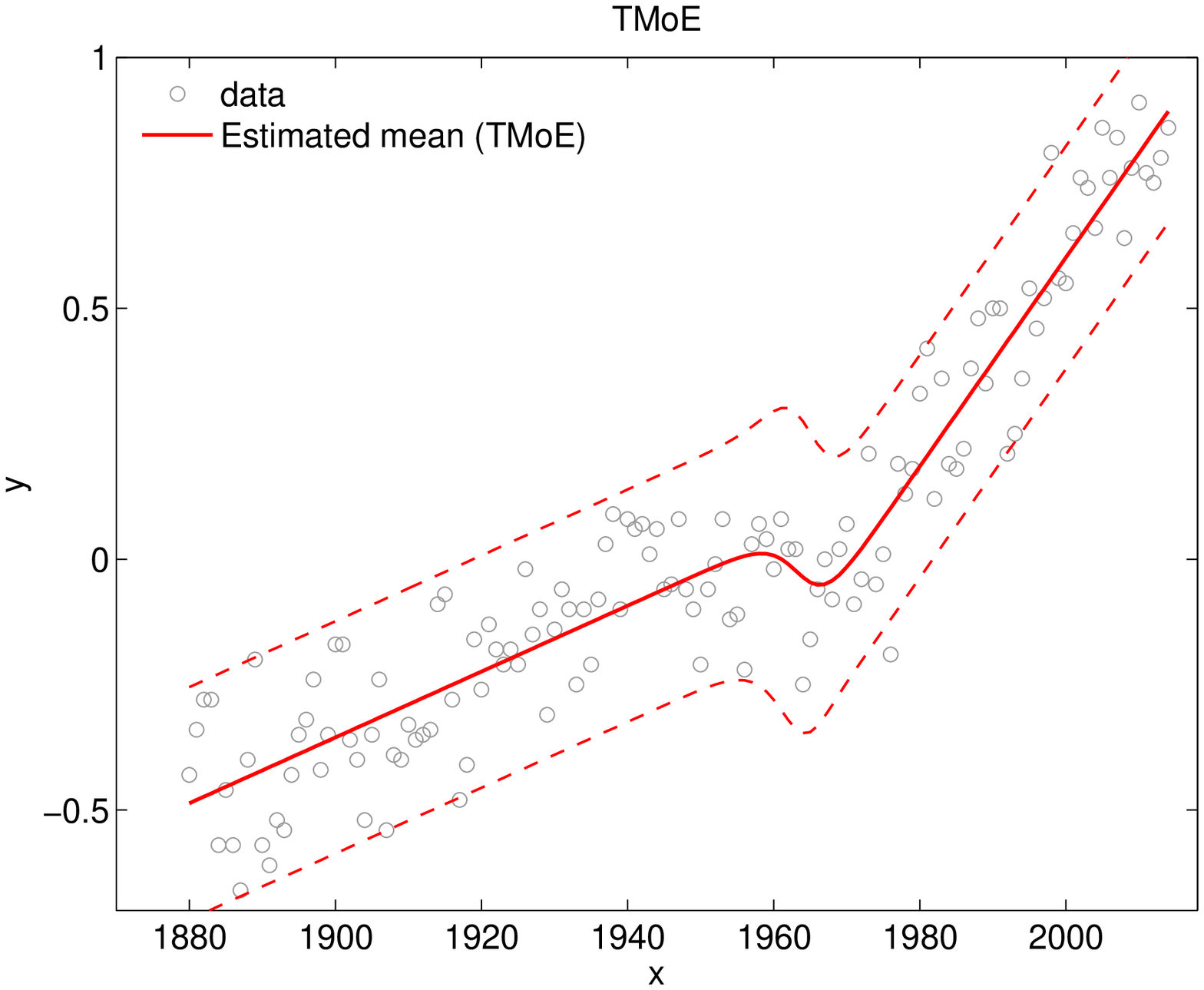} & \includegraphics[width=7.5cm]{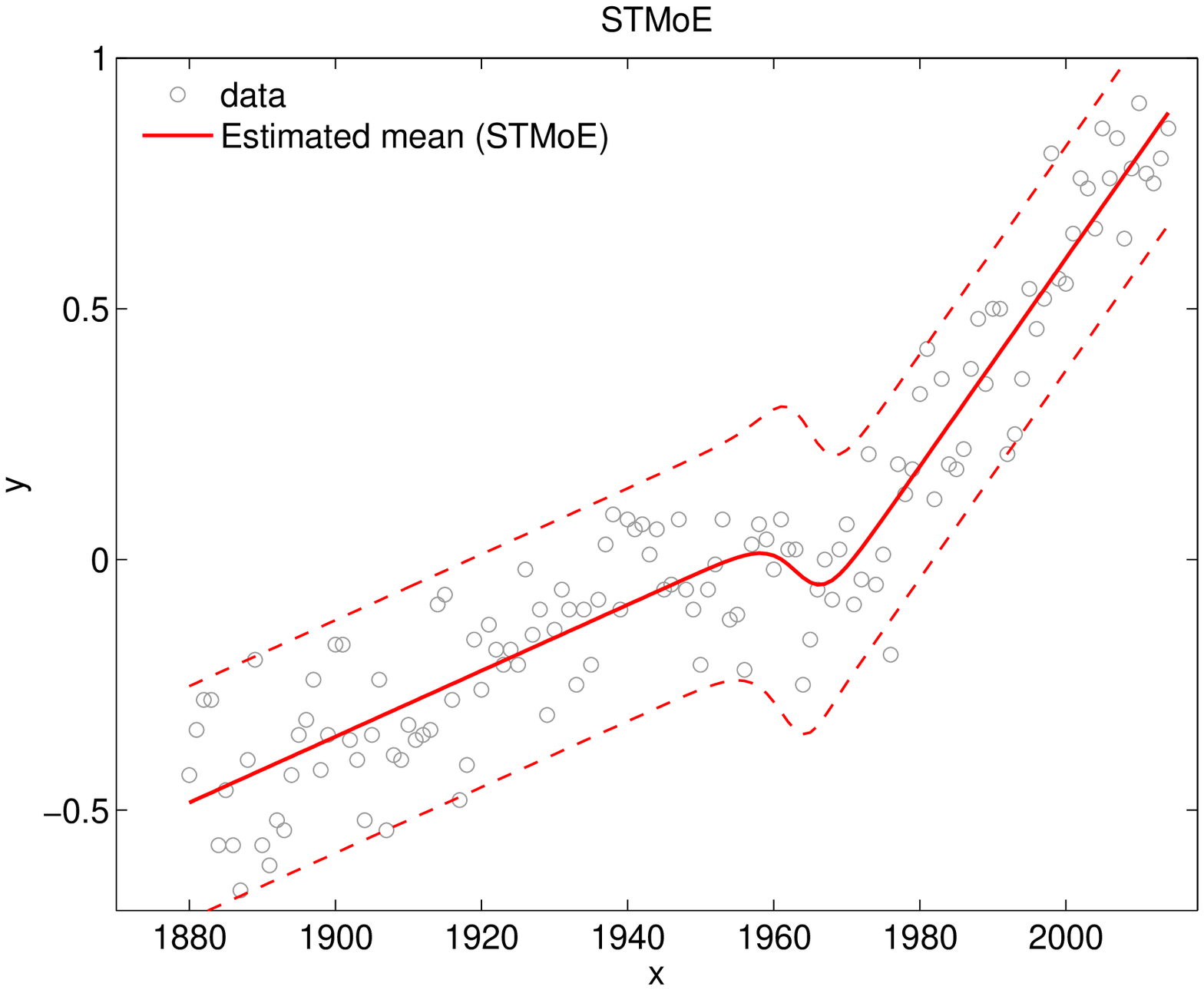}
   \end{tabular}
      \caption{\label{fig. temperature anomalies data and models means}The fitted MoLE models to the temperature anomalies data set. Upper-left: NMoE model, Upper-right: SNMoE model, Bottom-left: TMoE model, Bottom-right: STMoE model. The predictor $x$ is the year and the response $y$ is the temperature anomaly.  The shaded region represents plus and minus twice the estimated (pointwise) standard deviation as presented in Section \ref{sec: Prediction using the NNMoE}.}
\end{figure}
\begin{figure}[htbp]
   \centering 
   \begin{tabular}{cc}
   \includegraphics[width=7.5cm]{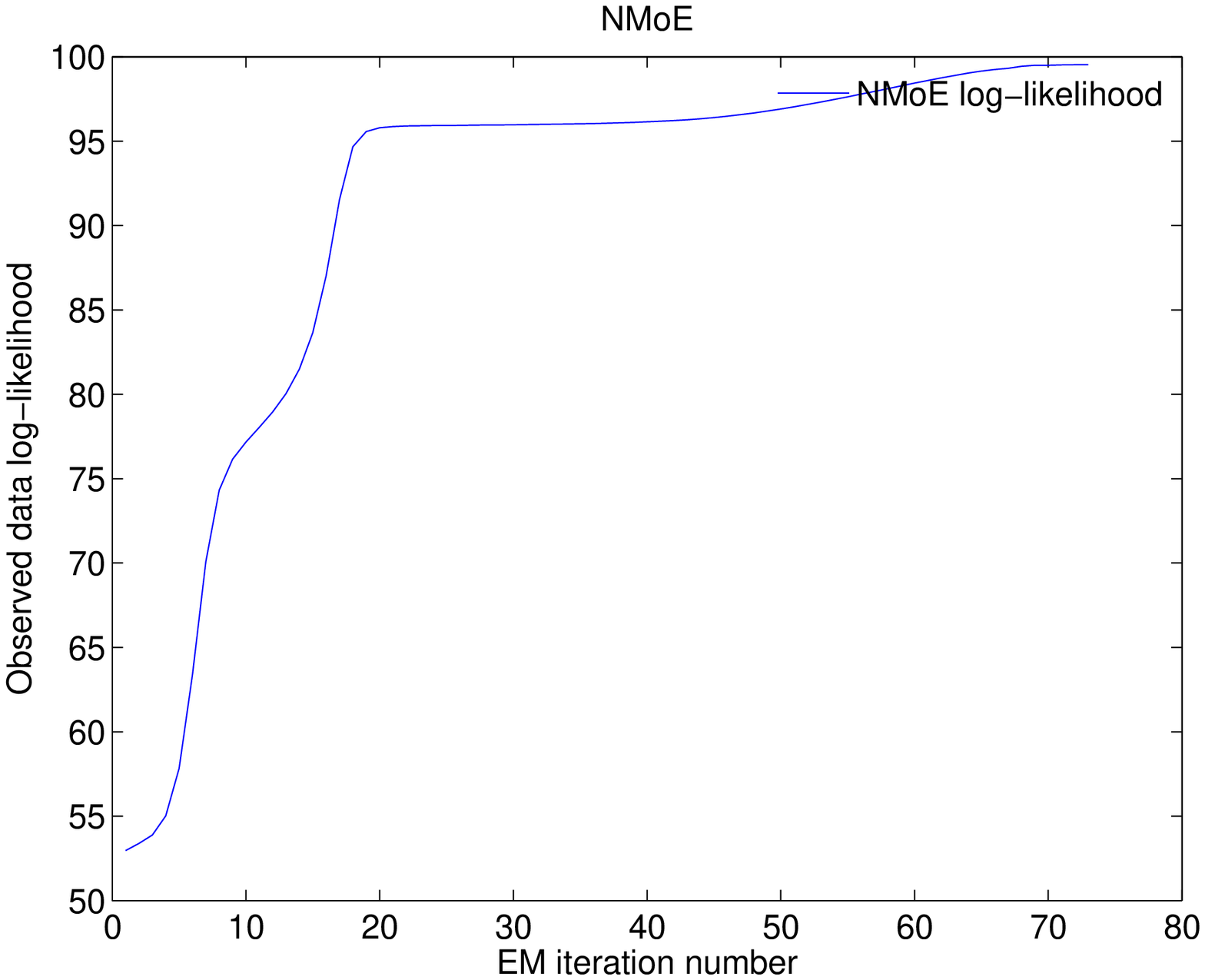} & 
   \includegraphics[width=7.5cm]{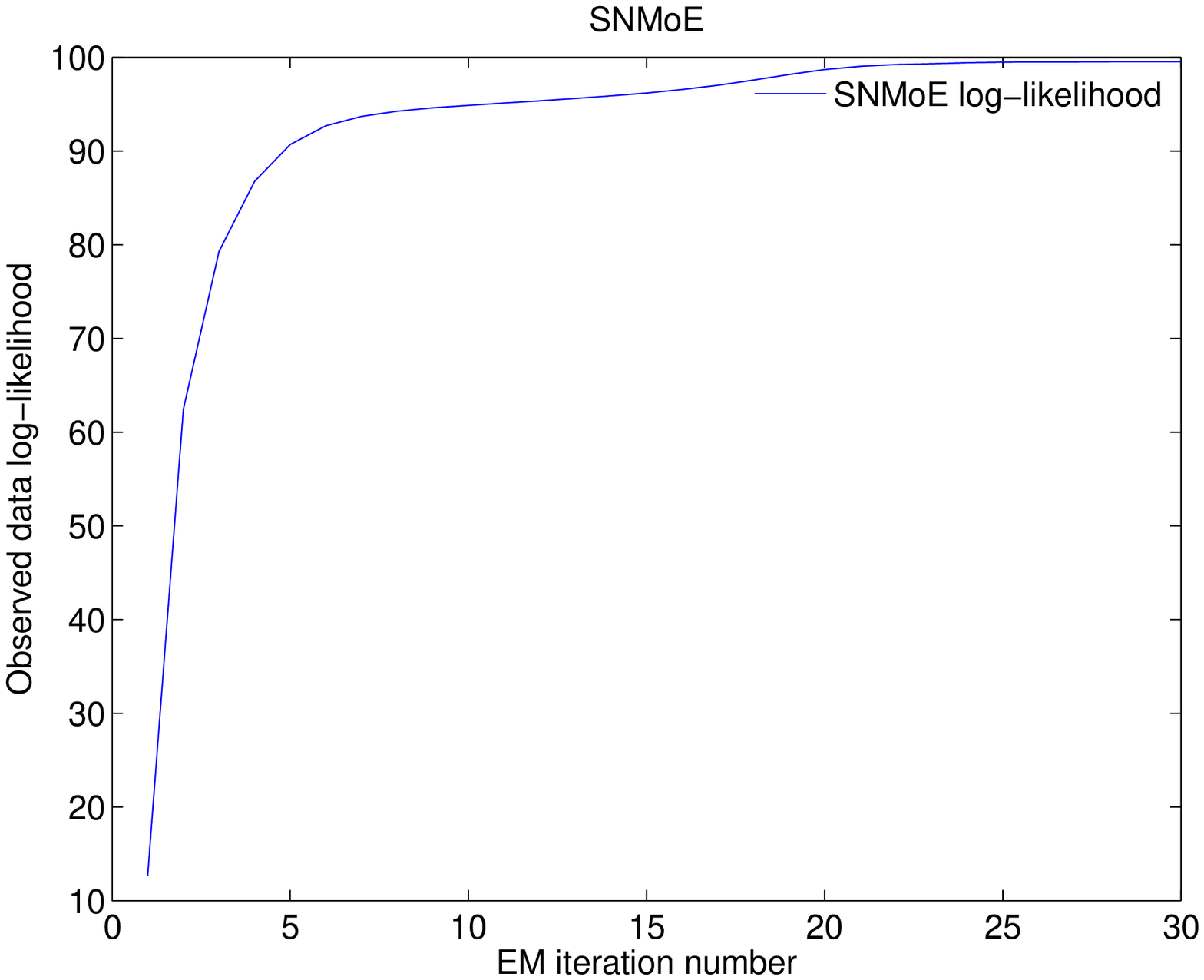}\\
  \includegraphics[width=7.5cm]{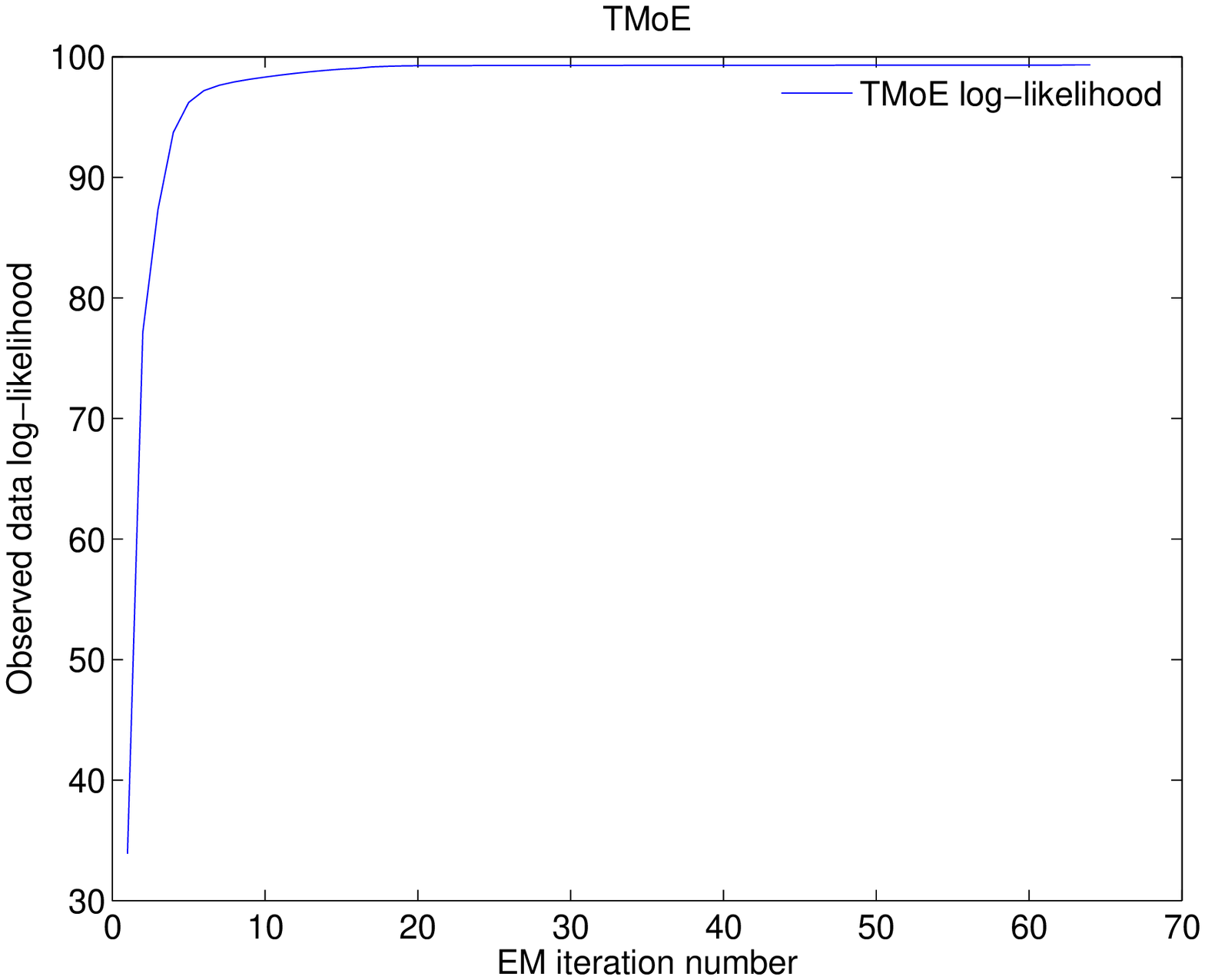} & 
  \includegraphics[width=7.5cm]{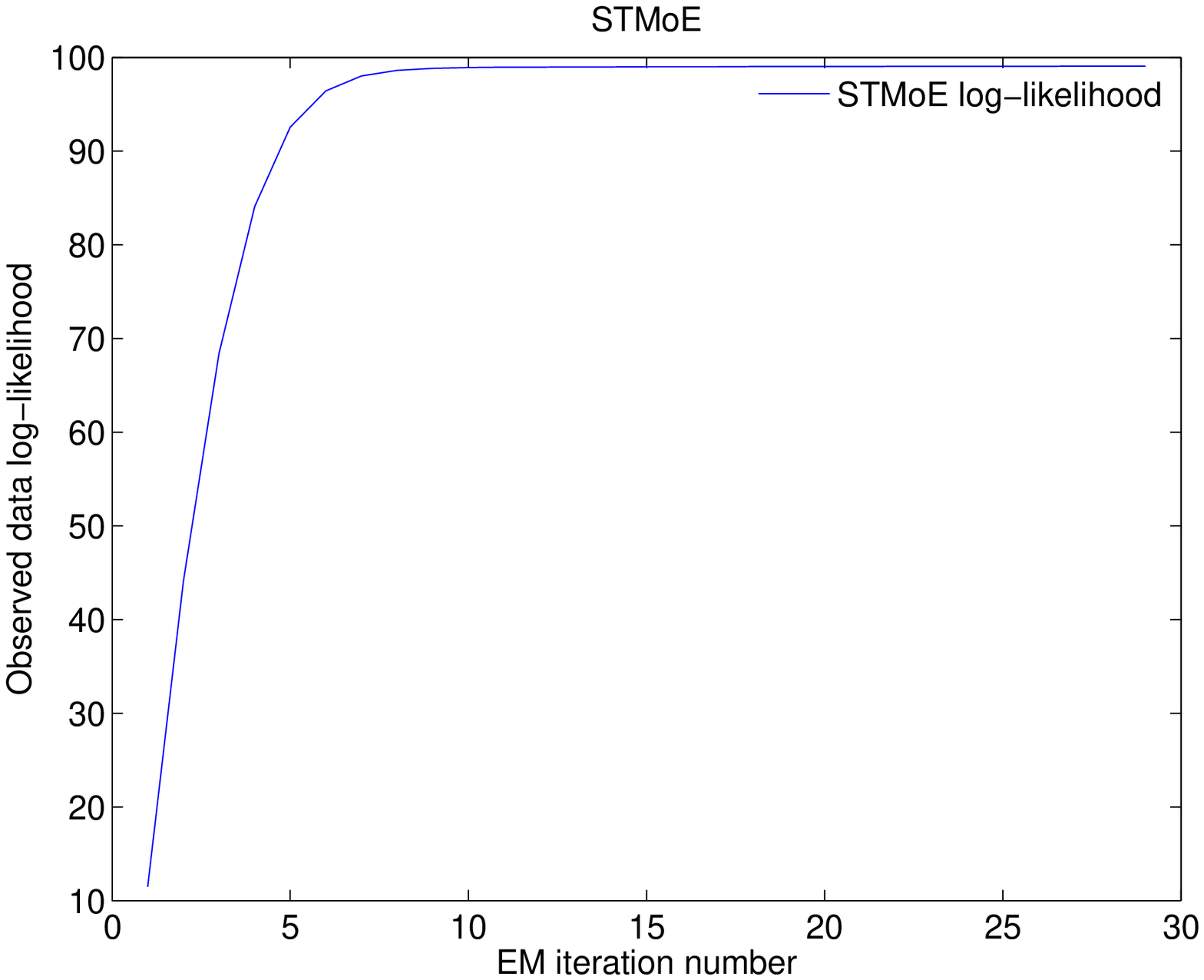}\\
   \end{tabular}
      \caption{\label{fig. temperature anomalies data and models loglik}The log-likelihood during the EM iterations when fitting the MoLE models to the temperature anomalies data set. Upper-left: NMoE model, Upper-right: SNMoE model, Bottom-left: TMoE model, Bottom-right: STMoE model.}
\end{figure}
The values of estimated MoE parameters for the temperature anomalies data set are given in Table \ref{tab. estimated parameters for the temperature anomalies data set}.
%
One can see that the parameters common for the all models are quasi-identical. 
It can also be seen that the SNMoE model provides s a fit with a skewness very close to zero. Similarly, the STMoE model provide a solution with a skewness close to zero. This may support the hypothesis of non-asymmetry for this data set.
Then, both the TMoE and STMoE fits provide a degrees of freedom more than 17, which tends to approach a normal distribution.
On the other hand, the regression coefficients are also similar to those found by \citet{Nguyen2014-MoLE} who used a Laplace mixture of linear experts.
{\setlength{\tabcolsep}{3pt
\begin{table}[htbp]
\centering
{\small
\begin{tabular}{l c c  c c c c c c c c c c}
\hline
param. & $\alpha_{10}$ & $\alpha_{11}$ & $\beta_{10}$ & $\beta_{11}$ & $\beta_{20}$ & $\beta_{21}$ & $\sigma_{1}$& $\sigma_{2}$ & $\lambda_{1}$ & $\lambda_{2}$ & $\nu_{1}$ & $\nu_{2}$ \\ 
model		& & & & & & & & & & & & \\
 \hline
 \hline
NMoE  	& 946.483 & -0.481  & -12.805 & 0.006 & -41.073 & 0.020 & 0.115 & 0.110 & - & -  & -  & -  \\
SNMoE	& 950.950  & -0.484 & -12.805 & 0.006 & -41.074 & 0.020 & 0.115 & 0.110 & -8.7e-13 & -9.2e-13 & - & - \\
TMoE		& 947.225 & -0.482 & -12.825 &  0.006 & -41.008 & 0.020 & 0.114 & 0.108 & - & - & 70.828 & 54.383\\ 
STMoE	& 931.966 & -0.474 & -12.848 & 0.006  & -40.876 & 0.020 & 0.113 & 0.105 & 0.024 & -0.015 & 41.048 & 17.589\\
 \hline
\end{tabular}}
\caption{\label{tab. estimated parameters for the temperature anomalies data set}Values of the estimated MoE parameters for the temperature anomalies data set.}
\end{table}
}

We performed a model selection procedure on the temperature anomalies data set to choose the best number of MoE components from values between 1 and 5. Table  \ref{tab. Model selection temprature anomalies data} gives the obtained values of the used model selection criteria, that is BIC, AIC, and ICL. 
One can see that, except the result provided by AIC for the NMoE model which provide a high number of components, all the others results provide evidence for two components in the data.  
%
{\setlength{\tabcolsep}{3pt
\begin{table}[htbp]
\centering
{\small 
\begin{tabular}{l |ccc | ccc |  ccc | ccc}
\hline
	& \multicolumn{3}{c|}{NMoE}	&	\multicolumn{3}{c|}{SNMoE}	&	\multicolumn{3}{c|}{TMoE}  & \multicolumn{3}{c}{STMoE}\\
\cline{2-13}
K			 &   BIC 	 & 	AIC 	&    ICL		 &   BIC 	 & 	AIC 	&    ICL		 &   BIC 	 & 	AIC 	&    ICL		 &   BIC 	 & 	AIC 	&    ICL \\
\hline
\hline
1 & 46.0623 &  50.4202 &  46.0623 &	43.6096 &  49.4202 &  43.6096 &	43.5521 &  49.3627 &  43.5521 & 40.9715 &  48.2347 &  40.9715 \\
2 & \underline{79.9163} &  91.5374 &  \underline{79.6241} &	\underline{75.0116} &  \underline{89.5380} &  \underline{74.7395} & 	\underline{74.7960} &  \underline{89.3224} &  \underline{74.5279} & \underline{69.6382} & \underline{87.0698}  &  \underline{69.3416} \\
3 & 71.3963 &  90.2806 &  58.4874 &	63.9254 &  87.1676 &  50.8704 &	63.9709 &  87.2131 &  47.3643 & 54.1267 & 81.7268  & 30.6556 \\
4 & 66.7276 &  92.8751 &  54.7524 &	55.4731 &  87.4312 &  41.1699	& 56.8410  & 88.7990  & 45.1251  & 42.3087 &  80.0773 &  20.4948 \\
5 & 59.5100 &  \underline{92.9206} &  51.2429 &	45.3469 &  86.0207 &  41.0906	& 43.7767  & 84.4505  & 29.3881 	& 28.0371 &  75.9742 &  -8.8817	 \\ 
\hline 
\end{tabular}
}
\caption{\label{tab. Model selection temprature anomalies data}Choosing the number of expert components $K$ for the temperature anomalies data by using the information criteria BIC, AIC, and ICL. Underlined numbers indicate the highest value for each criterion.}
\end{table}}

\section{Conclusion and future work}
\label{sec: Conclusion}

 In this paper, we proposed new non-normal MoE models, which generalize the normal MoE. They are based on the skew-normal, $t$ and skew $t$ distribution and are respectively the SNMoE, TMoE, and STMoE.
The SNMoE model is suggested for non-symmetric data, the TMoE for data with possibly outliers and heavy tail, and the STMoE is suggested for both possibly non-symmetric, heavy tailed and noisy data.
 We developed EM-type algorithms to infer each of the proposed models and  described the use of the models in non-linear regression and prediction as well as in model-based clustering.
The developed models are successfully applied on simulated and real data sets.
The results obtained on simulated data confirm the good performance of the models in terms of density estimation, non-linear regression function approximation and clustering.
 In addition, the simulation results provide evidence of the robustness of the TMoE and STMoE models to outliers, compared to the normal alternative models.
The proposed models were  also successfully applied to two different real data sets, including a situation with outliers. 
The model selection using information criteria tends to promote using BIC against in particular AIC which may perform poorly in the analyzed data. 
The obtained results support  the potential benefit of the proposed approaches for practical applications.

In this paper, we only considered the MoE in their standard (non-hierarchical) version. One interesting future direction is therefore to extend the proposed models to the hierarchical mixture of experts framework \citep{jordanHME}. Furthermore, a  natural  future extension of this work is to consider the case of   MoE for multiple regression on multivariate data rather than simple regression on univariate data. 

\bibliographystyle{plainnat}
 \addcontentsline{toc}{section}{References}
{\small \bibliography{NNMoE}}

\end{document}

%% file: mymathsym.tex
\newcommand{\bse}{\boldsymbol{e}}

\newcommand{\bsr}{\boldsymbol{r}}

\newcommand{\bsx}{\boldsymbol{x}}
\newcommand{\bsX}{\boldsymbol{X}}
\newcommand{\bsy}{\boldsymbol{y}}

\newcommand{\bsZ}{\boldsymbol{Z}}

\newcommand{\bsalpha}{\boldsymbol{\alpha}}
\newcommand{\bsbeta}{\boldsymbol{\beta}}

\newcommand{\bsOmega}{\boldsymbol{\Omega}}

\newcommand{\bssigma}{\boldsymbol{\sigma}}
\newcommand{\bslambda}{\boldsymbol{\lambda}}

\newcommand{\bsPsi}{\boldsymbol{\Psi}}
\newcommand{\bsvPsi}{\boldsymbol{\varPsi}}

\newcommand{\Indicatrice}{\mathds{1}}

\newcommand{\E}{\mathbb{E}}
\newcommand{\V}{\mathbb{V}}
\newcommand{\Pro}{\mathbb{P}}
\newcommand{\R}{\mathbb{R}}

\newcommand{\ICL}{\text{ICL}}